\documentclass[11pt,openright,oneside]{book}

\usepackage{snapshot}

\usepackage{color}
\usepackage{setspace}
\usepackage[cmex10]{amsmath}
\usepackage{psfrag}

\usepackage{subfigure} %
\usepackage{hyperref} %
\usepackage{multirow}
\usepackage{amsmath}
\usepackage{multicol}
\usepackage{setspace}
\usepackage{amsmath}
\usepackage{subfigure} 
\usepackage{multicol}
\usepackage{algorithmic}
\usepackage{algorithm}

\usepackage{epsfig}

\usepackage{cite}
   \usepackage{graphicx}
   \usepackage[lmargin=2.5cm,rmargin=2.5cm,tmargin=3cm,bmargin=3cm]{geometry}
\begin{document}
\date{}
\setcounter{page}{0}
\doublespacing
\title{\huge Realistic and Efficient Channel Modeling\\ for Vehicular Networks}

\author{
Submitted in partial fulfillment of the requirements for\\
the degree of\\
Doctor of Philosophy\\
in\\
Electrical and Computer Engineering\\\\\\
Mate Boban\\\\
Dipl.-Inform., University of Zagreb\\\\\\\\\\
Carnegie Mellon University\\
Pittsburgh, PA\\\\
December, 2012
}

\maketitle

\pagenumbering{roman}
\addcontentsline{toc}{chapter}{Abstract}
\chapter*{\centering Abstract}
Vehicular Ad Hoc Networks (VANETs) are envisioned to support three types of applications: safety, traffic management, and commercial applications. By using wireless interfaces to form an ad hoc network, vehicles will be able to inform other vehicles about traffic accidents, potentially hazardous road conditions, and traffic congestion. Commercial applications %
are expected to provide incentive for faster adoption of the technology. 

To date, %
VANET research efforts have relied heavily on simulations, due to prohibitive costs of deploying real world testbeds. Furthermore, the characteristics of VANET protocols and applications, particularly those aimed at preventing dangerous situations, require that the initial testing and evaluation be performed in a simulation environment before they are tested in the real world where their malfunctioning can result in a hazardous situation. 

Existing channel models implemented in discrete-event VANET simulators are by and large simple stochastic radio models, %
 based on the statistical properties of the chosen environment, thus not accounting for the specific obstacles in the region of interest. It was shown in \cite{dhoutaut06} and~\cite{koberstein09} that such models  are unable to provide satisfactory accuracy for typical VANET scenarios. 

While there have been several VANET studies recently  %
that %
introduced static objects (e.g., buildings) into the channel modeling process, modeling of mobile objects (i.e., vehicles) has been neglected. We performed extensive measurements in different environments (open space, highway, suburban, urban, parking lot) to characterize in detail the impact that vehicles have on communication in terms of received power, packet delivery rate, and effective communication range. Since the impact of vehicles was found to be significant, we developed a model that accounts for vehicles as three-dimensional obstacles and takes into account their impact on the line of sight obstruction, received signal power, packet reception rate, and message reachability. The model is based on the empirically derived vehicle dimensions, accurate vehicle positioning, and realistic mobility patterns. We validate the model against measurements and it exhibits realistic propagation characteristics while maintaining manageable complexity. 

In highway environments, vehicles are the most significant source of signal attenuation and variation. In urban and suburban environments, apart from vehicles, static objects such as buildings and foliage have a significant impact on inter-vehicle communication. Therefore, to enable realistic modeling in urban and suburban environments, we developed a model that incorporates static objects as well.
The model requires minimum geographic information: the location and the dimensions of modeled objects (vehicles, buildings, and foliage). 
We validate the model against measurements and show that it successfully captures both small-scale and large-scale propagation effects in different environments (highway, urban, suburban, open space). 

Finally, we performed experiments which showed that selecting tall vehicles as next-hop relays is beneficial in terms of higher power at the receiver, smaller number of hops to reach the destination, and increased per-hop communication range. Based on these results, %
we designed a technique that utilizes the advantageous position of antennas on tall vehicles (e.g., buses, trucks) to relay the messages. %
The technique improves communication performance by decreasing the number of hops needed to reach the destination, thus reducing the end-to-end delay and  increasing the effective communication range.
\newpage
\addcontentsline{toc}{chapter}{Resumo}
\chapter*{\centering Resumo}

As redes veiculares (VANETs) est\~{a}o preparadas para suportar pelo menos
tr\^{e}s tipos de aplica\c{c}\~{o}es: dissemina\c{c}\~{a}o de informa\c{c}\~{o}es de seguran\c{c}a,
gest\~{a}o de tr\'{a}fego, e aplica\c{c}\~{o}es comerciais. Recorrendo a interfaces
r\'{a}dio sem fios para formar uma rede ad hoc, os ve\'{i}culos ser\~{a}o capazes de
informar outros ve\'{i}culos sobre acidentes, condi\c{c}\~{o}es de estrada
potencialmente perigosas, e congestionamentos. Ao mesmo tempo, espera-se 
que as aplica\c{c}\~{o}es comerciais forne\c{c}am os incentivos econ\'{o}micos para uma
r\'{a}pida ado\c{c}\~{a}o da tecnologia. 

At\'{e} agora, as avalia\c{c}\~{a}o do desempenho de redes veiculares basearam-se 
principalmente em simula\c{c}\~{o}es, devido aos custos proibitivos na implementa\c{c}\~{a}o
de testbeds reais. Al\'{e}m disso, as caracter\'{i}sticas dos protocolos e aplica\c{c}\~{o}es
para redes veiculares, em particular aquelas destinadas \`{a} preven\c{c}\~{a}o de 
situa\c{c}\~{o}es perigosas, exigem que os testes iniciais se realizem por simula\c{c}\~{a}o 
antes de serem aplicadas no mundo real, uma vez que o seu mau funcionamento pode
causar situa\c{c}\~{o}es de perigo.

Os modelos actuais de canais de comunica\c{c}\~{a}o implementados em simuladores s\~{a}o, 
no geral, modelos probabil\'{i}sticos simplificados que se baseiam numa
caracteriza\c{c}\~{a}o estat\'{i}stica do cen\'{a}rio a ser analisado e, por consequente, 
n\~{a}o incorporam os obst\'{a}culos espec\'{i}ficos presentes na
regi\~{a}o de interesse. Estudos em \cite{dhoutaut06} e~\cite{koberstein09} mostraram que tais
modelos s\~{a}o incapazes de fornecer uma precis\~{a}o satisfat\'{o}ria em
cen\'{a}rios VANET t\'{i}picos.

Embora v\'{a}rios estudos recentes em VANETs tenham introduzido objetos
est\'{a}ticos (por exemplo, edif\'{i}cios) no processo de modela\c{c}\~{a}o do
canal, a modela\c{c}\~{a}o de objetos m\'{o}veis, nomeadamente ve\'{i}culos, tem sido
negligenciada. Neste trabalho foram realizadas medi\c{c}\~{o}es em diferentes ambientes
(espa\c{c}o aberto, autoestrada, suburbano, urbano, parque de estacionamento) para
caracterizar em detalhe o impacto que os ve\'{i}culos t\^{e}m na comunica\c{c}\~{a}o em termos 
do n\'{i}vel de pot\^{e}ncia do sinal recebido, da taxa
de entrega de pacotes, e do raio de alcance efetivo da comunica\c{c}\~{a}o.
Uma vez que o impacto dos ve\'{i}culos se mostrou significativo, propomos
um modelo que abstrai os ve\'{i}culos como obst\'{a}culos tridimensionais e
leva em considera\c{c}\~{a}o o seu impacto sobre a linha de vista entre o
transmissor e o receptor, a pot\^{e}ncia do sinal que chega ao receptor, a
taxa de entrega de pacotes, e o alcance dos mesmos na rede. O modelo baseia-se nas dimens\~{o}es do ve\'{i}culo, no posicionamento do ve\'{i}culo, e em padr\~{o}es de
mobilidade realistas, todos obtidos empiricamente. Os resultados previstos pelo modelo s\~{a}o
comparado com os resultados obtidos atrav\'{e}s de medi\c{c}\~{o}es, sendo observada
uma boa concord‰ncia entre ambos. 

Em ambientes de estrada, os ve\'{i}culos
s\~{a}o a fonte mais significativa de atenua\c{c}\~{a}o do sinal. Em ambientes
urbanos e suburbanos, para al\'{e}m de ve\'{i}culos, objetos est\'{a}ticos como
edif\'{i}cios e a vegeta\c{c}\~{a}o t\^{e}m um impacto sobre a comunica\c{c}\~{a}o
inter-ve\'{i}culos. Portanto, e para permitir uma modela\c{c}\~{a}o do canal
realista em ambientes urbanos e suburbanos, este trabalho apresenta um modelo que
incorpora os objetos est\'{a}ticos supracitados. O modelo requer informa\c{c}\~{a}o geogr\'{a}fica m\'{i}nima: 1) informa\c{c}\~{a}o sobre a localiza\c{c}\~{a}o do ve\'{i}culos;
e 2) informa\c{c}\~{a}o sobre os contornos dos objetos est\'{a}ticos. O modelo \'{e} validado por compara\c{c}\~{a}o
com resultados experimentais, e demonstra-se que \'{e} capaz de capturar com sucesso os efeitos 
da atenua\c{c}\~{a}o e propaga\c{c}\~{a}o em larga escala em diferentes cen\'{a}rios (autoestrada, urbano, suburbano, espa\c{c}o aberto). 

Finalmente, realizaram-se testes que mostraram que a
escolha de ve\'{i}culos altos como n\'{o}s principais para reencaminhamento de mensagens
\'{e} ben\'{e}fico em termos de uma maior for\c{c}a de sinal no receptor, menor nœmero de saltos
at\'{e} atingir o destino, e mais alcance de comunica\c{c}\~{a}o por salto.
Com base nesses resultados, \'{e} proposta uma t\'{e}cnia que tira vantagem da 
posi\c{c}\~{a}o elevada das antenas em ve\'{i}culos altos (como autocarros e cami\~{o}es) para o encaminhamento de
mensagens. A t\'{e}cnica proposta melhora o desempenho das comunica\c{c}\~{o}es,
diminuindo o nœmero de saltos necess\'{a}rio para chegar ao destino,
reduzindo assim o atraso total fim-a-fim e aumentando o alcance efetivo das comunica\c{c}\~{o}es.

\newpage

\vspace*{\fill} 
\begin{quote} 
\centering 
\emph{This is for Shesha}
\end{quote}
\vspace*{\fill}

\newpage

\chapter*{\centering Acknowledgements}

This thesis was made possible through the selfless support of a great number of people.

First, I would like to thank my advisors, Prof. Ozan K. Tonguz and Prof. Jo\~{a}o Barros. I joined Ozan's group at CMU back in 2007 as a Fulbright scholar. I was supposed to spend eight months at CMU -- five years later, here I am writing the acknowledgements for my thesis. While my journey has been long and winding, Ozan was always there to give me valuable advice. He instilled in me the scientific method and the rigorous approach to research, while also encouraging me to pursue out-of-the-box ideas. His ability to scope the problem and express it in simple terms is the ideal I strive for in my own research. 
When I officially started my Ph.D. in early 2009, I joined Jo\~{a}o's group. With his unique enthusiasm and energy, Jo\~{a}o brought a new dimension to my research. His unparalleled ability to get things done enabled me to pursue research directions that would otherwise remain inaccessible to me. Jo\~{a}o's support was key in enabling me to perform the experimental part of my work. Furthermore, his open approach to research allowed me to collaborate with some brilliant people that enriched both my knowledge and this thesis.
Ozan and Jo\~{a}o have also taught me the subtle differences between well-written, focused, and ambiguous, uninspiring scientific prose.\

\noindent The ability to identify research problems is the skill that defines an independent researcher. %
Since the goal of the Ph.D. is to educate independent researchers,
 I am most grateful to my advisors for giving me the complete freedom to find and go after the research problems that I considered interesting, at the same time keeping a keen eye on me so I do not get sidetracked. %

Next, I would like to thank Prof.~Peter Steenkiste, Prof.~Ana Aguiar and Prof.~Vijayakumar Bhagavatula for agreeing to serve on my defense committee. They provided invaluable advice which considerably improved the quality of this thesis.
I am grateful to Prof.~Steenkiste for his insightful advice during our collaboration over the past two and a half years. %
This thesis as a whole, and the experimental work contained within it in particular, has benefited greatly from his deep knowledge of wireless networking and experimental methods. %
 Prof.~Aguiar's expertise in the areas of simulation and measurements resulted in a more rigorous performance evaluation of the models developed in this thesis. Additionally, Prof. Aguiar participated in one of the measurement campaigns, results of which are used in this thesis. Despite his many obligations, including acting as a Dean of the College of Engineering, Prof.~Bhagavatula kindly accepted to serve on my committee. I am deeply grateful for his valuable time and technical advice.

I was fortunate to collaborate closely with Dr.~Tiago T. V. Vinhoza and Rui Meireles. %
I learned a great deal from Tiago about the physical layer, wireless communication, and the mathematical underpinnings of communications. 
Tiago is a careful listener with an ability to explain, in so many words, when the ideas are worth pursuing and also when they are off target. His incredible attention to detail and encyclopedic knowledge of all things EE made it a real pleasure to hang out with him at the white board and collaborate on projects that form the core of this thesis. Also, I am most grateful to him for proofreading the early drafts of this thesis.
Rui was my companion in virtually all measurement campaigns and the cowriter of the ensuing papers. Driving countless times around the same route during rush hours, in the middle of the night, and on weekends in cheap rental cars to collect enough measurement data is nobody's idea of a good time. %
Rui's composure, organizational ability, and extensive knowledge of protocols and scripting enabled us to stay on track both literally and figuratively. 

I would like to express my gratitude to Prof.~Michel Ferreira, who provided the aerial photography which enabled a thorough investigation into the impact of vehicular obstructions. He also participated in the design of the model for vehicles as obstacles, as well as in the initial discussions during which the tall vehicle relaying idea was formed.

I am indebted to Prof.~Bla\v{z}enka Divjak and Antun Brumni\'{c} at the University of Zagreb, whose encouragement and support in my decision to study abroad was unwavering. %
Bla\v{z}enka helped me to apply for the Fulbright scholarship and has supported me ever since in more ways than I could possibly recollect. Even when it meant confronting the powers that be, she stood by my side.
Throughout the past seven years, Antun provided me with constant advice when I needed it the most. Every time I was at the crossroads and needed to make a difficult decision, he told me exactly what I needed to hear to make the right decision. His integrity and uncompromising dedication have been an inspiration for me. %

My labmates at the University of Porto were a joy to be around. For countless conversations next to the coffee machine, ranging from complaints about the daily grind to space travel, but most often converging to sports and politics, I would like to thank Tiago Vinhoza, Jo\~{a}o Almeida, Saurabh Shintre, Ian Marsh, Rui Costa, Paulo Oliveira, Pedro Santos, S\'{e}rgio Cris\'{o}stomo, and Diogo Ferreira. For all the pleasant conversations not involving sports and politics, I would like to thank  Jo\~{a}o Paulo Vilela, Maricica Nistor, Gerhard Maierbacher, Traian Abrudan, and Hana Khamfroush. 
Special thanks are also due to the my labmates at CMU for helping me to ease the pressure of graduate life at CMU. In particular, I would like to thank Wantanee Viriyasitavat (for helping out on numerous occasions, including proofreading my thesis and asking the right questions during my practice talks), Jiun-Ren Lin (for taking time to provide useful feedback on my practice talks and helping me with the thesis defense logistics, among many other things), Hsin-Mu (Michael) Tsai (for his wise advice and brilliantly sharp questions that improved my research), and Vishnu Naresh Boddeti (for our bike trips to and from DC and for numerous enjoyable chats over the years). Amir Moghimi, Paisarn Sonthikorn, Yi Zhang, and Xindian Long were always there when I needed advice during my first year at CMU.  
I would also like to thank Paulo Oliveira, Xiaohui (Eeyore) Wang, Carlos Pereira, and Shshank Garg,  fellow students who set time apart from their busy schedules to help perform the experiments by driving a silly number of times around the same locations. To all the soccer folks that are either listed above or remain unnamed, I thank for playing the beautiful game, thus keeping me sane by occasionally taking my mind off research. %

My Ph.D. studies were funded by a grant from the Portuguese Foundation for Science and Technology under the Carnegie Mellon $\mid$ Portugal program (grant SFRH/BD/33771/2009) and the DRIVE-IN project (CMU-PT/NGN/0052/2008).

I would have not been able to focus on my graduate studies if it were not for the unconditional love and support of my family. %
My sincerest gratitude goes to my mother and father, for instilling in me the courage and allowing me to take my own path, even at times when they did not agree with my choice. %
Despite the physical distance between us during the last five years, %
 I always felt their support. I would also like to thank my grandparents, in particular my late grandfather Ljubo. %
 Through the vivid example of his own life, he has showed me the value of hard work, perseverance, and integrity.    

Finally, I would like to thank my wife Sanja. Words cannot express the gratitude I feel for the love and support she has given me.
She was my companion, best friend, and a source of inspiration throughout the last six and a half years. %
Every time I was down, she found a way to lift me up, always with a kind word of support, even during times when she was the one suffering much more.
As a small token of appreciation, with all my love I dedicate this thesis to Sanja. %

\clearpage

\tableofcontents
\listoffigures 
\listoftables
\newpage
\pagenumbering{arabic}
\chapter{Introduction} \label{ch:intro}
\section{Motivation} \label{sec:Intro}

Current VANET simulators have gone a long way from the simulation environments used in early VANET research, which often assumed unrealistic models such as random waypoint mobility, circular transmission range, or interference-free environment \cite{kotz04}. However, significant efforts still remain in order to improve the realism of VANET simulators, at the same time providing a computationally inexpensive and efficient platform for the evaluation of proposed VANET applications~\cite{bai06}. We distinguish three key building blocks of VANET simulators:
\begin{itemize}
\item	Mobility models;
\item	Networking (data exchange) models;
\item	Channel (Signal propagation) models.
\end{itemize}
Mobility models deal with realistic representation of vehicular movement, including mobility patterns (i.e., constraining vehicular mobility to the actual roadway), interactions between the vehicles (e.g., speed adjustment based on the traffic conditions) and traffic rule enforcement (e.g., intersection control through traffic lights and/or road signs). Networking models are designed to provide realistic data exchange, including simulating the medium contention, routing protocols, and upper layer protocols.

\begin{figure}[htb]	
\centering
\includegraphics[width=0.99\textwidth]{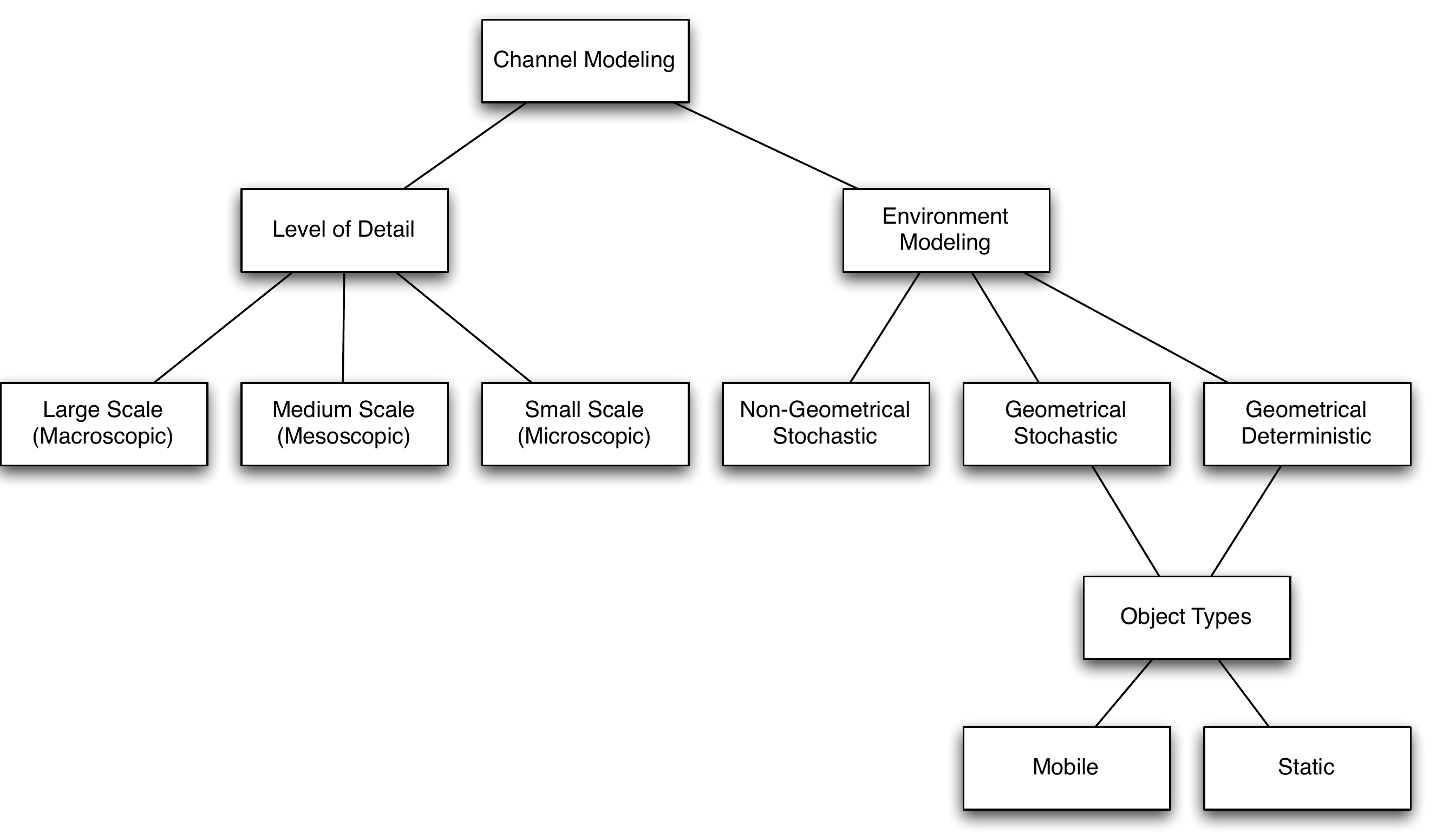} %
\caption{Structure of the channel modeling subsection of VANET simulation environments}\label{fig:graph}
\end{figure}

In this thesis, we are concerned with realistic channel modeling and simulation for large scale, discrete-event, packet-level VANET simulators. Modeling the signal propagation environment realistically is one of the pillars of successful evaluation of VANET protocols and applications. Physical characteristics of the (simulated) received signal directly affect the upper layers of the protocol stack, such as effectiveness of medium access, routing, and transport protocols, as well as characteristics important for applications (e.g., throughput, message delay, etc). Modeling the VANET channels realistically requires incorporation of the complex environment surrounding the communicating vehicles, in terms of both static objects (e.g., buildings, foliage), as well as mobile objects (other vehicles on the road).

Depending on the level of detail, channel models can be divided to small, medium, and large-scale models~\cite{boban11_3}. With respect to the method of calculation, the models are characterized as deterministic and stochastic. Furthermore, depending on whether site-specific geometrical information is used, the models are categorized as geometrical and non-geometrical. Detailed surveys on the existing vehicle-to-vehicle (V2V) channel models, including classification of models, can be found in~\cite{cheng09},~\cite{molisch09},~\cite{matolak09}, and~\cite{matolak08}. 
Figure~\ref{fig:graph} shows the classification of different model types and their relationships. 

We first perform extensive set of measurements to prepare the ground for designing the large scale packet-level V2V channel model. We measure the channel characteristics (received power and packet delivery rate) in a number of real-world scenarios. We perform experiments %
 in and around Pittsburgh, PA, USA, and Porto, Portugal and in distinct environments where VANETs will be deployed: highway, suburban, urban, open space, and isolated parking lot. We characterize the impact of both mobile objects (vehicles) and static objects (buildings and foliage) on the received power, packet delivery rate, and effective range.

Based on the measurements, in the second part of the thesis we design a computationally efficient packet-level V2V channel model for simulating vehicular communications in discrete-event simulators. We first describe existing channel models implemented in large scale packet-level VANET simulators and motivate the need for more accurate models that are able to capture the behavior of the signal on a per-link basis, rather than relying solely on the overall statistical properties of the environment. More specifically, as shown in~\cite{dhoutaut06} and \cite{koberstein09}), simplified stochastic radio models (e.g., free space \cite{goldsmith05}, log-distance path loss \cite{rappaport96}, etc.), which are based on the statistical properties of the chosen environment and do not account for the specific obstacles in the region of interest, are unable to provide satisfactory accuracy for typical VANET scenarios. Contrary to this, topography-specific, highly realistic channel models (e.g., based on ray-tracing \cite{maurer04}) yield results that are in very good agreement with the real world. However, these models are: 1) computationally too expensive due to prohibitive computation costs as the network grows; and 2) most often bound to a specific location for which a detailed geographic database is required, thus making them impractical for extensive simulation studies. For these reasons, such models have not been implemented in VANET simulators. We explore how site-specific models can be used so that both of these caveats are minimized, at the same time retaining the realism of the model. 

With regards to Fig.~\ref{fig:graph}, we propose a model that can be considered a hybrid between geometrical stochastic and geometrical deterministic models. It models the path loss deterministically using the geographical descriptors of the objects (namely, object outlines). Additionally, based on the location and the number of objects, the model stochastically assigns the additional signal variation (fading) on top of the path loss. The model takes into account both mobile  %
and static objects. %

The model leverages a limited amount of geographical information that is easily available in order to produce results comparable to those in the real world. Specifically, we use locations of the vehicles along with the information on the location and shape of the buildings and foliage. Vehicle locations are available through either real world traces (e.g., via GPS) or traffic mobility models, whereas the building and foliage outlines and locations are available freely from projects such as the OpenStreetMap (www.openstreetmap.org). 
The premise of the model is that line of sight (LOS) and non-LOS (NLOS) links exhibit considerably different channel characteristics. This is corroborated by numerous studies, both experimental and analytical (e.g., \cite{usdot06_2}, \cite{Otto2009}, \cite{boban11}, \cite{meireles10}, and \cite{masui02}), which have shown %
that the resulting channel characteristics for LOS and non-LOS links are fundamentally different. %
For this reason, our approach to modeling is to use %
simple geographical descriptors of the simulated environment (outlines of buildings, foliage, and vehicles on the road) to determine the large-scale signal shadowing effects and classify the communication links into three groups:
\begin{itemize}
\item Line of sight (\textbf{LOS}) -- links that have an unobstructed optical path between the transmitting and receiving antennas;
\item Non-LOS due to vehicles (\textbf{NLOSv}) -- links whose LOS is obstructed by other vehicles;
\item Non-LOS due to buildings/foliage (\textbf{NLOSb}) -- links whose LOS is obstructed by static objects (buildings or foliage).
\end{itemize}
We take a specific approach in estimating the properties for each of these link types.
The model is intended for implementation in packet-level, discrete-event VANET simulators (notable examples are Jist/SWANS~\cite{jist}, NS-2~\cite{ns2}, NS-3~\cite{ns3}, QualNet~\cite{qualnet}, etc.). We employ computational geometry concepts suitable for representation of the geographic data required in simulating VANETs. We form a bounding volume hierarchy (BVH) structures~\cite{klosovski98}, in which we store the information about outlines of both the vehicles and buildings. %
VANET-related geometric data lends itself to an efficient BVH implementation, due to its inherent geometrical structure (namely, relatively simple object outlines and no overlapping of building and vehicle outlines). The model can be used with any discrete-event VANET simulator, provided that: a) location of vehicles are known (which is a must for VANET simulations); and b) location and the outlines of buildings and streets are known (this information is freely available from numerous geographical databases). 

In the final part of the thesis, we discuss an application %
is aimed at alleviating the impact of vehicular obstructions by selecting the tall vehicles as relaying nodes. We evaluate this application using the model we introduced. Specifically,   we show that, using small-scale experiments, significant benefits can be obtained by opting for tall vehicles as next hop relays, as opposed to short (personal) vehicles. We perform simulations with the proposed model and we validate the results with experiments involving tall and short vehicles. The proposed technique matches the existing techniques in low vehicle density scenarios and outperforms them in high density scenarios.%

\section{Thesis Organization}
The rest of the thesis is organized as follows. In Chapter~\ref{ch:experiments}, we present a set of measurements which provide insights into the impact of vehicular obstructions on inter-vehicle communication. Based on the insights of the measurement study, Chapter~\ref{ch:vehModel} presents a model that incorporates the vehicles as three-dimensional objects in the channel modeling. The model takes into account the impact of vehicles on the line of sight obstruction, received signal power, packet reception rate, and message reachability. In order to enable realistic channel simulation in urban areas for VANET simulators, in Chapter~\ref{ch:completeModel} we introduce a channel model that incorporates both vehicles and static objects (namely, buildings and foliage). 
Furthermore, in Chapter~\ref{ch:TVR} we perform experiments to determine how much of the adverse effects of vehicular obstructions can be alleviated using the tall vehicles' raised antennas to achieve a better channel. Based on the measurements, we develop a relaying technique that utilizes tall vehicles to improve effective communication range and packet delivery. 
Finally, concluding remarks and future research directions are given in Chapter~\ref{ch:conclusion}.

\chapter{Experimental Evaluation of Vehicles as Obstructions} \label{ch:experiments}

Channel models for vehicular networks typically disregard the effect of vehicles as physical obstructions for the wireless signal. We tested the validity of this simplification by quantifying the impact of obstructions through a series of wireless experiments reported in~\cite{meireles10}.
Using two cars equipped with Dedicated Short Range Communications (DSRC) hardware~\cite{dsrc09} designed for vehicular use, we perform experimental measurements in order to collect received signal power and packet delivery ratio information in a multitude of relevant scenarios: parking lot, highway, suburban and urban canyon.
Upon separating the data into line of sight (LOS) and non-line of sight (NLOS) categories, our results show that obstructing vehicles cause significant impact on the channel quality. A single obstacle can cause a drop of over 20 dB in received signal strength when two cars communicate at a distance of 10 m. At longer distances, NLOS conditions affect the usable communication range, effectively halving the distance at which communication can be achieved with 90\% chance of success. 
The presented results motivate the inclusion of vehicles in the radio propagation models  %
used for VANET simulation in order to increase the level of realism. %

\section{Motivation} \label{sec:Introduction}

Based on the parties involved, two main communication paradigms exist in Vehicular Ad Hoc Networks (VANETs): Vehicle-to-Vehicle (V2V) communication, where vehicles on the road communicate amongst themselves; and Vehicle-to-Infrastructure (V2I) communication, where vehicles communicate with nearby roadside equipment. The relatively low heights of the antennas on the communicating entities in V2V communication imply that the optical line of sight (LOS) can easily be blocked by an obstruction, either static (e.g., buildings, hills, foliage) or mobile (other vehicles on the road). 

There exists a wide variety of experimental studies dealing with the propagation aspects of V2V communication. Many of these studies deal with static obstacles, often identified as the key factors affecting signal propagation (e.g., ~\cite{Otto2009,Cheng2007,andersen95}). %
However, it is reasonable to expect that a significant portion of the V2V communication will be bound to the road surface, especially in highway environments, thus making the LOS between two communicating nodes susceptible to interruptions by other vehicles. Even in urban areas, it is likely that other vehicles, especially large public transportation and commercial vehicles such as buses and trucks, will often obstruct the LOS. %

Despite this, as noted in \cite{martinez09}, virtually all of the state of the art VANET simulators neglect the impact of vehicles as obstacles on
signal propagation, mainly due to the lack of an appropriate methodology capable of incorporating the effect of vehicles realistically and efficiently. %
This motivated us to perform extensive measurements to precisely determine the impact of vehicles on the signal power and packet reception rate in different real world scenarios.
We focused on measuring the impact of NLOS conditions on received signal strength and packet delivery ratio. Our goal was to isolate the following three variables:

\begin{itemize}
\item \textbf{Environment} ---
We distinguish one parking lot and three on-the-road scenarios: urban, suburban, and highway. The parking lot experiments allowed us to control factors such as the distance between the vehicles and the number and location of vehicles obstructing the LOS. The on-the-road experiments allowed us to analyze the effect of NLOS conditions in the typical real world environments where VANETs will be used.

\item \textbf{Line of sight conditions} ---
To isolate the impact of moving vehicles on the channel quality, we distinguished between the following situations: LOS, NLOS due to vehicular obstacles (NLOSv), and NLOS due to static obstructions (NLOSb)\footnote{In our experiments, buildings were the predominant static objects that obstructed LOS. Therefore, in the rest of the text we refer to this condition as NLOSb (NLOS due to buildings).}. %

\item \textbf{Time of day} --- We introduce this variable to help determine how often the vehicles encounter NLOSv conditions at different times of day (since NLOSv obstruction is temporally variable) and how this affects the signal.
\end{itemize}

Using these variables and following the work reported in \cite{boban11}, we designed a set of experiments using two vehicles equipped with Dedicated Short Range Communication (DSRC) devices to characterize the impact of vehicles as obstacles on V2V communication at the communication link level. We aimed at quantifying the additional attenuation and packet loss due to vehicular obstructions.

The rest of the chapter is organized as follows. The experimental setup is described in Section~\ref{sec:ExperimentalSetup}. Section~\ref{sec:Results} discusses the results and Section~\ref{sec:RelatedWork} describes previous work on experimental evaluation and modeling of V2V communication. Section~\ref{sec:Conclusions} concludes the chapter.

\section{Experiment Setup} \label{sec:ExperimentalSetup}

\subsection{Network Configuration}\label{subsec:NetworkConfiguration}

\begin{figure*}[t]
  \begin{center}
    \includegraphics[width=0.91\textwidth]{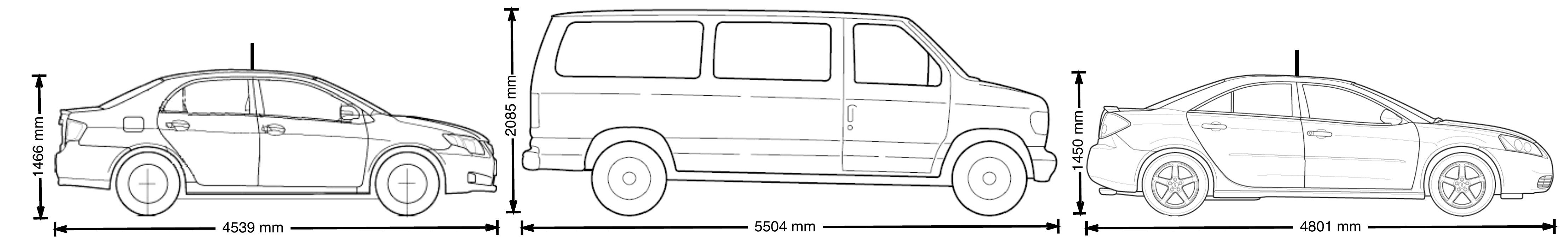}
     \caption[Scaled drawing of the vehicles used in the experiments]{Scaled drawing of the vehicles used in the experiments. Left to right: 2009 Toyota Corolla, a 2010 Ford E-Series, and a 2009 Pontiac G6. Blueprints courtesy of carblueprints.info~\cite{carblueprints}.}
      \label{vehicle-dimensions}
   \end{center}
\end{figure*}

\begin{table}
	\begin{center}
		\begin{tabular}{|l|c|c|}
			\hline
			\textbf{Parameter} & \textbf{802.11p} & \textbf{802.11b/g} \\ 
			\hline
			Channel & 180 & 1 \\ 
			\hline
			Center frequency (MHz) & 5900 & 2412\\ 
			\hline
			Bandwidth (MHz) & 20 & 20\\ 
			\hline
			Data rate (Mbps) & 6 & 1\\ 
			\hline
			Tx power (setting, dBm) & 18 & 18\\ 
			\hline
			Tx power (measured, dBm) & 10 & 16\\ 
			\hline
			Antenna gain (dBi) & 5 & 3\\ 
			\hline
			Beacon frequency (Hz) & 10 & 10\\ 
			\hline
			Beacon size (Byte) & 36 & 64\\ 
			\hline
		\end{tabular}
		\caption{Hardware configuration parameters}
		\label{tab:hw-config}
	\end{center}
	
\end{table}

We equipped each car with a NEC LinkBird-MX, a custom-built development platform for vehicular communications~\cite{festag08}. These DSRC devices operate at the 5.85-5.925~GHz band and implement the IEEE 802.11p wireless standard, specifically designed for automotive use~\cite{80211p2010}. The radios were connected to vertically polarized Mobile Mark ECOM6-5500 omnidirectional antennas, which measure 26~centimeters in height.
Adding a GPS receiver to each Linkbird-MX and taking advantage of the built-in beaconing functionality, we recorded the locations of the vehicles, the packet delivery ratio (PDR) and the received signal strength indicator (RSSI) throughout the experiments.

To get a sense of the difference between the IEEE 802.11p and the off-the-shelf WiFi (IEEE 802.11b/g) equipment, we also performed experiments with %
Atheros WiFi cards and GPS receivers. We used the \emph{ping} application and the Wireshark network protocol analyzer~\cite{wireshark} to collect the same location, PDR, and RSSI information as with the Linkbirds.

The hardware configuration parameters used in the experiments are summarized in Table~\ref{tab:hw-config}. We used the lowest available data rate for each standard to get the largest possible communication range. The actual power at the antenna outputs was measured using a real time spectrum analyzer and no significant power fluctuations were observed. We used 20~MHz channels for both standards to have a closer comparison of the two. Relatively small packet sizes (see Table~\ref{tab:hw-config}) were used in order to reflect the message size for proposed safety applications \cite{bai06}. Since larger packets would be more susceptible to fading, our results provide a lower bound on the effect of non-line of sight conditions.

The experiments were performed with a simple vehicular ad-hoc network comprised of two vehicles, both sedans of similar and average height: a Toyota Corolla and a Pontiac G6. In order to directly affect the line of sight between these two vehicles, we used a larger, non-networked vehicle as a LOS obstacle: a Ford E-Series van. The relevant dimensions of all three vehicles are depicted in Fig.~\ref{vehicle-dimensions}. With 26~cm antennas centrally mounted on the roof for the best possible reception (as experimentally shown by \cite{kaul07}), the van sits around 37~cm taller than the tip of the antennas on the sedans, effectively blocking the LOS while positioned between them.

\subsection{Scenarios}\label{subsec:Scenarios}

A set of parking lot and on-the-road experiments were designed to isolate the effect of vehicles as obstacles from other variables and to provide insights into the effect of vehicles in different environments where VANETs will be used. %
All of the experiments were performed in, or near, Pittsburgh PA, USA in good weather conditions, with clear skies and no rain.

\begin{figure*}[t]
  \begin{center}
\subfigure[Parking lot environment: experiment with the obstructing van]
{
\label{exp-setup:a}
\includegraphics[height=0.15\textheight]{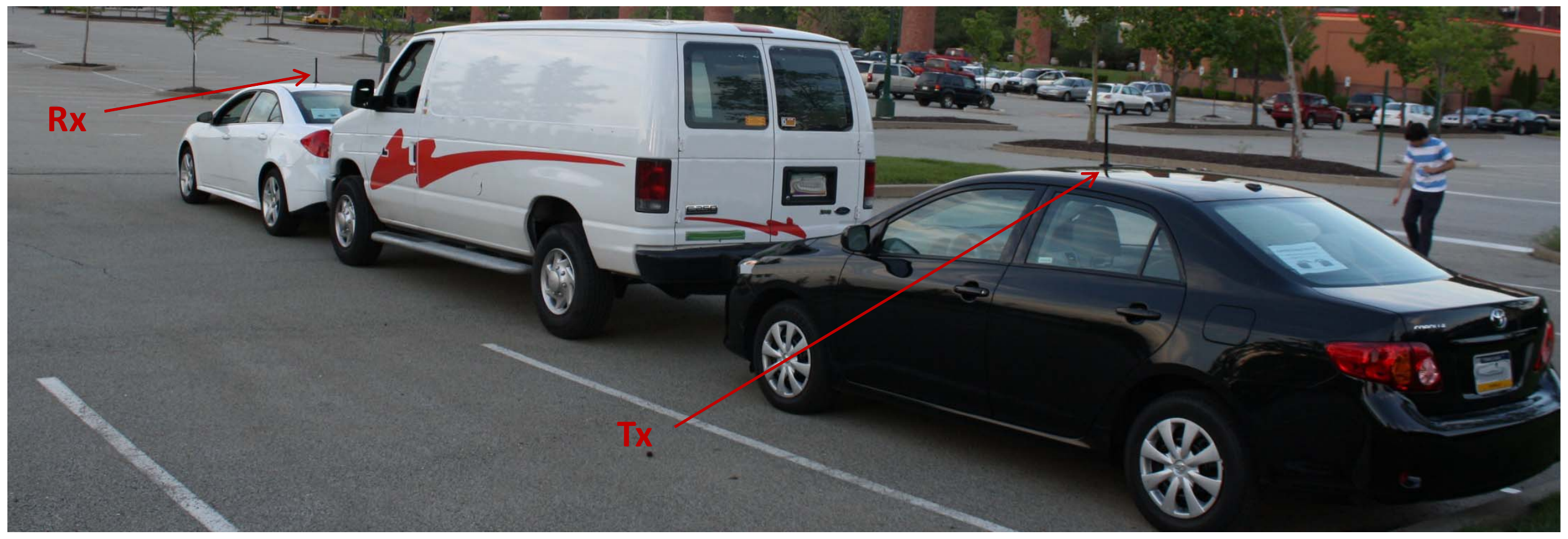}
}
\subfigure[Urban canyon in Downtown Pittsburgh]
{
\label{exp-setup:c}
\includegraphics[height=0.15\textheight]{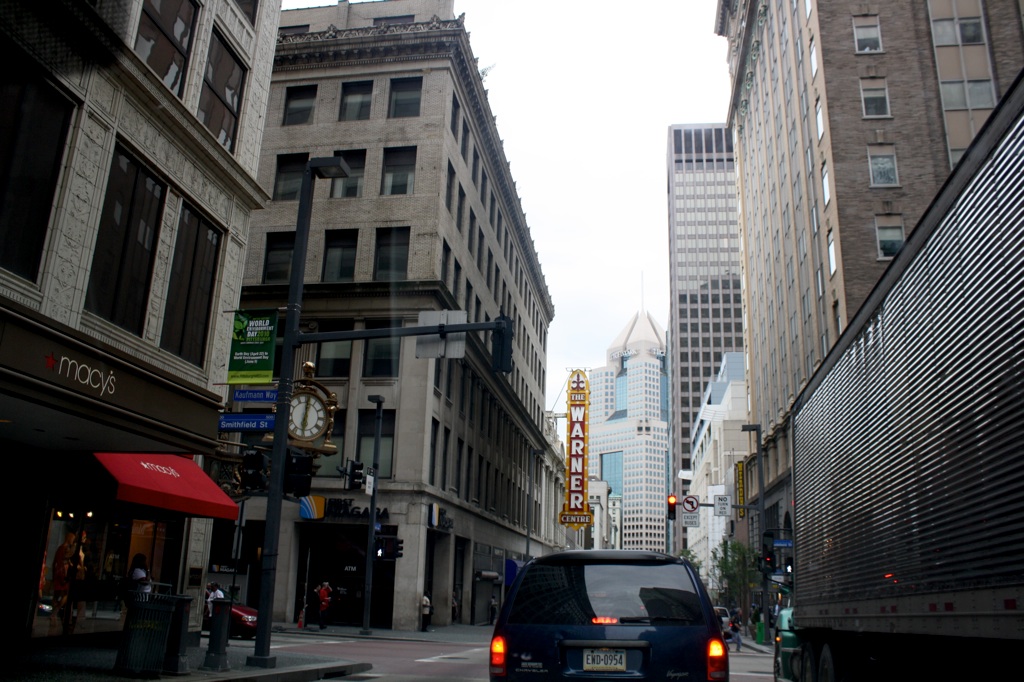}
} 
\subfigure[Parking lot environment: experiment with the obstructing truck]
{
\label{exp-setup:d}
\includegraphics[height=0.1454\textheight]{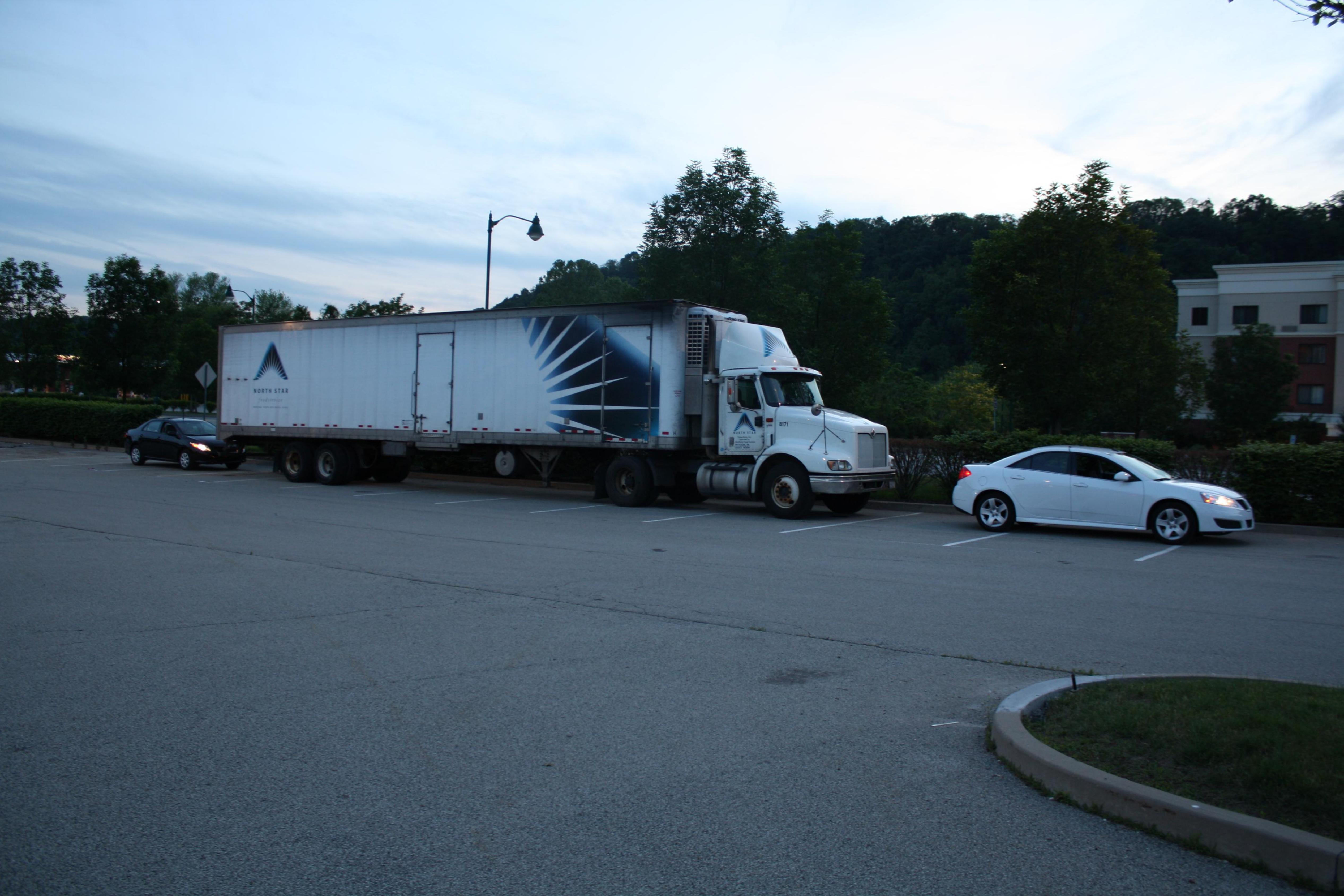}
}
\subfigure[Hardware]
{
\label{exp-setup:b}
\includegraphics[height=0.1454\textheight]{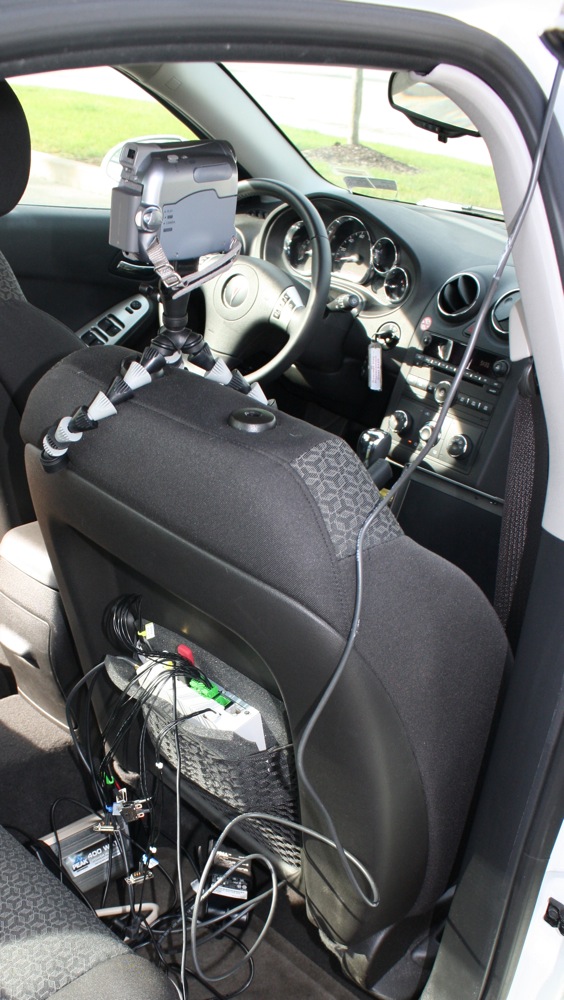}
} 
     \caption{Experimental setup.}
      \label{exp-setup}
   \end{center}
\end{figure*}

The \textbf{parking lot experiments} were performed in the Loews Complex parking lot (lat:~40.405139, long:~-79.91925), which is open, large (200~m by 200~m), mostly flat and during the day, practically empty. We collected signal information for the following scenarios:
\begin{itemize}
\item Cars parked 10, 50 and 100~m apart, with and without the van placed halfway across the gap.
\item Cars starting next to each other and slowly moving apart, with and without an obstruction in between them. In this experiment, we replaced the obstructing van with a 4 meter tall semi-trailer truck shown in Fig.~\ref{exp-setup:d}.
\end{itemize}

For the \textbf{on-the-road experiments}, we identified three typical environments where VANETs will be used:

\begin{itemize} 

\item \textbf{Highway} --- 
In this environment, the obstructions are caused by the terrain profile, e.g., crests and corners. We performed experiments on a 85~km stretch of the U.S. Interstate 79 between the Pittsburgh Airport (Coordinates: 40.4516, -80.1099) and Grove City, PA (Coordinates: 41.14174, -80.15498).

\item \textbf{Suburban} --- 
In this environment, wide streets are typically lined with small buildings and trees. There are also occasional crests, dips, and blind corners. We used a residential, 4 lane, 5~km stretch of Fifth Ave. in Pittsburgh, PA (Coordinates: 40.45008, -79.92768) for this scenario. 

\item \textbf{Urban canyon} --- 
In this environment, streets cut through dense blocks of tall buildings which significantly affect the reception of radio signals. We performed experiments on a two~km trapezoidal route around Grant Street (Coordinates: 40.44082, -79.99579) in downtown Pittsburgh (Fig.~\ref{exp-setup:c}).

\end{itemize}

For each environment, we performed the experiments by driving the cars for approximately one hour, all the time collecting GPS and received signal information. Throughout the experiment, we videotaped the view from the car following in the back for later analysis of the LOS/NLOSv/NLOSb conditions.

We performed two one-hour experiment runs for each on-the-road scenario: one at a rush hour period with frequent NLOSv conditions, and the other late at night, when the number of vehicles on the road (and consequently, the frequency of vehicle-induced -- NLOSv -- conditions) is substantially lower. This, by itself, worked as a heuristic for the LOS conditions. Furthermore, to more accurately distinguish between LOS conditions, we used the recorded videos to separate the LOS, NLOSv, and NLOSb data.

\section{Results}\label{sec:Results}

\subsection{Parking lot experiments}\label{subsec:parkingLotExperiments}

\begin{figure}[t!]
\centering
\subfigure[802.11g]
{
   \label{fig:parkinglot-rssi:11g}
   \includegraphics[width=0.33\textwidth,angle=0]{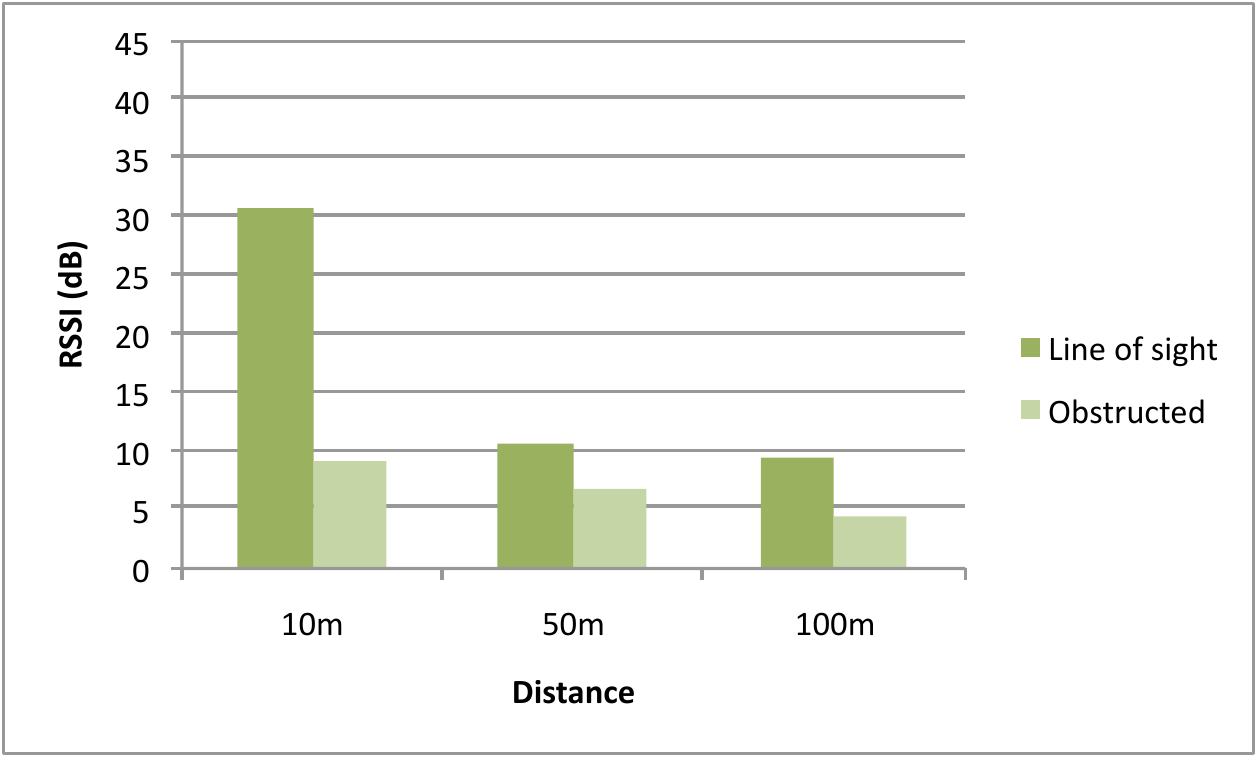}
}
\subfigure[802.11p]
{
   \label{fig:parkinglot:11p}
   \includegraphics[width=0.33\textwidth,angle=0]{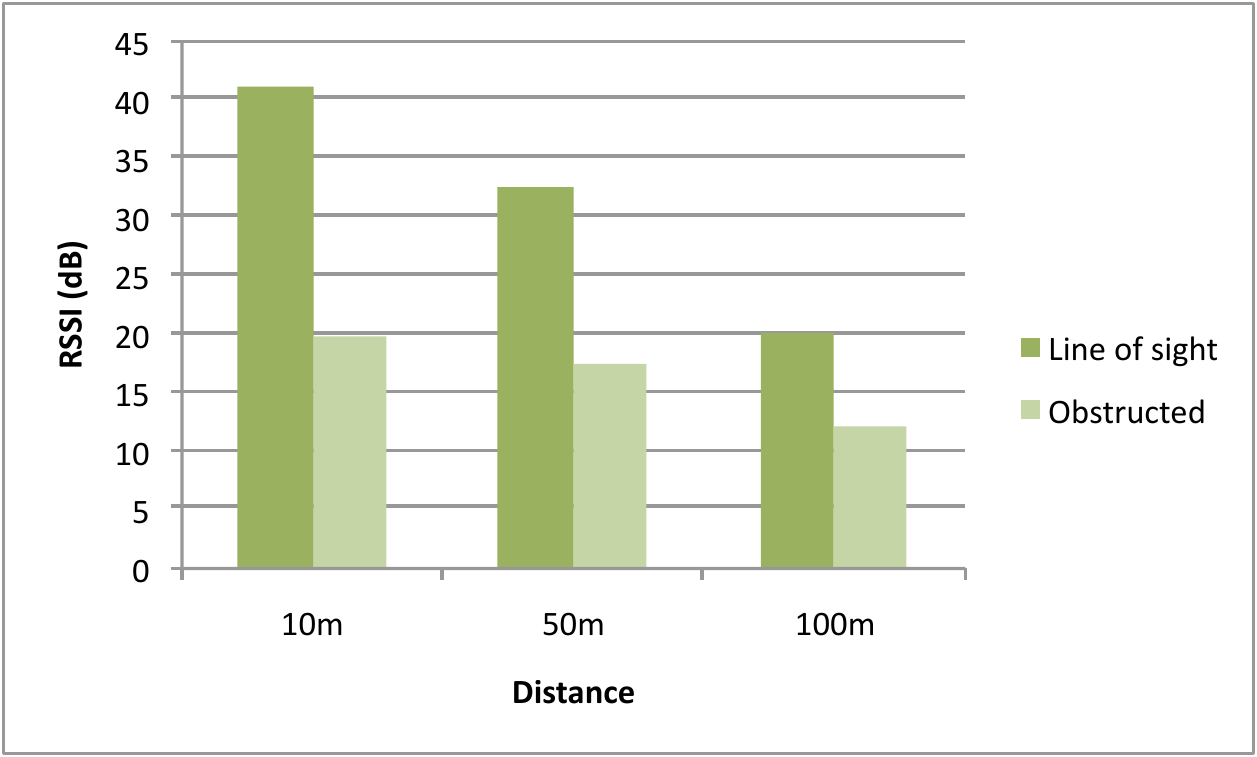}
}
\caption[Parking lot experiment results]{Parking lot experiment: average received signal strength measured at fixed distances with and without the obstructing van for both 802.11g and 802.11p standards.   }
\label{fig:parkinglot-rssi}
\end{figure}

All of the parking lot experiments were performed at relatively short distances, meaning the packet delivery ratio was almost always 100\%. We therefore focus on RSSI to analyze the effect of LOS conditions on channel quality. For ease of presentation, we report the RSSI values in dB as provided by the Atheros cards. The RSSI values can be converted to dBm by subtracting 95 from the presented values.

First, we consider the experiments where the cars were placed at a fixed distance from each other. Figure~\ref{fig:parkinglot-rssi} shows the RSSI results. The standard deviation was under 1 dB and the 95\% confidence intervals were too small to represent; we thus focus on the average values. The difference in absolute RSSI values between the 802.11b/g and 802.11p standards is mainly due to the difference in antenna gains, hardware calibrations, and the quality of the radios.

Blocking the LOS has clear negative effects on the RSSI. Even though the absolute values differ between the standards, the overall impact of NLOSv conditions is quite similar. At 10~m, the van reduced the RSSI by approximately 20~dB in both cases. As the distance between communicating nodes increased, the effect of the van was gradually reduced. At 100~m, the RSSI in the NLOSv case was approximately 5 and 7~dB below the LOS case for 802.11b/g and 802.11p, respectively.

\begin{figure}
\centering
   \includegraphics[width=0.33\textwidth,angle=0]{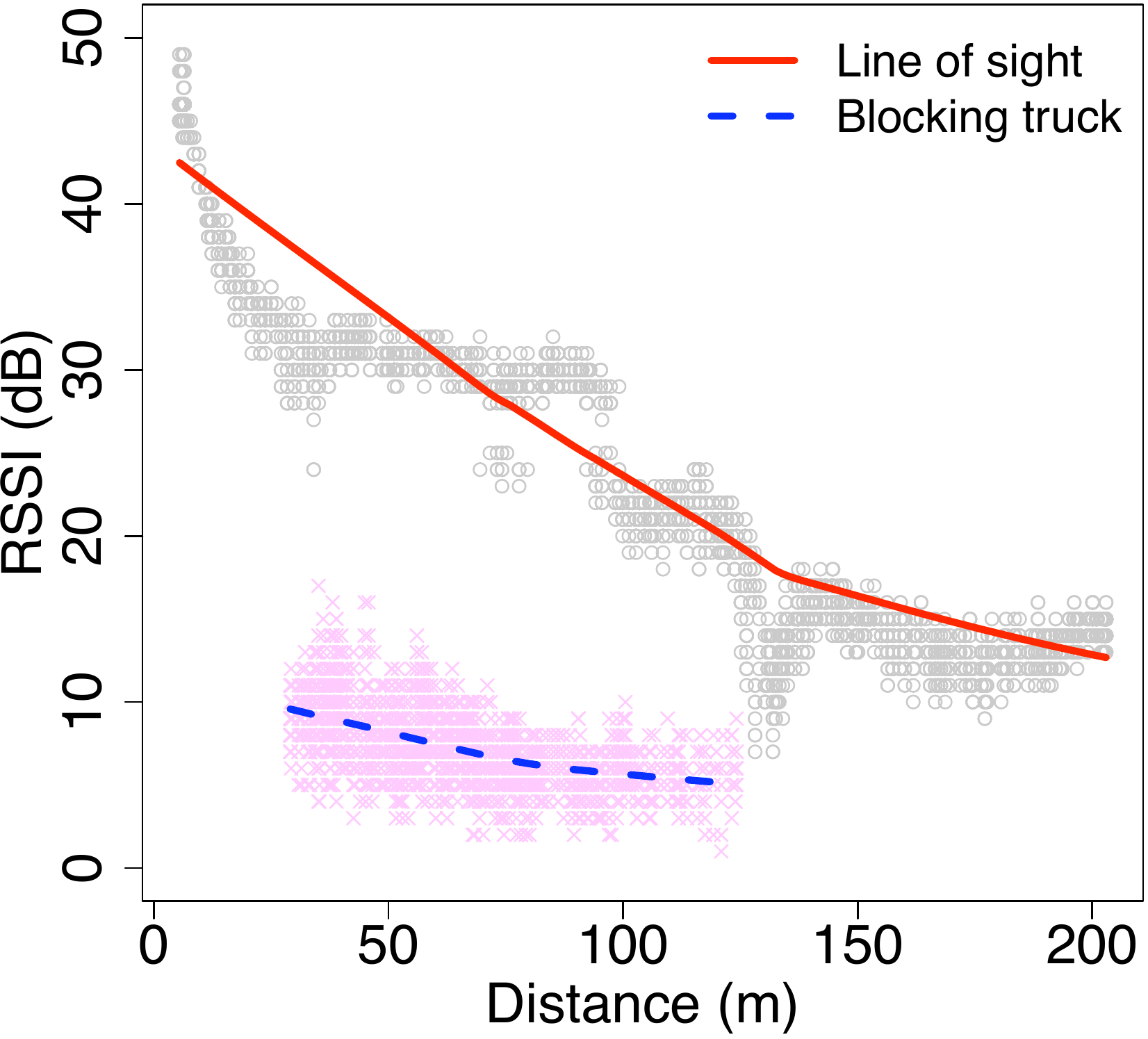}
\caption[RSSI as a function of distance for LOS and NLOSv conditions]{RSSI as a function of distance in 802.11p for LOS and NLOSv conditions due to the obstructing truck shown in Fig.~\ref{exp-setup:d}.}
\label{fig:parkinglot-carsMovingToMax}
\end{figure}

Furthermore, we performed an experiment where, starting with the cars next to each other, we slowly moved them apart. We did this experiment without any LOS obstruction and with a 4~m tall semi-trailer truck parked halfway between the vehicles (Fig.~\ref{exp-setup:d}). %
Figure~\ref{fig:parkinglot-carsMovingToMax} shows the RSSI as a function of distance. The dots represent individual samples, while the curves show the result of applying locally weighted scatter plot smoothing (LOWESS) to the individual points. %
The truck had a large impact on RSSI, with a loss of approximately 27~dB at the smallest recorded distance of 26~m (the length of the truck) when compared with the LOS case. For comparison, the van attenuated the signal by 12~dB at 20~m. The RSSI drop caused by the truck decreased as the cars move further away from it, an indication that the angle of the antennas' field of view that gets blocked makes a difference.

\subsection{On-the-road experiments}\label{subsec:realWorldExperiments}

For the on-the-road experiments, we drove the test vehicles in the three scenarios identified in Section~\ref{subsec:Scenarios} and collected RSSI and PDR information to use as indicators of channel quality. To accurately analyze the LOS and NLOS conditions, we placed each data point in one of the following categories, according to the information we obtained by reviewing the experiment videos:

\begin{itemize} 
\item \textbf{Line of sight (LOS)} --- no obstacles between the sender and receiver vehicles.

\item \textbf{Vehicular obstructions (NLOSv)} --- LOS blocked by other vehicles on the road.

\item \textbf{Static obstructions (NLOSb)} --- LOS blocked by immovable objects, such as buildings or terrain features, like crests and hills.

\end{itemize}

\begin{figure*}[thb!]
\centering
\subfigure[Highway]
{
   \label{fig:mobile-pdr:highway}
   \includegraphics[width=0.33\textwidth,angle=0]{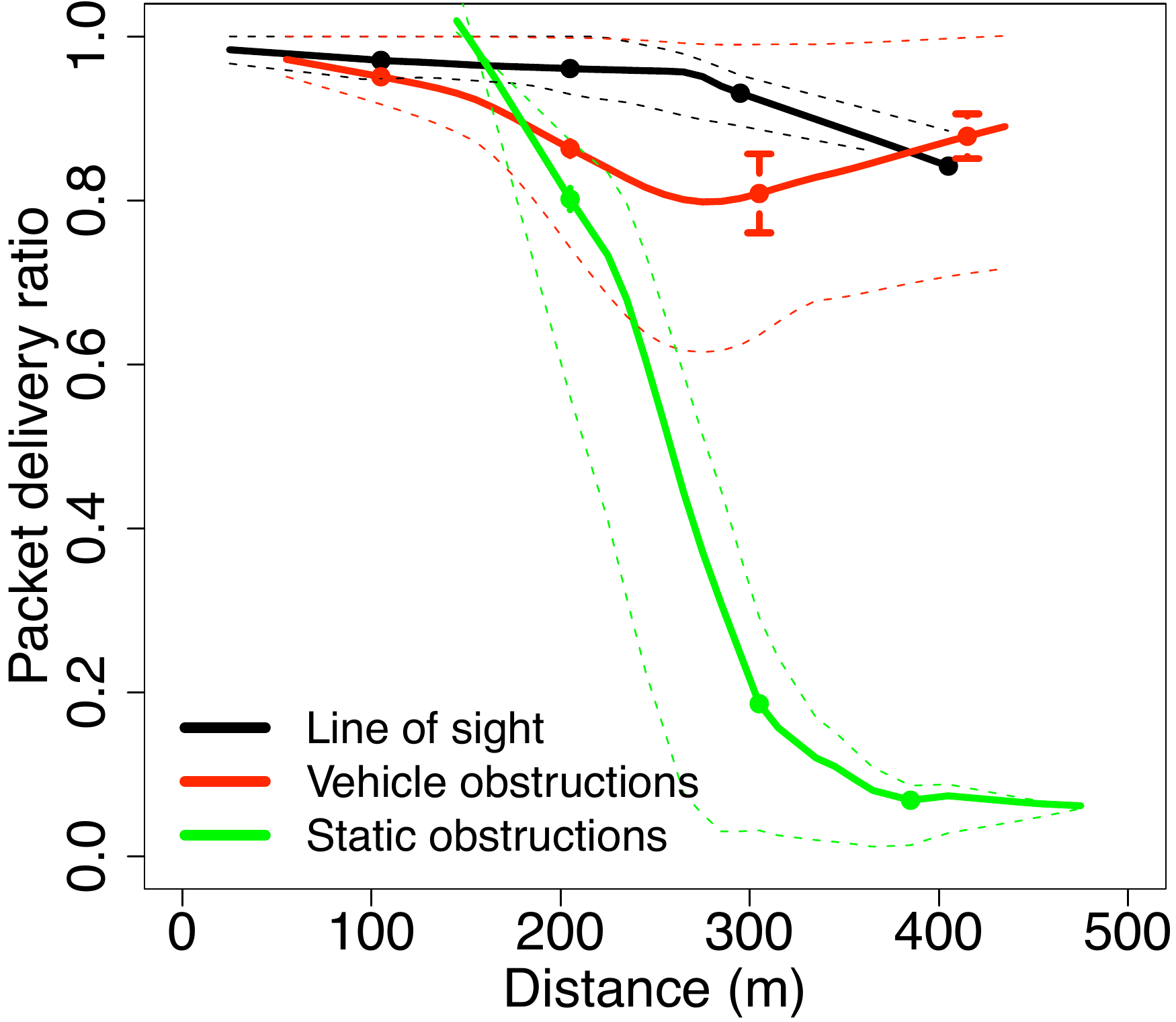}
}
\subfigure[Suburban]
{
   \label{fig:mobile-pdr:suburban}
   \includegraphics[width=0.33\textwidth,angle=0]{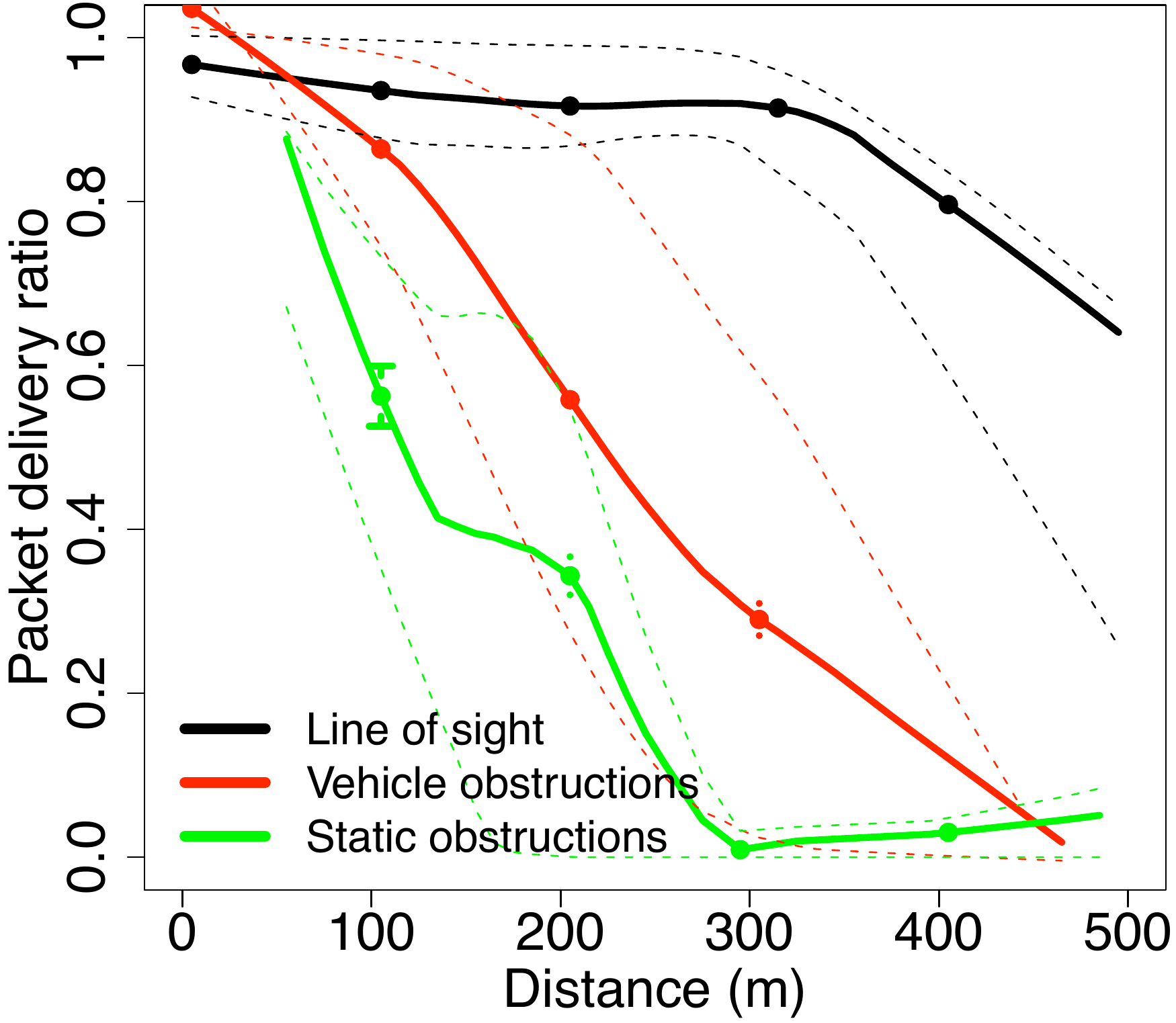}
}
\\
\subfigure[Urban canyon]
{
   \label{fig:mobile-pdr:urban-canyon}
   \includegraphics[width=0.33\textwidth,angle=0]{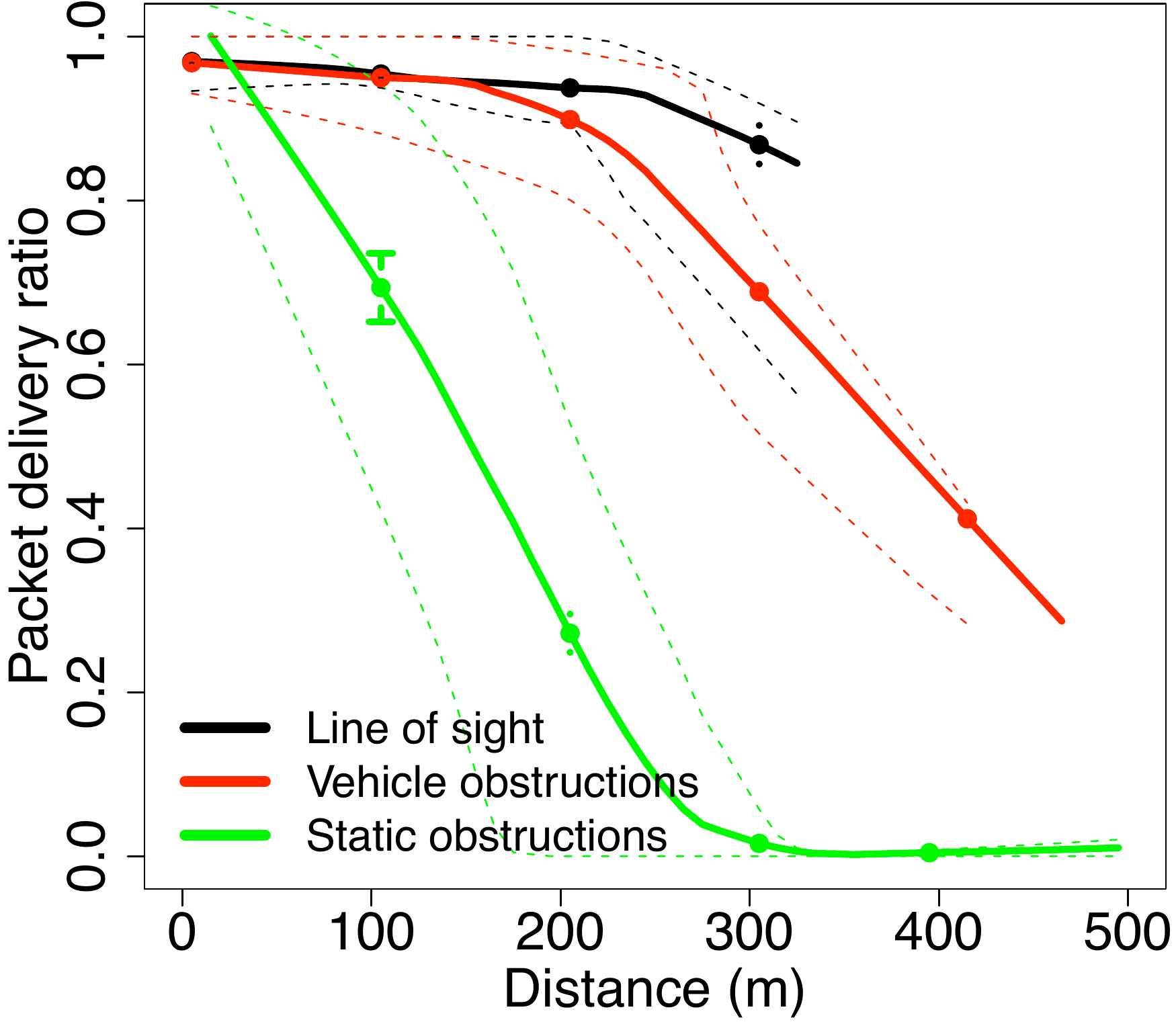}
}
\subfigure[Overall]
{
   \label{fig:mobile-pdr:overall}
   \includegraphics[width=0.33\textwidth,angle=0]{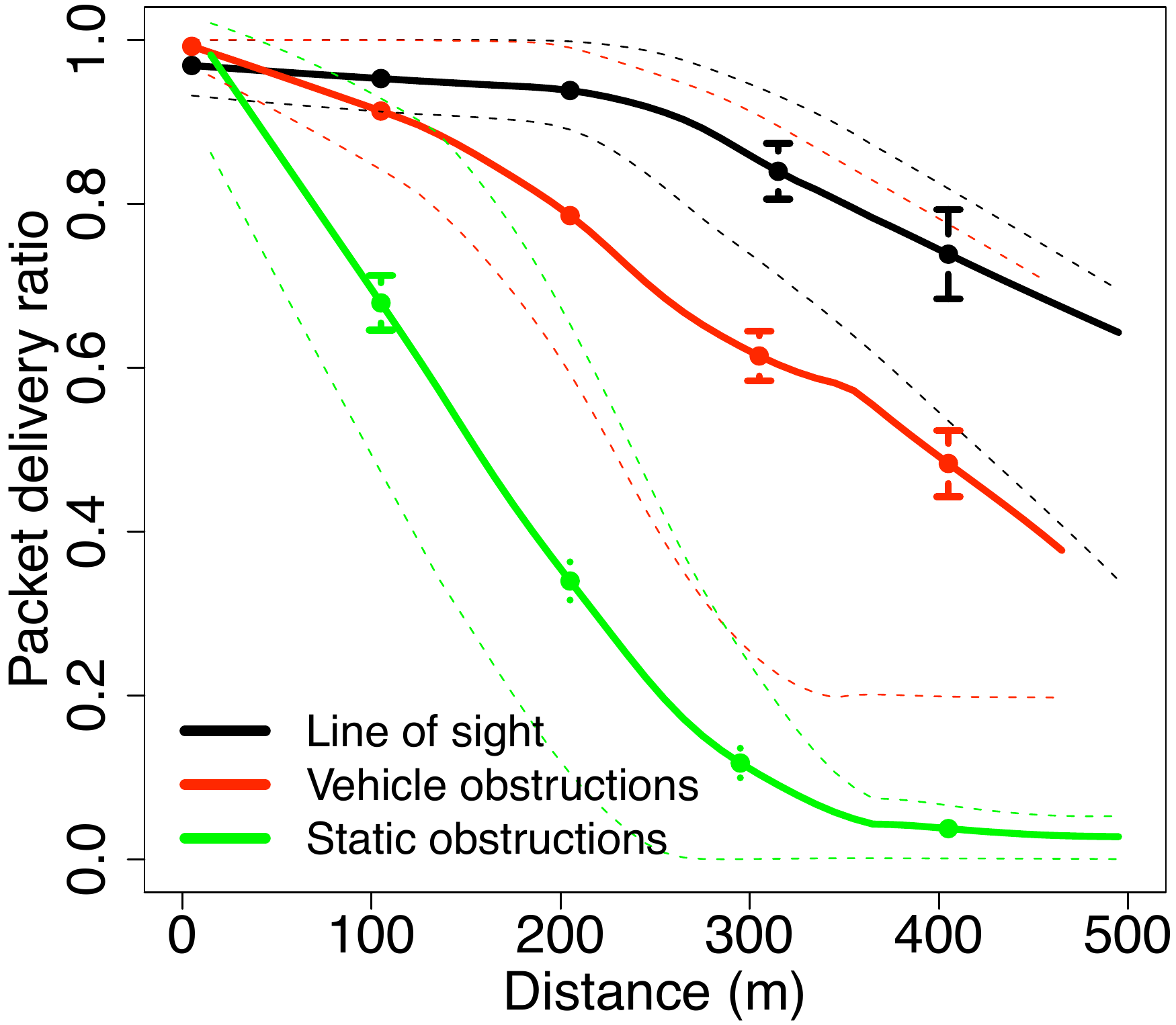}
}
\caption[Packet delivery ratio: on-the-road experiments]{Packet delivery ratio as a function of distance for the on-the-road experiments. The dashed lines represent the 20\% and 80\% quantiles.}
\label{fig:mobile-pdr}
\end{figure*}

To compute the PDR, we counted the number of beacons sent by the sender and the number of beacons received at the receiver in a given time interval. We used a granularity of 5 seconds (50 beacons) for the calculations. We use 10~m bins for the distance and show: the mean, its associated 95\% confidence intervals and the 20 and 80\% quantiles (dashed lines). To make the data easier to read, we use LOWESS to smooth the curves.

Figure~\ref{fig:mobile-pdr} shows the PDR as a function of distance separately for each on-the-road scenario, as well as aggregated over all three. For all scenarios, the PDR for the LOS case is above 80\% even at long distances, only dropping below that threshold in the suburban scenario and only after 400~m. At short distances, the difference between the PDR for LOS and NLOSv is almost non-existent. However, above 100~m there is a significant increase in the number of dropped packets in the NLOSv case. In the suburban scenario, the NLOSv PDR drops to zero at 500~m. In the urban canyon case, it drops to 30\% at roughly the same distance. %
Interestingly, in the highway scenario the NLOSv PDR stays high at long distances. One possible explanation could be that in the long sweeping highway curves the angle of the antennas' field of view blocked by vehicular obstructions is smaller than in other environments. Looking at the data for the static obstructions, we see marked differences in PDR, even when compared to the NLOSv case. In all environments, the PDR drops to 20\% or less at approximately 300~m, including the highway environment.

\begin{figure}
\centering
   \includegraphics[width=0.33\textwidth,angle=0]{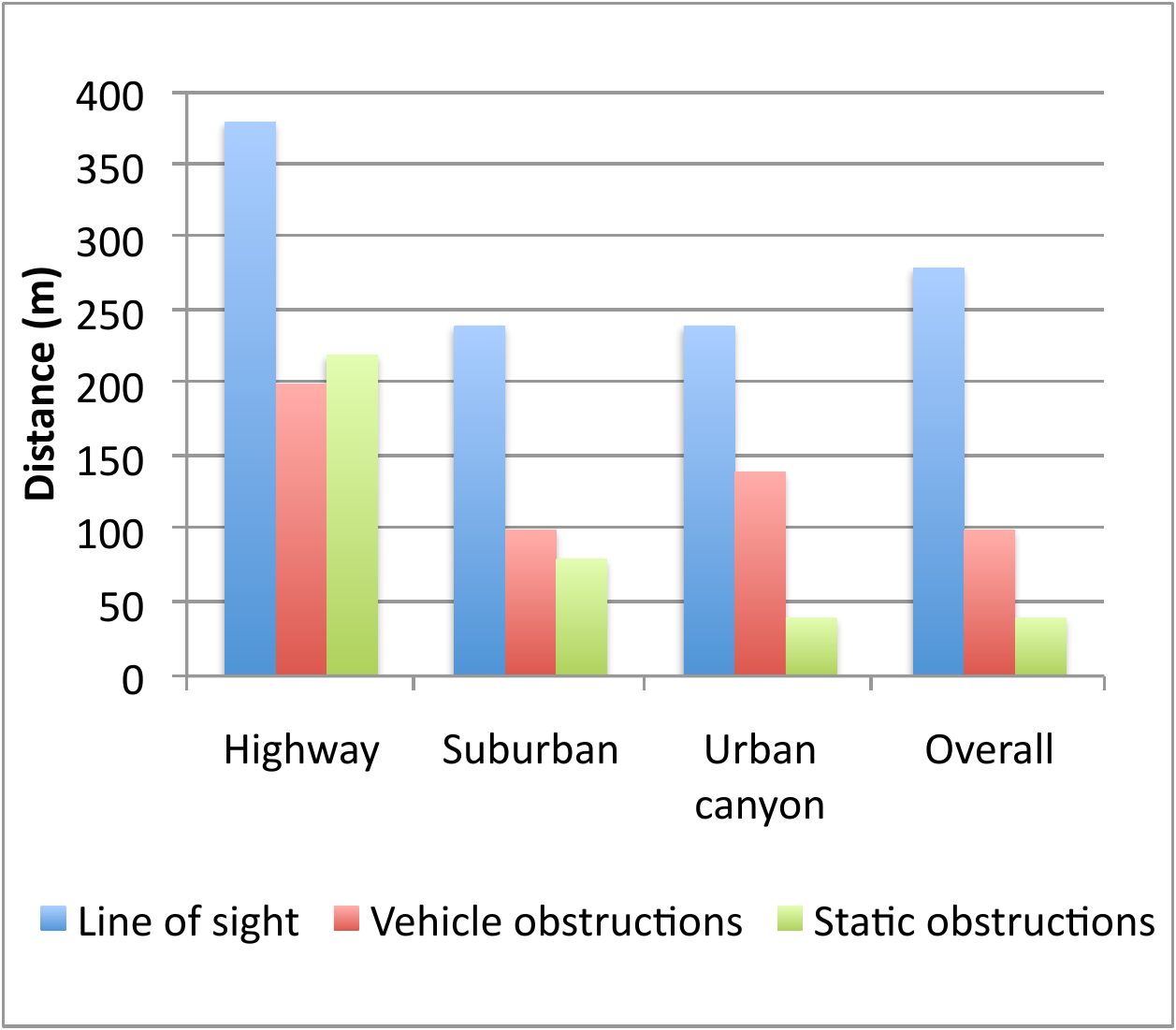}
\caption[Reliable communication range]{The reliable communication range calculated as the maximum distance at which the PDR was above 90\%.}
\label{fig:mobile-pdr:range}
\end{figure}

\begin{figure}
\centering
   \includegraphics[width=0.33\textwidth,angle=0]{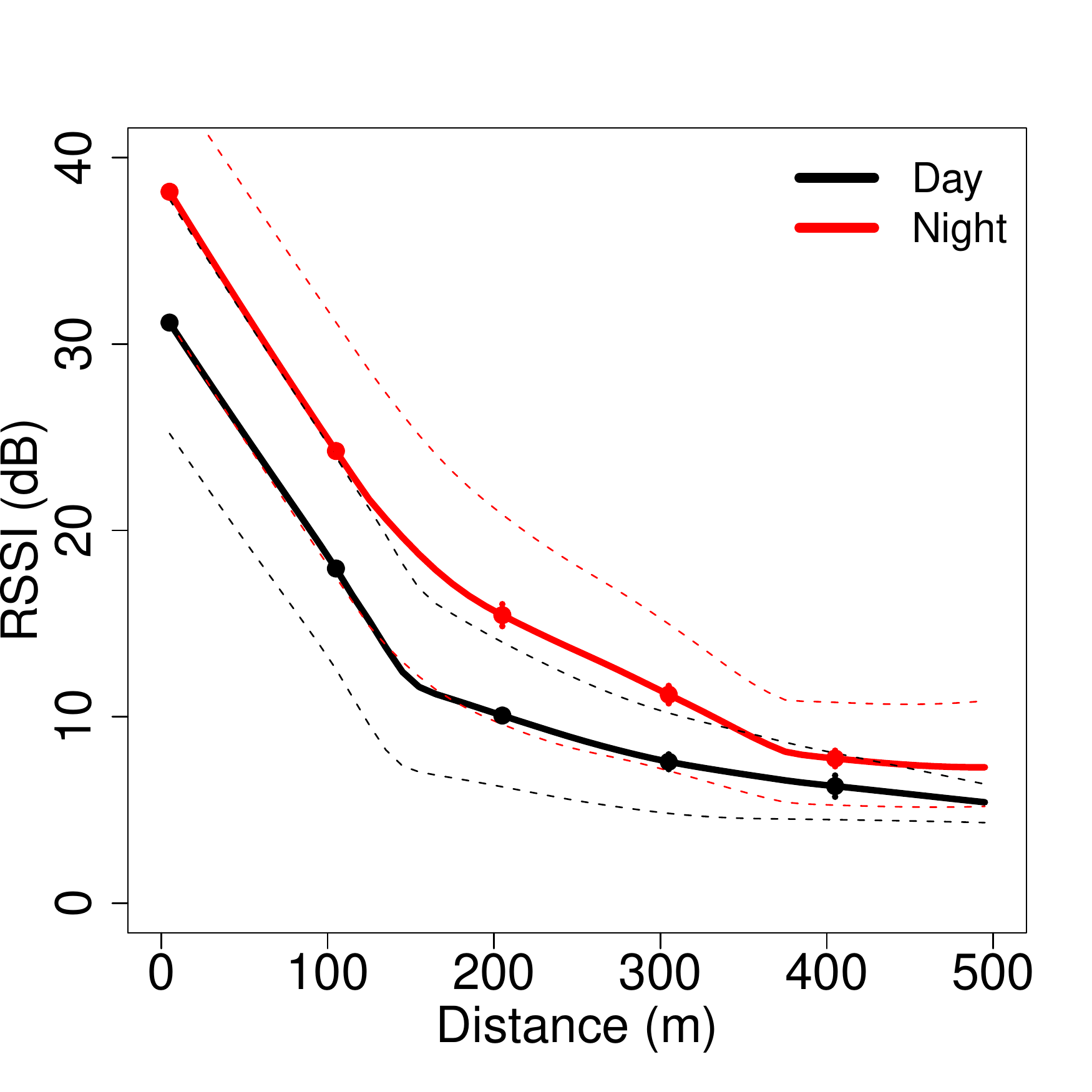}
\caption[Difference between the daytime and nighttime experiments]{The overall difference between the daytime experiments (frequent NLOSv conditions) and nighttime experiments (predominantly LOS).}
\label{fig:day_vs_night}
\end{figure}

\begin{figure*}
\centering
\subfigure[Suburban (35.000 LOS, 20.000 NLOSv and 7.000 NLOSb data points)]
{
   \label{fig:mobile-rssi:suburban}
   \includegraphics[width=0.33\textwidth,angle=0]{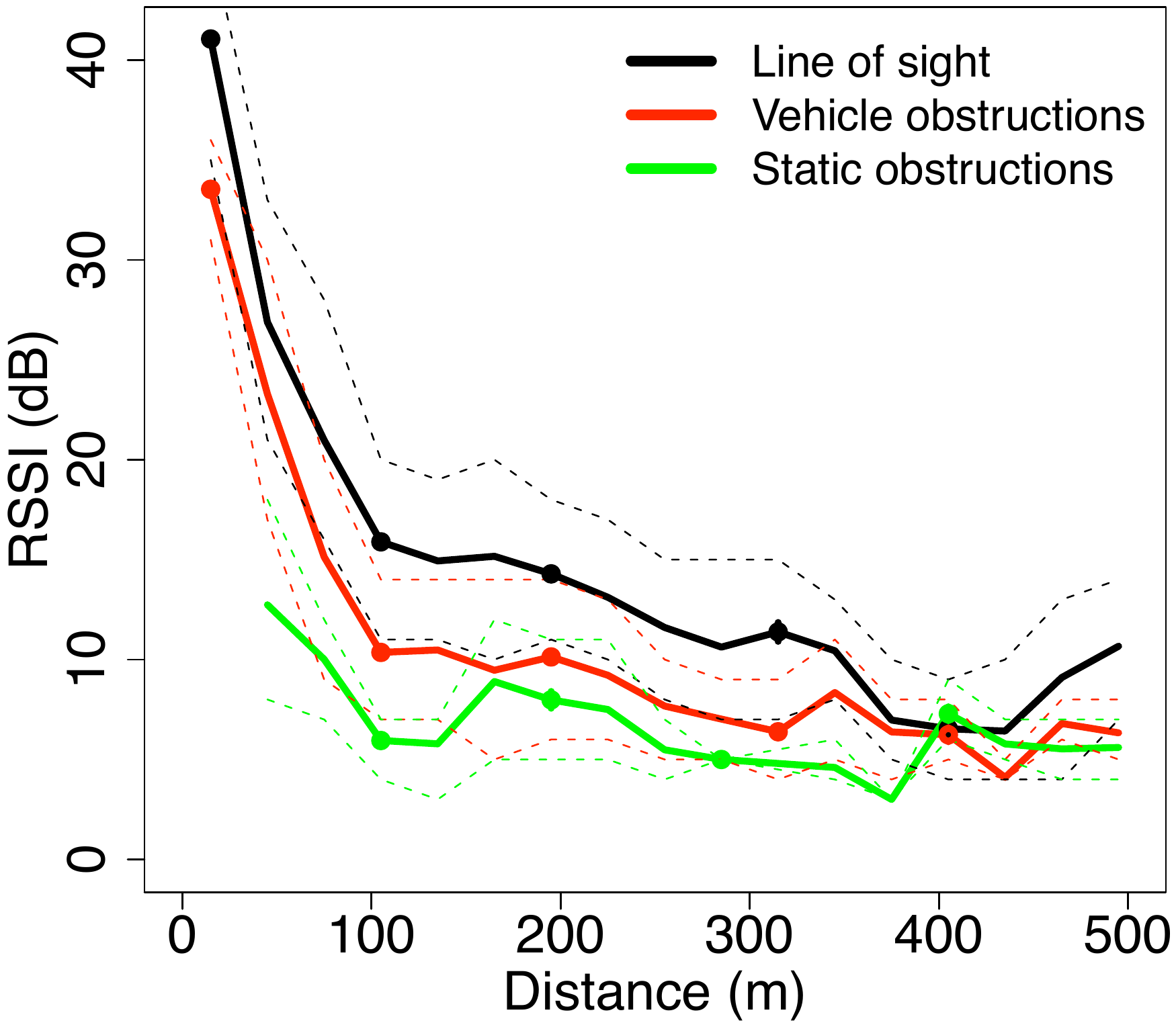}
}
\subfigure[Highway (14.000 LOS, 17.000 NLOSv and 1.000 NLOSb data points)]
{
   \label{fig:mobile-rssi:highway}
   \includegraphics[width=0.33\textwidth,angle=0]{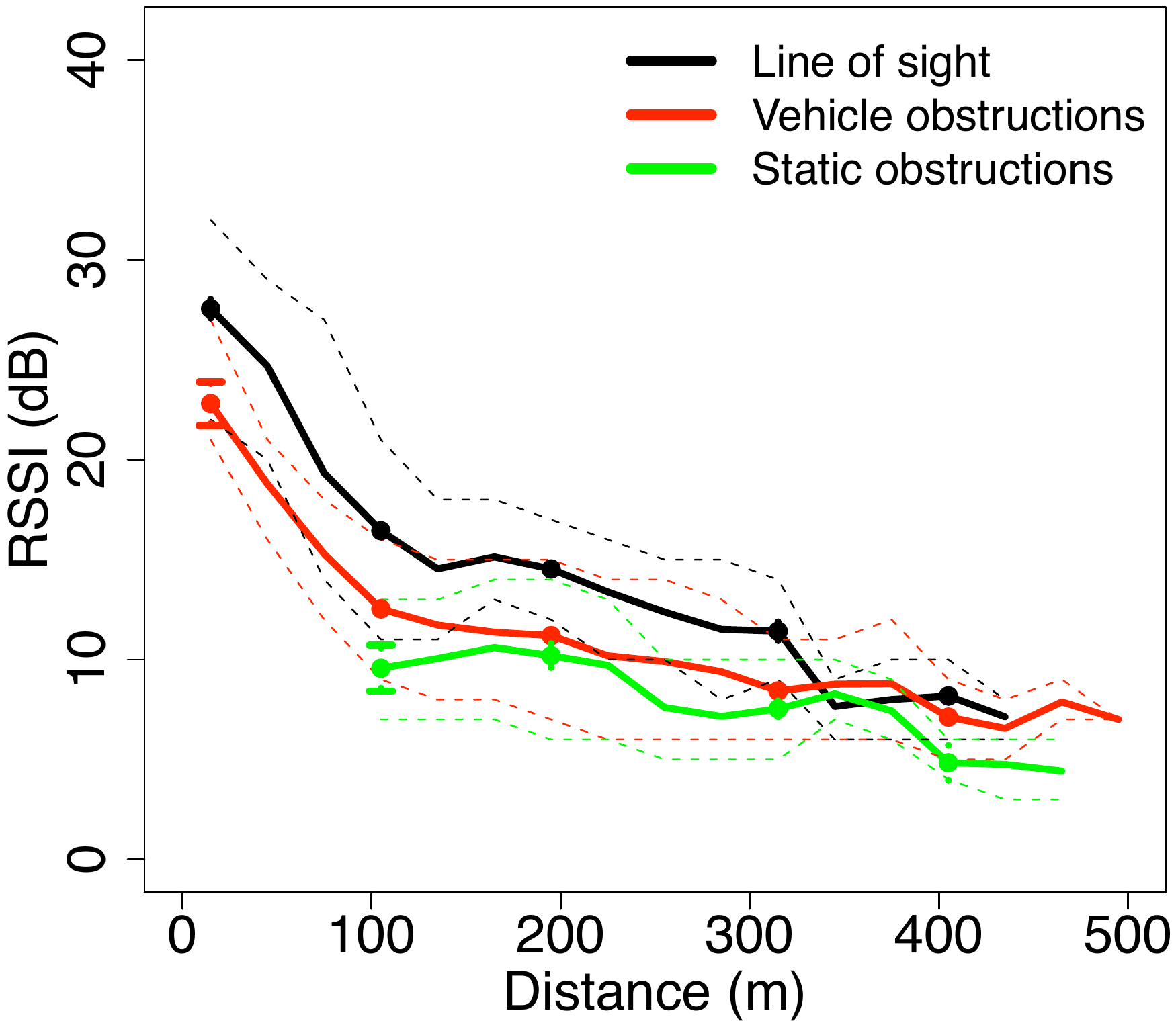}
}
\\
\subfigure[Urban canyon (62.000 LOS, 25.000 NLOSv and 1.000 NLOSb data points)]
{
   \label{fig:mobile-rssi:urban-canyon}
   \includegraphics[width=0.33\textwidth,angle=0]{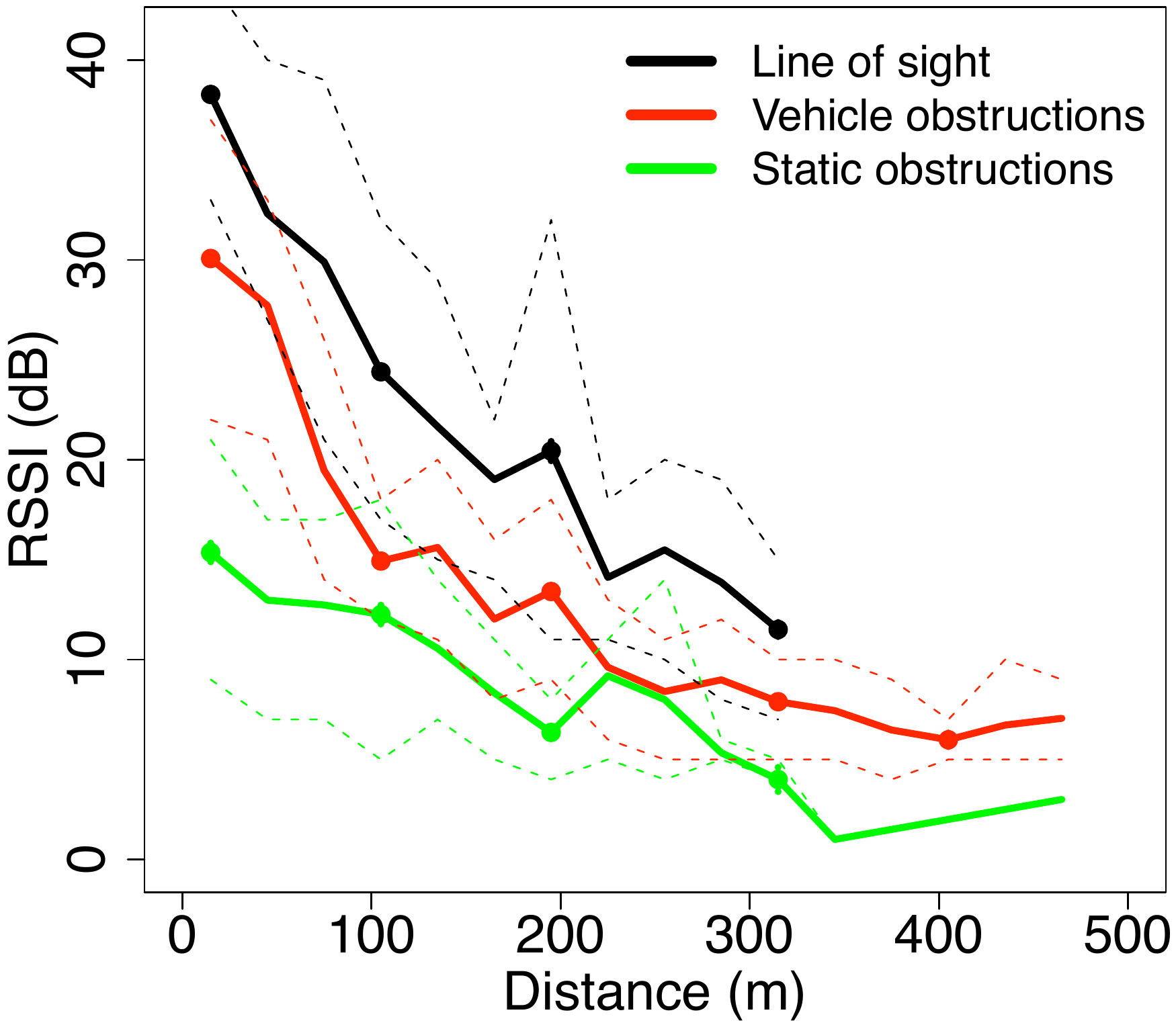}
}
\subfigure[Overall (35.000 LOS, 62.000 NLOSv and 9.000 NLOSb data points)]
{
   \label{fig:mobile-rssi:overall}
   \includegraphics[width=0.33\textwidth,angle=0]{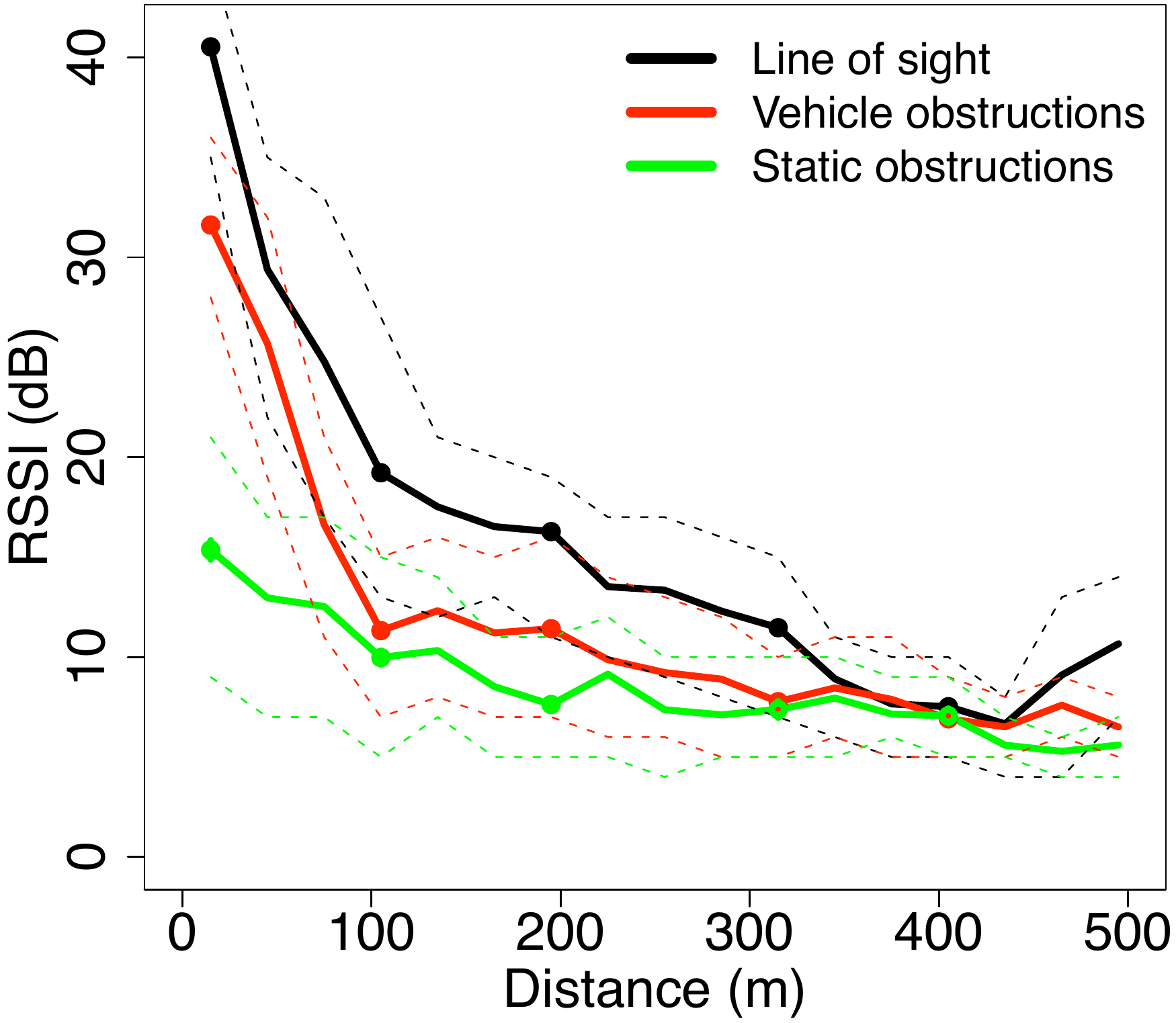}
}
\caption[Received signal strength: on-the-road experiments]{Received signal strength as a function of distance for the on-the-road experiments. The dashed lines represent the 20\% and 80\% quantiles.}
\label{fig:mobile-rssi}
\end{figure*}

To shed some light on the practical implications of these results, Fig.~\ref{fig:mobile-pdr:range} shows the reliable communication range under different LOS conditions. This range was calculated as the maximum distance at which the mean PDR was above or equal to 90\%. In all of the environments, the obstructing vehicles significantly decreased the effective communication range. The largest relative difference was observed in the suburban environment, with a 60\% reduction in range, and the smallest in the urban environment, with a 40\% reduction. The static obstructions have an even more negative impact, decreasing the overall communication range by 85\% on average. Using other target success probabilities (from 95\% to 50\%), we observed the following trends:
\begin{itemize}
\item For targets above 90\%, the importance of the LOS conditions is reduced. For a target PDR of 95\%, NLOSv conditions cause a 25\% decrease of the usable range.
\item Gradually decreasing the target PDR from 90\% to 50\% we observed a trend where the effective range in NLOSv conditions converges to around 50\% of what is achievable in the LOS case.
\end{itemize}

Regarding RSSI, we analyze each successfully received packet and plot the mean RSSI as a function of distance using 30 meter bins. We also plot 20\% and 80\% quantiles and 95\% confidence intervals at selected points. 

Figure~\ref{fig:day_vs_night} shows the overall RSSI as a function of distance for daytime (frequent NLOSv) and nighttime (infrequent NLOSv) experiments. Since the same routes were used in both experiments, the obstructing vehicles were the only variable changing between day and night. The difference between the plots shows the significant impact of the obstructing vehicles on the received signal power. %

Figure~\ref{fig:mobile-rssi} shows the resulting RSSI plots for each of the individual on-the-road experiment scenarios (Figs.~\ref{fig:mobile-rssi}(a)-(c)) and for the general case where we aggregate all data in each LOS category (Fig.~\ref{fig:mobile-rssi:overall}). The difference between LOS and NLOSv conditions varies in magnitude across scenarios but the overall trends are roughly similar and indicative of the significant impact that both vehicles and static obstacles had. Generally, we can observe the following trends in the difference between LOS and NLOSv conditions as we move from short to longer distances:

\begin{enumerate}
\item There is a large average difference of up to 10~dB between LOS and NLOSv conditions at short distances. This is most likely due to the vehicles blocking a large angle of the antennas' field of view. In the parking lot experiments the difference was up to 20~dB at these distances (see Fig.~\ref{fig:parkinglot-rssi}). The smaller difference in the road experiments is due to averaging over all vehicular obstructions, regardless of their height or angle relative to the antennas.
Interestingly, the absolute RSSI values at short distances in the highway scenario were significantly lower than in the other scenarios. 

\item As the distance increases, the difference between LOS an NLOSv conditions decreases slightly and then roughly stabilizes. %

\item At longer distances (above approximately 400~m), the difference gradually decreased to the point of being non-existent. This can be explained by two factors. First, the successful packet reception requires a minimum SINR. If the attenuation is strong enough that this threshold is crossed, the packet is dropped. At long distances, the successfully received packets are close to this minimum SINR threshold, so the difference between LOS conditions can only be observed in  terms of PDR. 

Also, for 5.9~GHz frequency and the given heights of the antennas, the first Fresnel ellipsoid becomes obstructed by the ground level at 400~m \cite[Chap. 3]{rappaport96}. Therefore, the road itself starts effectively blocking the LOS between the communicating vehicles. This finding is in line with the results reported in \cite{boban11}.

\end{enumerate}

It is interesting to observe the large difference in RSSI observed in the urban canyon scenario. %
This difference is perhaps best explained by the multipath effects caused by the buildings. The tunneling effect created reflected rays with relatively low phase difference to the LOS ray, which in turn acted constructively on the received power.

We also captured data pertaining to the effect of static obstructions on the channel quality. In the urban canyon the obstructions were mainly buildings, which had a profound impact on RSSI. A loss of around 15~dB compared with the NLOSv case at shorter distances and around 4~dB at larger distances was observed. In the suburban and highway scenarios, obstructions were mostly created by crests on the road. The results indicate that they can make a difference of up to 3~dB of additional attenuation atop the NLOSv attenuation.

The results presented in this section inevitably point to the fact that obstructing vehicles have to be accounted for in channel modeling. Not modeling the vehicles results in overly optimistic received signal power, PDR and communication range.

\section{Related Work} \label{sec:RelatedWork}

Regarding V2V communication, Otto et al.~in~\cite{Otto2009} performed V2V experiments in the 2.4~GHz frequency band in an open road environment and reported a significantly worse signal reception during a traffic heavy, rush hour period in comparison to a no traffic, late night period. 
A similar study presented in \cite{takahashi03} analyzed the signal propagation in ``crowded'' and ``uncrowded'' highway scenarios (based on the number of vehicles on the road) for the 60~GHz frequency band, and reported significantly higher path loss for the crowded scenarios. %
With regards to experimental evaluation of the impact of vehicles and their incorporation in channel models, a lightweight model based on Markov chains was proposed in \cite{dhoutaut06}. Based on experimental measurements, the model extends the stochastic shadowing model and aims at capturing the time-varying nature of the V2V channel based on a set of predetermined parameters describing the environment. Tan et al.~in~\cite{tan08} performed experimental measurements in various environments (urban, rural, highway) at 5.9~GHz to determine the suitability of DSRC for vehicular environments with respect to delay spread and Doppler shift. The paper distinguishes LOS and NLOS communication scenarios by coarsely dividing the overall obstruction levels. The results showed that DSRC provides satisfactory performance of the delay spread and Doppler shift, provided that the message is below a certain size. A similar study was reported in \cite{paier07}, where experiments were performed at 5.2~GHz. Path loss, power delay profile, and Doppler shift were analyzed and statistical parameters, such as path loss exponent, were deduced for given environments.
Based on measurements, a realistic model based on optical ray-tracing was presented in \cite{maurer04}. The model encompassed all of the obstructions in a given area, including the vehicles, and yielded results comparable with the real world measurements. However, the high realism that the model exhibits is achieved at the expense of high computational complexity.

Experiments in urban, suburban, and highway environments with two levels of traffic density (high and low) were reported in \cite{sen08}. The results showed significantly differing channel properties in low and high traffic scenarios. Based on the measurements, several V2V channel models were proposed. The presented models are specific for a given environment and vehicle traffic density.
Several other studies~\cite{Jerbi2007,Wu2005,matolak05,usdot06_2} point out that other vehicles apart from the transmitter and receiver could be an important factor in modeling the signal propagation by obstructing the LOS between the communicating vehicles. %

Virtually all of the studies mentioned above emphasize that LOS and NLOS for V2V communication have to be modeled differently, and that vehicles act as obstacles and affect signal propagation to some extent. However, these studies at most quantify the macroscopic impact of the vehicles %
by defining V2V communication environments as uncrowded (LOS) or crowded (NLOS), depending on the relative vehicle density, without analyzing the impact that obstructing vehicles have on a single communication link. %

\section{Conclusions} \label{sec:Conclusions} %
In this work we set out to experimentally evaluate the impact of obstructing vehicles on V2V communication. %
For this purpose, we ran a set of experiments with near-production 802.11p hardware in a multitude of relevant scenarios: parking lot, highway, suburban and urban canyon.

Our results indicate that vehicles blocking the line of sight significantly attenuate the signal when compared to line of sight conditions across all scenarios. Also, the effect appears to be more pronounced the closer the obstruction is to the sender, with over 20~dB attenuation at bumper-to-bumper distances. The additional attenuation decreased the packet delivery ratio at longer distances, halving the effective communication range for target average packet delivery ratios between 90\% and 50\%. The effect of static obstacles such as buildings and hills was also analyzed and shown to be even more pronounced than that of vehicular obstructions. %

With respect to channel modeling, even the experimental measurements proposed for certification testing of DSRC equipment \cite{acosta07} do not directly address the effect of vehicles in the V2V environment, thus potentially underestimating the attenuation and packet loss. Our work shows that not modeling vehicles as physical obstructions takes away from the realism of the channel models, thus affecting the simulation of both the physical layer and the upper layer protocols.

\chapter{Modeling Vehicles as Obstacles in VANETs} \label{ch:vehModel}
A thorough understanding of the communications channel between vehicles is essential for realistic modeling of Vehicular Ad Hoc Networks (VANETs) and the development of related technology and applications. The impact of vehicles as obstacles on vehicle-to-vehicle (V2V) communication has been largely neglected in VANET research, especially in simulations. Useful models accounting for vehicles as obstacles must satisfy a number of requirements, most notably accurate positioning, realistic mobility patterns, realistic propagation characteristics, and manageable complexity. We present a model that satisfies all of these requirements~\cite{boban11}. Vehicles are modeled as physical obstacles affecting the V2V communication. The proposed model accounts for vehicles as three-dimensional obstacles and takes into account their impact on the LOS obstruction, received signal power, and the packet reception rate. We utilize the experiments reported in Chapter~\ref{ch:experiments} along with two highway datasets containing vehicle locations collected via stereoscopic aerial photography to test our proposed model. We confirm the importance of modeling the effects of obstructing vehicles through experimental measurements. %
By obstructing the LOS, vehicles induce significant attenuation and packet loss. The algorithm behind the proposed model allows for computationally efficient implementation in VANET simulators. It is also shown that by modeling the vehicles as obstacles, significant realism can be added to existing simulators with relevant implications on the design of upper layer protocols. 

\section{Motivation} \label{sec:Introduction}
Vehicle to vehicle (V2V) communication is proposed as the communication paradigm for a number of traffic safety (\cite{bai06, boban12}, traffic management (\cite{nadeem04}), and infotainment applications (\cite{bai06,chen05,tonguz10}). 
In the previous chapter we have shown that in V2V communication, due to the relatively low elevation of the antennas on the communicating vehicles, other vehicles act as obstacles to the signal, often affecting propagation even more than static obstacles (e.g., buildings or hills), especially in the case of an open road. %
As noted in a recent survey on Vehicular Ad Hoc Network (VANET) simulators \cite{martinez09}, state of the art VANET simulators such as NS-2 \cite{ns2}, JiST/SWANS/STRAW \cite{choffnes05}, and NCTU-NS \cite{nctuns},  
consider the vehicles as dimensionless entities that have no influence on signal propagation. 
One reason lies in the fact that realistic propagation models for such highly dynamic networks are generally deemed to be computationally expensive (e.g., ray-tracing \cite{glassner89}), and mobile obstacles increase the complexity even further.  
Simplified stochastic radio models, which rely on the statistical properties of the chosen environment and do not account for the specific obstacles in the region of interest, are thus preferred for use in simulators, and are believed to offer reasonable approximations at low computational cost \cite{dhoutaut06}. A recent study showed, however, %
that stochastic radio models do not provide satisfying accuracy for typical VANET scenarios \cite{koberstein09}.  
On the other extreme, topography-specific, highly realistic channel models (such as the one presented in \cite{maurer04}) yield results that are in very good agreement with the real world, albeit at a price. These models are computationally too expensive and bound to a specific location (e.g., a particular neighborhood in a city, such as in \cite{maurer04}) to be practically useful for simulations. For these reasons, such models are not implemented in VANET simulators.

Hence, there exists a need for accurate and efficient V2V channel models. 
To provide such a model, we incorporate vehicles as obstacles and present a method to analyze the existence of the LOS 
component of the signal for each communicating pair. The focus on the existence of the LOS component was motivated by the recent experimental V2V studies reported in \cite{acosta06} and \cite{paier07}. 
These studies showed that, when existent, the LOS component of the signal carries orders of magnitude more energy than 
the remaining components (e.g., due to reflection or diffraction). 
This effect was shown to be particularly pronounced in highway environments. We therefore analyze the data collected on Portuguese highways 
to show that, as physical obstacles, vehicles have a significant impact on signal propagation, by frequently obstructing the LOS between the communicating vehicles.  
Based on the (non-)existence of LOS, we implemented a simple and efficient model for vehicles as obstacles and showed that for the proposed VANET communication standard, the Dedicated Short Range Communication (DSRC) \cite{dsrc09}, the %
signal attenuation due to the obstructing vehicles is significant. To further verify the predictions of the proposed model, we conducted empirical measurements which corroborated the results regarding the signal attenuation due to vehicles. Not modeling the vehicles as obstacles thus leads to unrealistic assumptions about the physical layer, and this was shown to have significant implications on the behavior of the upper layers of the protocol stack (e.g., \cite{dhoutaut06}, \cite{Otto2009}, and \cite{kaul07}).

Our main contributions are as follows. 
\begin{enumerate}
\item By analyzing the real-world data, we %
\emph{quantify} %
the impact of vehicles as obstacles on V2V communication in terms of LOS obstruction. The results show that the obstructing vehicles have a significant impact on LOS in both sparse and dense vehicular networks and should therefore be included in V2V channel modeling in order to obtain more realistic simulation results. 
\item Based on the LOS obstruction analysis, we develop a model for incorporating the vehicles as obstacles in VANET simulators. The model encompasses calculation of LOS obstruction, as well as a simple signal propagation model to characterize the effects of obstructing vehicles on the received signal power and the packet reception ratio. We confirm the results of the model by performing empirical V2V measurements.
\item In order to make the proposed model suitable for implementation in VANET simulation environments, we designed it %
with the following characteristics in mind:
\begin{itemize}
\item suitability for any VANET environment (e.g., urban, suburban, highway) with any vehicle density; 
\item topology/location independence; 
\item feasible implementation in VANET simulators; and 
\item complementarity and compatibility with VANET signal propagation models for static obstacles (e.g., buildings, foliage, etc). %
\end{itemize}
The results obtained by employing the proposed model show that significant improvements can be obtained with regards to the realism of the simulation results, at the same time maintaining relatively low computational cost. For this reason, we argue that such model needs to be implemented in VANET simulators in order to increase the credibility of simulation results.
\end{enumerate}

The rest of the chapter is organized as follows. Section~\ref{sec:Related} describes previous work on channel characterization in V2V communication. 
The methodology for evaluating the impact of vehicles on LOS and signal behavior is described in Section~\ref{sec:Evaluating}, whereas Section~\ref{sec:Model} presents the requirements of the proposed model and the means to obtain the required data. %
Section~\ref{sec:Results} highlights the obtained results. %
Finally, Section~\ref{sec:Conclusions} concludes the chapter.

\section{Related Work} \label{sec:Related}

Our approach towards V2V communication is based on the following hypothesis: the low heights of the antennas in V2V communication system suggests that other vehicles can act as obstacles for signal propagation, most notably by obstructing the LOS between the communicating vehicles. %
Numerous studies, both experimental and analytical (e.g., \cite{usdot06_2} %
 and \cite{masui02}), have shown that %
 LOS and non-LOS (NLOS) scenarios must be separately modeled in VANETs, because the resulting channel characteristics are fundamentally different. 

Several other experimental studies point out that other vehicles apart from the transmitter and receiver could be an important factor for the signal propagation (mainly by obstructing the LOS, thus decreasing the received signal power) and therefore should be included in channel modeling  (e.g., \cite{Jerbi2007, matolak05,takahashi03}). 

Wang et al.~in~\cite{cheng09} analyzed the state of the art in V2V channel measurement and modeling.  
Based on the approach of modeling the environment (geometrically or non-geometrically), and the distribution of objects in the environment (stochastic or deterministic), three main types of models were identified: non-geometrical stochastic models, geometry-based deterministic models, and geometry-based stochastic models. 
Using this classification, we present an overview of the existing research on V2V communication and channel modeling with respect to vehicles as obstacles. 

\subsection{Non-Geometrical Stochastic Models}

Otto et al.~in~\cite{Otto2009} performed experiments at the 2.4~GHz frequency band in urban, suburban, and open road environments. Although the study focused on static obstacles such as buildings, the results showed a significantly worse signal reception on the same open road during the traffic heavy, rush hour period when compared to a no traffic, late night period. The measurements for the rush hour period showed a mean path loss exponent of 3.31 and a shadowing deviation of 4.84 dB, whereas in the late night period the mean path loss exponent was 3.1 with a shadowing deviation of 3.23 dB. The observed difference can only be attributed to other vehicles obstructing the signal, since all other system parameters remained the same.
 
Cheng et al.~in~\cite{Cheng2007} performed measurements of the V2V channel in the 5.9~GHz frequency band and pointed out that vehicles as obstacles are the most probable cause for the difference in received signal power between the obtained experimental measurements and the 
dual slope piecewise linear channel model used in that study. Extensive measurement campaigns reported in \cite{sen08} analyzed urban, suburban, and highway environments with two levels of traffic density (high and low). The measurements showed significantly differing channel properties in low and high traffic scenarios. Based on the measurements, several V2V channel models were proposed. The presented models are specific for a given environment and vehicle traffic density.
A simple error model for V2V communication was presented in \cite{zang05}, where the authors differentiate the LOS and NLOS communication due to vehicles using a highly abstracted model where a threshold distance is used to separate the LOS and NLOS communication. As noted in the same paper, higher realism requires a more detailed channel model that differentiates between the LOS and NLOS communication induced by vehicles. 

\subsection{Geometry-Based Deterministic Models}
A highly realistic model, based on optical ray-tracing was presented in \cite{maurer04}. The model encompasses all objects in the analyzed environment (both static and mobile) and evaluates the signal behavior by analyzing the 50 strongest propagation paths between the communicating pair. The model was compared against experimental measurements and showed close agreement. However, the realism of the model is achieved at the cost of high computational complexity and location-specific modeling. Even with the recent advances in optimizing the execution of ray-tracing models \cite{kim09}, the method remains computationally too expensive to be implemented in VANET simulators. Additionally, detailed knowledge about the topology of the analyzed environment is necessary in order to accurately model the channel.

\subsection{Geometry-Based Stochastic Models}

Karedal et al.~in~\cite{karedal09} designed a %
model for the V2V channel based on extensive measurements performed in highway and suburban environment at the 5.2 GHz frequency band. The model distributes the vehicles as well as other objects at random locations and analyzes four distinct signal components: LOS, discrete components from mobile objects, discrete components from static objects, and diffuse scattering. Based on the obtained measurement data, a set of model parameters for the two environments is prescribed, and the non-stationarity of the V2V channel can be captured by employing a mobility model for the vehicles (it was shown in \cite{paier07_2} that the wide-sense stationary uncorrelated scattering assumption does not hold for the V2V channel).
Cheng et al.~in~\cite{cheng09_3} presented a MIMO channel model that takes into account the LOS, single-bounced rays, and double-bounced rays by employing a combined two-ring and ellipse model. By properly defining the parameters, the model can be used in various V2V environments with varying vehicle densities. Due to the static nature of the employed geometric model, the non-stationarity of the V2V channel can not be captured.

With regards to the implementation of vehicles as obstacles in simulators, virtually all of the state of the art VANET simulators neglect the impact of vehicles as obstacles on signal propagation, mainly due to the lack of an appropriate methodology capable of incorporating the effect of vehicles both realistically and efficiently.

To the best of our knowledge, up to now there has been no study that focused on vehicles as obstacles by systematically quantifying their impact on LOS and consequently on the received signal power. Apart from quantifying the impact of vehicles, we present a computationally efficient model for the implementation of vehicles as obstacles in VANET simulators. Our model can be seen as a simplified geometry-based deterministic model.

\section{Model Analysis}%
\label{sec:Evaluating}

\subsection{The Impact of Vehicles on Line of Sight}

In order to isolate and quantify the effect of vehicles as obstacles on signal propagation, we do not consider the effect of other obstacles such as buildings, overpasses, vegetation, or other roadside objects on the analyzed highways. Since those obstacles can only further reduce the probability of LOS, our approach leads to a  
best case analysis for probability of LOS.

\begin{figure*}[t]
	\centering
	\fbox{\psfig{figure=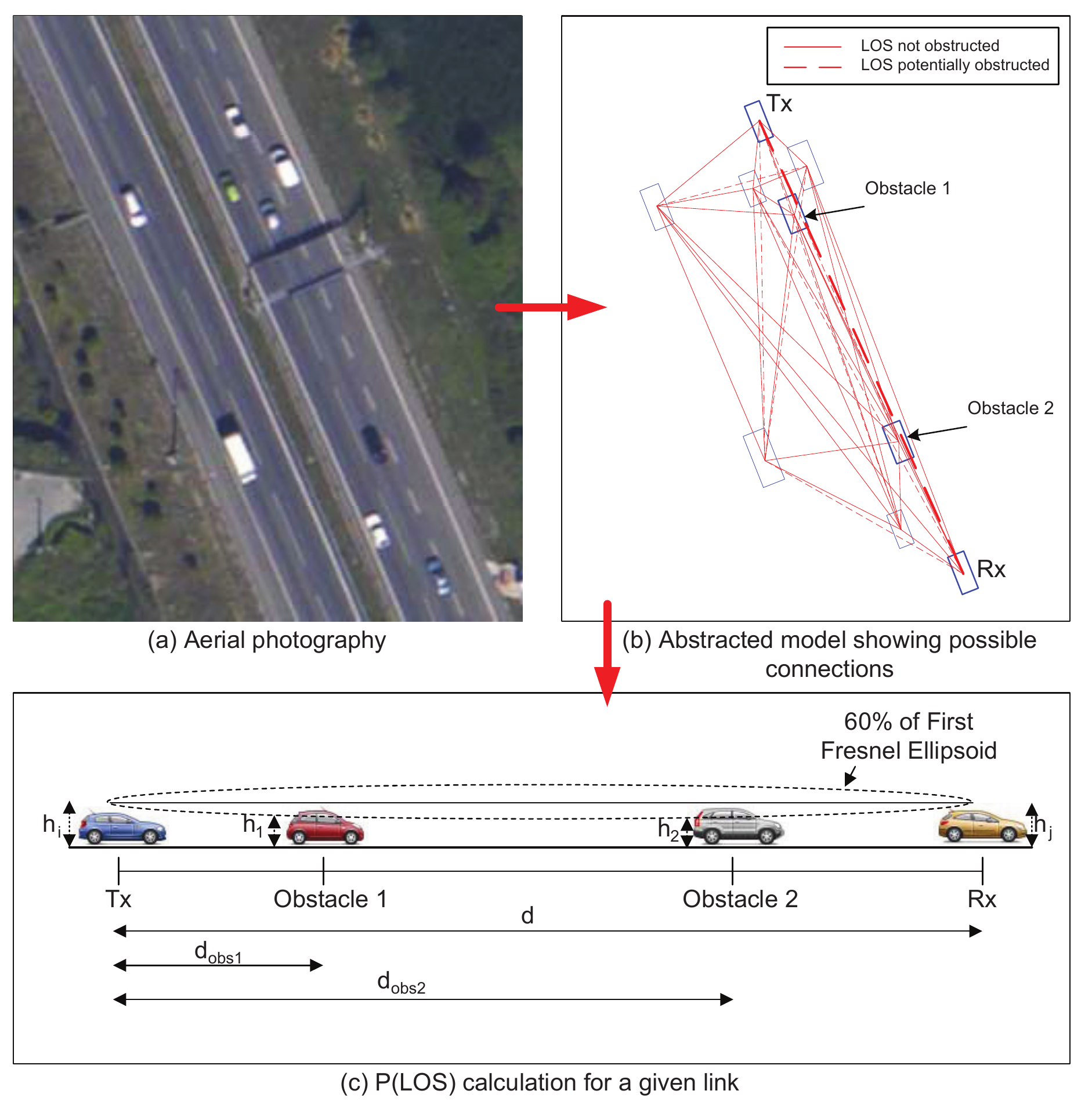,width=0.75\textwidth}} %
	\caption[Model for evaluating the impact of vehicles as obstacles on LOS]{\small Model for evaluating the impact of vehicles as obstacles on LOS (for simplicity, vehicle antenna heights ($h_a$) are not shown in subfigure (c)).}%
  \label{explanation}
\end{figure*}
Figure~\ref{explanation} describes the methodology we use to quantify the impact of vehicles as obstacles on LOS in a V2V environment. Using aerial imagery (Fig.~\ref{explanation}a) to obtain the location and length of vehicles,  
we devise a model that is able to analyze all possible connections between vehicles within a given range (Fig.~\ref{explanation}b). For each link  
-- such as the one between the vehicles designated as transmitter (Tx) and receiver (Rx) in Fig.~\ref{explanation}b -- the model determines the existence or non-existence of the LOS based on the number and dimensions of vehicles potentially obstructing the LOS (in case of the aforementioned vehicles designated as Tx and Rx, the vehicles potentially obstructing the LOS are those designated as Obstacle 1 and Obstacle 2 in Fig.~\ref{explanation}b). 

The proposed model calculates the (non-)existence of the LOS for each link (i.e., between all communicating pairs) in a deterministic fashion, based on the dimensions of the vehicles and their locations. However, in order to make the model mathematically tractable, we derive the expressions for the microscopic (i.e., per-link and per-node) and macroscopic (i.e., system-wide) probability of LOS. It has to be noted that, from the electromagnetic wave propagation perspective, the LOS is not guaranteed with the existence of the visual sight line between the Tx and Rx. It is also required that the Fresnel ellipsoid is free of obstructions \cite[Chap. 3]{rappaport96}. Any obstacle that obstructs the Fresnel ellipsoid might affect the transmitted signal. As the distance between the transmitter and receiver increases, the diameter of the Fresnel ellipsoid increases accordingly. Besides the distance between the Tx and Rx, the Fresnel ellipsoid diameter is also a function of the wavelength. 

As we will show later in Section~\ref{sec:Model}, the vehicle heights follow a normal distribution. To calculate $P(LOS)_{ij}$, i.e., the probability of LOS for the link between vehicles $i$ and $j$, with one vehicle as a potential obstacle between Tx and Rx (of height $h_i$ and $h_j$, respectively), we have: %
\begin{equation}
P(LOS|h_i,h_j)= 1 - Q\left(\frac{h - \mu}{\sigma}\right)
\label{eq:first}
\end{equation}
and
\begin{equation}
h = (h_j-h_i)\frac{d_{obs}}{d} + h_i - 0.6r_f + h_a,
\end{equation}
where the $i,j$ subscripts are dropped for clarity, and $h$ denotes the effective height of the straight line that connects Tx and Rx at the obstacle location when we consider the first Fresnel ellipsoid. %
Furthermore, $Q(\cdot)$ represents the $Q$-function, $\mu$ is the mean height of the obstacle, $\sigma$ is the standard deviation of the obstacle's height, $d$ is the distance between the transmitter and receiver, $d_{obs}$ is the distance between the transmitter and the obstacle, $h_a$ is the height of the antenna, and $r_f$ is the radius of the first Fresnel zone ellipsoid which is given by
\[
r_f = \sqrt{\frac{\lambda d_{obs}(d-d_{obs})}{d}},
\]
with $\lambda$ denoting the wavelength. We use the appropriate $\lambda$ for the proposed standard for VANET communication (DSRC), which operates in the 5.9~GHz frequency band. 
In our studies, we assume that the antennas are located on top of the vehicles in the middle of the roof (which was experimentally shown to be the overall optimum placement of the antenna \cite{kaul07}), and we set the $h_a$ to 10 cm.
As a general rule commonly used in literature, LOS is considered to be unobstructed if intermediate vehicles obstruct the first Fresnel ellipsoid by less than 40\% \cite[Chap. 3]{rappaport96}.
Furthermore, for $N_o$ vehicles as potential obstacles between the Tx and Rx, we get (see Fig.~\ref{explanation}c)

\begin{equation}
P(LOS|h_i,h_j) = \prod_{k=1}^{N_o} \left[1 - Q\left(\frac{h_k - \mu_k}{\sigma_k}\right)\right],
\end{equation}
where $h_k$ is the effective height of the straight line that connects Tx and Rx at the location of the $k$-th obstacle considering the first Fresnel ellipsoid, $\mu_k$ is the mean height of the $k$-th obstacle, and $\sigma_k$ is the standard deviation of the height of the $k$-th obstacle.

Averaging over the transmitter and receiver antenna heights with respect to the road, we obtain the unconditional $P(LOS)_{ij}$
\begin{equation}
P(LOS)_{ij} = \int\int P(LOS|h_i,h_j) p(h_i) p(h_j) dh_i dh_j,
\end{equation}
where $p(h_i)$ and $p(h_j)$ are the probability density functions for the transmitter and receiver antenna heights with respect to the road, respectively.

The average probability of LOS for a given vehicle $i$, $P(LOS)_i$, and all its $N_i$ neighbors is defined as
\begin{equation}
P(LOS)_i= \frac{1}{N_i}\sum_{j=1}^{N_i}P(LOS)_{ij} 
\label{eq:per_node_plos}
\end{equation}
To determine the system-wide ratio of LOS paths blocked by other vehicles, we average $P(LOS)_i$ over all $N_v$ vehicles in the system, yielding %
\begin{equation}
\overline{P(LOS)}= \frac{1}{N_v}\sum_{i=1}^{N_v}P(LOS)_i.
\label{eq:average_plos}
\end{equation}

Furthermore, we analyze the behavior of the probability of LOS for a given vehicle $i$ over time. Let us denote the $i$-th vehicle probability of LOS at a given time $t$ as $P(LOS)^t_{i}$. 
We define the change in the probability of LOS for the $i$-th vehicle over two snapshots at times $t_1$ and $t_2$ as
\begin{equation}
\Delta P(LOS)_{i} = |P(LOS)^{t_2}_{i} - P(LOS)^{t_1}_{i}|,
\label{eq:delta_plos}
\end{equation}
where $P(LOS)^{t_1}_{i}$ and $P(LOS)^{t_2}_{i}$ are obtained using \eqref{eq:per_node_plos}.

It is important to note that equations \eqref{eq:first} to \eqref{eq:delta_plos} depend on the distance between the node $i$ and the node $j$ (i.e., transmitter and receiver) in a \emph{deterministic manner}. More specifically, the snapshot obtained from aerial photography provides the exact distance $d$ (Fig.~\ref{explanation}c) between the nodes $i$ and $j$. While in our study we used aerial photography to get this information, any VANET simulator would also provide the exact location of vehicles based on the assumed mobility model (e.g., car-following \cite{rothery92}, cellular automata \cite{tonguz09_2,shih11}, etc.), hence the distance $d$ between the nodes $i$ and $j$ would still be available. This also explains why the proposed model is independent of the simulator used, since it can be incorporated into any VANET simulator, regardless of the underlying mobility model, as long as the locations of the vehicles are available. Furthermore, even though we used the highway environment for testing, the proposed model can be used for evaluating the impact of obstructing vehicles on any type of road, irrespective of the shape of the road (e.g., single or multiple lanes, straight or curvy) or location (e.g., highway, suburban, or urban\footnote{To precisely quantify the impact of obstructing vehicles in complex urban environments, the interplay between the vehicle-induced obstruction and the obstruction caused by other objects (e.g., buildings, overpasses, etc.) needs to be determined. Chapter~\ref{ch:completeModel} focuses on such environments.}). 

\subsection{The Impact of Vehicles on Signal Propagation}

The attenuation on a radio link increases if one or more vehicles intersect the ellipsoid corresponding to 60\% of the radius of the first Fresnel zone, independent of their positions on the Tx-Rx link (Fig.~\ref{explanation}c). This increase in attenuation is due to the diffraction of the electromagnetic waves. The additional attenuation due to diffraction depends on a variety of factors: the obstruction level, the carrier frequency, the electrical characteristics, the shape of the obstacles, and the amount of obstructions in the path between transmitter and receiver.
To model vehicles obstructing the LOS, %
we use the knife-edge attenuation model. It is reasonable to expect that more than one vehicle can be located between transmitter (Tx) and receiver (Rx). Thus, we employ the multiple knife-edge model described in ITU-R recommendation \cite{itu07}. %

\subsubsection{Single Knife-Edge}
The simplest obstacle model is the knife-edge model, which is a reference case for more complex obstacle models (e.g., cylinder and convex obstacles). Since the frequency of DSRC radios is 5.9~GHz, the knife-edge model theoretically presents an adequate approximation for the obstacles at hand (vehicles), as the prerequisite for the applicability of the model, namely a significantly smaller wavelength than the size of the obstacles \cite{itu07}, is fulfilled (the wavelength of the DSRC is approximately 5~cm, which is significantly smaller than the size of the vehicles).

The obstacle is seen as a semi-infinite perfectly absorbing plane that is placed perpendicular to the radio link between the Tx and Rx. Based on the Huygens principle, the electric field is the sum of Huygens sources located in the plane above the obstruction and can be computed by solving the Fresnel integrals \cite{00parsons}. A good approximation for the additional attenuation (in dB) due to a single knife-edge obstacle $A_{sk}$ can be obtained using the following equation \cite{itu07}:

\begin{equation}
A_{sk} = \left\{ \begin{array}{cl}
6.9 + 20\log_{10} \left[\sqrt{(v-0.1)^2+1} + v - 0.1\right]; \\ \textrm{ for $v > -0.7$} \\
0;  \textrm{ otherwise,}
\end{array} \right.
\label{eq:singleknifeedge}
\end{equation}

where $v = \sqrt{2}H/r_f$, $H$ is the difference between the height of the obstacle and the height of the straight line that connects Tx and Rx, and $r_f$ is the Fresnel ellipsoid radius. 

\subsubsection{Multiple Knife-Edge}

The extension of the single knife-edge obstacle case to the multiple knife-edge is not immediate. All of the existing methods in the literature are empirical and the results vary from optimistic to pessimistic approximations \cite{00parsons}. 
The Epstein-Petterson method \cite{epstein53} presents a more optimistic view, whereas the Deygout \cite{deygout66} and Giovanelli \cite{giovaneli84} are more pessimistic approximations of the real world. Usually, the pessimistic methods are employed when it is desirable to guarantee that the system will be functional with very high probability. On the other hand, the more optimistic methods are used when analyzing the effect of interfering sources in the communications between transmitter and receiver. To calculate the additional attenuation due to vehicles, we employ the ITU-R method \cite{itu07}, which can be seen as a modified version of the Epstein-Patterson method, where correcting factors are added to the attenuation in order to better approximate reality.

\section{Model Requirements} \label{sec:Model} 

The model proposed in the previous section is aimed at evaluating the impact of vehicles as obstacles using %
geometry concepts and relies heavily on realistic modeling of the physical environment. In order to employ the proposed model accurately, realistic modeling of the following physical properties is necessary: determining the exact position of vehicles and the inter-vehicle spacing; determining the speed of vehicles; and determining the vehicle dimensions.

\begin{table} 
	\centering
		\caption{Analyzed highway datasets}
		\begin{tabular}{|c c c c c|}
\hline \textbf{Dataset} & \textbf{Size} & \textbf{\# vehicles} & \textbf{\# tall vehicles} & \textbf{Veh. density} \\
		 	\hline 
\hline A28 & 12.5~km & 404 & 58 (14.36\%) & 32.3~veh/km\\ 
\hline A3 & 7.5~km & 55 & 10 (18.18\%) & 7.3~veh/km\\ 
\hline
\end{tabular}
	\label{datasets}
\end{table}
 
\subsection{Determining the exact position of vehicles and the inter-vehicle spacing}

The position and the speed of vehicles can easily be obtained from any currently available VANET mobility model. However, in order to test our methodology with the most realistic parameters available, %
we used \emph{aerial photography}. This technique is used by the traffic engineering community as an alternative to ground-based traffic monitoring \cite{mccasland65}, and was recently applied to VANET connectivity analysis \cite{ferreira09}. It is well suited to characterize the physical interdependencies of signal propagation and vehicle location, because %
it gives the exact position of each vehicle.
We analyzed two distinct datasets, namely two Portuguese highways near the city of Porto, A28 and A3, both with four lanes (two per direction). 
Detailed parameters for the two datasets are presented in Table~\ref{datasets}. 
For an extensive description of the method used for data collection and analysis, we refer the reader to \cite{ferreira09}. 

\subsection{Determining the speed of vehicles}

For the observed datasets, besides the exact location of vehicles and the inter-vehicle distances, stereoscopic imagery was once again used to determine the speed and heading of vehicles. Since the successive photographs were taken with a fixed time interval (5 seconds), by marking the vehicles on successive photographs we were able to measure the distance the vehicle traversed, and thus infer the speed and heading of the vehicle. The measured speed and inter-vehicle spacing is used to analyze the behavior of vehicles as obstacles while they are moving. %

\begin{figure*}
\begin{center}
	\subfigure[\footnotesize{Inter-vehicle spacing on A28}]{\fbox{\psfig{figure=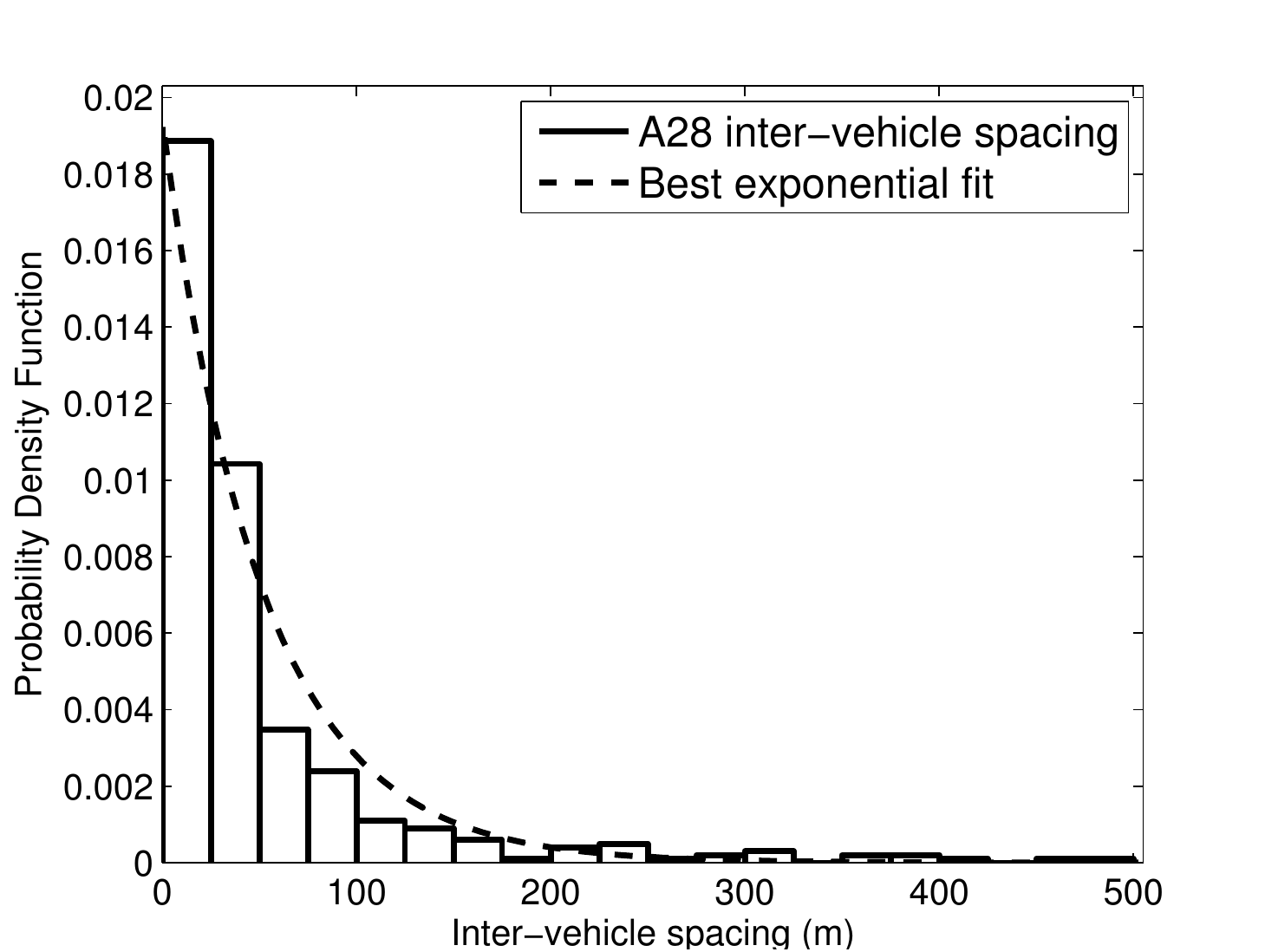,width=0.45\textwidth}}}
	\subfigure[\footnotesize{Inter-vehicle spacing on A3}]{\fbox{\psfig{figure=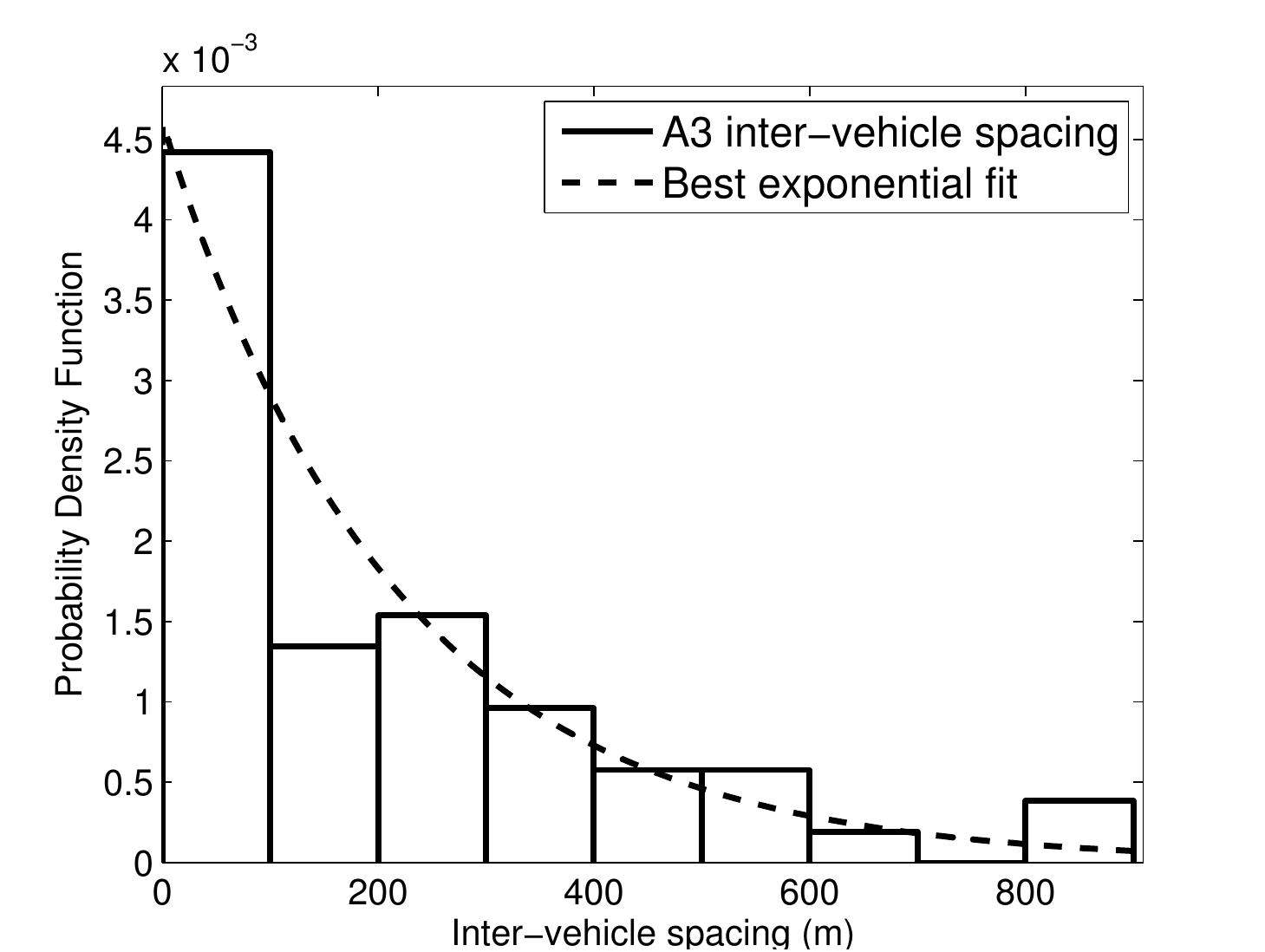,width=0.45\textwidth}}}
	\subfigure[Speed distribution on A28]{\fbox{\psfig{figure=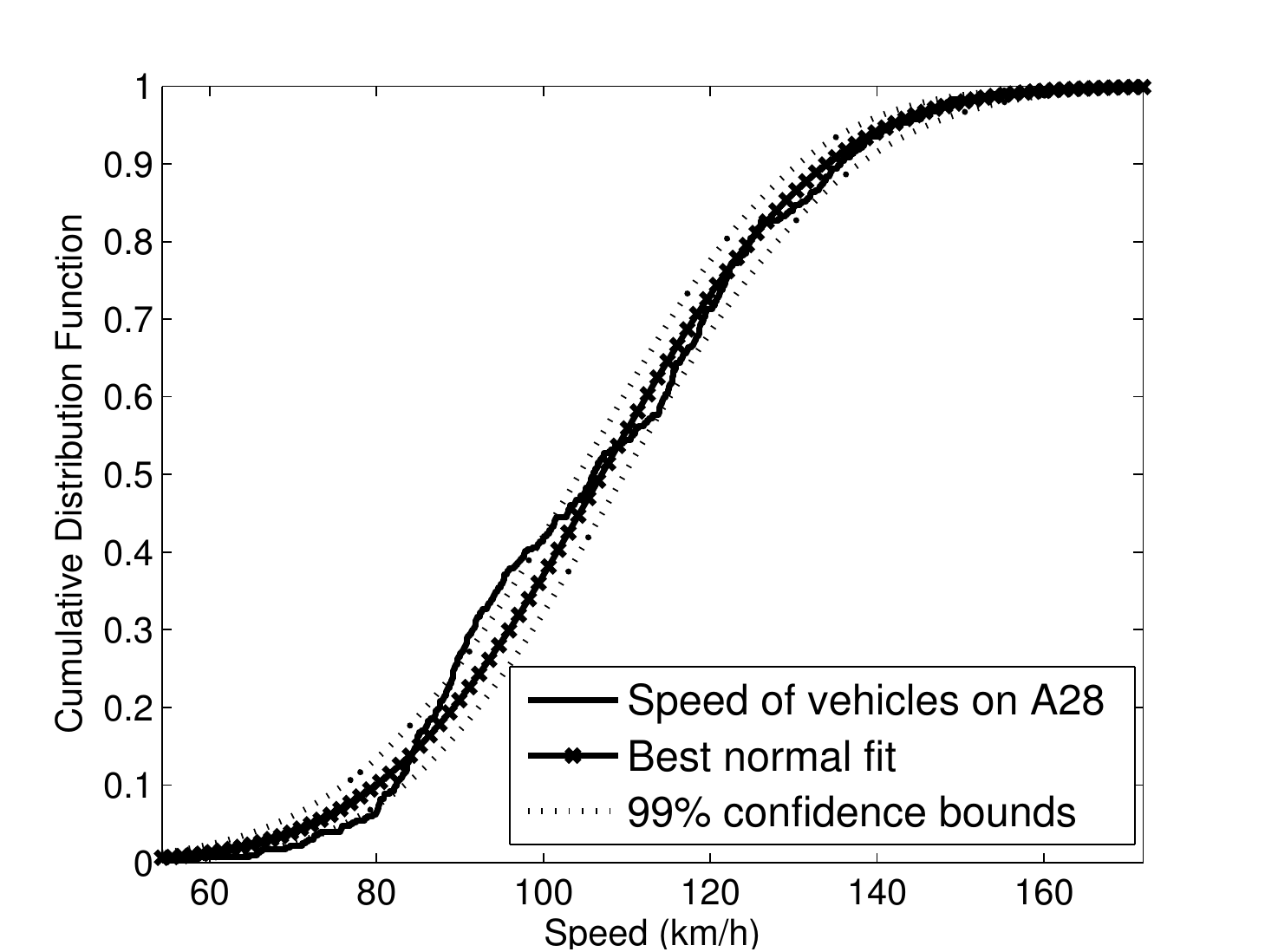,width=0.45\textwidth}}}
	\subfigure[Speed distribution on A3]{\fbox{\psfig{figure=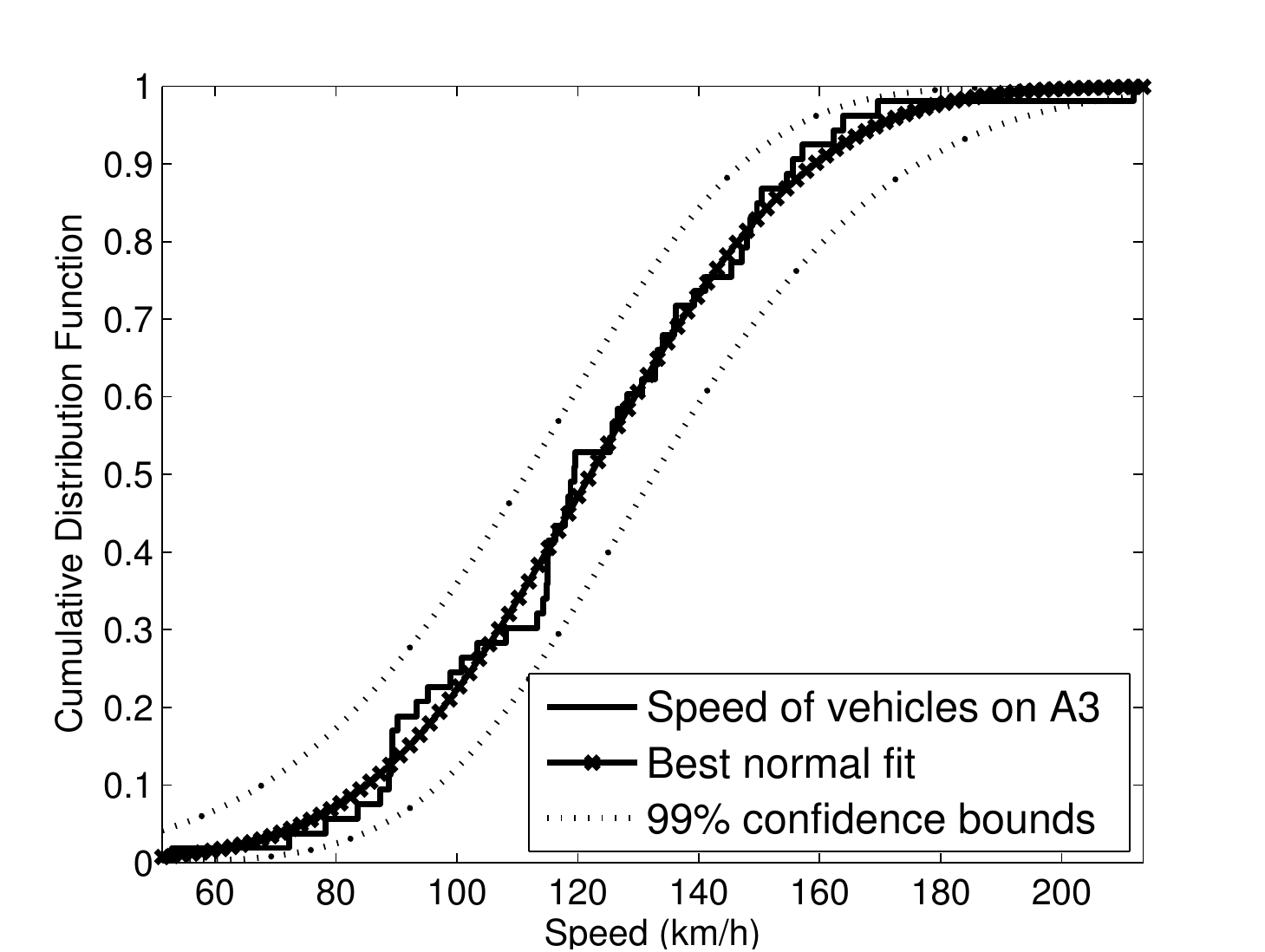,width=0.45\textwidth}}}
  \caption{Speed and inter-vehicle spacing distribution on highways.}
  \label{speedSpacing}
\end{center}
\end{figure*}

Figures \ref{speedSpacing}a and \ref{speedSpacing}b show the distribution of inter-vehicle spacing (defined as the distance between a vehicle and its closest neighbor) for the A28 and A3, respectively.
The distribution of inter-vehicle spacing for %
both cases can be well fitted with an \emph{exponential probability distribution}. This agrees with the empirical measurements made on the I-80 interstate in California reported in \cite{wisit07}.  
Figures \ref{speedSpacing}c and \ref{speedSpacing}d show the speed distribution for the A28 and A3, respectively. The speed distribution on both highways is well approximated by a normal probability distribution.
Table~\ref{FitParameters} shows the parameters of best fits for inter-vehicle distances and speeds.

\begin{table}
	\centering
		\caption{Parameters of the Best Fit Distributions for Vehicle Speed and Inter-vehicle spacing}
		\begin{tabular}{|c c c|}
\hline \multicolumn{3}{|c|}{\textbf{Data for A28}} \\			
\hline \textbf{Parameter} & \textbf{Estimate} & \textbf{Std. Error}\\
		 	\hline 
\hline \textbf{Speed: normal fit} & &\\
\hline mean (km/h)	& 106.98 & 1.05\\ 
\hline std. deviation (km/h)	& 21.09 & 0.74\\ \hline
\hline \textbf{Inter-vehicle spacing : exponential fit} & &\\
\hline mean (m)	& 51.58 & 2.57\\ \hline
\hline \multicolumn{3}{|c|}{\textbf{Data for A3}} \\
\hline \textbf{Parameter} & \textbf{Estimate} & \textbf{Std. Error}\\
		 	\hline 
\hline \textbf{Speed: normal fit} & &\\
\hline mean (km/h)	& 122.11 & 3.97\\ 
\hline std. deviation (km/h) & 28.95 & 2.85\\ \hline
\hline \textbf{Inter-vehicle spacing : exponential fit} & &\\
\hline mean (m)	& 215.78 & 29.92\\ \hline
\end{tabular}
	\label{FitParameters}
\end{table}

\subsection{Determining the vehicle dimensions}

In order to implement the model in VANET simulators, apart from the information available in the current VANET simulators, very few additional pieces of information are necessary. Specifically, the required information pertains to the physical dimensions of the vehicles. 
From the photographs, we were able to obtain the length of each vehicle accurately, however the width and height could not be determined with satisfactory accuracy 
due to resolution constraints and vehicle mobility. 
To assign proper widths and heights to vehicles, %
we use the data made available by the Automotive Association of Portugal \cite{acap}, which issued an official report about all vehicles currently in circulation in Portugal. From the report we extracted the eighteen most popular personal vehicle brands which comprise 92\% of all personal vehicles circulating on Portuguese roads, and consulted an online database of vehicle dimensions \cite{carfolio} to arrive at the distribution of height and width required for our analysis. The dimensions of the most popular personal vehicles showed that both the vehicle widths and heights can be modeled as a normal random variable. Detailed parameters for the fitting process for both personal and tall vehicles are presented in Table~\ref{FitParametersWH}. For both width and height of personal vehicles, the standard error for the fitting process remained below 0.33\% for both the mean and the standard deviation. The data regarding the specific types of tall vehicles (e.g., trucks, vans, or buses) currently in circulation was not available. Consequently, the precise dimension distributions of the most representative models could not be obtained. For this reason, we infer tall vehicle height and width values from the data available on manufacturers' websites, which can serve as rough dimension guidelines that show significantly different height and width in comparison to personal vehicles.

\begin{table}
	\centering
		\caption{Parameters of the Best Fit Distributions for Vehicle Width and Height}
		\begin{tabular}{|c c|}
\hline \multicolumn{2}{|c|}{\textbf{Personal vehicles}} \\			
\hline \textbf{Parameter} & \textbf{Estimate}\\
		 	\hline 
\hline \textbf{Width: normal fit} & \\
\hline mean (cm)	& 175\\ 
\hline std. deviation (cm)	& 8.3\\ \hline
\hline \textbf{Height: normal fit} & \\
\hline mean (cm)	& 150\\ 
\hline std. deviation (cm)	& 8.4\\ \hline
\hline \multicolumn{2}{|c|}{\textbf{Tall vehicles}} \\			
\hline \textbf{Parameter} & \textbf{Estimate}\\
		 	\hline 
\hline \textbf{Width: constant} & \\
\hline mean (cm)	& 250\\ 
\hline \textbf{Height: normal fit} & \\
\hline mean (cm)	& 335\\ 
\hline std. deviation (cm)	& 8.4\\ \hline
\end{tabular}
	\label{FitParametersWH}
\end{table}

\section{Results} \label{sec:Results}
We implemented the model described in previous sections %
in Matlab. In this section we present the results based on testing the model using the A3 and A28 datasets. We also present the results of the empirical measurements that we performed in order to characterize the impact of the obstructing vehicles on the received signal strength. We emphasize that the developed model is not dependent on these datasets, but can be used in any environment by applying the analysis presented in Section~\ref{sec:Evaluating}. Furthermore, the observations pertaining to the inter-vehicle and speed distributions on A3 and A28 (Fig.~\ref{speedSpacing}) are used only to characterize the behavior of the highway environment over time. We do not use these distributions in our model; rather, we use actual positions of the vehicles. Since the model developed in Section~\ref{sec:Evaluating} is intended to be utilized by VANET simulators, the positions of the vehicles can easily be 
obtained through the employed vehicular mobility model.

We first give evidence that vehicles as obstacles have a significant impact on LOS communication in both sparse (A3) and more dense (A28) networks. 
Next, we analyze the microscopic probability of LOS to determine the variation of the LOS conditions over time for a given vehicle.
Then, we used the speed and heading information %
to characterize both the microscopic and macroscopic behavior of the probability of LOS on highways over time in order to determine how often the proposed model needs to be recalculated in the simulators, and to infer the stationarity of the system-wide probability of LOS.
Using the employed multiple knife-edge model, we present the results pertaining to the decrease of the received power and packet loss for DSRC due to vehicles. %
Finally, we corroborate our findings on the impact of the obstructing vehicles and discuss the appropriateness of the knife-edge model by performing empirical measurements of the received signal strength in LOS and non-LOS conditions.

\begin{table}
	\centering
		\caption{$\overline{P(LOS)}$ for A3 and A28}
\begin{tabular}{|c c c c|}\hline
		 \multicolumn{4}{|c|}{\bf Highways} \\ \hline\hline%
		 & \multicolumn{3}{c|}{Transmission Range (m)} \\ %
		 Highway & 100 & 250 & 500 \\ \hline
		A3 $\overline{P(LOS)}$& 0.8445 & 0.6839 & 0.6597 \\ \hline
		A28 $\overline{P(LOS)}$& 0.8213 & 0.6605 & 0.6149  \\ \hline 
		\end{tabular} 
\label{tab:PLOS_A3A28}
\end{table}

\subsection{Probability of Line of Sight}
\subsubsection{Macroscopic probability of line of sight}
\begin{figure}
\begin{center}
	\fbox{\psfig{figure=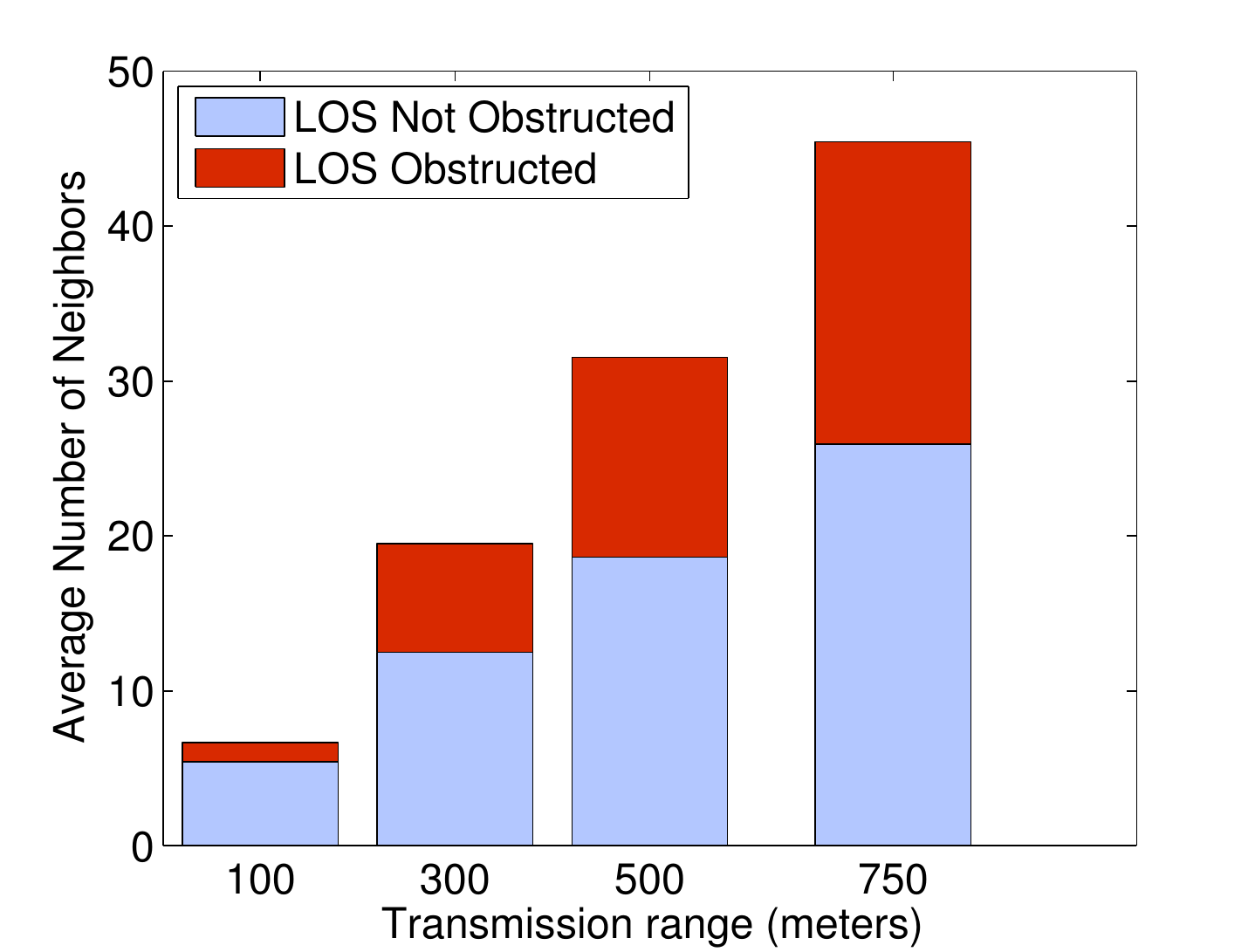,width=0.45\textwidth}}
  \caption{ Average number of neighbors with unobstructed and obstructed LOS on A28 highway.}
  \label{neighbors}
\end{center}
\end{figure}

Table~\ref{tab:PLOS_A3A28} %
presents the values of $\overline{P(LOS)}$ with respect to the observed range on highways. %
The highway results show that even for the sparsely populated A3 highway the impact of vehicles on $\overline{P(LOS)}$ is significant. This can be explained by the exponential inter-vehicle spacing, which makes it more probable that the vehicles are located close to each other, thus increasing the probability of having an obstructed link between two vehicles. 
For both highways, %
it is clear that the impact of other vehicles as obstacles can not be neglected even for vehicles that are relatively close to each other (for the observed range of 100~m, $\overline{P(LOS)}$ is under 85\% for both highways, which means that there is a non-negligible 15\% probability that the vehicles will not have LOS while communicating). 
To confirm these results, Fig.~\ref{neighbors} shows the average number of neighbors with obstructed and unobstructed LOS for the A28 highway. 
The increase of obstructed vehicles in both absolute and relative sense is evident. 

\subsubsection{Microscopic probability of line of sight}
In order to analyze the variation of the probability of LOS for a vehicle and its neighbors over time, 
we observe the $\Delta P(LOS)_i$ (as defined in equation \eqref{eq:delta_plos}) on A28 highway for the maximum communication range of 750~m. 
Table~\ref{tab:deltaPLOS_A28} shows the $\Delta P(LOS)_i$. %
The variation of probability of LOS is moderate for periods of seconds (even for the largest offset of 2 seconds, only 15\% of the nodes have the $\Delta P(LOS)_i$ greater than 20\%). This result suggests that the LOS conditions between a vehicle and its neighbors will remain largely unchanged for a period of seconds. Therefore, a simulation time-step of the order of seconds can be used for calculations of the impact of vehicles as obstacles. From a simulation execution standpoint, the time-step of the order of seconds is quite a long time when compared with the rate of message transmission, measured in milliseconds; this enables a more efficient and scalable design and modeling of vehicles as obstacles on a microscopic, per-vehicle level. With the proper implementation of the LOS intersection model discussed in Sections~\ref{sec:Evaluating}, %
the modeling of vehicles as obstacles should not induce a large overhead in the simulation execution time.

\begin{table}
	\centering
		\caption{Variation of $P(LOS)_i$ over time for the observed range of 750~m on A28.}		
	\begin{tabular}{|c| c c c c|}\hline
		 & \multicolumn{4}{c|}{$\Delta P(LOS)_i$ in \%} \\ %
	  Time offset	& $<$ 5\%& 5-10\% & 10-20\% & $>$20\% \\ \hline
	1ms  & 100\%& 0\% &0\% & 0\% \\ \hline
	10ms & 99\%& 1\% &0\%& 0\% \\   \hline
	100ms& 82\%& 15\% &3\% & 0\% \\   \hline 
	1s 	 &35\%& 33\% & 22\%& 10\% \\ \hline
	2s 	 &31\%& 25\% & 29\% & 15\% \\ \hline
		\end{tabular} 	
\label{tab:deltaPLOS_A28}
\end{table}

\subsubsection{Stochastic properties of line of sight in mobile vehicular network}
Figs.~\ref{speedSpacing}a and \ref{speedSpacing}b show that a Poisson process with parameter $\alpha$ can be used to describe the distribution of vehicles on highways at a given time $t$.  It is reasonable to assume that, for a vehicular traffic in the free-flow phase, the rate of change of the parameter $\alpha$ over time is quite slow, thus the Poisson process can be considered homogeneous %
for a certain amount of time. 
This allows us to utilize one of the key properties of homogeneous Poisson processes, namely stationary increments, which says that if two road segments %
are of the same length, the probability distribution function of the number of vehicles over those segments is equal \cite{papoulis84}. Therefore, we can conclude that for a certain period of time, %
the probability distribution function of the number of vehicles on two road segments will only depend on the size of the segments. Based on the homogeneity assumption, applying this property on the same segment of the road but at different times results in identical probability distribution for the number of vehicles. Therefore, it is expected that the $\overline{P(LOS)}$ over the observed road segment will not change over time, as long as the arrival rate $\alpha$ remains constant.

In order to confirm these results, %
 we performed tests using two snapshots of A28 highway %
taken on the same road interval %
5 seconds one after another. 
By inferring the speed and heading of the vehicles from the snapshots, it was possible to accurately interpolate the positions of 
the vehicles for 1~ms, 10~ms, 100~ms, 1~s, and 2~s offsets from the first snapshot.

\begin{figure}
	\centering
	\fbox{\psfig{figure=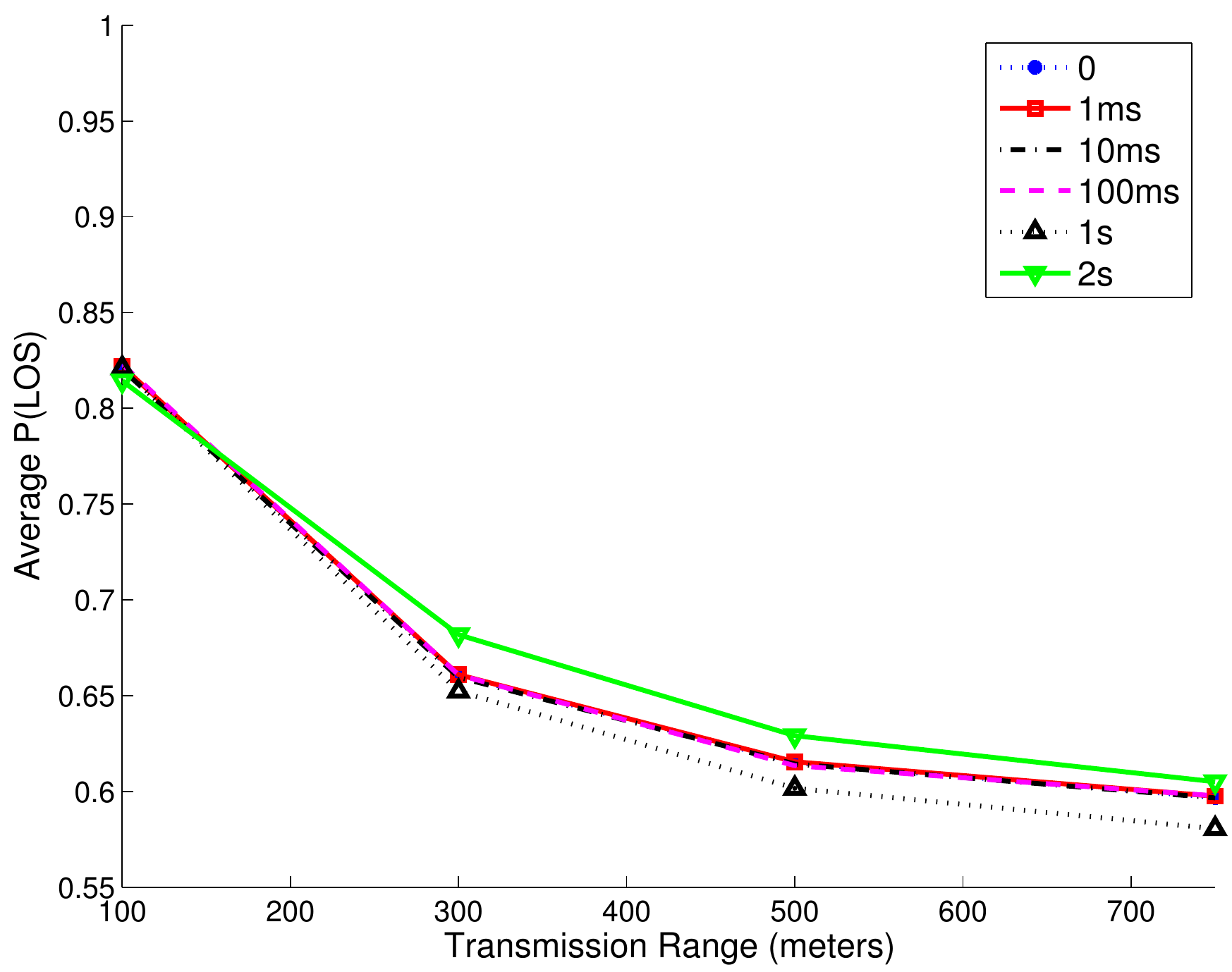,width=0.45\textwidth}} %
	\caption{ $\overline{P(LOS)}$ vs. observed range for different time offsets}%
  \label{plosRange}
\end{figure}

The obtained results showed that the average inter-vehicle spacing remains invariant for the observed time offsets, thus confirming the first-order stationarity %
of the underlying Poisson process. 
Similarly, Fig.~\ref{plosRange} shows that for various communication radii (100 - 750~m), the $\overline{P(LOS)}$ does not change for the observed time offsets. Therefore, we can conclude that $\overline{P(LOS)}$ remains constant on the observed road interval as long as the arrival rate of the generating Poisson process remains constant. 
Thus, the presented $\overline{P(LOS)}$ results hold for both the instantaneous V2V communication as well as for V2V communication over time (i.e., for moving vehicular network).

\subsection{Received Power}

Based on the methodology developed in Section~\ref{sec:Evaluating}, %
we utilize the multiple knife-edge model to calculate the additional attenuation due to vehicles. We use the obtained attenuation to calculate the received signal power for the DSRC. We employed the knife-edge model for its simplicity and the fact that it is well studied and often used in the literature. 
However, we point out that the LOS analysis and the methodology developed in Section~\ref{sec:Evaluating} can be used in conjunction with any channel model that relies on the distinction between the LOS and NLOS communication (e.g., \cite{zang05} or \cite{wang04}).

\begin{figure}
	\centering
	\fbox{\psfig{figure=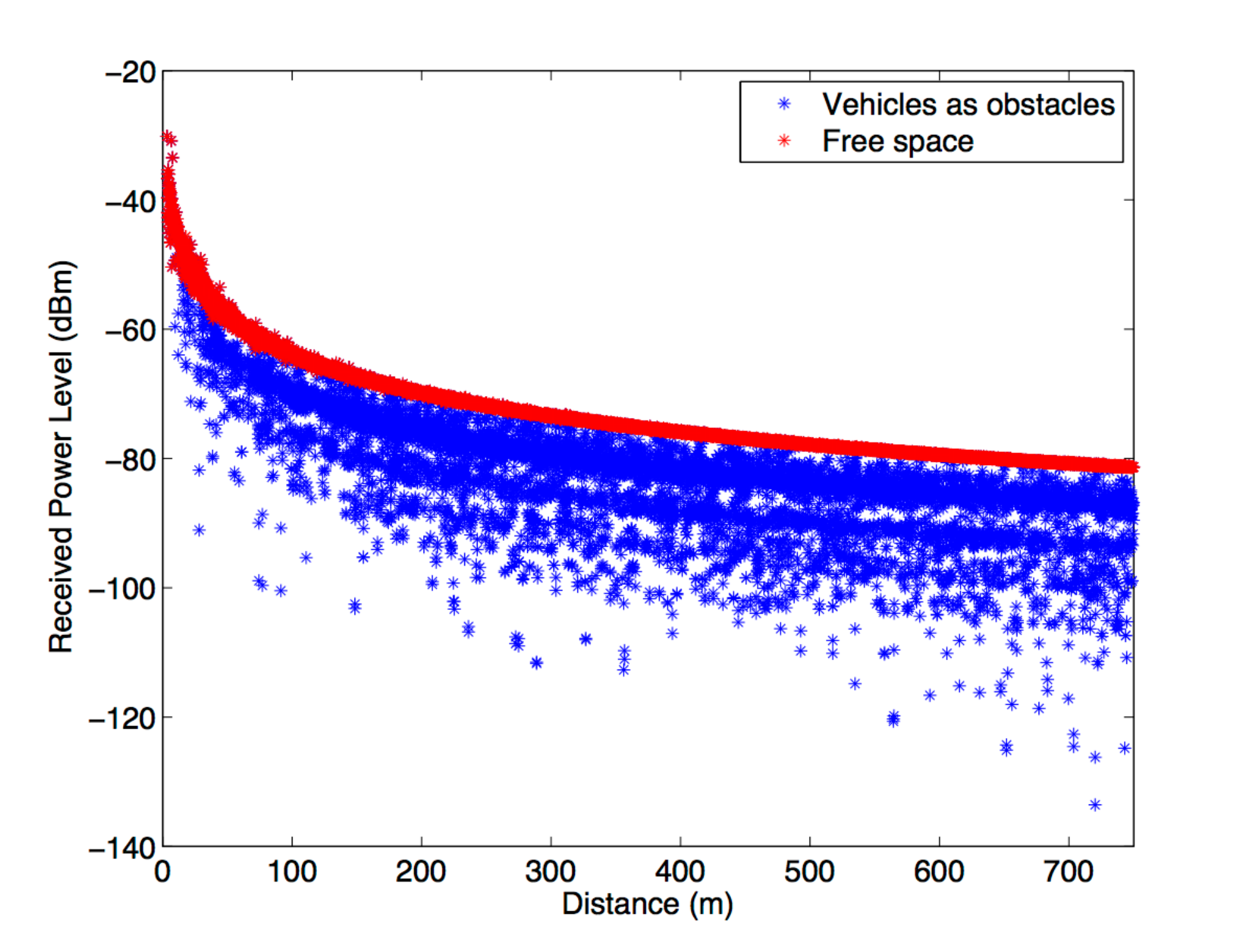,width=0.45\textwidth}} %
	\caption{ The impact of vehicles as obstacles on the received signal power on highway A28.}%
  \label{power}
\end{figure}

For the A28 highway and the observed range of 750~m, with the transmit power set to 18~dBm, 3~dBi antenna gain for both transmitters and receivers, at the 5.9 GHz frequency band, the results for the free space path loss model \cite{goldsmith05} (i.e., not including vehicles as obstacles) and our model that accounts for vehicles as obstacles are shown in Fig.~\ref{power}. The average additional attenuation due to vehicles was 9.2~dB for the observed highway. 

\begin{table}
	\centering
		\caption{Requirements for DSRC Receiver Performance}		
	\begin{tabular}{|c c c|}\hline
	  Data Rate (Mb/s)& Modulation & Minimum sensitivity (dBm)  \\ \hline
	  3   & BPSK   & $-85$ \\\hline
	  4.5 & BPSK   & $-84$ \\\hline
	  6   & QPSK   & $-82$\\\hline
	  9   & QPSK   & $-80$ \\\hline
	  12  & QAM-16 & $-77$ \\\hline
	  18  & QAM-16 & $-70$ \\\hline
	  24  & QAM-64 & $-69$ \\\hline
	  27  & QAM-64 & $-67$ \\\hline
		\end{tabular} 	
\label{tab:DSRCrates}
\end{table}

\begin{figure}
	\centering
	\fbox{\psfig{figure=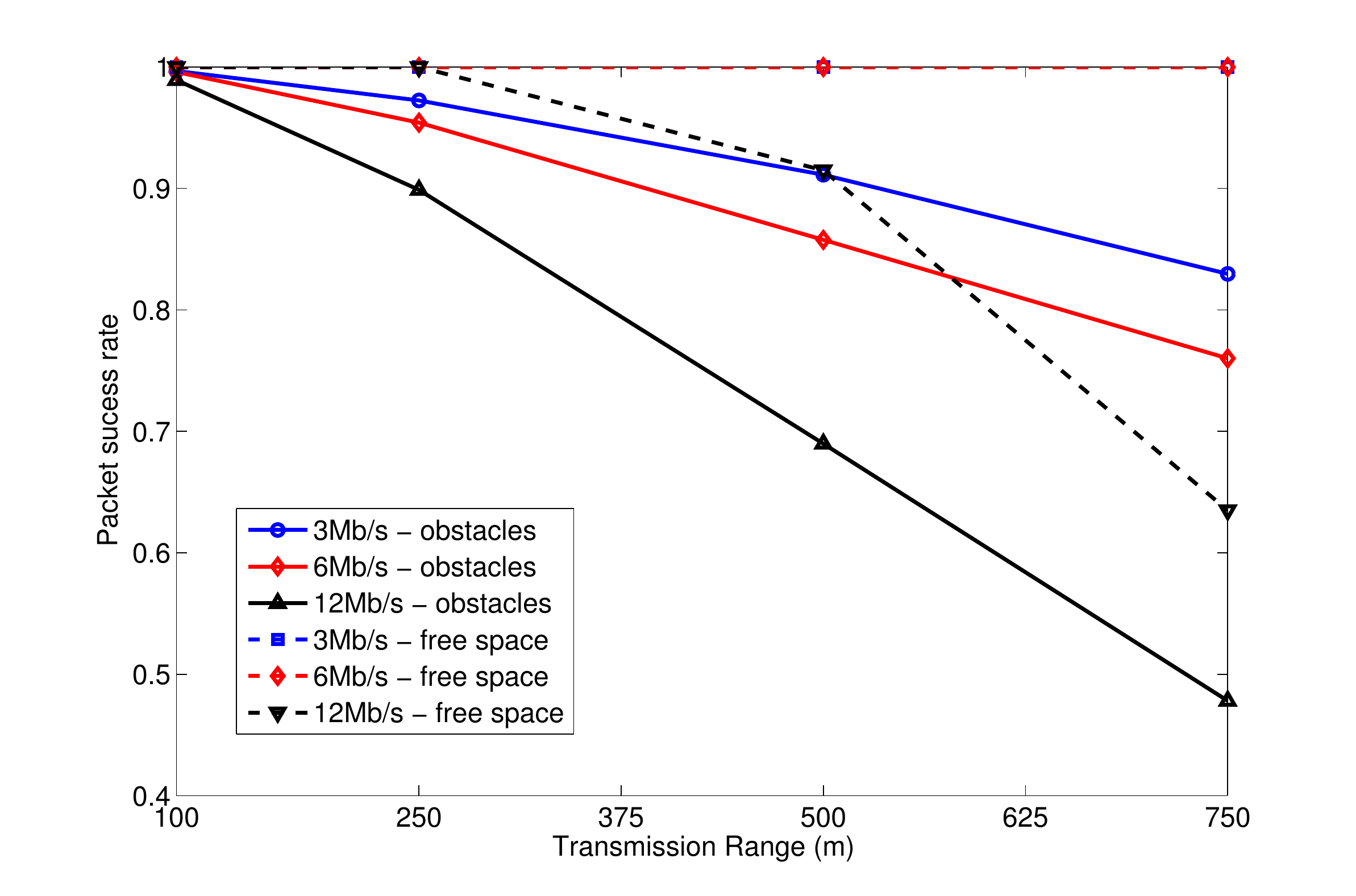,width=0.45\textwidth}} %
	\caption[The impact of vehicles as obstacles on A28 highway]{ The impact of vehicles as obstacles on packet success rate for various DSRC data rates on A28 highway.}%
  \label{per}
\end{figure}

Using the minimum sensitivity thresholds as defined in the DSRC standard (see Table~\ref{tab:DSRCrates}) \cite{astm03}, we calculate the packet success rate (PSR, defined as the ratio of received messages to sent messages) as follows. We analyze all of the communicating pairs within an observed range, and calculate the received signal power for each message. Based on the sensitivity thresholds presented in Table~\ref{tab:DSRCrates}, we determine whether a message is successfully received. For the A28 highway, Fig.~\ref{per} shows the PSR difference between the free space path loss and the implemented model with vehicles as obstacles for rates of 3, 6, and 12 Mb/s.
The results show that the difference is significant, as the percentage of lost packets can be up to 25\% higher when vehicles are accounted for. 

These results show that not only do the vehicles significantly decrease the received signal power, but the resulting received power is highly variable even for relatively short distances between the communicating vehicles, thus calling for a microscopic, per-vehicle analysis of the impact of obstructing vehicles. Models that try to average the additional attenuation due to vehicles could fail to describe the complexity of the environment, thus yielding unrealistic results. Furthermore, the results show that the distance itself can not be solely used for determining the received power, since even the vehicles close by can have a number of other vehicles obstructing the communication path and therefore the received signal power becomes worse than for vehicles %
further apart that do not have obstructing vehicles between them.  

\subsection{Experimental Evaluation}

\begin{figure}
	\centering
	\fbox{\psfig{figure=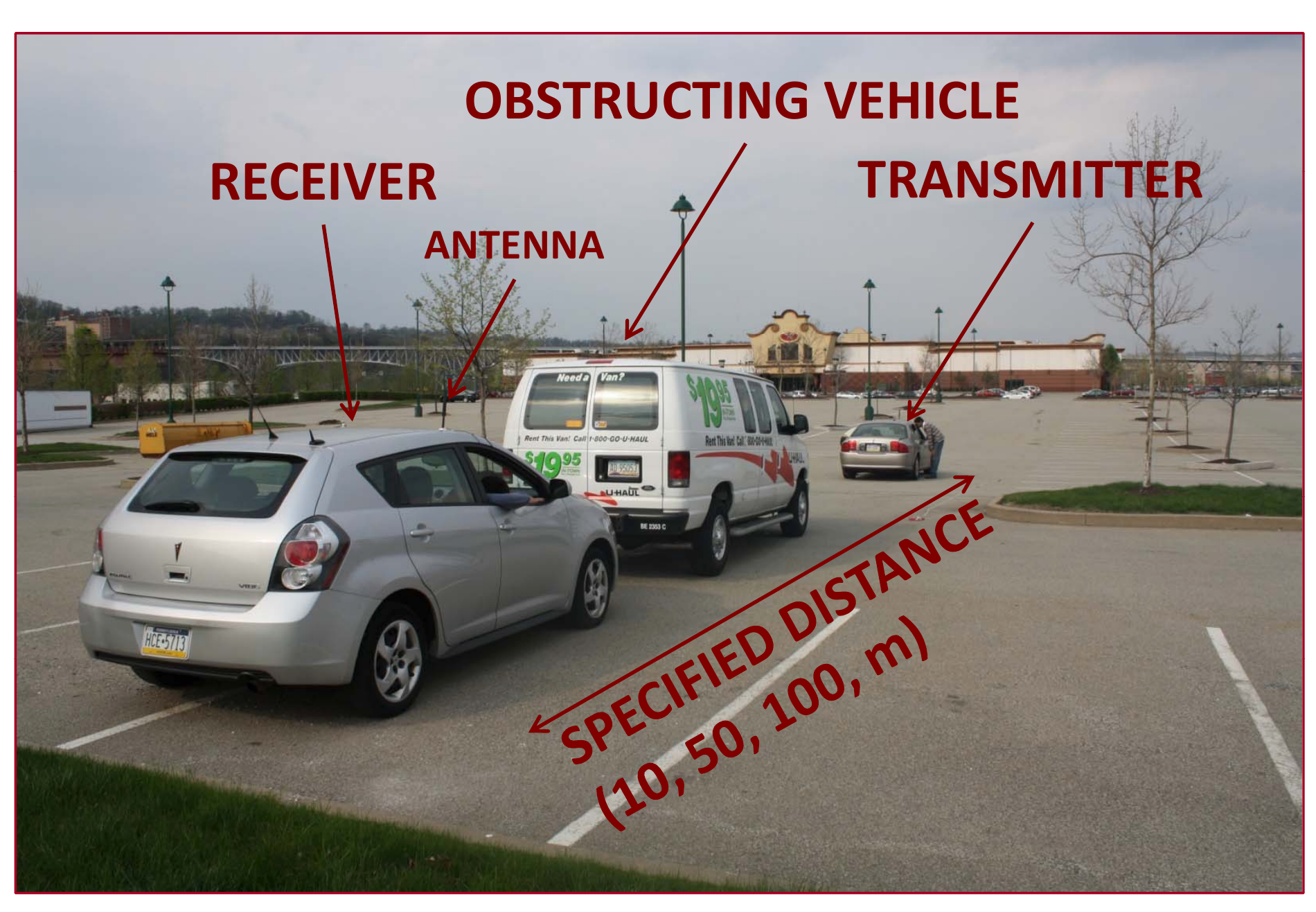,width=0.45\textwidth}} %
	\caption{ Experiment setup.}%
  \label{experiment}
\end{figure}

\begin{table}
	\centering
		\caption{Dimensions of Vehicles} 
\begin{tabular}{|c c c c|} \hline
		 & \multicolumn{3}{c|}{\bf Dimensions (m)} \\ %
		 \bf Vehicle & \bf Height & \bf Width & \bf Length \\ \hline \hline
		2002 Lincoln LS (TX) & 1.453 & 1.859 & 4.925\\ \hline
		2009 Pontiac Vibe (RX) & 1.547 & 1.763 & 4.371\\ \hline 
		2010 Ford E-250 (Obstruction) & 2.085 & 2.029 & 5.504\\ \hline
		\end{tabular} 
\label{tab:dimensions}
\end{table}

We performed measurements in order to validate the model and to further confirm the validity of the importance of modeling the effect of vehicles on the received signal strength.  
To isolate the effect of the obstructing vehicles, we aimed at setting up a controlled environment without other obstructions and with minimum impact of other variables (e.g., other moving objects, electromagnetic radiation, etc). For this reason, we performed experiments in an empty parking lot in Pittsburgh, PA (Fig.~\ref{experiment}). We analyzed the received signal strength for the no obstruction, LOS case, and the non-LOS case where we introduced an obstructing vehicle (the van shown in Fig.~\ref{experiment}) between the transmitter (Tx) and the receiver (Rx) vehicles. The received signal strength was measured for the distances of 10, 50, and 100~m between the Tx and the Rx. In case of the non-LOS experiments, the obstructing van was placed in the middle between the Tx and the Rx. 
We performed experiments at two frequency bands: 2.4~GHz (used by the majority of commercial WiFi devices) and 5.9~GHz (the band at which spectrum has been allocated for automotive use worldwide \cite{dsrc09}).
For 2.4~GHz experiments, we equipped the Tx and Rx vehicles with laptops that had Atheros 802.11b/g wireless cards installed and we used 3 dBi gain omnidirectional antennas. 
For 5.9~GHz experiments, we equipped the Tx and Rx vehicles with NEC Linkbird-MX devices, described in section~\ref{subsec:NetworkConfiguration}, whereas hardware configuration parameters are shown in Table~\ref{tab:hw-config}.
The data rate for 2.4~GHz experiments was 1~Mb/s, with 10 messages (140~B in size) sent per second using the ping command, whereas for 5.9~GHz experiments the data rate was 6~Mb/s (the lowest data rate in 802.11p for 20~MHz channel width) with 10 beacons \cite{festag08} (36~B in size) sent per second. Each measurement was performed for at least 120 seconds, thus resulting in a minimum of 1200 data packets transmitted per measurement. We %
collected the per-packet Received Signal Strength Indication (RSSI) information. %

\begin{figure*}
\begin{center}
	\subfigure[2.4~GHz.]{\psfig{figure=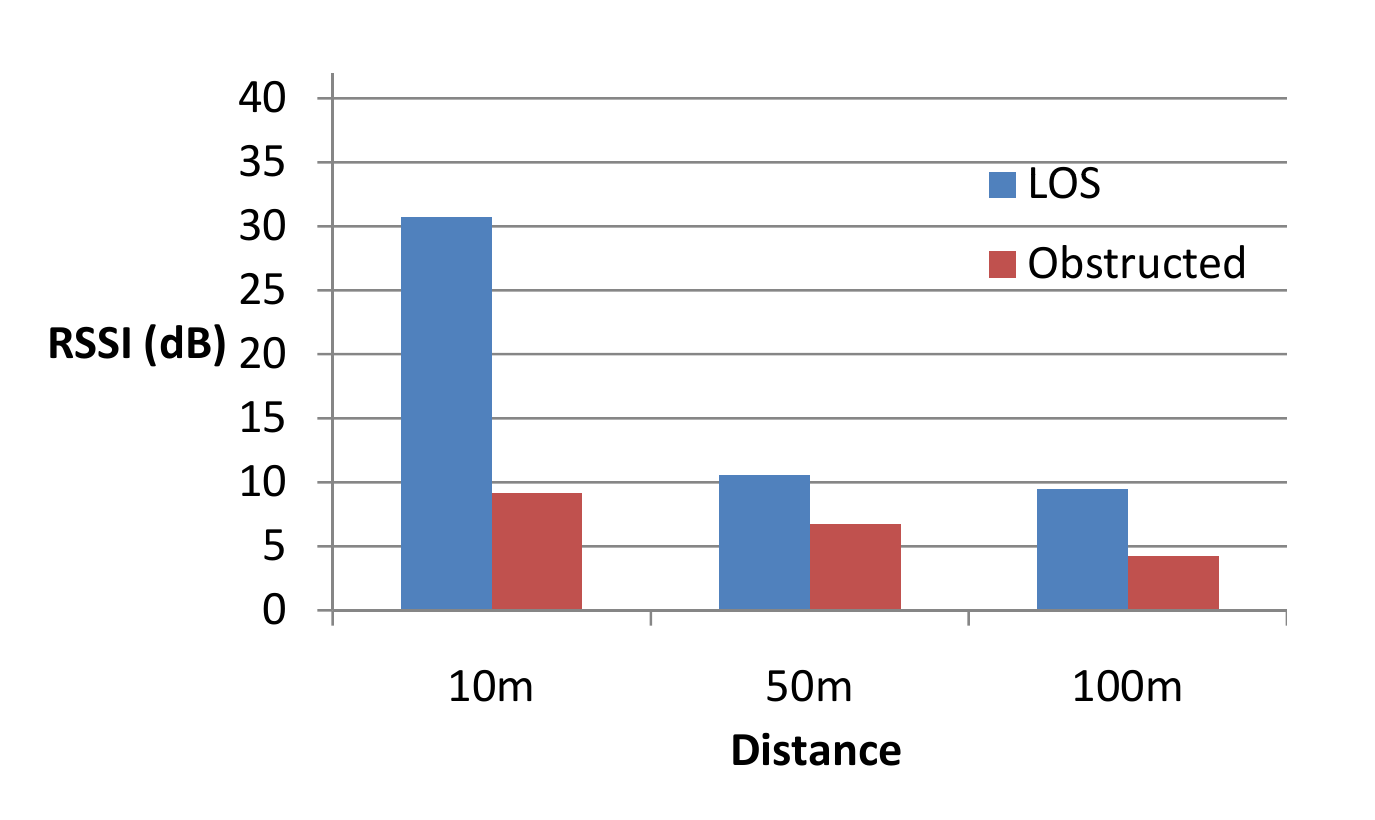,width=0.45\textwidth}}
	\subfigure[5.9~GHz]{\psfig{figure=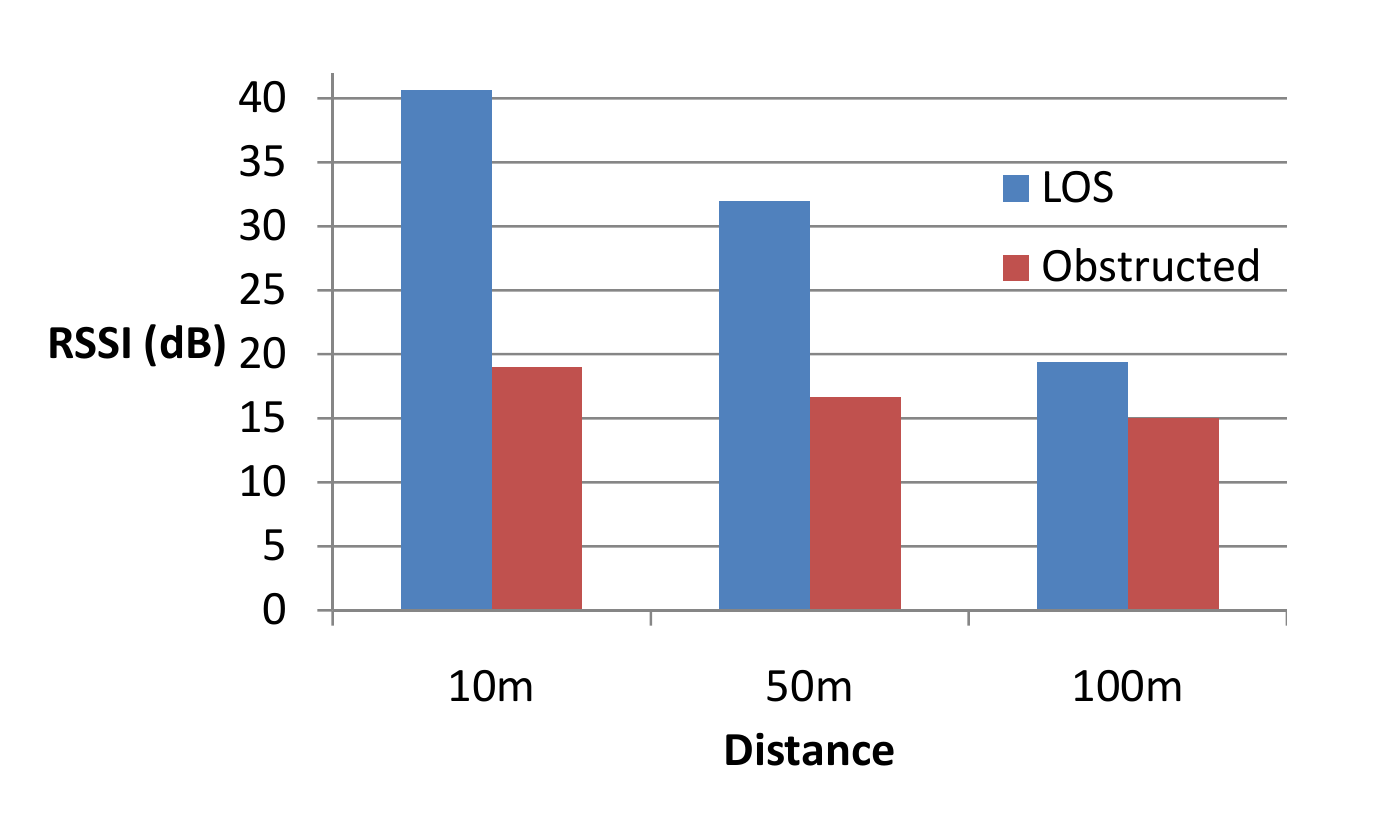,width=0.45\textwidth}}
  \caption{ RSSI measurements: average RSSI with and without the obstructing vehicle.}
  \label{measurements}
\end{center}
\end{figure*}

\begin{figure}[b]
	\centering
	\fbox{\psfig{figure=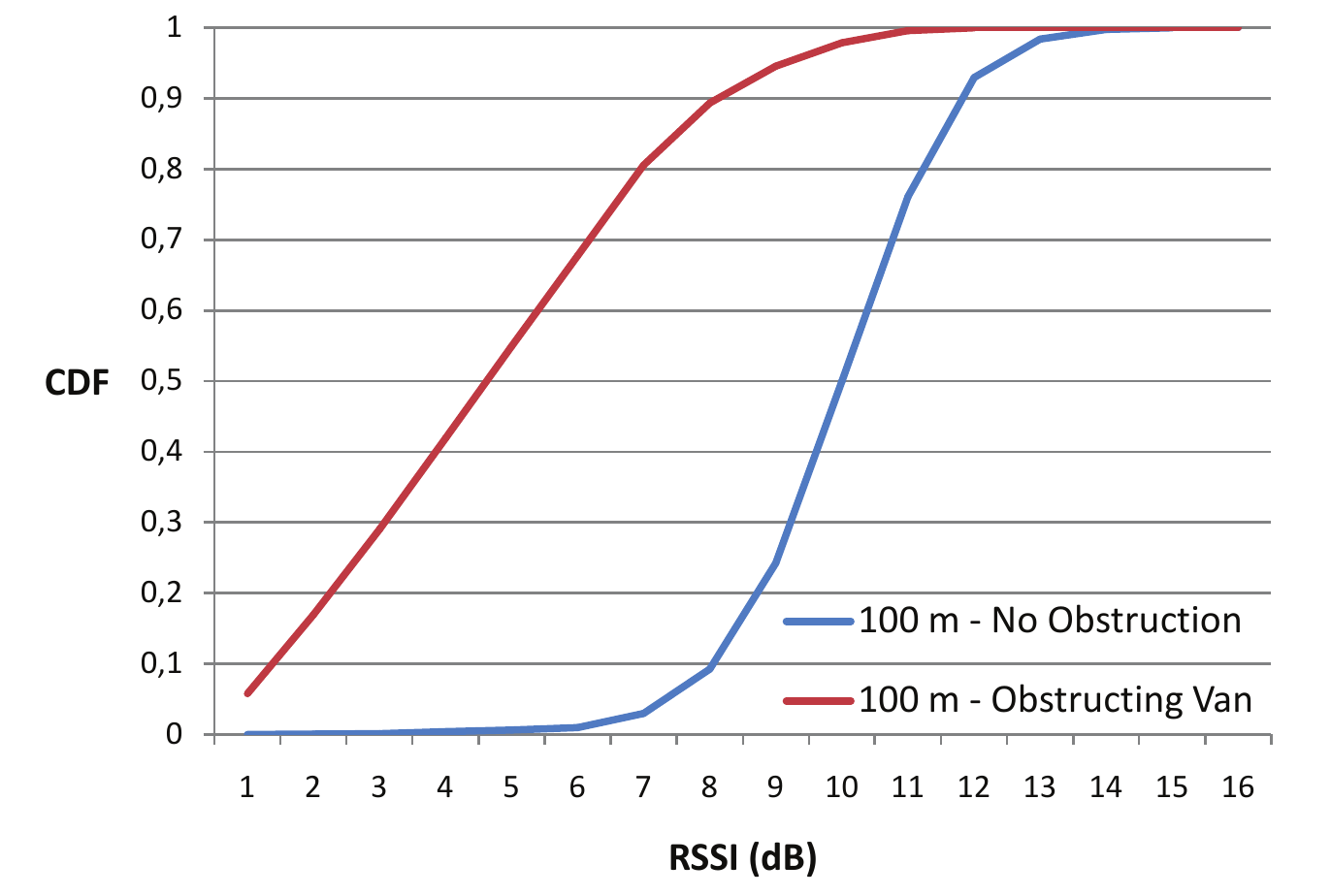,width=0.45\textwidth}} 
	\caption[Distribution of RSSI in case of LOS non-LOS (obstructing van)]{ Distribution of the RSSI for 100~m in case of LOS (no obstruction) and non-LOS (obstructing van) at 2.4~GHz.}
  \label{measurements_distribution}
\end{figure}

Figures \ref{measurements}a and \ref{measurements}b show the RSSI for the LOS (no obstruction) and non-LOS (van obstructing the LOS) measurements at 2.4~GHz and 5.9~GHz, respectively. The additional attenuation at both central frequencies ranges from approx. 20~dB at 10~m distance between Tx and Rx to 4~dB at 100~m. Even though the absolute values for the two frequencies differ (resulting mainly from the different quality radios used for 2.4~GHz and 5.9~GHz experiments), the relative trends indicate that the obstructing vehicles attenuate the signal more significantly the closer the Tx and Rx are. 
To provide more insight into the distribution of the received signal strength for LOS and non-LOS measurements, Fig.~\ref{measurements_distribution} shows the cumulative distribution function (CDF) of the RSSI measurements for 100~m in case of LOS and non-LOS at 2.4~GHz. The non-LOS case exhibits a larger variation and the two distributions are overall significantly different, thus clearly showing the impact of the obstructing van. Similar distributions were observed for other distances between the Tx and the Rx.

In order to determine how well the knife-edge model fits our measurements, we calculated the additional attenuation due to the knife-edge diffraction model for the given parameters: distances between the Tx and the Rx of 10, 50, and 100 m, location of the obstructing van, the dimensions of the vehicles, 2.4~GHz and 5.9~GHz frequency band, 3~dBi and 5~dBi antenna gains, and 18~dBm transmit power. The difference between the measurements and the knife-edge model was negligible at 100~m (e.g., 0.17~dB for 100~m at 2.4~GHz) and increased with the decrease of distance between the Tx and Rx (e.g., 1.2~dB for 50~m distance at 2.4~GHz and 10+~dB at 10~m). The knife-edge model approximates the real world measurements fairly well at longer distances between the Tx and Rx; however, it is too optimistic with regards to the additional attenuation at shorter distances (10~m). Therefore, more accurate models should be used to characterize additional attenuation due to vehicles at short distances.

\section{Conclusions} \label{sec:Conclusions} %
We proposed a new model for incorporating vehicles as obstacles in VANET simulation environments. 
First, we analyzed the real world data collected by means of %
stereoscopic aerial photography and showed that vehicles as obstacles have a significant impact on LOS obstruction in both dense and sparse vehicular networks, and should therefore be included in V2V channel modeling.
Then, based on the concepts of computational geometry, we modeled the vehicles as three-dimensional objects that can act as LOS obstructions between other communicating vehicles. Next, we designed a mechanism for calculating additional attenuation due to vehicles as obstacles, and we showed that the obstructing vehicles significantly decrease the received signal power and the packet success rate. We also performed experimental measurements in order to confirm the significance of the impact of obstructing vehicles on the received signal strength.
clearly indicate that vehicles as obstacles have a significant impact on signal propagation (see Fig.~\ref{power} and \ref{measurements}); therefore, in order to properly model V2V communication, it is imperative to account for vehicles as obstacles. Furthermore, the effect of vehicles as obstacles can not be neglected even in the case of relatively sparse vehicular networks, as the analyzed A3 highway dataset showed. 

Furthermore, neglecting vehicles as obstacles in VANET simulation and modeling %
has profound effects on the performance evaluation of upper layers of the communication stack. The expected effects on the data link layer are twofold: a) the medium contention is overestimated in models that do not include vehicles as obstacles in the calculation, thus potentially representing a more pessimistic situation than the real-world with regards to contention and collision; and b) the network reachability is bound to be overestimated, due to the fact that the signal is considered to reach more neighbors and at a higher power than in the real world. These results have important implications for vehicular Medium Access Control (MAC) protocol design; MAC protocols will have to cope with an increased number of hidden vehicles due to other vehicles obstructing them. %

The algorithm behind the proposed model, even though microscopically evaluating the attenuation due to vehicles (i.e., calculating additional attenuation due to vehicles for each communicating pair separately), remains computationally efficient, location independent, and compatible with models that evaluate the effect of other types of obstacles. By implementing the proposed model in VANET simulators, significant benefits can be obtained with respect to increased credibility of simulation results, at %
the expense of a relatively small computational overhead.

\chapter{Vehicle-to-Vehicle Channel Model for VANET Simulations} \label{ch:completeModel}

\section{Motivation}\label{motivationCompleteModel}

In the previous chapter, we proposed a model for incorporating the impact of vehicles as obstructions in vehicle-to-vehicle (V2V) communication. %
In highway environments, we showed that vehicles are the predominant source of shadowing~\cite{meireles10,boban11_2}. To realistically characterize the channel in environments where other objects have a significant impact on the signal (e.g., buildings and foliage in urban and suburban areas), we need to model these objects as well. To that end, in this chapter we design a model that, apart from vehicles, also takes into account buildings and foliage\footnote{Throughout the text, due to the lack of a more appropriate all-encompassing term, we use the term ``foliage'' for %
vegetation such as
 trees, %
bushes, shrubbery, etc.}. 

Our main goal is to design a computationally manageable channel model for implementation in discrete-event packet-level Vehicular Ad Hoc Network (VANET) simulators. To better estimate the channel conditions, %
we account for specific objects in the region around the communicating vehicles. %
Previous studies have shown that simplified statistical channel models are unable to simulate VANET channels accurately~\cite{dhoutaut06, koberstein09}. On the other hand, location-specific channel models such as those based on ray-tracing \cite{maurer04} yield results that are in a very good agreement with the real world. However, these models are computationally too expensive due to prohibitive computation costs as the network grows. %
Another notable problem of ray-tracing models %
is the need for detailed object database (e.g., location of doors and windows on buildings), as well as the sensitivity to the inaccuracies of the object database, which  make it difficult to correctly predict the path of the reflecting and diffracting rays interacting with the building.. %
For these reasons, ray-tracing models have not been implemented in large-scale, packet-level VANET simulators.

We aim to bridge the gap between overly simplified statistical models and computationally expensive ray-tracing models by performing location-specific channel modeling with respect to large objects in the vicinity of the communicating vehicles, at the same time limiting the calculations by using only the simple representation of the objects (i.e., outlines).
We use the real-world locations and dimensions of the buildings, foliage, and vehicles to determine the line of sight conditions for each link. To provide the baseline for the validation of our model, we start by performing a set of experiments in different environments (highway, urban, suburban, open space). Using the measurements, we characterize the most important factors impacting V2V links in each of the environments, which we then use to design a more efficient and accurate channel model. We implement spatial tree structures for efficient manipulation of geographic data to discriminate between three different link types: line of sight (LOS), non-LOS due to vehicles (NLOSv), and non-LOS due to static objects (NLOSb). Apart from large-scale signal variations due to shadowing, we also calculate the small-scale effects using the information about the number and size of the objects around the communicating vehicles. We validate our model using the collected experimental data. The results generated by the model match the measurements well in terms of both the large-scale effects (shadowing) as well as small-scale (multipath). 
We provide the complete simulation recipe for the implementation of the model in simulators and we implement the model in Matlab. %
We show that the model scales well by simulating networks of different size, with up to tens of thousands of objects in the scene and hundred thousand communicating pairs. Since the model requires only the locations and outlines of the modeled objects (buildings, foliage, and vehicles), all of which are easy to obtain, it is well suited for the implementation in discrete-event VANET simulators.

The rest of the chapter is organized as follows. The details of the experimental setup are shown in Section~\ref{sec:ExpSetupComplete}. Section~\ref{sec:SpatialTreeStructures} explains the spatial tree structures we use to efficiently implement the model. Section~\ref{sec:modelDescription} describes the proposed channel model, along with the recipe for implementation of the model in VANET simulators. %
Results validating the proposed model against measurements are shown in Section~\ref{resultsComplete}, whereas the computational performance of the model %
is discussed in Section~\ref{sec:Performance}. Section~\ref{sec:relWorkComplete} describes the related work, while Section~\ref{sec:conclusionsComplete} concludes the chapter.

\section{Experiment Setup}\label{sec:ExpSetupComplete}%

\begin{figure}
  \begin{center}
  \subfigure[ Porto Downtown.]{\label{fig:DowntownPortoRoute}\includegraphics[height=0.4\textwidth]{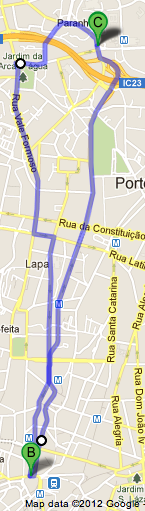}}
  \subfigure[ Porto Open Space (Leca).]{\label{fig:LecaRoute}\includegraphics[height=0.4\textwidth]{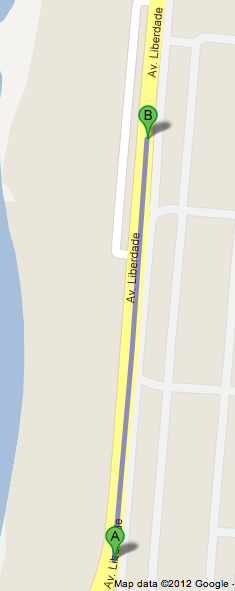}}
\subfigure[Porto Urban Highway (VCI).]{\label{fig:VCIImgRoute}\includegraphics[height=.4\textwidth]{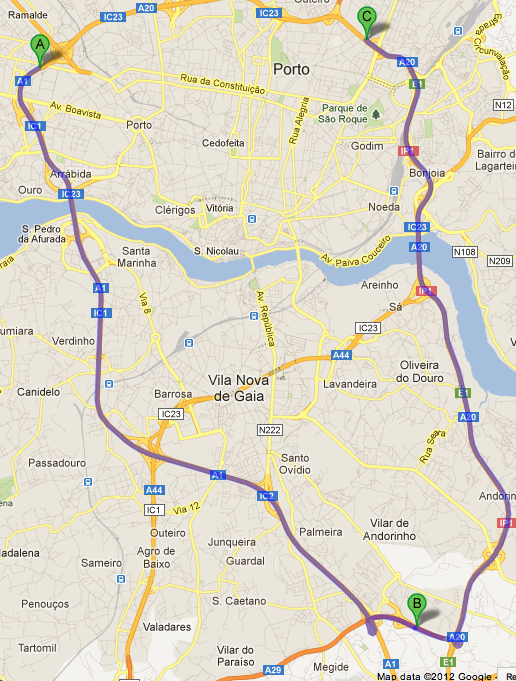}}  
  \subfigure[Porto Higway (A28).]{\label{fig:A28ImgRoute}\includegraphics[height=.4%
\textwidth]{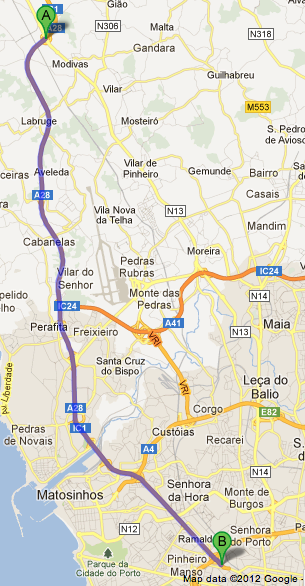}}
  \subfigure[ Pittsburgh Suburban (5th Ave).]{\label{fig:5thRoute}\includegraphics[height=0.25\textwidth]{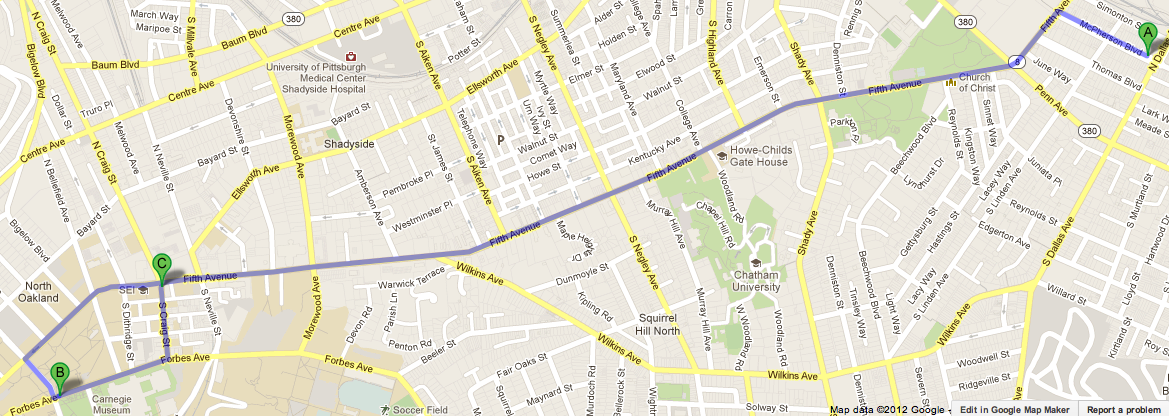}}
  \subfigure[ Pittsburgh Open Space (Homestead Grays Bridge).]{\label{fig:HomesteadRoute}\includegraphics[height=0.48\textwidth]{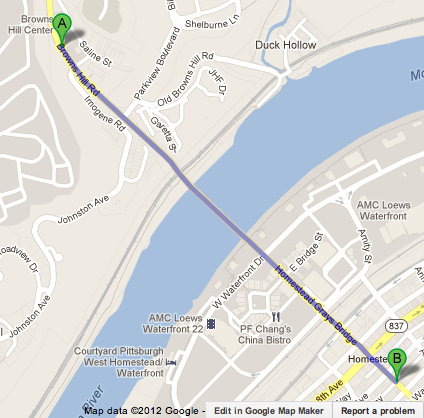}}
     \caption[Locations of the experiments]{Experiment locations with indicated routes. }%
      \label{fig:ExperimentRoutes}
   \end{center}
\end{figure}

\begin{figure}
  \begin{center}
  \subfigure[ Porto Downtown]%
  {\label{fig:DowntownPorto}\includegraphics[width=0.41\textwidth]{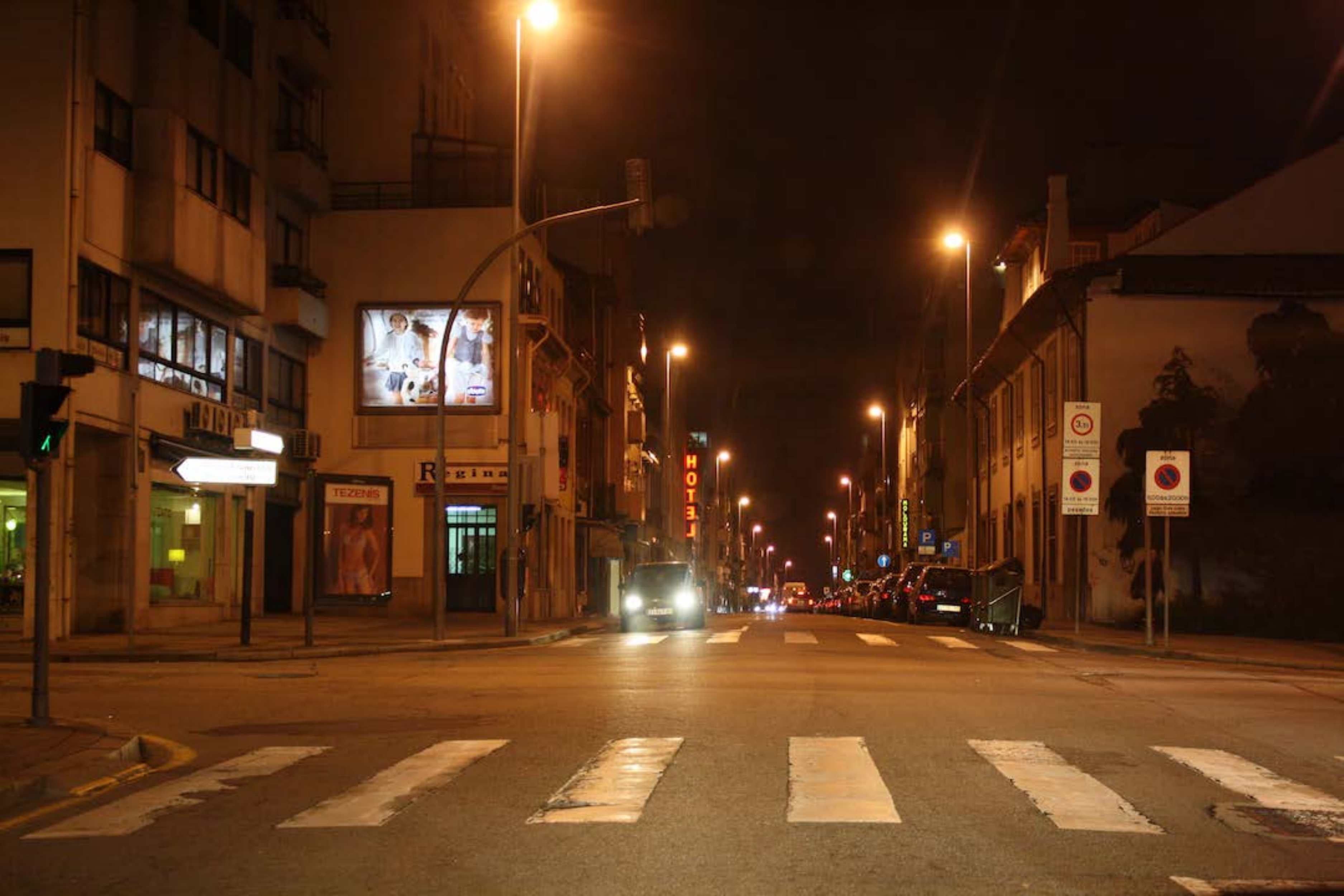}}
  \subfigure[ Porto Open Space (Leca)]%
  {\label{fig:Leca}\includegraphics[width=0.58\textwidth]{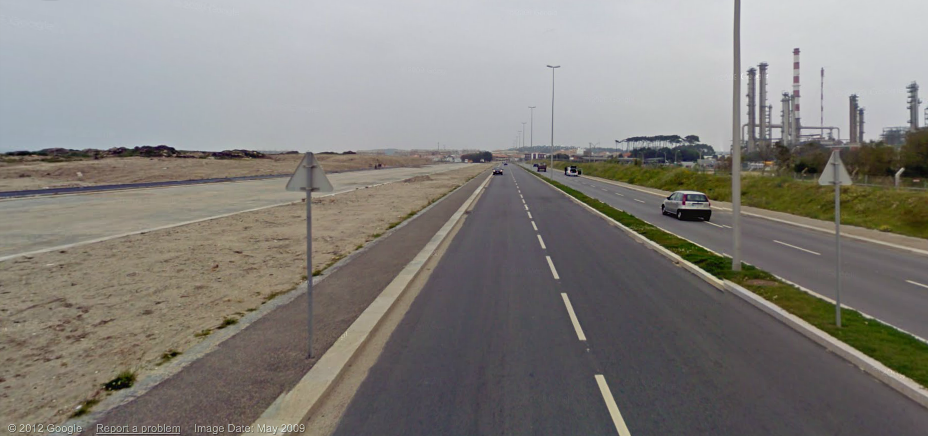}}
\subfigure[Porto Urban Highway (VCI)] %
{\label{fig:VCIImg}\includegraphics[width=.455\textwidth]{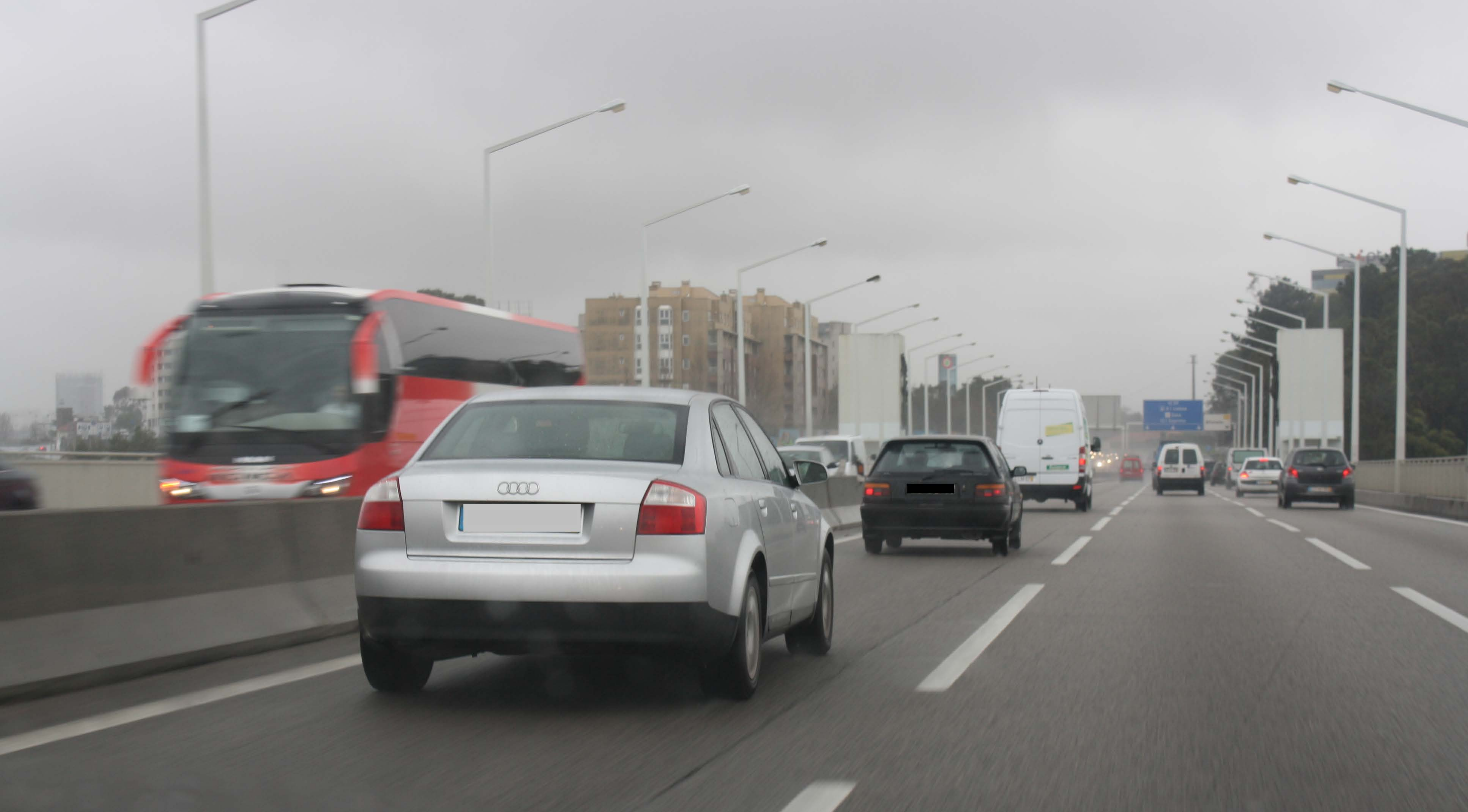}}  
  \subfigure[Porto Higway (A28)] %
  {\label{fig:A28Img}\includegraphics[width=.535%
\textwidth]{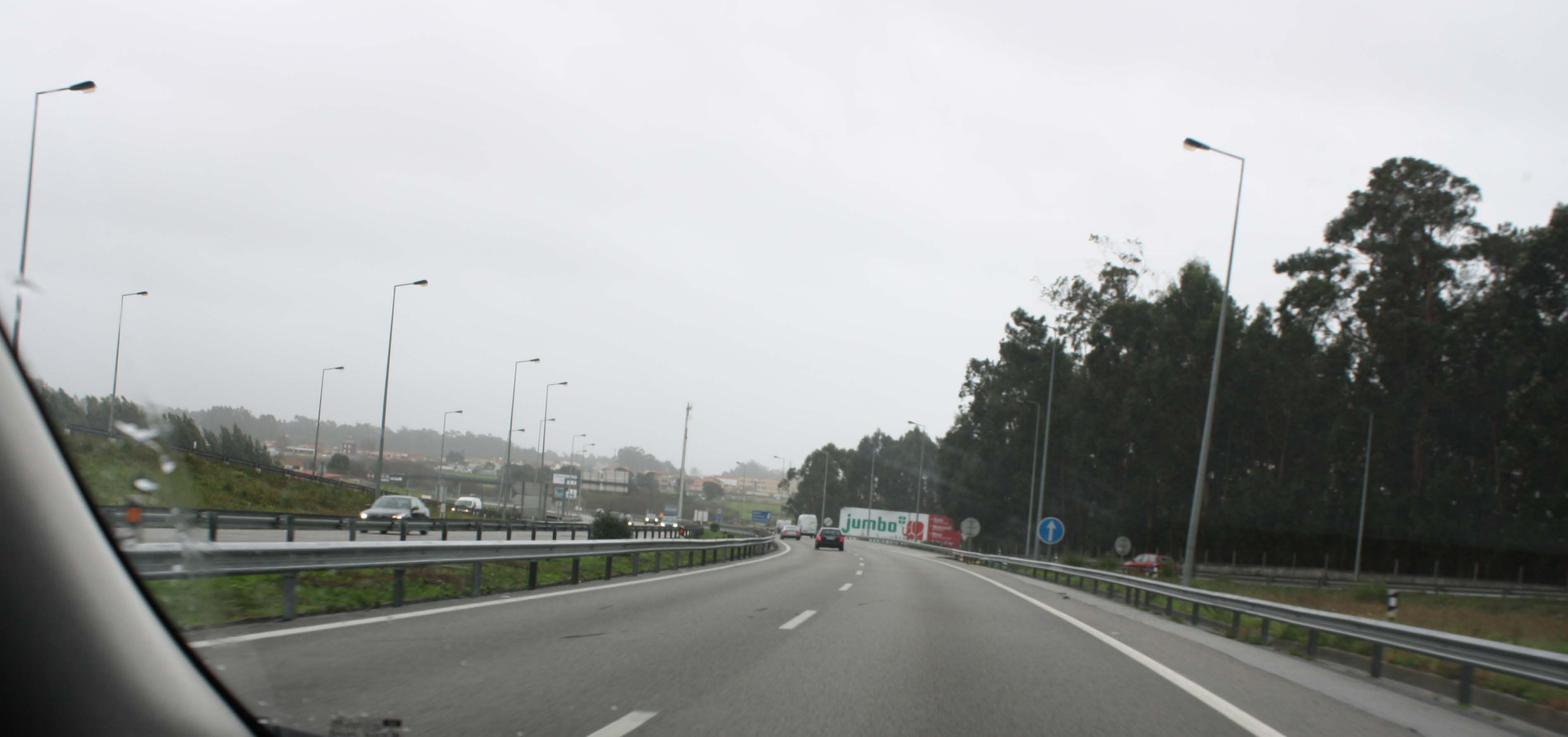}}
  \subfigure[ Pittsburgh Suburban (5th Ave)] %
  {\label{fig:5th}\includegraphics[width=0.5\textwidth]{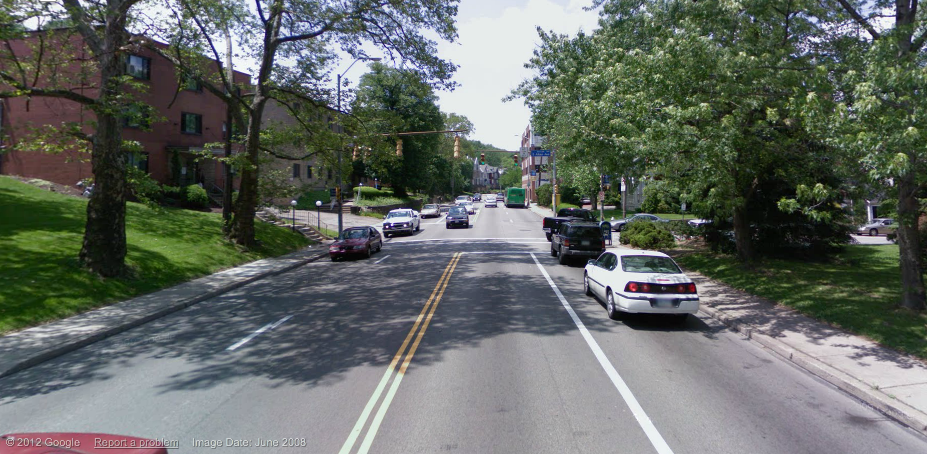}}
  \subfigure[ Pittsburgh Open Space (Homestead Grays Bridge)] %
  {\label{fig:Homestead}\includegraphics[width=0.48\textwidth]{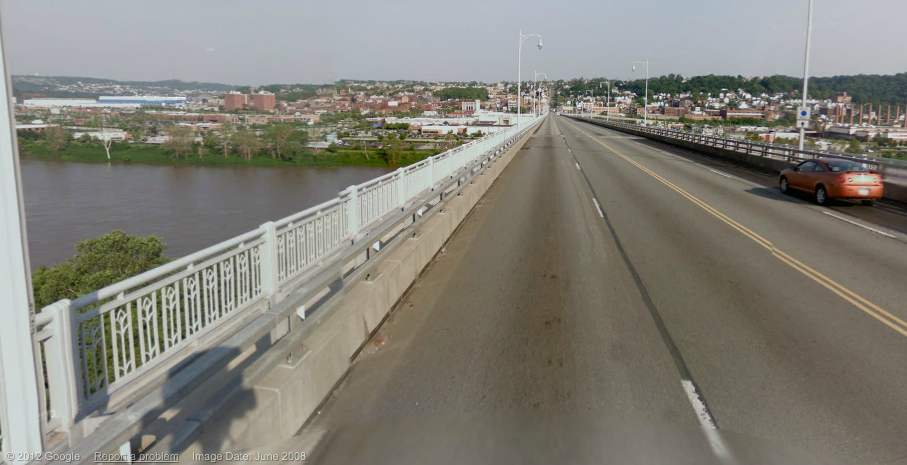}}
     \caption[Snapshots of the experiment locations]{Snapshots of the experiment locations.} %
      \label{fig:ExperimentLocations}
   \end{center}
\end{figure}

As a baseline for the model  validation and to extract statistical parameters to be used in the model, %
we performed experiments in the following locations: %
\begin{itemize}
\item Porto Downtown  -- 9~km route shown in Fig.~\ref{fig:DowntownPortoRoute}, going from the Paranhos parish to the Avenida dos Aliados in downtown Porto and back. Approximate coordinates (lat,lon): 41.153673, -8.609913; %
\item Porto Open Space (Leca) -- 1~km route shown in Fig.~\ref{fig:LecaRoute}. Approximate coordinates (lat,lon): 41.210615, -8.713418;
\item Porto Urban Highway (VCI) -- 24~km route shown in Fig.~\ref{fig:VCIImgRoute}. Approximate coordinates (lat,lon): 41.1050224 -8.5661420; 
\item Porto Higway (A28) -- 13.5~km route shown in Fig.~\ref{fig:A28ImgRoute}. Approximate coordinates (lat,lon): 41.22776, -8.695148; 
\item Porto Outlet -- shown in Fig.~\ref{fig:OutletOverlayReflDiffr}. Approximate coordinates (lat,lon): 41.300137, -8.707385; 
\item Pittsburgh Suburban (5th Ave) -- 7~km route shown in Fig.~\ref{fig:5thRoute}. %
Approximate coordinates (lat,lon):  40.4476089, -79.9398574; 
\item Pittsburgh Open Space (Homestead Grays Bridge) -- 2~km route shown in Fig.~\ref{fig:HomesteadRoute}. Approximate coordinates (lat,lon):  40.4103279, -79.9181137). 
\end{itemize}
Photographs of each of the measurement locations can be seen in Fig.~\ref{fig:ExperimentLocations}. Measurements were performed multiple times at each of these locations. Experiments were performed between May 2010 and December 2011. Each vehicle was equipped with a NEC LinkBird-MX V3, a development platform for vehicular communications~\cite{festag08}. Details regarding the devices and DSRC parameter setup are identical to those described in Section~\ref{subsec:NetworkConfiguration} and Table~\ref{tab:hw-config}. Identical hardware setup and radio parameters were used in all experiments.
We also performed experiments in downtown Pittsburgh. However, due to many high-rises taller than 100~meters, the GPS reception suffered from multipath that occasionally generated location errors in excess of 30 meters. %
Therefore, we do not include these results in our analysis. The buildings in downtown Porto are significantly lower, thus the GPS data is more accurate.

\begin{table} 
	\centering
		\caption[Porto Downtown Buildings and Vehicle Dataset]{Porto Downtown Buildings and Vehicle Dataset (more details available in~\cite{ferreira09})}
		\begin{tabular}{|c c c c c c|}
 \hline \textbf{City area} & \textbf{\# buildings} & \textbf{Area} & \textbf{\# vehicles} & \textbf{\# tall vehicles} & \textbf{Veh. density} \\
			& &  \textbf{of buildings} & & &  \textbf{(km$^2$)}\\
			\hline
\hline 41.3~km$^2$ & 17346 & 8.6~km$^2$ & 10566 & 595 (5.6\%) & 255 veh/km$^2$\\ 
\hline
\end{tabular}
	\label{tab:PortoDataset}
\end{table}

\begin{table}
	\centering
		\caption{Dimensions of Vehicles Used in the Experiments} 
\begin{tabular}{|c c c c|} \hline
		 & \multicolumn{3}{c|}{\bf Dimensions (m)} \\ %
		 \bf Vehicle & \bf Height & \bf Width & \bf Length \\ \hline \hline
		\emph{Portugal} & & & \\ \hline
		2007 Kia Cee'd & 1.480  & 1.790 & 4.260\\ \hline
		2002 Honda Jazz & 1.525 & 1.676 & 3.845\\ \hline 
		2010 Mercedes Sprinter & 2.591 & 1.989 & 6.680\\ \hline
		2010 Fiat Ducato & 2.524 & 2.025 & 5.943\\ \hline
		\emph{USA} & & & \\ \hline
		2009 Toyota Corolla & 1.466 & 1.762 & 4.539\\ \hline
		2009 Pontiac G6 & 1.450 & 1.793 & 4.801\\ \hline 
		\end{tabular} 
\label{tab:dimensionsCompleteModel}
\end{table}

\begin{figure}
  \begin{center}
    \includegraphics[width=0.55\textwidth]{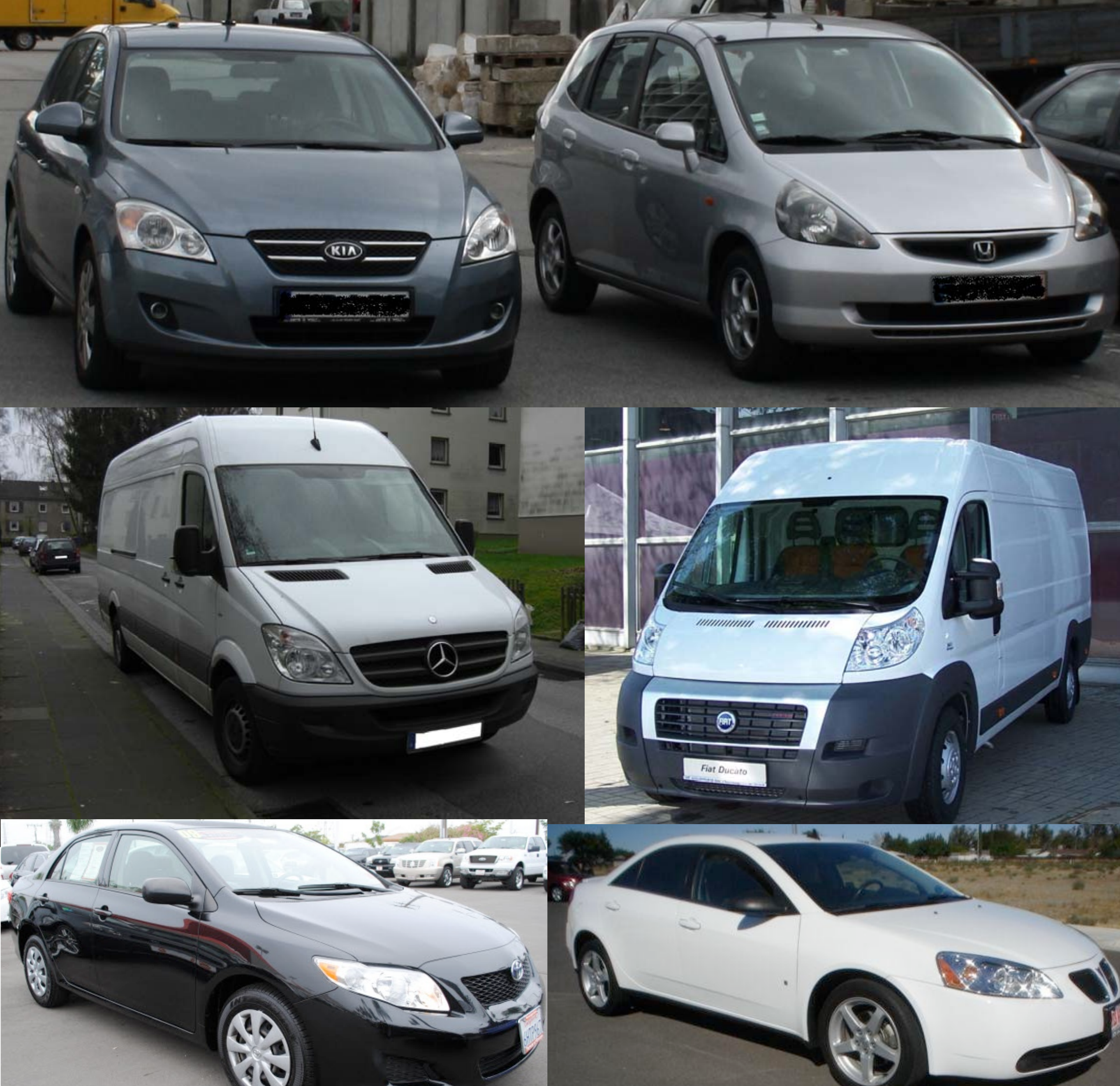}
     \caption[Vehicles used in the experiments]{Vehicles used in the experiments. First row: Kia Cee'd and Honda Jazz; second row: Mercedes Sprinter and Fiat Ducato; third row: Toyota Corolla and Pontiac G6.}
      \label{vehiclesCompleteModel}
   \end{center}
\end{figure}

Furthermore, we used building and vehicle outlines of the city of Porto, Portugal, described in Table.~\ref{tab:PortoDataset} and freely available from the Drive-In project website~\cite{drivein}. %
A snapshot of the data is shown in Fig.~\ref{portoBuildings}. We also used the building and foliage outlines from OpenStreetMap~\cite{openstreetmap}. %
As of yet, for the locations where we performed the measurements, the foliage data is scarcely represented in the OpenStreetMap database. Therefore, the results we obtain mostly pertain to buildings and vehicles. However, as described by Wang et al.~in~\cite{wang12}, high-precision foliage maps (1m x 1m resolution) can be extracted using image classification techniques \cite{blaschke10} on freely available aerial photography data~\cite{ferreira09,naip}.

We used regular passenger cars and commercial vehicles depicted in Fig.~\ref{vehiclesCompleteModel}. %
Dimensions of the vehicles are listed in Table~\ref{tab:dimensionsCompleteModel}. %
All passenger cars we used have a height of approximately 1.5 meters, which coincides with the statistical mean height for personal vehicles~\cite{boban11}, %
whereas both vans are approximately 2.5~meters tall. %

\section{Spatial Tree Structure for Efficient Object Manipulation}\label{sec:SpatialTreeStructures}

\begin{figure}[t]	
\centering
\includegraphics[width=0.95\textwidth]{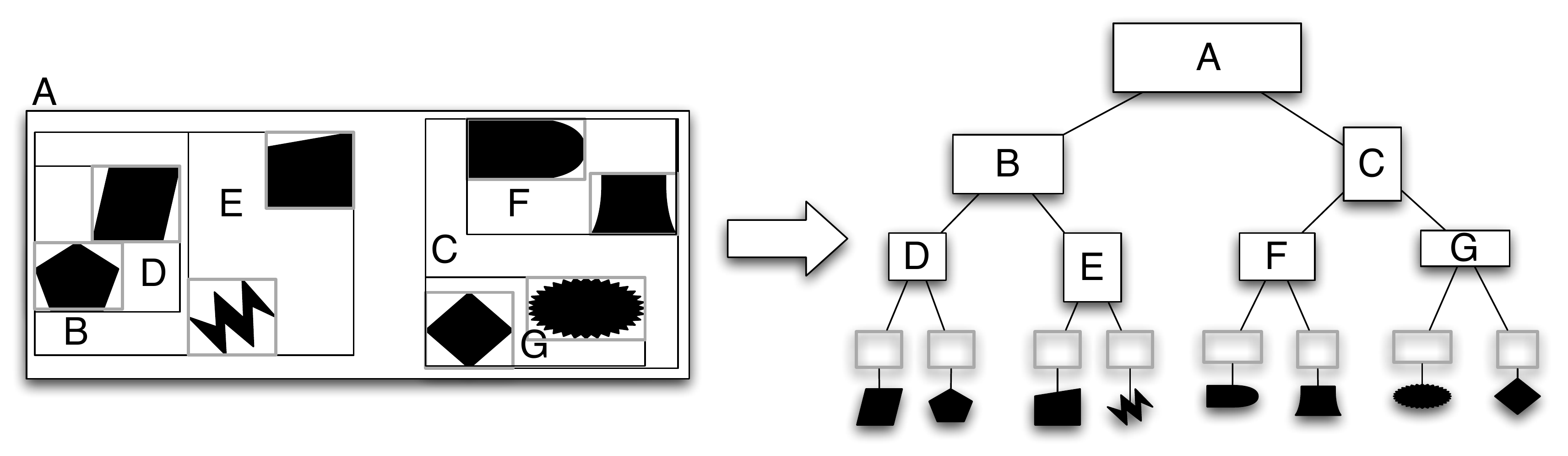} %
\caption[Bounding Volume Hierarchy]{Example of a bounding volume hierarchy (BVH), using rectangles as bounding volumes (adapted from~\cite{haverkort04}).
}\label{bvh}
\end{figure}

\begin{figure}[htb]	
\centering
\includegraphics[width=0.7\textwidth]{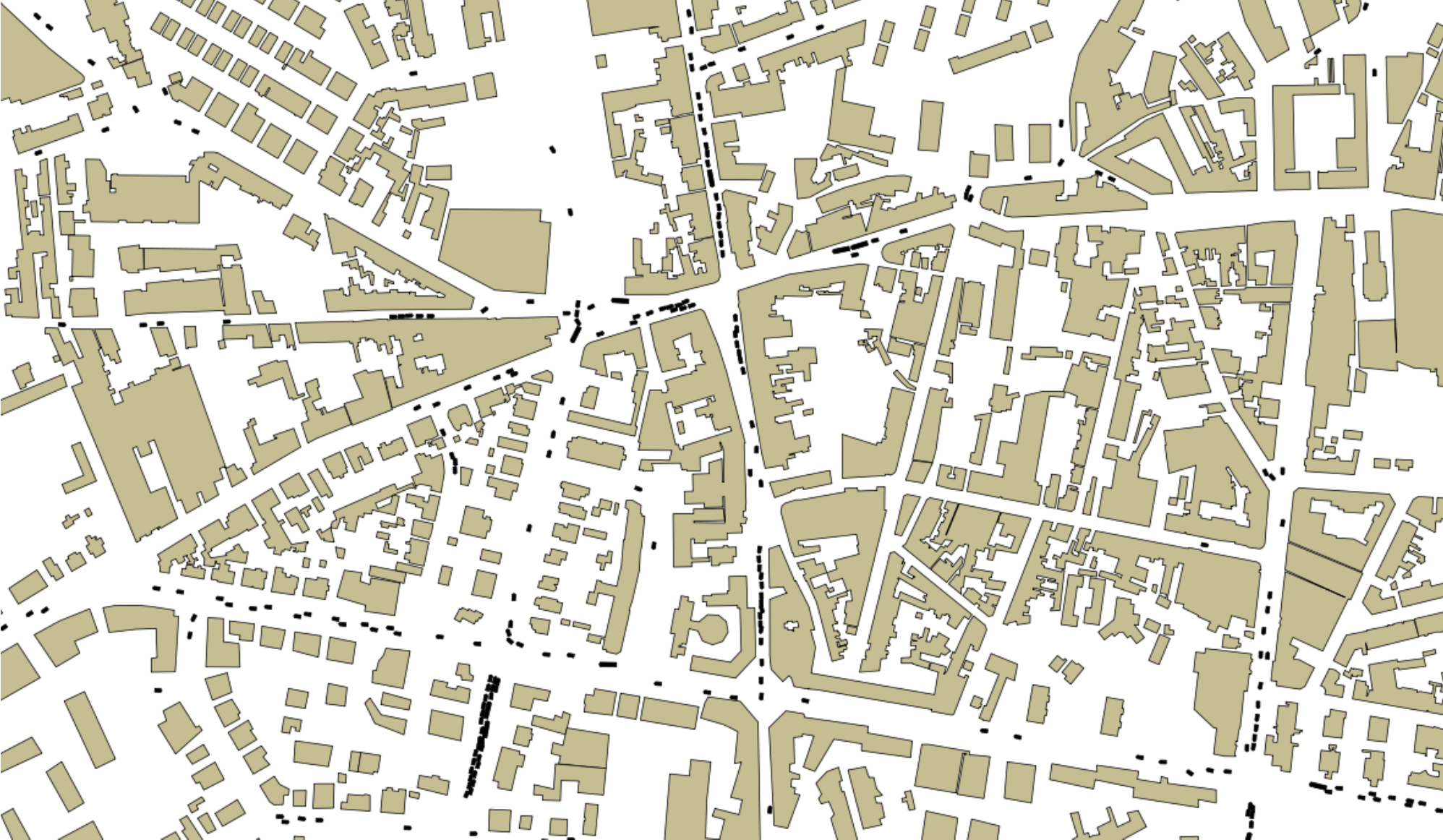} %
\caption[Outlines of the buildings and vehicles]{Outlines of the buildings and vehicles (vehicles colored black) extracted from aerial imagery in a neighborhood of Porto, Portugal.}\label{portoBuildings}
\end{figure}

Before we discuss the structure of our channel model, we introduce the spatial tree structure used for efficient VANET object manipulation. 
For a description of the modeled area, 
we use the outlines of buildings and foliage available through free geographic databases such as OpenStreetMap~\cite{openstreetmap}. Such sources of %
 geographical descriptors have become available recently, with a crowdsourced approach to geographic data collection and processing. %
Apart from the outlines of buildings and foliage available in such databases, we use the dimensions and locations of the vehicles, which can be obtained through GPS logs, vehicular mobility model, or aerial photography~\cite{boban11}.
Based on the outlines of the objects, we form bounding volume hierarchy (BVH) structures~\cite{klosovski98}.
BVHs are tree structures in which objects in the field are structured hierarchically based on their location in space. An example of a BVH is given in Fig.~\ref{bvh}. VANET-related geometric data lends itself to an efficient BVH implementation, due to its inherent geometrical structure (namely, relatively simple object outlines and no overlapping of object outlines).
More specifically, we utilize R-trees (where R stands for ``rectangle'', since it is used as the bounding shape)~\cite{guttman84} to store the vehicle, building, and foliage outlines. R-trees are often used to store spatial objects (streets, buildings, geographic regions, counties, etc.) in geographic databases. Even though they do not have good worst-case performance\footnote{When bounding rectangles of all objects overlap in a single point/area, the operation of checking the object intersection %
is quadratic in the number of objects in the R-tree (i.e., it is the same as the na\"{\i}ve approach that checks for intersection of every object with every other object). However, such extreme situations do not occur when modeling vehicular environments.}, in practice they were shown to have good tree construction and querying performance, particularly when the stored data has certain properties, such as limited object overlap~\cite{theodoridis96}. 
Figure~\ref{portoBuildings} shows the outlines of the vehicles and buildings we use in the model. We utilize R-trees to store the vehicle, building, and foliage outlines. We store vehicle outlines in a separate R-tree. The main difference in storing the outline of vehicles when compared to buildings and foliage is that, unlike vehicles, buildings and foliage do not move, therefore the model only needs to compute their R-tree once, after which it does not change. %
On the other hand, the vehicle R-tree changes at each simulation time-step. %

We construct each tree using a top-down approach, whereby the algorithm starts with all objects (i.e., vehicles, buildings, or foliage) and splits them into child nodes (we use binary R-tree, i.e., each non-leaf node has two child nodes). To keep the tree balanced, we sort the current objects at each node splitting based on the currently longer axis so that each created child node contains half of the objects. We note that similar tree data structures, such as \emph{k}-d tree and quadtree/octree, could be used instead of R-tree, with consideration to the specific application at hand and limitations and advantages of a specific data structure (for details, see de Berg et.~al~\cite{berg97}).

\section{Description of the Channel Model}\label{sec:modelDescription}

\begin{figure}[htb]	
\centering
\includegraphics[width=0.7\textwidth]{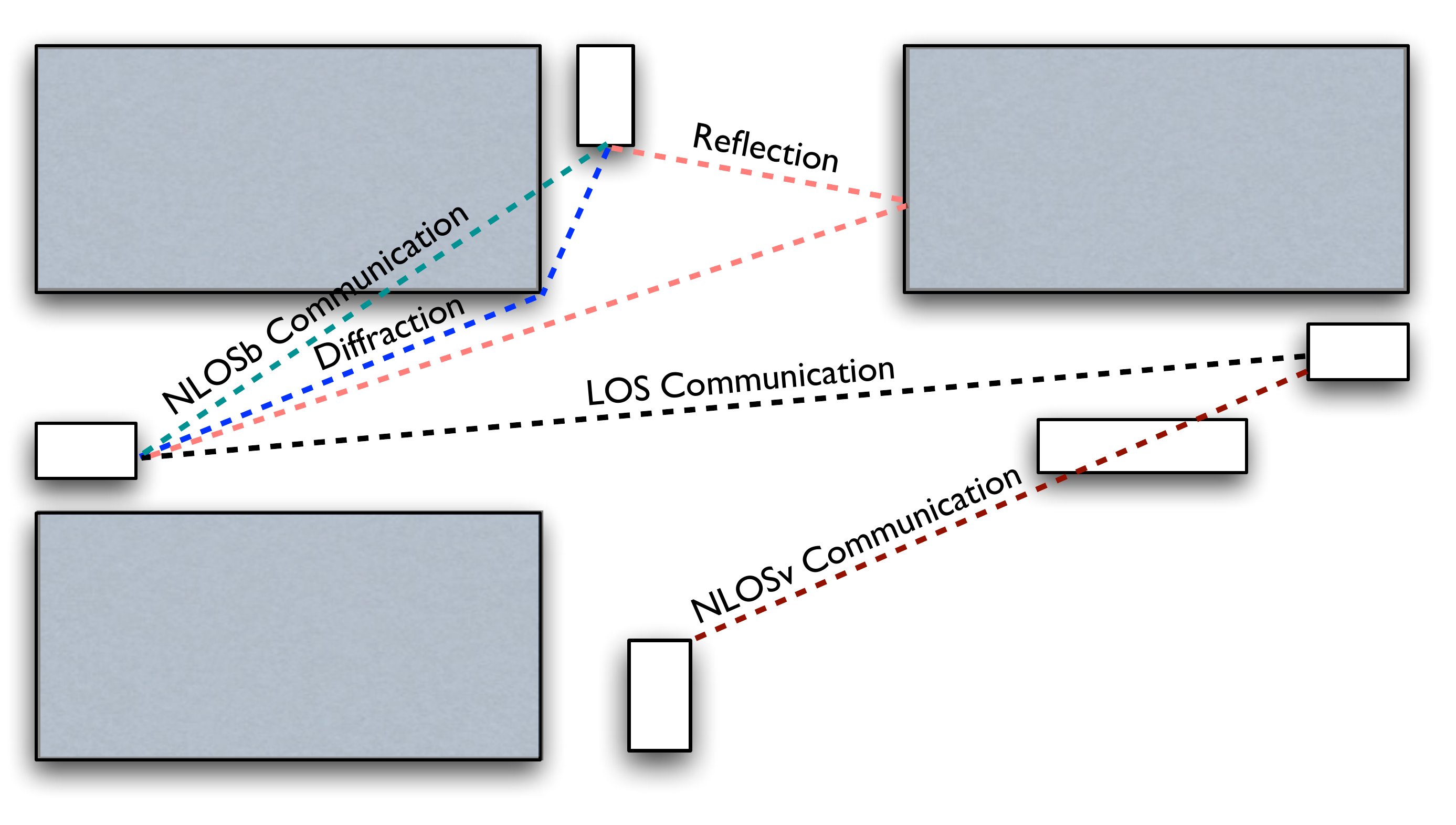} %
\caption{Link types and propagation effects captured by the model. White rectangles represent vehicles; gray rectangles represent buildings.}\label{modelDescription}
\end{figure}

\begin{figure}[htb]	
\centering
\includegraphics[width=0.7\textwidth]{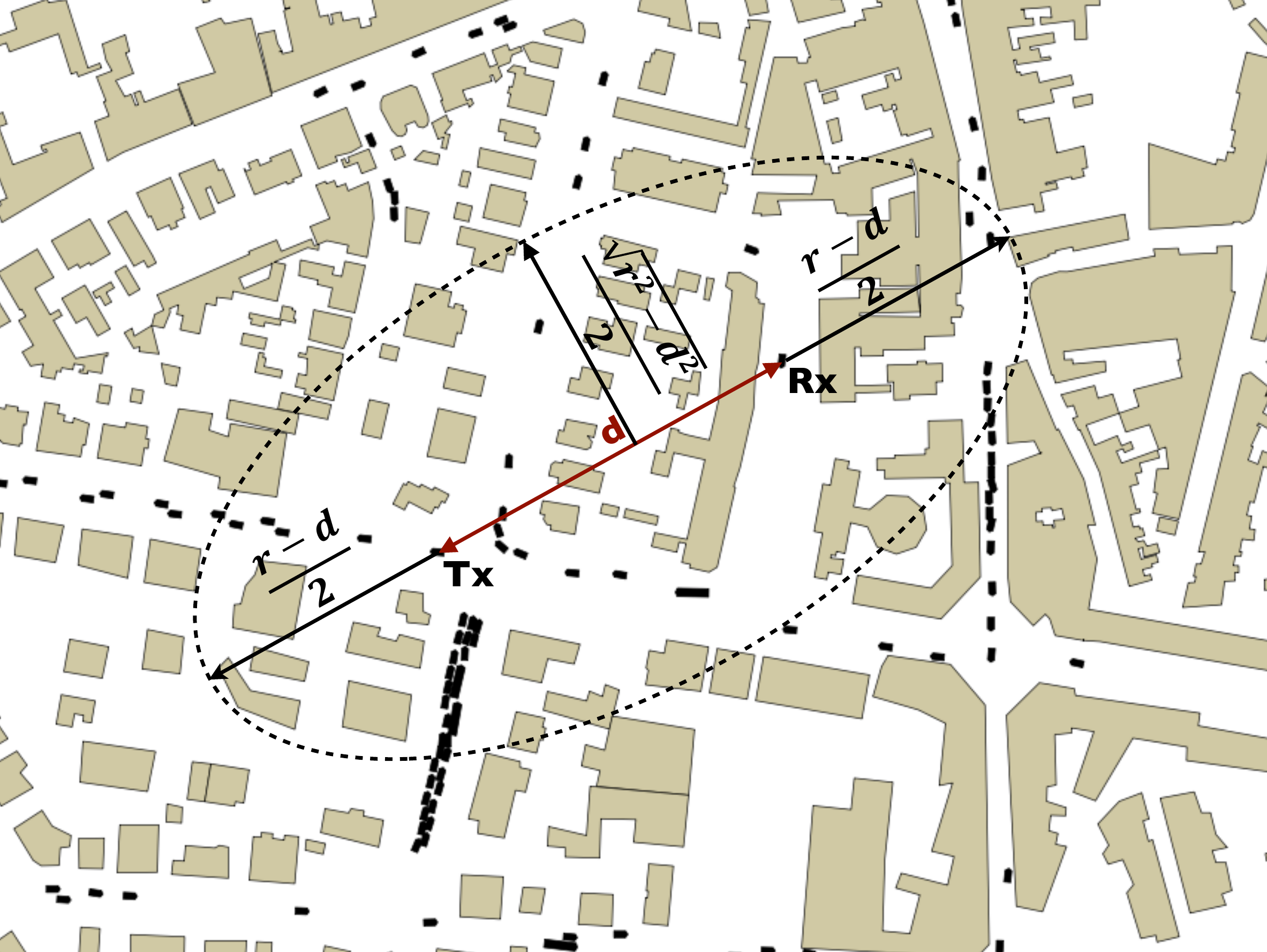} %
\caption[Explanation of the area used to determine fading, diffractions, and reflections]{Explanation of the area that the model analyzes to determine fading, diffractions, and reflections. The search space encompasses the ellipse whose foci are the transmitting (Tx) and receiving (Rx) vehicles. This ensures that all objects whose sum of distances to Tx and to Rx (i.e., from Tx to object and from object to Rx) is less than $r$ (maximum communicating distance for a given environment) are checked for diffractions and reflections. Note that the length of the major diameter of the ellipse is $r$, irrespective of the distance $d$ between Tx and Rx. The minor diameter's length is $\sqrt{r^2-d^2}$. The area of the ellipse is largest when Tx and Rx are close together, and the area goes to zero as the distance between Tx and Rx $d$ goes to $r$.
}\label{NLOSExplanation}
\end{figure}

We designed a model that, in addition to the LOS component, incorporates the following propagation effects (shown in Fig.~\ref{modelDescription}): 1) transmission (propagation through material); 2) diffraction; and 3) reflection. %
We focus on modeling the impact of vehicles, buildings, and foliage (as opposed to smaller objects such as traffic signs, traffic lights, etc.) for two reasons. First, on highways, obstructing vehicles are the most important objects for modeling the V2V channel, %
as the roads are predominantly straight and the largest portion of communication happens over the face of the road~\cite{boban11,meireles10}. In urban areas, obstructing vehicles have a significant impact for communicating pairs that are on the same street~\cite{meireles10}. Furthermore, the 2-D nature of the roads in suburban and urban areas implies that communication also happens outside the road surface. In such cases, static obstructions such as buildings and foliage play an important role. Buildings and foliage are the main source of obstructions for communication on the intersections and across different streets~\cite{karedal10}. Buildings and vehicles are also the main sources of reflections and diffractions~\cite{paier07_2}. Furthermore, other static objects such as lamp posts, street signs, railings, etc.,  are neither readily available in geographic databases, nor  would it be computationally feasible to model them due to their number, shape, and size. %
We validate the model against a set of experiments performed in different environments. Based on the measurements, we limit the complexity of the geometric model to a point where it is representable of the real world, but requires orders of magnitude less computations than complex ray-tracing models. %

\subsection{Classification of link types}
As mentioned previously, %
we distinguish three types of links: 1) line of sight (\textbf{LOS}); 2) non-LOS due to vehicles (\textbf{NLOSv}); and 3) non-LOS due to buildings/foliage (\textbf{NLOSb}).
Using the insights from the measurements in different environments, we apply different propagation models used for each of the three link types.  Table~\ref{tab:linkTypes} shows the employed models. Specifically, through measurements in open space, urban, suburban, and highway environments,  as well as consulting existing V2V measurements (e.g.,~\cite{meireles10,sommer2011using}), we concluded that LOS links are well approximated with a two-ray ground reflection model. Similarly, the NLOSv links are well modeled using the vehicles-as-obstacles model developed in Chapter~\ref{ch:vehModel}. Finally, for NLOSb links, we calculate single-interaction reflections and diffractions to account for the ``around the corner'' communication, and log-distance path loss~\cite{rappaport96} for cases where single-interaction rays are either non-existent or carry low power. We elaborate on the models used for the three link types in more detail in Section~\ref{resultsComplete}.

\subsection{Rules for reducing the computational complexity of the model}
As shown in the Fig.~\ref{modelDescription}, in addition to the LOS signal component, we model the following propagation effects:
1) transmission (propagation through material); 2) diffraction; and 3) reflection. However, if we were to calculate all significant rays between the transmitting and receiving vehicle, the model would not be different from the existing ray-tracing models (e.g., \cite{maurer04}, \cite{glassner89}).
Therefore, we exploit additional information available in VANETs, along with the specific and known geometric properties of the environment, in order to increase the performance of the model and make it suitable for implementation in large-scale, packet-level VANET simulators. 

\begin{table} 
	\centering
\caption[Modeling different link types]{Models used for different link types}
		\begin{tabular}{|c c|}
\hline \textbf{Link Type} &  \textbf{Propagation Mechanisms} \\
		 	\hline 
\hline \textbf{LOS} & Two-ray ground~\cite{rappaport96} \& fading (Section~\ref{subsec:fadingModel})\\
\hline \textbf{NLOSv} & Vehicles-as-obstacles (Ch.~\ref{ch:vehModel}) \& fading (Section~\ref{subsec:fadingModel}) \\
\hline \textbf{NLOSb} & Reflections and diffractions (Ch.~\ref{subsec:efield}) \& log-distance path loss \& fading (Section~\ref{subsec:fadingModel}) \\
\hline
\end{tabular}
	\label{tab:linkTypes}
\end{table}

\begin{enumerate}

\item Apart from checking which objects are blocking the LOS of the link and classifying the link into LOS, NLOSv, or NLOSb category, the model searches for the objects inside the ellipse shown in the Fig.~\ref{NLOSExplanation}. These objects are later used to calculate both large and small-scale signal variations. %

\item %
We calculate the small-scale signal variations based on the variations measured in real-world environments and using the number, relative location, and density of potentially reflecting and diffracting objects (other vehicles, buildings, foliage) around the communicating pair. We discuss the implemented model in detail in Section~\ref{subsec:fadingModel}.
\item For each link, we first check the blockage of LOS by any of the buildings of foliage. If there is LOS blockage, we do \emph{not} check the vehicle R-tree for LOS blockage. The reason for this is that LOS-obstructing buildings and foliage incur significantly more loss than LOS-obstructing vehicles (see, e.g.,~\cite{meireles10, mangel11, karedal10, paier07, durgin98}). For links whose LOS is not blocked by buildings or foliage, we check the R-tree containing vehicles.
\item R-trees enable efficient intersection testing and neighbor querying~\cite{guttman84}. Apart from using R-trees for link type classification, to determine reflected and diffracted rays, we use them %
 to efficiently implement a variation of the method of images~\cite[Chap. 7]{00parsons} -- a technique used to geometrically determine the reflected and diffracted rays.

\item For LOS, NLOSv, and NLOSb links, we define the maximum communication range $r$ as shown in Fig.~\ref{NLOSExplanation}, which determines the threshold distance above which the received power is assumed to be insufficient to correctly decode the message at the receiver, irrespective of the channel conditions. Specifically, we define $r_{LOS}$, $r_{NLOSv}$, and $r_{NLOSb}$ for LOS, NLOSv, and NLOSb links, respectively. %
In general, these radii are functions of transmit power, receiver sensitivity, antenna gains, and the surrounding environment. For a given set of radio parameters (reception threshold, transmit power, etc.), the ranges can be obtained either through field measurements or analytically.  
\item We used the insights from the experiments to finalize the structure of the model. Specifically, we tested the benefits we obtain when considering reflections and diffractions in all three types of links (LOS, NLOSv, and NLOSb). The measurement results for LOS and NLOSv links in different environments showed that these link types are well represented using path loss propagation mechanisms along with the fading as estimated by accounting for objects inside the appropriate ellipse as shown in Fig.~\ref{NLOSExplanation} (models used for link types are shown in Table~\ref{tab:linkTypes}, whereas details on the match between the model and measurements for these links are shown in Section~\ref{resultsComplete}). Furthermore, adding reflections and diffractions resulted in minimal benefits in terms of accuracy, while incurring a high computational overhead. %
On the other hand, for NLOSb links, reflected and diffracted rays account for a significant portion of the received power. Therefore, we explicitly model the single interaction reflections and diffractions for NLOSb links only. 

\end{enumerate}

\subsection{Transmission through Foliage}

For foliage, we use the attenuation-through-transmission model based on the measurements described in~\cite{goldhirsh98,ulaby90,benzair91,durgin98}. Specifically, we use the empirically-derived formulation from~\cite{goldhirsh98} where attenuation for deciduous trees is calculated per meter of transmission using %

\begin{equation}
MEL = 0.79 f^{0.61},
\end{equation}
where $MEL$ is mean excess loss per meter of transmission through trees and $f$ is frequency in GHz~\cite{benzair91}. For DSRC frequency centered at 5.9~GHz, this results in attenuation of 2.3~dB per meter of transmission through trees, which is in line with the measurement results in the 5.85~GHz band reported in~\cite{durgin98}. Similar calculations can be performed for coniferous trees as well as for seasonal changes when trees are not in full foliage~\cite{goldhirsh98}. Decision on which kind of trees to model (deciduous or coniferous) and the level of foliage (e.g., due to the time of the year) needs to be determined for the location where the simulations are carried out. Furthermore, geographic databases such as OpenStreetMap have tagging systems that allow for specification of such characterstics~\cite{openstreetmap}; provided that the different types of vegetation are tagged properly, they can be distinguished and modeled accordingly. Finally, we do not model any reflections or diffractions off foliage (i.e., only transmission attenuation is accounted for); rather, we implicitly encompass their scattering effects in the fading calculations.

\subsection{Combining multiple paths: E-field and received power calculations}\label{subsec:efield}

Once all contributing rays (LOS, reflected, diffracted, and transmitted) have been obtained, we calculate their contributions in terms of the E-field and the received power for each link. %
We obtain the resultant E-field envelope as follows~\cite[Chap. 3.]{rappaport96}:
\begin{align} 
|E_{TOT}| &= |E_{LOS} + \sum_jE_{Refl_j} + \sum_kE_{diffr_k}|,
\label{eq:Etot}
\end{align}
where $E_{LOS}$, $E_{Refl}$, and $E_{diffr}$ are E-fieds of line or sight, reflected, and diffracted rays, respectively. Expanding eq.~\ref{eq:Etot}, we get

\begin{align}
E_{TOT} &=\frac{E_0d_0}{d_{LOS}}\cos\left(\omega_c\left(t-\frac{d_{LOS}}{c}\right)\right) + \sum_{j}R_j\frac{E_0d_0}{d_{j}}\cos\left(\omega_c\left(t-\frac{d_{j}}{c}\right)\right) + \sum_{k}D_k\frac{E_0d_0}{d_{k}}\cos\left(\omega_c\left(t-\frac{d_{k}}{c}\right)\right),
\label{eq:EtotExpanded}
\end{align}

where $\frac{E_0d_0}{d_{LOS}}$ is the envelope E-field at a reference distance $d0$, $\omega_c$ is the angular frequency ($\omega_c = 2\pi f$), $t$ is the time at which the E-field is evaluated, $d_x$ represents distance traversed by ray $x$,  $R_j$ is the reflection coefficient of reflected ray $j$, $D_k$ is the diffraction coefficient of diffracted ray $k$. When the originating medium is free space, the reflected coefficient $R$ is calculated as follows for vertical and horizontal polarization, respectively~\cite[Chap. 3.]{rappaport96}:

\begin{equation}
R_{||} = \frac{-\epsilon_r \sin\theta_i + \sqrt{\epsilon_r - \cos^2\theta_i}}{\epsilon_r \sin\theta_i + \sqrt{\epsilon_r - \cos^2\theta_i}}
\end{equation}

and

\begin{equation}
R_{\perp} = \frac{\sin\theta_i - \sqrt{\epsilon_r - \cos^2\theta_i}}{\sin\theta_i + \sqrt{\epsilon_r - \cos^2\theta_i}},
\end{equation}

where $\theta_i$ is the incident angle, and $\epsilon_r$ is the relative permittivity of the material. 

Regarding diffractions, we do not calculate the diffraction coefficient directly; we approximate the E-field for diffracted rays using the knife-edge model~\cite{itu07}. However, the model contains all the geographical information to calculate the diffraction parameter for single diffractions using uniform theory of diffraction (UTD)~\cite{anderson98}.

The ensuing received power $Pr$ (in watts), assuming unit antenna gains, is calculated as follows:
\begin{equation}
Pr = \frac{|E_{TOT}|^2 \lambda^2}{480\pi^2},%
\label{eq:totPower}
\end{equation}
where $\lambda$ is the wavelength. Note that $Pr$ accounts for the slow fading signal component of LOS links, whereas for NLOSv and NLOSb links there are also contributions in terms of multipath generated by multiple diffractions around vehicles in case of NLOSv (horizontal and vertical multiple knife-edge diffractions) and single interaction reflections and diffractions in case of NLOSb. 

Furthermore, in cases where the accuracy of the geographic database is not sufficiently high for correct calculation of the phase shift -- this might often be the case for DSRC systems, since the wavelength is approximately 5~cm, thus requiring centimeter-grade database precision -- the phase shift component in the eq.~\ref{eq:EtotExpanded} for different incoming rays can be approximated using a distribution that represents the given environment well (e.g., uniform in case of isotropic scattering or based on the predominant angles of arrival in case of non-isotropic scattering~\cite{abdi02,karedal09}).

\subsection{Practical considerations for different link types and propagation mechanisms}

\subsubsection{LOS communication}

For LOS links, we implement the complete two-ray ground reflection model given by the following equation~\cite[Chap. 3]{rappaport96}:

\begin{align}
E_{TOT} &=\frac{E_0d_0}{d_{LOS}}\cos\left(\omega_c\left(t-\frac{d_{LOS}}{c}\right)\right) + R_{ground}\frac{E_0d_0}{d_{ground}}\cos\left(\omega_c\left(t-\frac{d_{ground}}{c}\right)\right),
\end{align}

where the reflection coefficient $R_{ground}$ and distance $d_{ground}$ for the ground-reflected ray are calculated according to the exact antenna heights
(i.e., we do not assume that the distance between transmitter and receiver is large compared to heights of the vehicles, as is often done in simulators~\cite{ns2},~\cite{ns3}). As will become apparent from our results (Section~\ref{resultsComplete}), using the exact height of the antennas is important, since even a 10~cm difference in height of either Tx or Rx results in significantly different interference relationship between the LOS and ground-reflected ray. %

In calculating $R_{ground}$, we model the relative permittivity $\epsilon_r$ to obtain the ``effective'' range of the reflection coefficient for the road.  It was pointed out in~\cite{kunisch08} that the idealized two-ray model is an approximation of the actual V2V channel, since the reflection coefficient is affected by the antenna location, diffraction over the vehicle roof below antenna, and the roughness of the road, among other. %
Therefore, we set the $e_r$ value used to generate the LOS results to 1.003, as this value %
minimized the mean square error for Leca dataset (see Section~\ref{resultsComplete}). Then, we use the same $e_r$ value for LOS links in all environments. Similar concept of effective reflection coefficient range calculation was used in the following studies:~\cite{karedal11TVT},~\cite{sommer2011using}, and \cite{kunisch08}.

\subsubsection{Reflections}
With respect to the reflection coefficients off building walls, we apply similar reasoning on the ``effective'' range of reflection coefficients as with the two-ray ground reflection model. We match the reflection coefficient distribution to the values empirically derived in~\cite{landron96}, where the authors extract the reflection coefficients for brick building walls from controlled measurements. In the measurement locations, walls of reflecting buildings were predominantly made of brick and concrete. 

\begin{figure}
  \begin{center}
    \includegraphics[width=0.75\textwidth]{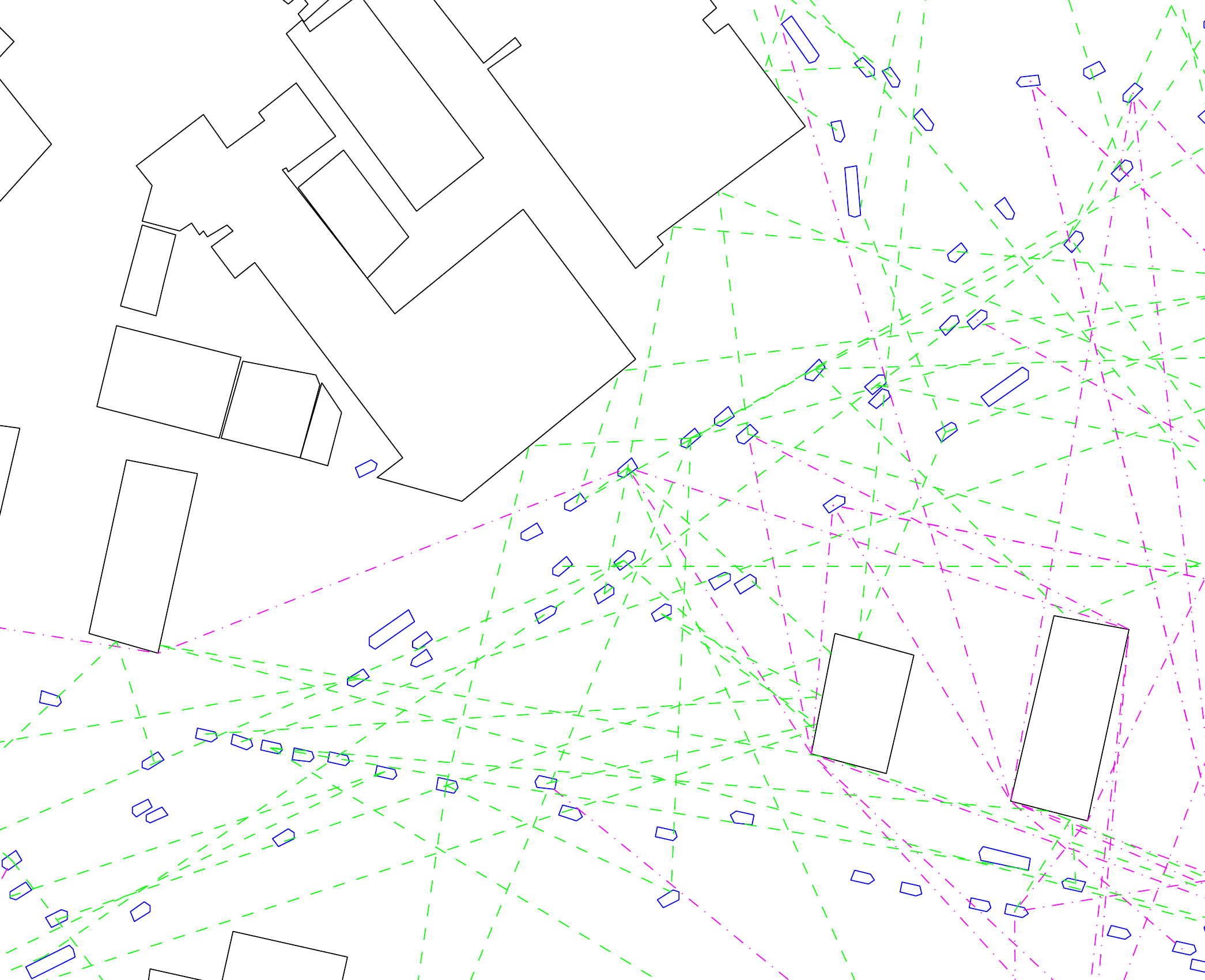}
     \caption[Downtown Porto with reflections and diffractions shown]{ Snapshot of a simulation in downtown Porto with reflections and diffractions shown for randomly selected communication pairs. Objects in the scene: buildings (black lines), vehicles (blue lines), reflected rays (green dashed lines), diffracted rays (magenta dash-dotted lines). Note that tall vehicles (elongated blue pentagons) obstruct reflected and diffracted rays, as they are most often taller than 60\% of the first Fresnel zone for the antennas of the communicating vehicles. %
     Short vehicles block the reflected/diffracted rays less frequently.} %
     
      \label{fig:PortoReflDiffr10kPairs}
   \end{center}
\end{figure}

The reflections are calculated from buildings and from vehicles. Since all buildings are significantly taller than any vehicle, any building can reflect the signal for any communicating pair. On the other hand, in order to be a reflector, a vehicle needs to be taller than both communicating vehicles' antennas, since otherwise the reflected ray does not exist. In practice, this means that tall vehicles are predominant among reflecting vehicles.  Furthermore, as can be seen in Fig.~\ref{fig:PortoReflDiffr10kPairs}, tall vehicles are more likely to block reflections coming off the building walls or other vehicles, whereas short vehicles are less likely to do so, since their height is often shorter than the height of the line between the communicating antennas  discounted for the 60\% of the first Fresnel zone.

\subsubsection{Diffractions}
For building diffractions, we use the same multiple knife-edge method on the building corners that we used for vehicle diffraction in Chapter~\ref{ch:vehModel} (eq.~\ref{eq:singleknifeedge}). The difference is that in the case of buildings, the only  diffraction in the \emph{horizontal plane} is accounted for (we assume that the buildings are too tall for diffraction over the rooftops), whereas in case of vehicles, the diffraction is calculated for both the \emph{vertical plane} (i.e., over the vehicle roofs) and the \emph{horizontal plane} (on the sides of the vehicles). In all cases, we perform the multiple knife-edge diffraction as described in~\cite{itu07}.  %

\subsubsection{Log-distance path loss in deep-fade areas}
Reflections and diffractions off buildings and vehicles are used for NLOSb links.
We limit the calculation of diffracted and reflected rays to single-interaction (single-bounce) rays, except for multiple diffraction due to vehicles. %
It was recently shown by Abbas et al.~in~\cite{abbas11} that single-interaction reflections and diffractions are most often the dominating propagation mechanisms in the absence of LOS. The authors conclude that ``single-bounce reflections with static objects e.g., buildings, roadsigns, and streetlights, often are the dominating propagation mechanisms in the absence of line of sight whereas the reflections from other vehicles contribute little unless these vehicles are tall enough.'' Similar findings are reported by  Paier et al.~in~\cite{paier09_2}. By determining the LOS conditions and modeling LOS and single-interaction rays, we aim to design a model that accounts for the most important rays, at the same time keeping the computational load manageable\footnote{The model is easily extensible to (recursively) account for the higher order interactions, however at a prohibitively increasing computational cost. Furthermore, a more detailed geographical database is required in order to model multiple interaction rays correctly.}.

However, communicating pairs that are not located on the same street or adjacent orthogonal streets (e.g., vehicles in parallel streets with contiguous buildings between the streets or vehicles several streets apart) most often do not have strong single-interaction reflected or diffracted rays, but are still occasionally able to communicate.
For such communicating pairs, multiple interaction reflections and scattering are the dominant contributors of the power at the receiver~\cite{00parsons}; calculating such rays  incurs prohibitively high computations and a geographical database with a high level of detail. Furthermore, our measurement results and those reported in similar studies (e.g.,~\cite{cardote11, giordano10, meireles10,mangel11}) show that communication range in NLOSb (i.e., building-obstructed) conditions using DSRC-enabled radios operating in the 5.9~GHz frequency band is limited to approximately 200~meters, even with the maximum transmit power. Thus, in order to avoid costly geometric computations which often yield power levels below reception threshold, %
at the same time allowing for communication in deeply faded areas, we determine the received power as follows. %
We calculate the received power using both the single-interaction diffractions and reflections through the described model and using the log-distance path loss model~\cite{00parsons}. We use the maximum received power of the two. %
The log-distance path loss $PL$ (in dB) for distance $d$ is given by~\cite{00parsons}
\begin{equation}
PL(d)\;=\;PL(d_0)\;+\;10\gamma\;\log_{10} \left( \frac{d}{d_0}\right),\
\label{eq:PathLoss}
\end{equation}
where $\gamma$ is the path loss exponent and $PL(d_0)$ is the path loss at a reference distance $d_0$.
For the log-distance path loss model, the received power $Pr_{PL}$ (in dB) at a distance $d$ is given by %
\begin{equation}
Pr_{PL}(d)\;=\;Pt\;-\;PL(d),\
\label{eq:PathLoss}
\end{equation}
where $Pt$ is the transmitted power in dB (where $Pt$ encompasses the transmit and receive antenna gains).

In the simulations, we used $\gamma$=2.9, which we extracted from the Porto urban dataset for the NLOSb conditions where there were no significant single-interaction reflections/diffractions. Previous studies reported similar values: $\gamma$=2.9 by Durgin et.~al~in~\cite{durgin98} (NLOSb environment), 2.3~$\leq \gamma \leq$~2.75 by Cheng et~al.~in~\cite{Cheng2007} (suburban environment), and 2.44~$\leq \gamma \leq$~3.39 by Paschalidis et~al.~in~\cite{paschalidis11} (urban environment -- various LOS conditions). %
Therefore, for NLOSb links, we determine the received power as the maximum of the received power calculated by the implemented model (using eq.~\ref{eq:totPower} in dB) and the %
log-distance path loss (eq.~\ref{eq:PathLoss}):
\begin{equation}
Pr_{NLOSb} = \max(Pr, Pr_{PL}).\
\end{equation}

\subsection{Small-scale signal variations}\label{subsec:fadingModel}

The model described above captures signal variations at different scales for different link categories (LOS, NLOSv, and NLOSb). %
For LOS links, the model accounts for large-scale signal variation due to distance and ground-reflection. 
NLOSv and NLOSb links, on the other hand, are by definition shadowed, albeit at different levels of obstruction within each category %
(in case of NLOSv, a small or a large blocking vehicle, or one or more vehicles; in case of NLOSb, deep or slight building obstruction). 
Thus, for NLOSv and NLOSb links, the variation captured by the proposed model accounts for shadowing variation as well as part of fast fading as follows. %
For NLOSv links, %
multiple diffraction paths around vehicles are accounted for (i.e., one path over the vehicle roofs and two potential paths on the sides of the vehicles). For NLOSb links, the multipath is partially accounted for by calculating the single-interaction reflections and diffractions. Therefore, the model accounts for the most significant rays in case of all three types of links (LOS, NLOSv, and NLOSb). %
To account for additional (smaller scale) signal variation inherent in V2V communication (e.g., due to effects such as scattering and higher order diffractions and reflections), %
using the insights obtained through experiments we designed a small-scale signal variation
 model that captures the richness of the propagation environment surrounding the communicating pair.

We first characterize the small-scale signal variations in the collected measurements; using the insights from the measurements, we design a simple model that complements the previously described components of the model that deal with large-scale signal variations. 

\subsubsection{Small-scale variations in the experimental datasets}
\begin{figure}
  \begin{center}
  
  \subfigure[\scriptsize LOS data from Open space Porto (Leca) dataset with best-fit normal distributions. %
  ]{\label{fig:LOSFading}\includegraphics[trim=0cm 6cm 0cm 6cm,clip=true,width=0.4\textwidth]{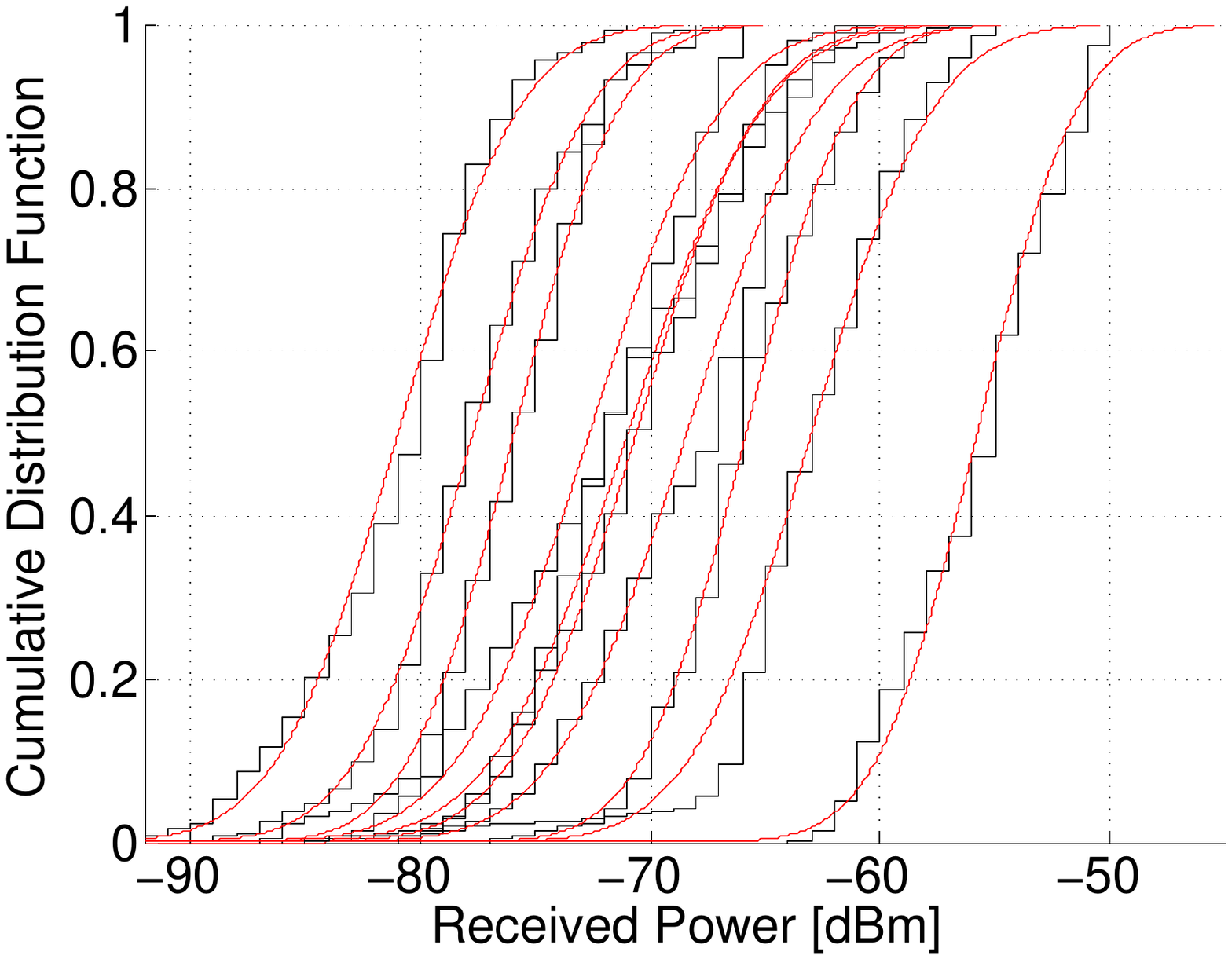}}
    \subfigure[\scriptsize LOS data from Urban (downtown Porto) dataset with best-fit normal distributions.]{\label{fig:LOSFadingUrb}\includegraphics[trim=0cm 6cm 0cm 6cm,clip=true,width=0.4\textwidth]{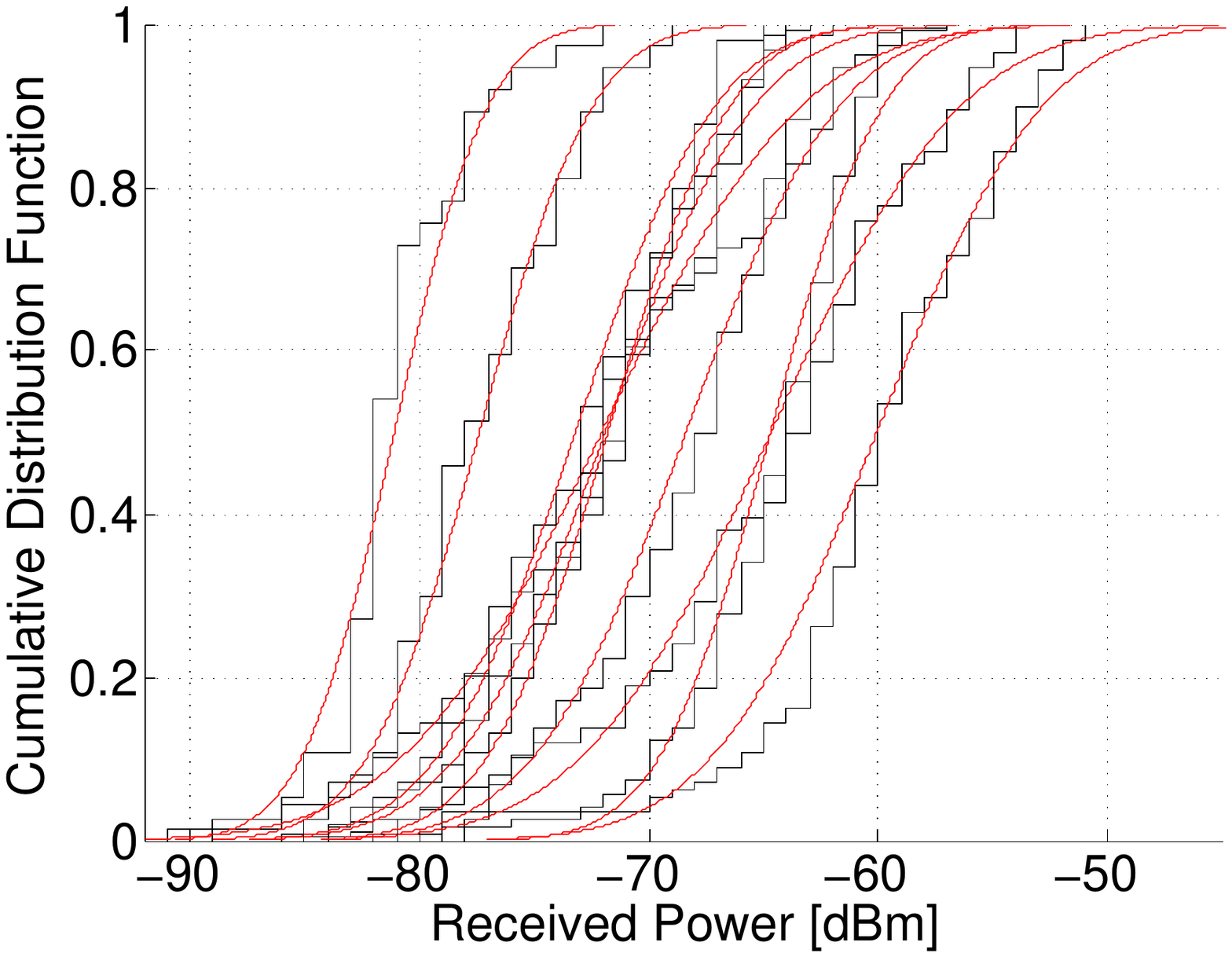}}

\subfigure[\scriptsize NLOSv data from Urban (downtown Porto) dataset with best-fit normal distributions.]{\label{fig:NLOSvFading}\includegraphics[trim=0cm 6cm 0cm 6cm,clip=true,width=0.4\textwidth]{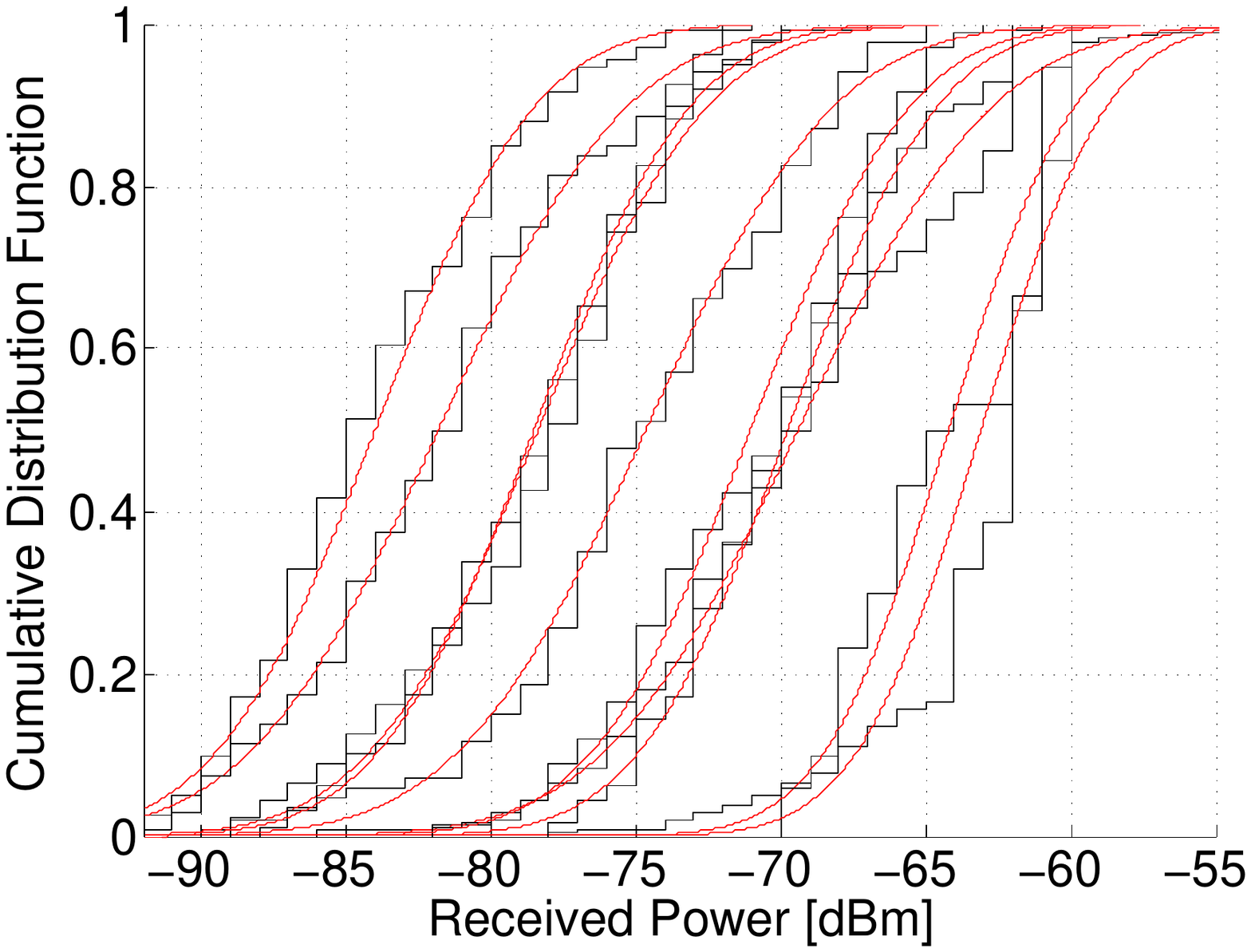}}
\subfigure[\scriptsize NLOSb data from Urban (downtown Porto) dataset with best-fit normal distributions.]{\label{fig:NLOSbFading}\includegraphics[trim=0cm 6cm 0cm 6cm,clip=true,width=0.4\textwidth]{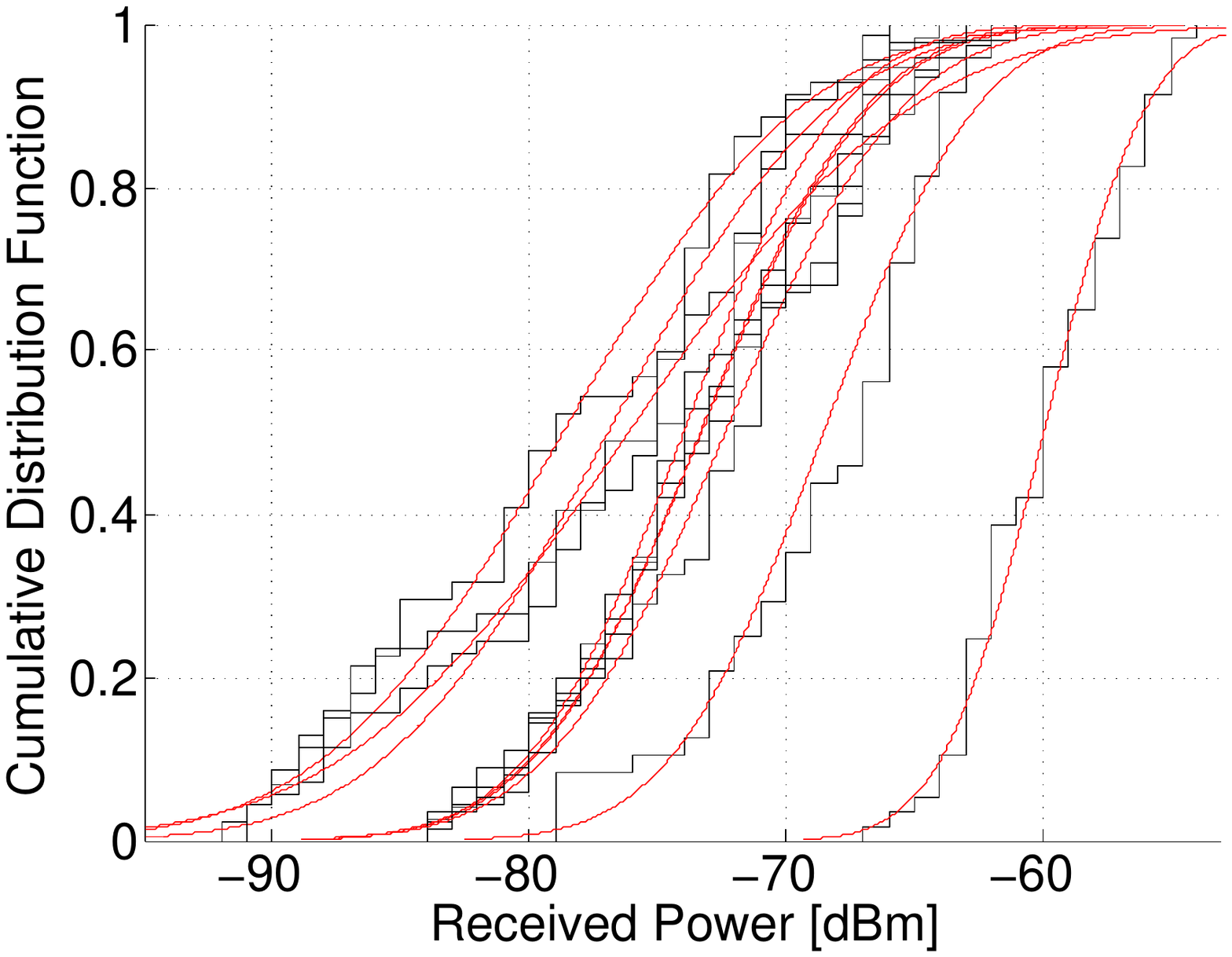}}
     \caption[CDF of the received power for two-meter distance bins]{Cumulative distribution functions of the received power for two-meter distance bins. All plotted bins contain at least 40 data points. For LOS (Figs.~\ref{fig:LOSFading} and~\ref{fig:LOSFadingUrb}) and NLOSv data (Fig.~\ref{fig:NLOSvFading}), the bins are centered at decades from 10 to 100 meters  (i.e., the curves represent the following bins, left to right: [99-101], [89-91], ..., %
[9-11] meters). For NLOSb data (Fig.~\ref{fig:NLOSbFading}), due to lack of data points at lower distances, the two-meter bins centered at the following distances are shown (left to right): 90, 85, 70, 65, 60, 55, 50, 45, and 15 meters. %
All plots are for passenger (short) vehicle experiments.}
      \label{fig:FadingTen}
   \end{center}
\end{figure}

We used the collected measurements to characterize small-scale variation in different LOS conditions, environments, and with different levels of vehicular traffic (i.e., temporal variation). %
For each collected measurement, we separate the data into LOS, NLOSv, and NLOSb category using the video recorded during the experiments. Then, we divide the collected data into two-meter distance bins. %
We selected two meter bins because we assume that the two meter window is small enough not to incur significant distance-related path loss dependence, at the same time containing enough data points to allow for a meaningful statistical characterization. %
Figure~\ref{fig:FadingTen} shows the distribution of received power for two-meter distance bins. For the LOS datasets, the normal distribution seems to fit the data reasonably well, with a better fit for the open space environment LOS data (Fig.~\ref{fig:LOSFading}) than the urban LOS data (Fig.~\ref{fig:LOSFadingUrb}) due to the richer reflection environment in the case of the latter.
Normal fit for the NLOSv and NLOSb data is less accurate due to the variety of conditions that are encompassed by the data (e.g., different number of obstructing vehicles in case of NLOSv, deep or slight building obstruction in case of NLOSb).
Based on the measured data, we elect to use normal distribution to describe the small-scale variation process in all three LOS conditions\footnote{This is in line with the results reported by recent V2V experimental studies described in~\cite{mangel11} and~\cite{abbas12}.}. Therefore, the empirically determined small-scale variation is a zero-mean %
normal distribution $N(0,\sigma)$ (normal in dB; log-normal in terms of power in Watts). %

\subsubsection{Accounting for additional small-scale signal variation} %

Apart from establishing the distribution of signal variation, we also need to determine its parameter -- i.e., the standard deviation $\sigma$ of the normal distribution -- since different environments and LOS conditions experience different levels of signal variation, as shown in Fig.~\ref{fig:FadingTen}. Therefore, we implement a simple model that accounts for the additional small-scale fading due to the objects in the area around the communicating pair as follows. Using the communication ellipse for each pair as explained in Fig.~\ref{NLOSExplanation}, we count the \emph{number of vehicles} and sum the \emph{area of static objects}  in the ellipse. We chose the area of the static objects, rather than their number because, unlike the size of vehicles, their area varies greatly (see Fig.~\ref{portoBuildings}). Since a large-area building/foliage is more likely to impact the communication than a smaller one, %
we use their area instead of their number in the calculations. In terms of different link types, the objects in the ellipse have the following effects: 1) for the communicating pairs located on the same street (i.e., LOS or NLOSv links), the objects inside the ellipse will include the vehicles along that street and buildings and foliage lining the street -- arguably, these are the most important sources of scattering, diffractions, and reflections that generate multipath fading for such links (see Fig.~\ref{fig:PortoReflDiffr10kPairs}); 2) similarly, for NLOSb links (i.e., links between vehicles on different streets and with buildings/foliage blocking the LOS), the ellipse will include buildings, foliage, and vehicles that generate significant reflecting, diffracting, and scattering rays (Fig.~\ref{fig:PortoReflDiffr10kPairs}).

Next, we set the minimum and maximum $\sigma$ for a given LOS condition based on the collected measurements. We do not extract the minimum and maximum for each experiment location, since we aim to determine a single minimum-maximum pair for the three LOS environments (LOS, NLOSv, and NLOSb), which could then be used across a number of different locations. Therefore, we utilize minimum and maximum $\sigma$ as calculated from the experiments and shown in Table~\ref{tab:fadingValues}. For simplicity, we use a single set of minimum/maximum values for both short and tall vehicles. The minimum $\sigma$ for NLOSv and NLOSb links is set to zero, since the most significant reflected and diffracted rays for these links are already accounted. For LOS links, on the other hand, we calculate the minimum $\sigma$ from the least variable environment in terms of small-scale fading (i.e., the open space). The maximum values for all three LOS link types have been calculated by finding the most variable fading environment from all the collected datasets. %
By averaging the standard deviation of the received power for all two-meter bins with more than 40 samples in that dataset, we obtained the values shown in Table~\ref{tab:fadingValues}. Note that minimum and maximum $\sigma$ values can be different for other environments; however, we believe the values we obtained are sound guidelines, since they are similar to small-scale variation observed in other experimental studies (e.g.,~\cite{abbas12,mangel11}).

\begin{table} 
	\centering
\caption[Minimum and fading $\sigma$ extracted from experimental data]{Minimum and maximum values of the fading deviation $\sigma$ extracted from experimental data }
		\begin{tabular}{|c c c|}
\hline \textbf{LOS Condition} & $\sigma_{min}$ (source) & $\sigma_{max}$ (source) \\
		 	\hline 
\hline \textbf{LOS} & 3.3~dB (Open space (Leca))& 5.2~dB (Urban Porto)\\
\hline \textbf{NLOSv} & 0~dB %
& 5.3~dB (Urban Porto) \\
\hline \textbf{NLOSb} &  0~dB %
 & 6.8~dB (Urban Porto)\\ 
\hline
\end{tabular}
	\label{tab:fadingValues}
\end{table}

The small-scale signal deviation $\sigma$ (in dB) for the communication pair $i$,  $\sigma_i$, is calculated as follows:
\begin{equation}
\sigma_i = \sigma_{min} + \frac{\sigma_{max}-\sigma_{min}}{2} \cdot \left( \sqrt{\frac{NV_i}{NV_{max}}} + \sqrt{\frac{AS_i}{AS_{max}}} \right),
\label{eq:FadingCalc}
\end{equation}

where $\sigma_{min}$ is the minimum small-scale signal deviation (in dB) for $i$'s LOS type (LOS, NLOSv, or NLOSb), $\sigma_{max}$ is the maximum deviation value for $i$'s LOS type, $NV_i$ is the number of vehicles per unit area in $i$'s ellipse, $NV_{max}$ is maximum number of vehicles per unit area, $AS_i$ is the area of static objects per unit area in $i$'s ellipse, and $AS_{max}$ is the maximum area of static objects per unit area. The value of $NV_{max}$ can be calculated a priori from historical data (e.g., maximum number of vehicles per area in a given city or a highway), whereas $AS_{max}$ can be calculated from geographical databases, such as~\cite{openstreetmap}.
In our calculations, we used $NV_{max}$ and $AS_{max}$ derived from the Porto dataset, with references defined on a square kilometer (i.e., maximum number of vehicles and maximum area of static objects in a square kilometer). For each of its constituents (vehicle-induced and static objects-induced signal variation), equation~\ref{eq:FadingCalc} is essentially a square interpolation between minimum small-scale variation in an environment (e.g., open space without any objects other than communicating vehicles) and maximum  variation (e.g., area with a high density of vehicles and buildings/foliage, such as the downtown of a city during rush hour). As shown in eq.~\ref{eq:FadingCalc} and due to the lack of a better classification, 
we give equal weights to the number of vehicles and the area of static objects when calculating the small-scale variation.
Finally, once the %
$\sigma$ is calculated for a given communication pair, we add a normally distributed random variable $N(0,\sigma)$ to the previously calculated received power (eq.~\ref{eq:totPower}):
\begin{equation}
Pr_{TOT_i} = 10\log_{10}(Pr_i) + N(0,\sigma_i).
\end{equation}

\subsection{Assumptions}
Since we are designing a channel model with a specific application in mind -- V2V communication using IEEE 802.11p (DSRC) radios~\cite{dsrc09} or similar technology -- the model relies on several IEEE 802.11p protocol characteristics in order to simplify the implementation of our model. Additionally, here we list other assumptions and simplifications we made when designing our model.
\begin{enumerate}	
\item %
IEEE 802.11p has been designed to cope with severe channel conditions~\cite{usdot06_2}. The channel bandwidth (10~MHz), symbol length (8~$\mu$s), guard time (1.6~$\mu$s), and Adjacent Channel Rejection (ACR) have all been designed so that the multipath fading/Doppler spread and Inter-Symbol Interference (ISI) do not affect the communication even in the harshest channel conditions~\cite{astm03}. In combination with Orthogonal Frequency-Division Multiplexing, it is envisioned to enable flat and ISI-free channels. For this reason, in our model we assume that the channel coherence bandwidth is sufficiently high so as not to cause frequency selective fading on OFDM subcarriers. Experimental studies by Acosta-Marum and Ingram in~\cite{acosta07} and by Paier et al.~in~\cite{paier07_2} confirm this assumption. However, a recent study by Fernandez et~al.~in~\cite{fernandez11} point out that the proposed IEEE 802.11p physical (PHY) layer (due to suboptimal equalization in particular) can suffer high packet error rates in highly-faded V2V environments. In case the PHY layer of IEEE 802.11p is unable to cope with the channel variations (thus resulting in the packet errors), the results generated by the model above would represent the upper bound of the performance in terms of packet delivery rate. %

\item We limit the calculation of diffracted and reflected rays for NLOSb links to single-interaction (single-bounce) rays. %
It was recently shown in~\cite{abbas11} and~\cite{paier09_2} that single-interaction reflections and diffractions are most often the dominating propagation mechanisms in the absence of LOS on urban intersections. However, even though communicating vehicles in parallel streets with contiguous buildings between the streets or vehicles several streets apart most often do not have strong single-interaction reflected or diffracted rays, they might be able to communicate. For this reason, we implement the log-distance path loss to compensate for the lack of higher-order reflections and diffractions.

\item We assume that buildings are too tall for any meaningful amount of power to be received over them. Since even the shortest buildings are at least 5~meters taller than the vehicles, simple calculations using knife-edge diffraction~\cite{itu07} show that the losses due to diffraction over the rooftops is in excess of 30~dB (with 40+~dB loss for buildings 15 or more meters taller than vehicles), thus making the power contribution over the rooftops negligible.  %
 Therefore, we do not model diffraction over the rooftops.
\item Currently, vehicles, buildings, and foliage are incorporated in our model currently. In environments where other objects have a significant impact (e.g., lamp posts, signs, railing, etc.), the model would need geographical information about these objects as well. However, such objects are currently not readily available in geographic databases. Furthermore, the additional gains in realism need to be compared against the increase in the computational complexity due to the additional objects, particularly if the number of such objects is large.

\item Due to the limited precision of the databases and the focus on packet-level simulation, we do not model scattering. Due to their potentially large number, scattering objects could significantly increase the complexity of the calculations that need to be performed by the model. Additionally, the precision of such calculations would be questionable.

\item Due to the small wavelength of DSRC (approx. 5~cm), in case when the geographical databases are inaccurate, the calculations of the incoming phase of the reflected/diffracted ray might be erroneous, thus randomly combining constructively or destructively at the receiver. 

\item Currently, we assume that the terrain is flat. For locations with significant elevation changes, the model can be adapted so that the elevation is included, %
provided such data is sufficiently accurate. %

\end{enumerate}
\subsection{Channel Model Simulation Structure}\label{sec:chModelStructure}

\begin{figure}
  \begin{center}
    \includegraphics[width=0.8\textwidth]{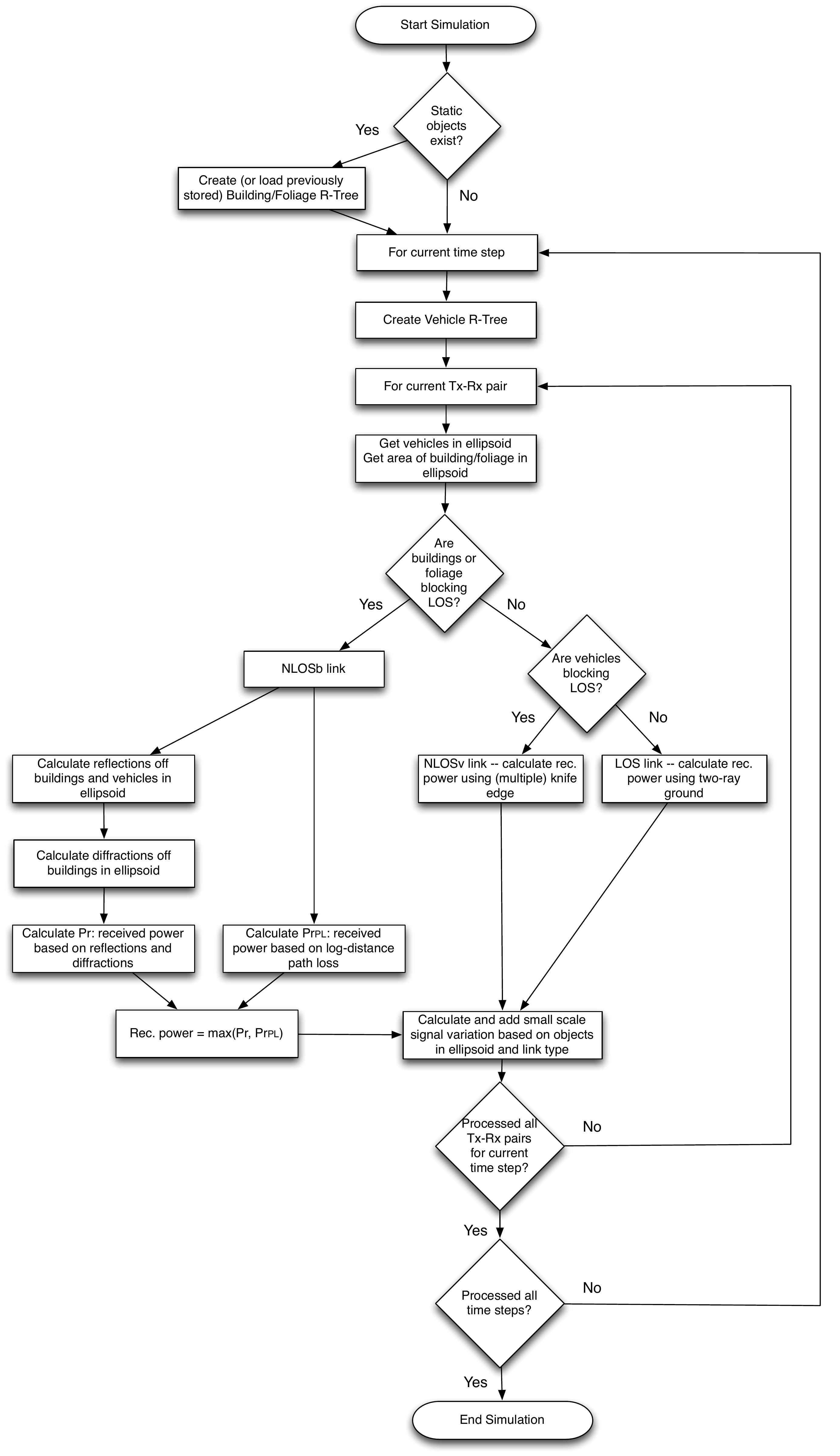}
     \caption{Channel model simulation flow.}
      \label{fig:simulationFlowchart}
   \end{center}
\end{figure}

Figure~\ref{fig:simulationFlowchart} shows the simulation execution flowchart of the model. The flowchart can be used to implement the proposed model in discrete-event packet-level VANET simulators.

\section{Results}\label{resultsComplete}
In this section we show the measurements and channel prediction results for the measurement locations shown in Fig.~\ref{fig:ExperimentRoutes} and~\ref{fig:ExperimentLocations}. We plug in the GPS locations of the measurement vehicles into the model and simulate the channel in the same locations for each of the transmitted packets. We use the actual dimensions of the vehicles (Table~\ref{tab:dimensionsCompleteModel}) and the corresponding static objects extracted from geographical databases. Furthermore, we use the communication ranges specified in Table~\ref{tab:rLOS}. These values are based on our own measurements, as well as results collected in \cite{meireles10,cardote11,karedal10,mangel11}.
Note that $r_{LOS}$ was set to 1000~meters outside urban areas and 500~meters in urban areas, whereas $r_{NLOSv}$ and $r_{NLOSb}$ did not change  in different environments.

\begin{table} 
	\centering
\caption{Max. communication ranges used for different link types}
		\begin{tabular}{|c c|}
\hline \textbf{Link Type} &  \textbf{Max. comm. range} \\
		 	\hline 
\hline \textbf{$r_{LOS}$} - highway & 1000\\
\hline \textbf{$r_{LOS}$} - urban & 500\\
\hline \textbf{$r_{NLOSv}$} & 400 \\
\hline \textbf{$r_{NLOSb}$} & 300 \\
\hline
\end{tabular}
	\label{tab:rLOS}
\end{table}

\begin{figure}
  \begin{center}
    \includegraphics[width=0.75\textwidth]{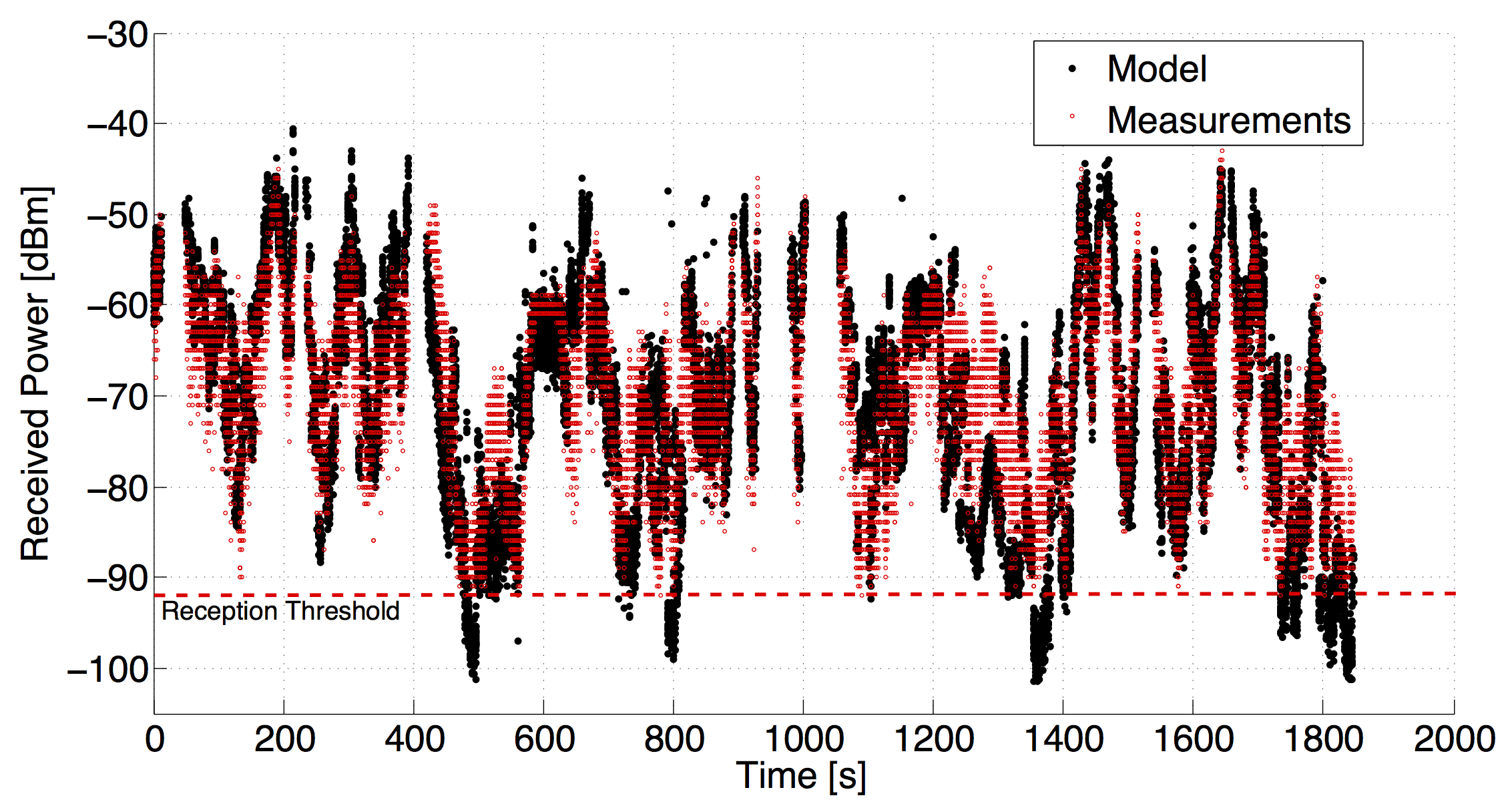}

     \caption[Received power for a 30-minute experiment in downtown Porto]{Received power for a single 30-minute experiment in downtown Porto, Portugal, along with the received power predicted by the model. Number of data points: 16500.}
      \label{fig:PortoComplete}
   \end{center}
\end{figure}

Figure~\ref{fig:PortoComplete} shows the received power for a 30-minute experiment conducted on a 10~km route in downtown Porto, Portugal (Fig.~\ref{fig:DowntownPorto}), along with the results generated by the model. Using the recorded videos of the experiments, we separated the data into line of sight (LOS), non-LOS due to vehicles (NLOSv) and non-LOS due to buildings (NLOSb). This allowed us to evaluate the ability of the model to deal with different types of LOS conditions. Since we empirically determined the fading for the measured data, %
we consider that, in practice, the best performance the model can have in terms of deviation around mean power for a given distance %
 is bounded by the fading value for the given data type, since it defines the fast-varying multipath component which remains after the large-scale effects have been correctly estimated. Therefore, for all results henceforth, apart from the mean difference in terms of received power, we also calculate the standard deviation around the mean difference for the results generated by the model and measurements (i.e., the standard error of the model). %

\begin{figure*}
\centering
\subfigure[\scriptsize Raw data from the Open space Porto (Leca) dataset collected through measurements and generated by the model.]{\label{fig:LOSFitsLecaRaw}\includegraphics[trim=0cm 5.5cm 0cm 5.5cm,clip=true,width=0.35\textwidth]{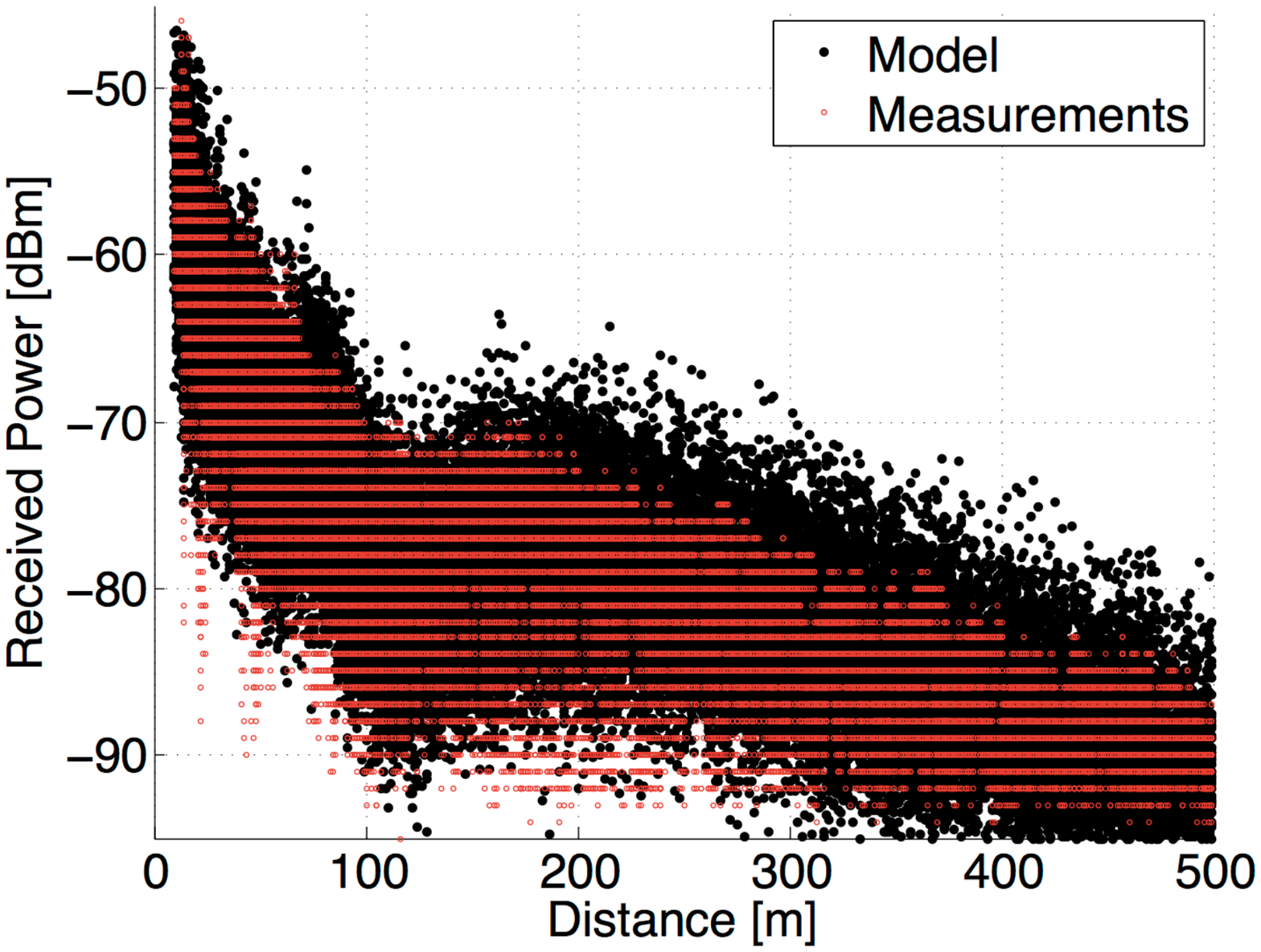}}
  \hspace{5mm}
\subfigure[\scriptsize Open space Porto (Leca) -- Coordinates: 41.210615, -8.713418.  Number of data points: 61000. \textbf{Measured averaged $\sigma$: 3.3~dB}]{\label{fig:LOSFitsLeca}\includegraphics[trim=0cm 5.5cm 0cm 5.5cm,clip=true,width=0.35\textwidth]{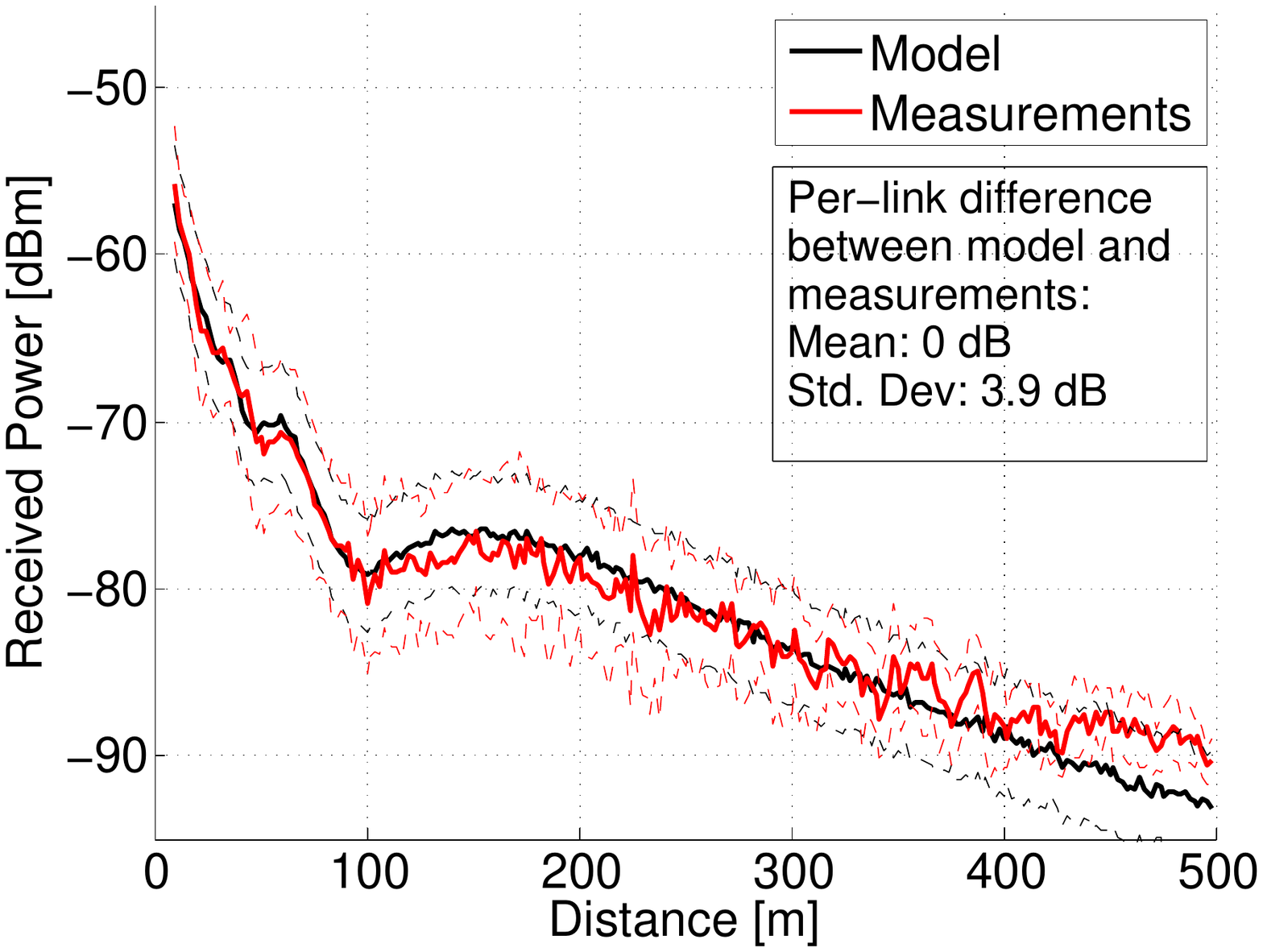}}
\subfigure[\scriptsize Open space Pittsburgh (Homestead Grays Bridge)  -- Coordinates:  40.4103279, -79.9181137. Number of data points: 10700. \textbf{Measured averaged $\sigma$: 4.6~dB}]{\label{fig:LOSFitsHomestead}\includegraphics[trim=0cm 5.5cm 0cm 5.5cm,clip=true,width=0.35\textwidth]{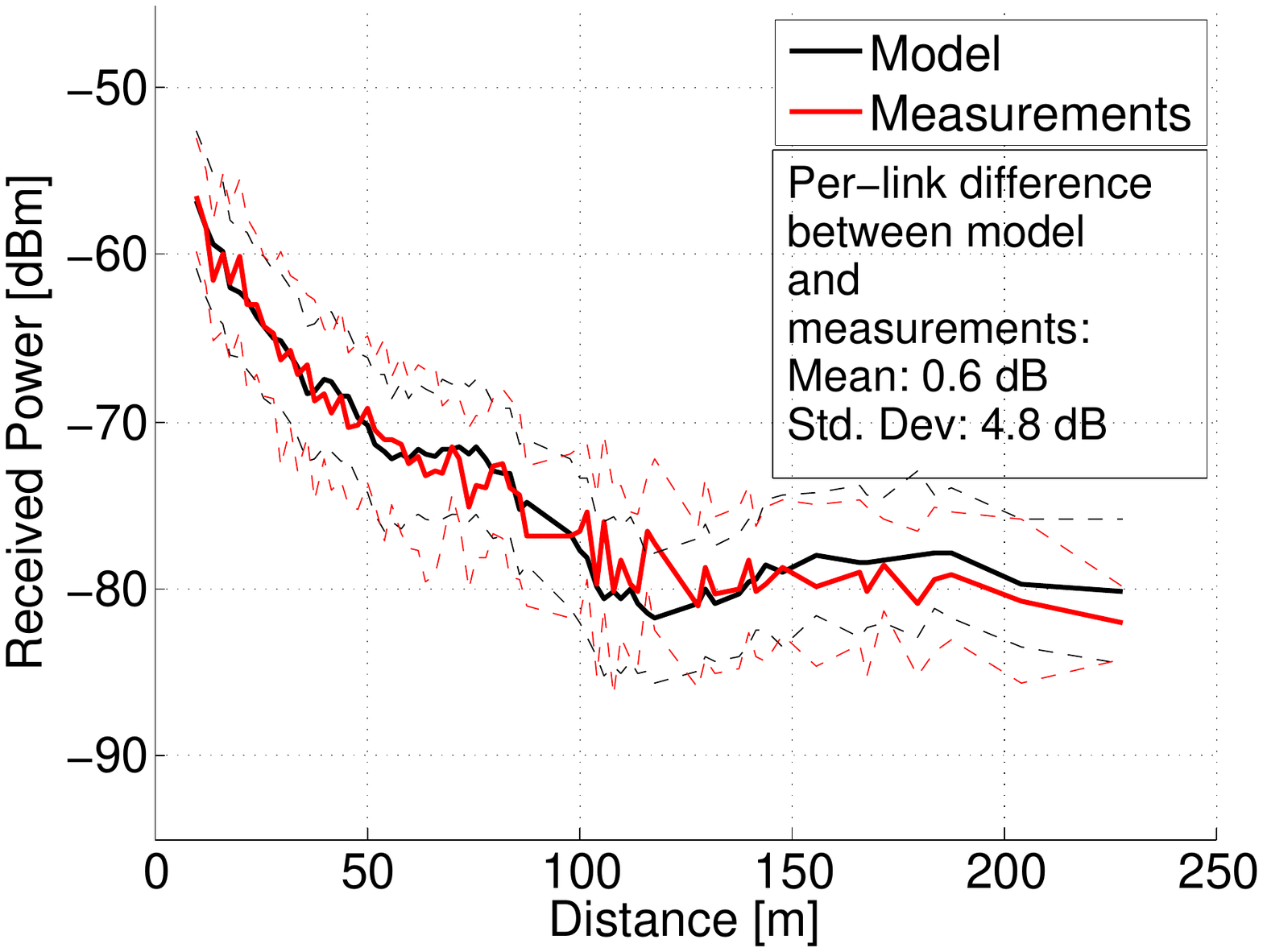}}
  \hspace{5mm}
\subfigure[\scriptsize Suburban Pittsburgh (5th Ave.) Nighttime Experiments  -- Coordinates:  40.4476089, -79.9398574. Number of data points: 11900.  \textbf{Measured averaged $\sigma$: 4.1~dB}]{\label{fig:LOSFits5thNight}\includegraphics[trim=0cm 5.5cm 0cm 5.5cm,clip=true,width=0.35\textwidth]{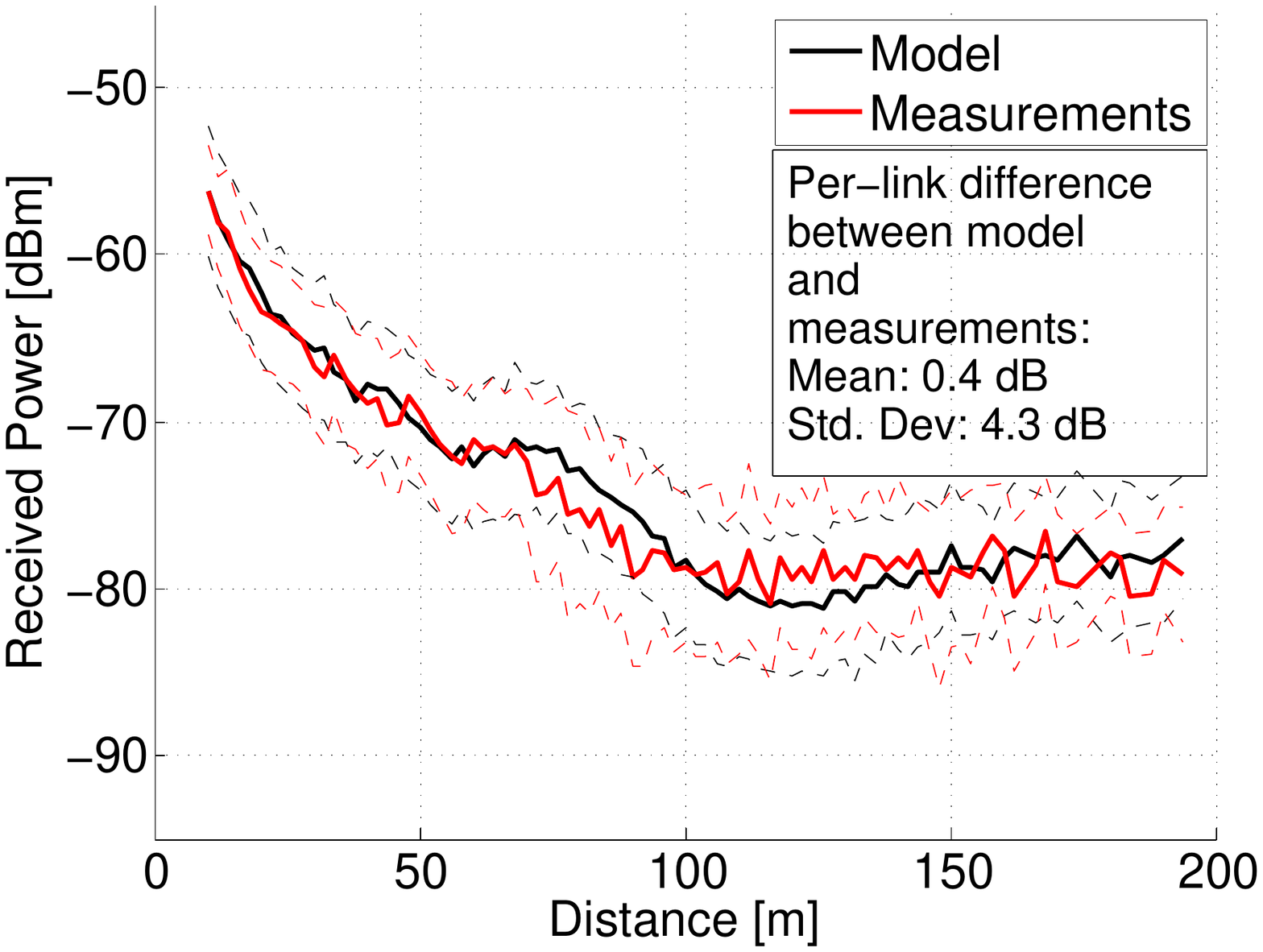}}
\subfigure[\scriptsize Suburban Pittsburgh (5th Ave.) Daytime Experiments  -- Coordinates:  40.4476089, -79.9398574.  Number of data points: 13000. \textbf{Measured averaged $\sigma$: 4.8~dB}]{\label{fig:LOSFitsFifthDay}\includegraphics[trim=0cm 5.5cm 0cm 5.5cm,clip=true,width=0.35\textwidth]{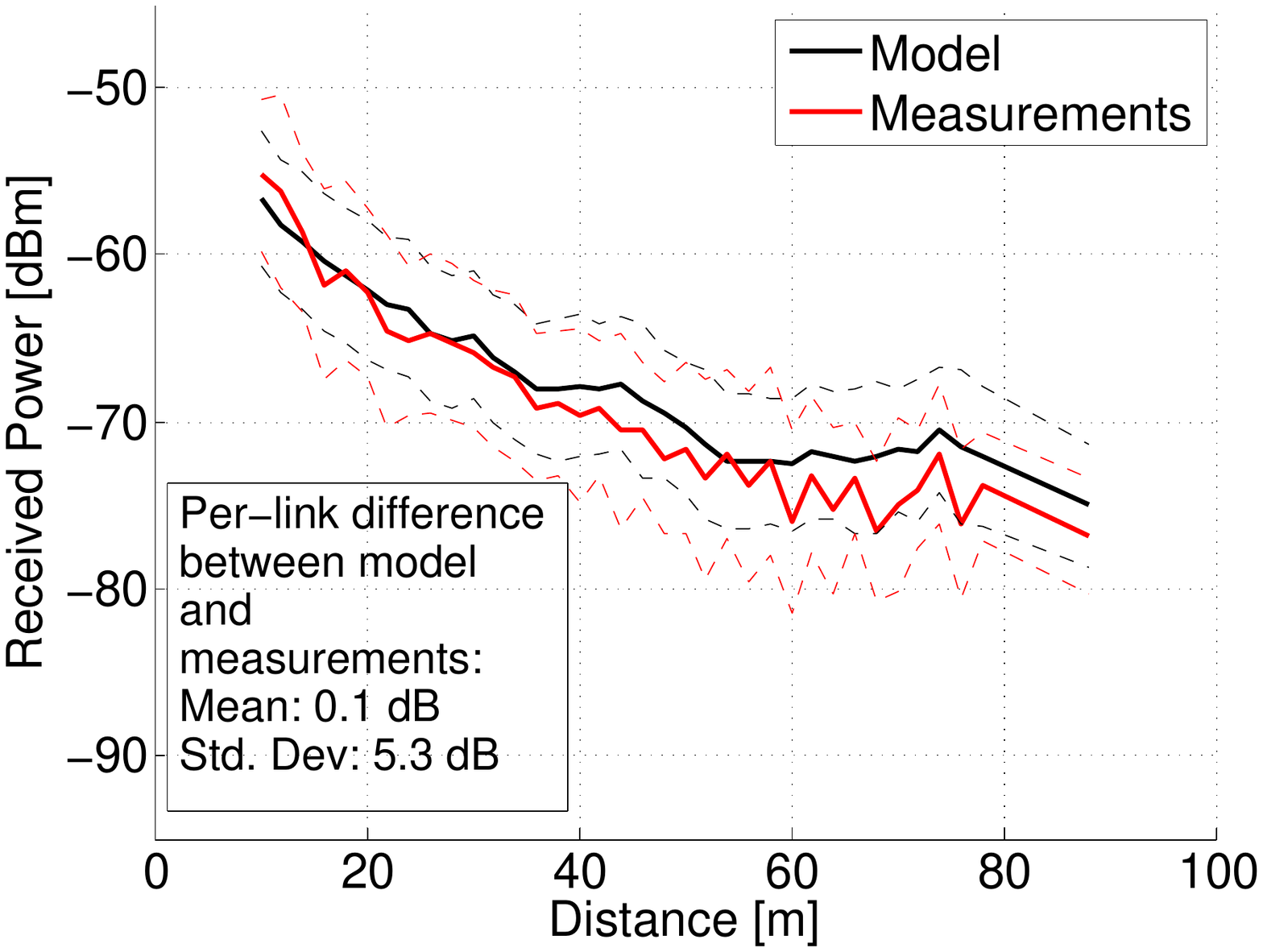}}
  \hspace{5mm}
\subfigure[\scriptsize Urban Porto -- Coordinates: 41.153673, -8.609913. Number of data points: 4400. \textbf{Measured averaged $\sigma$: 5.2~dB}]{\label{fig:LOSFitsUrban}\includegraphics[trim=0cm 5.5cm 0cm 5.5cm,clip=true,width=0.35\textwidth]{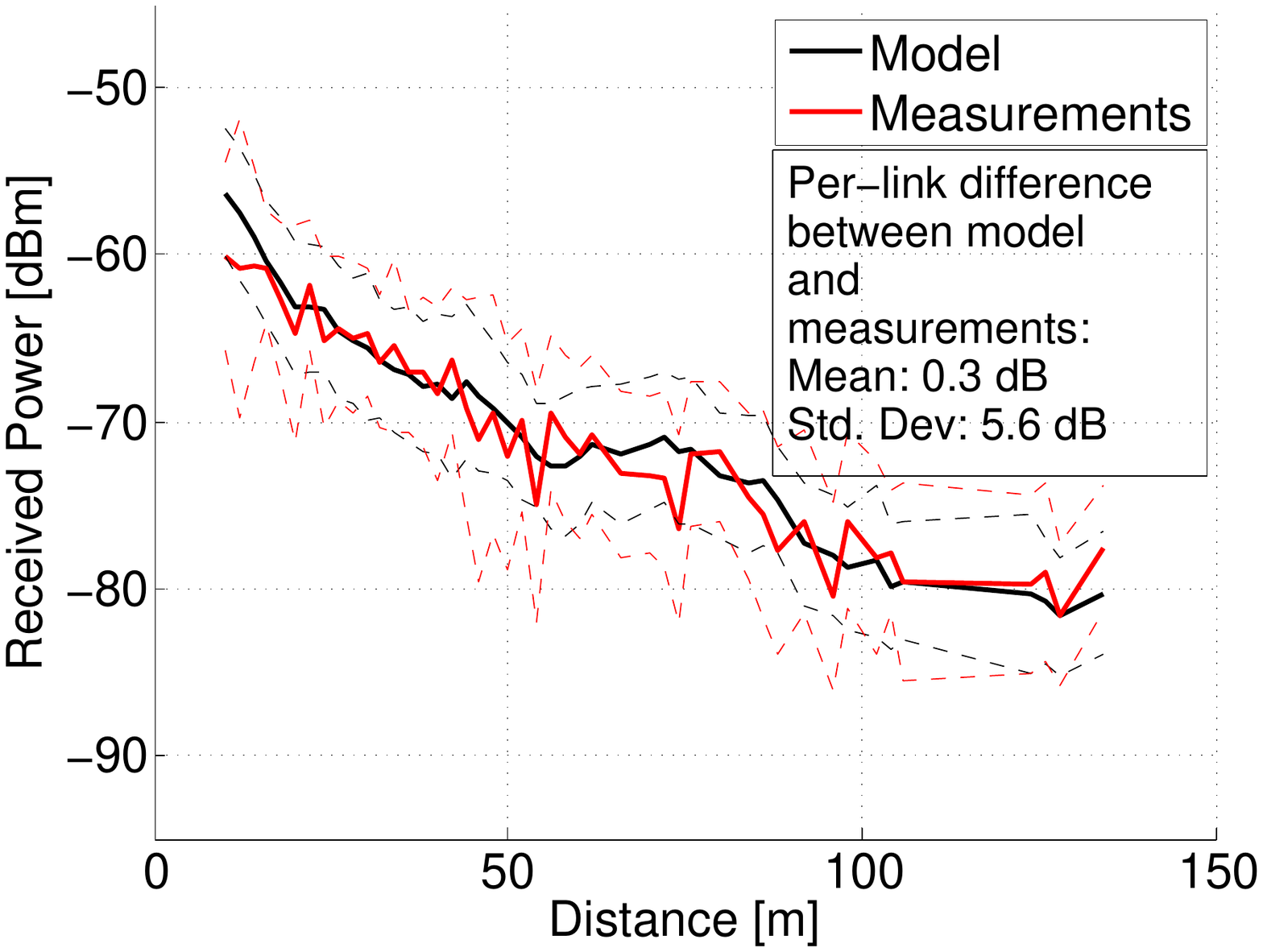}}
\caption[LOS data - model vs experimental measurements]{\small LOS data - model vs experimental measurements. Figure (a) shows the raw data collected through measurements and generated by the model. Figures (b) through (f) show the mean and the standard deviation around the mean received power for two-meter distance bins. Results are plotted only for bins with at least 40 data points. Error bars represent one standard deviation around the mean received power calculated for each distance bin separately. %
All results show the data collected with passenger (short) vehicles. The results with tall vehicles (vans) exhibited similar behavior.} %
\label{fig:LOSFits}
\end{figure*}

\subsection{LOS Links} \label{sec:LOSLinks}

Figure~\ref{fig:LOSFits} shows the results for the LOS data in different environments. The model fits the experimental data quite well in all environments, with the mean difference between model and measurements within 0.6~dB for each of the environments. Similarly, the standard error for all LOS datasets (shown in a text box in each of the subfigures of Fig.~\ref{fig:LOSFits}) is within 0.5~dB of the small-scale fading for that dataset (noted in the caption of each subfigure).
 Regarding the open space LOS results, we attribute the higher fading of the Open space Pittsburgh dataset (Fig.~\ref{fig:LOSFitsHomestead}) compared to the Open space Porto dataset (Fig.~\ref{fig:LOSFitsLeca}) to the guard rails and metal fence (visible in Fig.~\ref{fig:Homestead}), which did not exist in the Open space Porto location (Fig.~\ref{fig:Leca}). 
 The daytime suburban (5th Ave.) measurements (Fig.~\ref{fig:LOSFitsFifthDay}) and urban measurements in downtown Porto (Fig.~\ref{fig:LOSFitsUrban}) have a significantly richer reflection/diffraction environments due to the nearby vehicles in case of the former and both vehicles and buildings in case of the latter. This results in the increase of both the fading ($\sigma$) and the standard error. %
It is important to note that the mean and standard deviation of the difference in received power between the model and experiments is performed on a per-packet basis; for each collected measurement datapoint, we calculate the mean and standard deviation using the 
per-packet received power difference between the model and experiments. Furthermore, we use the same value of relative permittivity ($\epsilon_r$=1.003) in all environments to calculate the reflection coefficient for the ground reflection (i.e., we do not fit the value to a given dataset).

\subsection{NLOSv Links}

\begin{figure*}
\centering

\subfigure[\scriptsize Highway Porto (A28) Passenger Vehicles -- Coordinates: 41.22776, -8.695148. Number of data points: 14200. \textbf{Measured averaged $\sigma$: 4.5~dB}]{\label{fig:NLOSvHighA28CC}\includegraphics[trim=0cm 5.5cm 0cm 5.5cm,clip=true,width=0.4\textwidth]{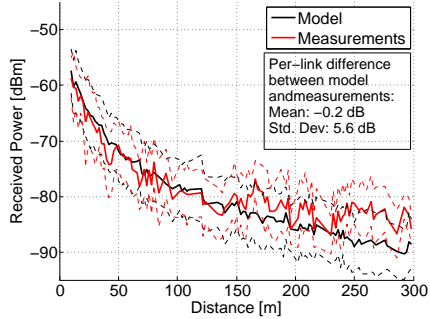}}
  \hspace{5mm}
\subfigure[\scriptsize Highway Porto (A28) Tall Vehicles -- Coordinates: 41.22776, -8.695148. Number of data points: 14700. \textbf{Measured averaged $\sigma$: 3.8~dB}]{\label{fig:NLOSvHighA28VV}\includegraphics[trim=0cm 5.5cm 0cm 5.5cm,clip=true,width=0.4\textwidth]{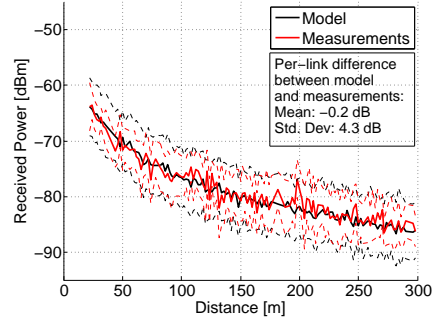}}

\subfigure[\scriptsize Urban Porto Passenger Vehicles -- Coordinates: 41.153673, -8.609913. Number of data points: 6300. \textbf{Measured averaged $\sigma$: 5.3~dB}]{\label{fig:NLOSvFitsUrbanCC}\includegraphics[trim=0cm 5.5cm 0cm 5.5cm,clip=true,width=0.4\textwidth]{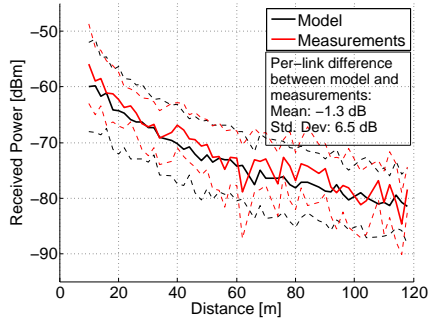}}
  \hspace{5mm}
\subfigure[\scriptsize Urban Porto Tall Vehicles -- Coordinates: 41.153673, -8.609913. Number of data points: 10500. \textbf{Measured averaged $\sigma$: 4.7~dB}]{\label{fig:NLOSvFitsUrbanVV}\includegraphics[trim=0cm 5.5cm 0cm 5.5cm,clip=true,width=0.4\textwidth]{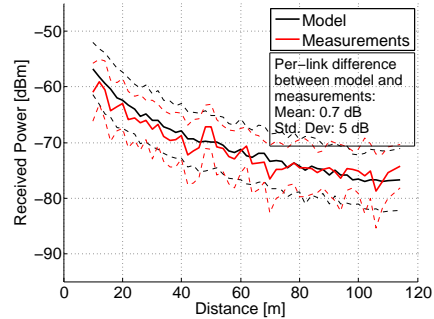}}

\subfigure[\scriptsize Suburban Pittsburgh (5th Ave.) Daytime Experiments  -- Coordinates:  40.4476089, -79.9398574. Number of data points: 9500. \textbf{Measured averaged $\sigma$: 4.5~dB}]{\label{fig:NLOSvFits5thDay }\includegraphics[trim=0cm 5.5cm 0cm 5.5cm,clip=true,width=0.4\textwidth]{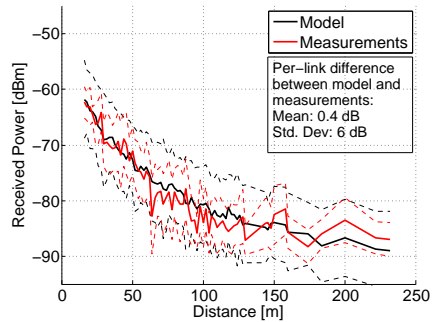}}

\caption[NLOSv data - model vs experimental measurements]{NLOSv data - model vs experimental measurements. Figures show the mean and the standard deviation around the mean received power for two-meter distance bins. Results are plotted only for bins with at least 40 data points. %
 }

\label{fig:NLOSvFits}
\end{figure*}

Figure~\ref{fig:NLOSvFits} shows the results for the NLOSv data in different environments and with both passenger (short) and tall vehicles. The model fits the experimental data well in all environments, with the mean difference between model and measurements within 1.3~dB in each of the environments. Again, the standard error for NLOSv datasets (shown in a text box in each of the subfigures of Fig.~\ref{fig:NLOSvFits}) is within 0.9~dB of the small-scale fading of that dataset (noted in the caption of each subfigure). It is interesting to see that NLOSv results for tall vehicles (vans) experience both lower fading and lower standard error. %
This is due to the advantageous position of the antennas on top of the vans, which are shadowed less frequently than the antennas on the shorter vehicles. This results in a more stable channel with less fading.

\subsection{NLOSb Links}

\begin{figure*}
\centering

\subfigure[\scriptsize Reflections (green lines) and diffractions (magenta) generated by the model and overlaid on the image of the Porto Outlet location. The vehicles started close to each other with clear LOS and slowly moved along paths indicated by the arrows, thus going from LOS to NLOSb conditions. During the measurements, the two large buildings that create reflections and diffractions were the only large protruding objects in the scene, with clearance in excess of 100~meters to the nearest objects (i.e., there were no parked vehicles). Coordinates of the location: 41.300137, -8.707385]{\label{fig:OutletOverlayReflDiffr}\includegraphics[width=0.75\textwidth]{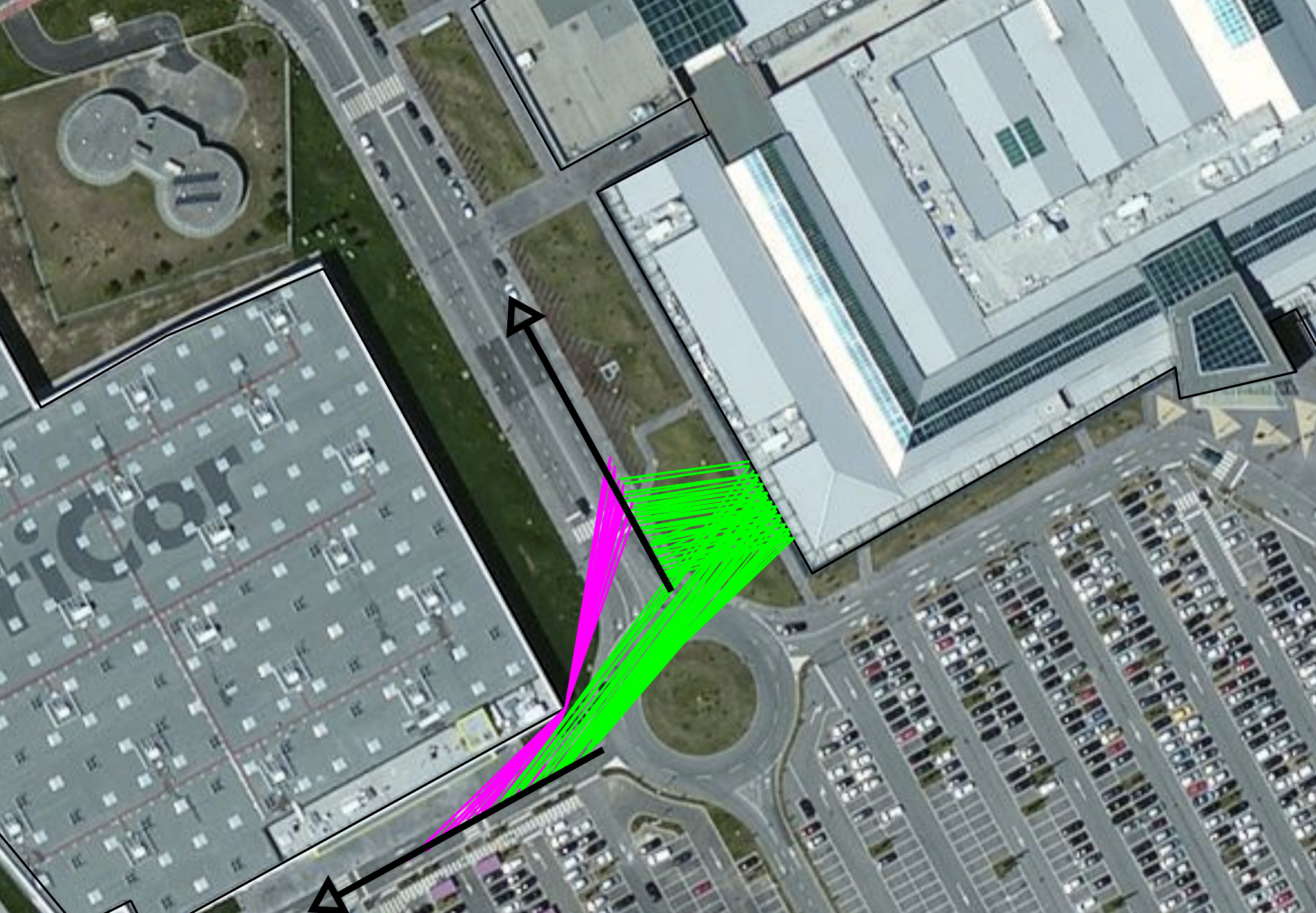}}
\subfigure[\scriptsize Transition of the model between LOS and NLOSb conditions at the Porto Outlet location for a single 30-second run. The distinct conditions are annotated. The outliers in the top right corner are due to GPS inaccuracy while vehicles were stationary. For comparison, we plot the log-distance path loss for the same location.]{\label{fig:outletAnnotated}\includegraphics[%
width=0.75\textwidth]{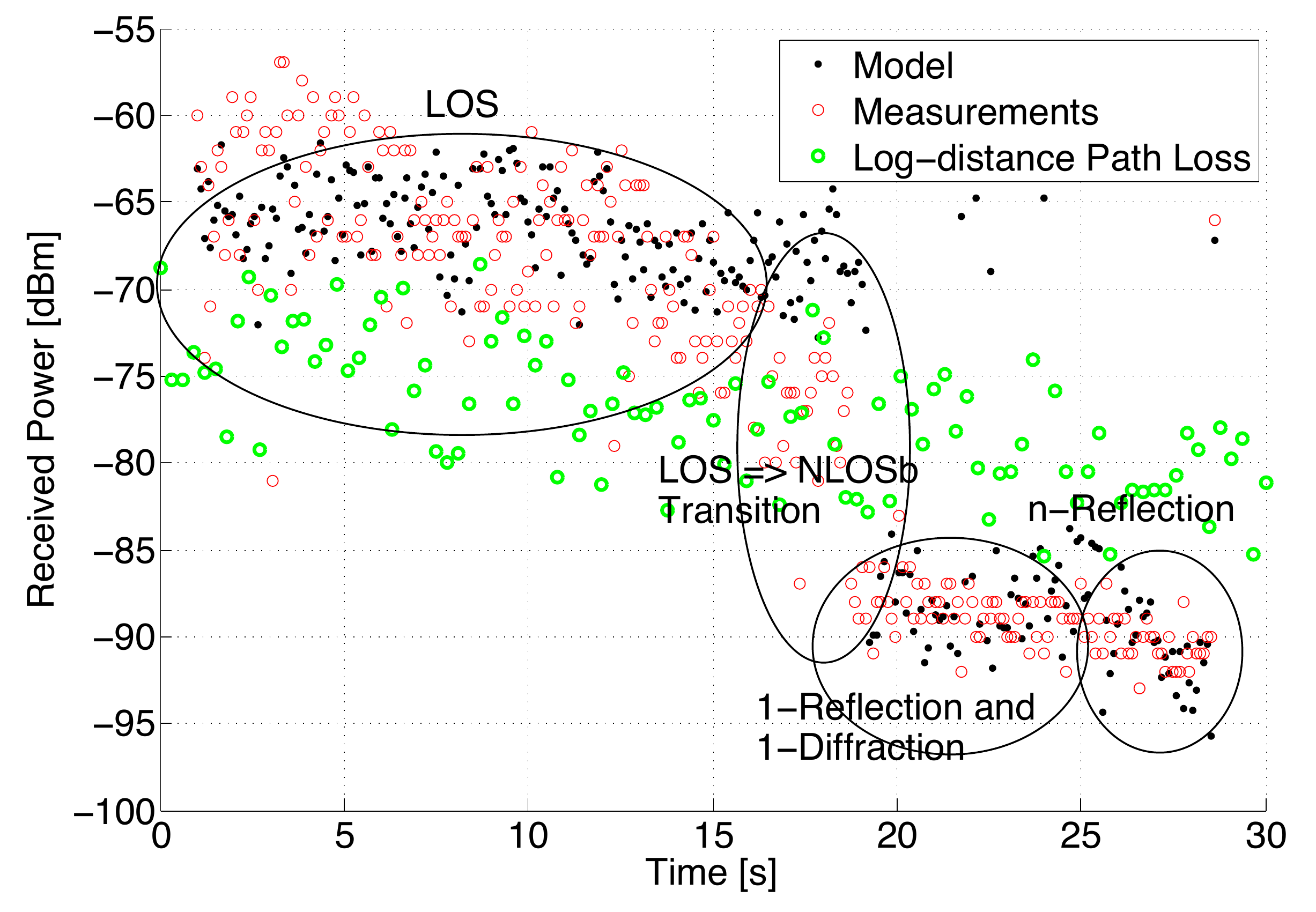}}
\caption{Outlet NLOSb experiment.}

\label{fig:Outlet}
\end{figure*}

\begin{figure}
  \begin{center}
    \includegraphics[trim=0cm 5cm 0cm 5cm,clip=true,width=0.5\textwidth]{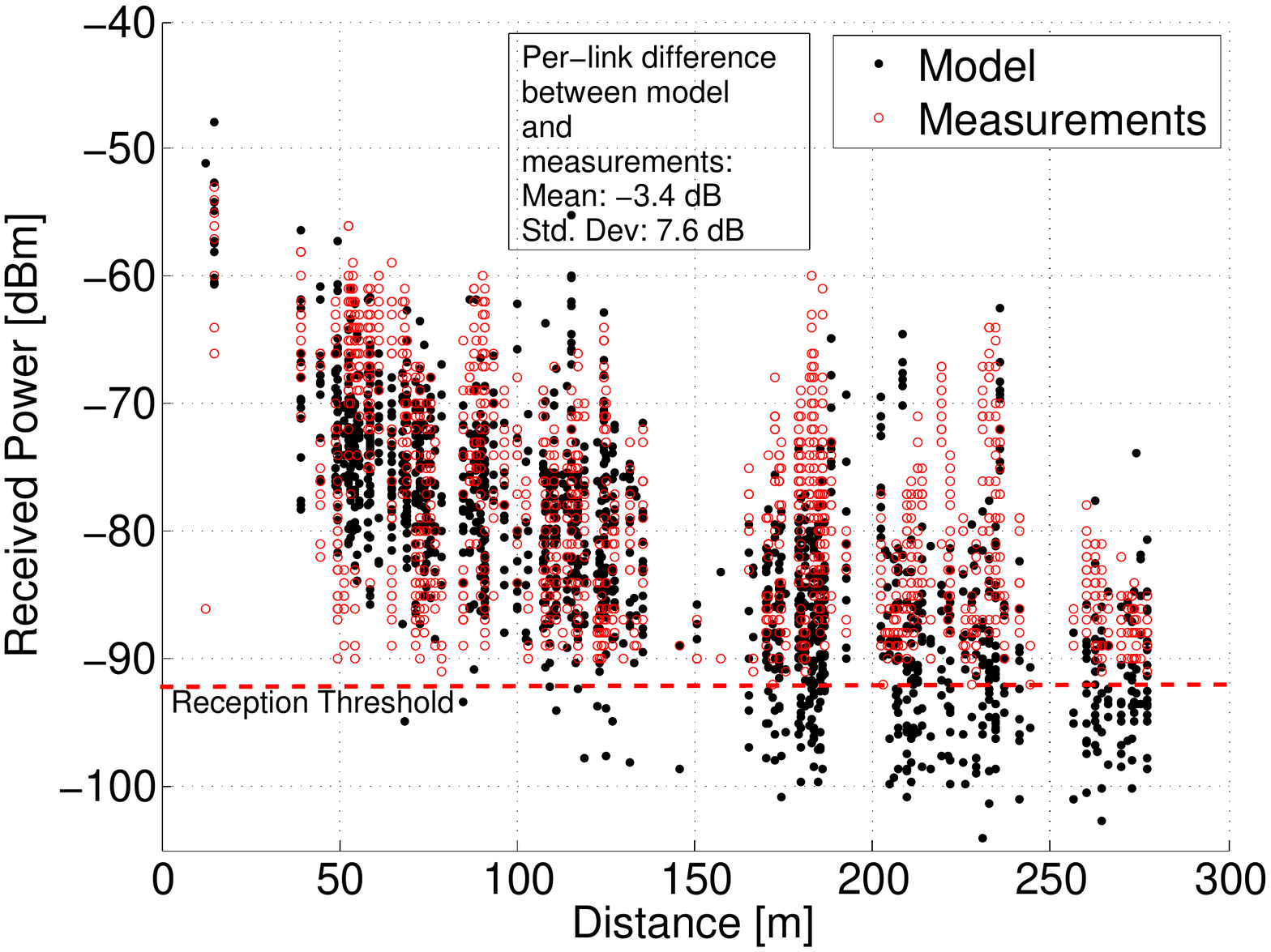}
     \caption[NLOSb data - model vs experimental measurements]{\small NLOSb data - model vs experimental measurements for downtown Porto dataset and passenger vehicles. Figure shows raw data collected in the measurements and generated by the model. The data received at distances above 130~m is predominantly due to ``around the corner'' situations, where the LOS is not severely obstructed (i.e., the diffraction loss is not high) and reflections off neighboring buildings often exist. \textbf{Measured averaged $\sigma$: 6.8~dB.} }
      \label{fig:NLOSbFits}
   \end{center}
\end{figure}

Figure~\ref{fig:OutletOverlayReflDiffr} shows the Porto Outlet location with the overlaid reflecting and diffracting rays as generated by the model. Once the vehicles are not in LOS, the predominant propagation mechanisms become single-interaction reflections and diffractions. Figure~\ref{fig:outletAnnotated} shows distinct transitions of the received power as the vehicles transition from LOS to NLOSb. The model is able to capture the steep drop in the received power once the LOS is obstructed by building. At the same time, the log-distance path loss, because it is unable to capture the transition between LOS and NLOSb, underestimates the received power in LOS conditions and overestimates it in NLOSb conditions. This result points out the importance of location-specific channel modeling: the transitions between different LOS conditions at exact times when they occur can only be accounted for by taking into account the objects in the location. On the other hand, models relying on the common parameters of a large environment (such as the overall path-loss exponent in the case of log-distance path loss) are unable to model such transitions, which results in ``averaging'' of the received power between different LOS conditions. Therefore, for such models to be useful for calculation of per-link channel characteristics, at minimum the different LOS conditions need to be identified and modeled separately.

Furthermore, a good match between experiments and model shows that transmission through the building does not play an important role. This confirms previous measurement studies that made similar observations: Anderson in~\cite{anderson98} performed experiments at 1.8~GHz and modeled the diffraction and reflection around an isolated building corner using uniform theory of diffraction (UTD). The author concludes that through-wall transmission is negligible compared to the corner diffraction and wall reflections. Similarly, Durgin et.~al in~\cite{durgin98} performed experiments at 5.85~GHz and point out that ``transmission through the house was not as important as outdoor multipath scattering''. For this reason, we do not consider the through-building transmission as an important effect and therefore do not include it in our model. 

Figure~\ref{fig:NLOSbFits} shows NLOSb data for the experiments performed in downtown Porto. The difference between the model and experiments is notably higher than in the case of LOS and NLOSv. Furthermore, NLOSb is the only data type where significant portion (13\%) of correctly received and decoded experimental data was classified as below the reception threshold (-92dBm) by the model, resulting in 3.4~dB mean power difference. This disparity is due to three reasons: 1) the large fading of the NLOSb dataset, where the variation around the signal reception threshold makes the model generate the data which is below threshold; 2) during experiments, only decodable data can be recorded (i.e., anything below the reception threshold of -92~dBm is not recorded); and 3) along the route, there was occasional foliage which we did not have in the geographical database, thus it was not modeled. To put 1) and 2) into perspective, the packet delivery rate for NLOSb data from 100~m to 500~m is below 20\% (i.e., more than 80\% of the packets sent above 100~m is not received).

\subsection{Combined large-scale and small-scale signal variation}

\begin{figure}
  \begin{center}
    \includegraphics[trim=8cm 20.5cm 8cm 20.5cm,clip=true,width=0.45\textwidth]{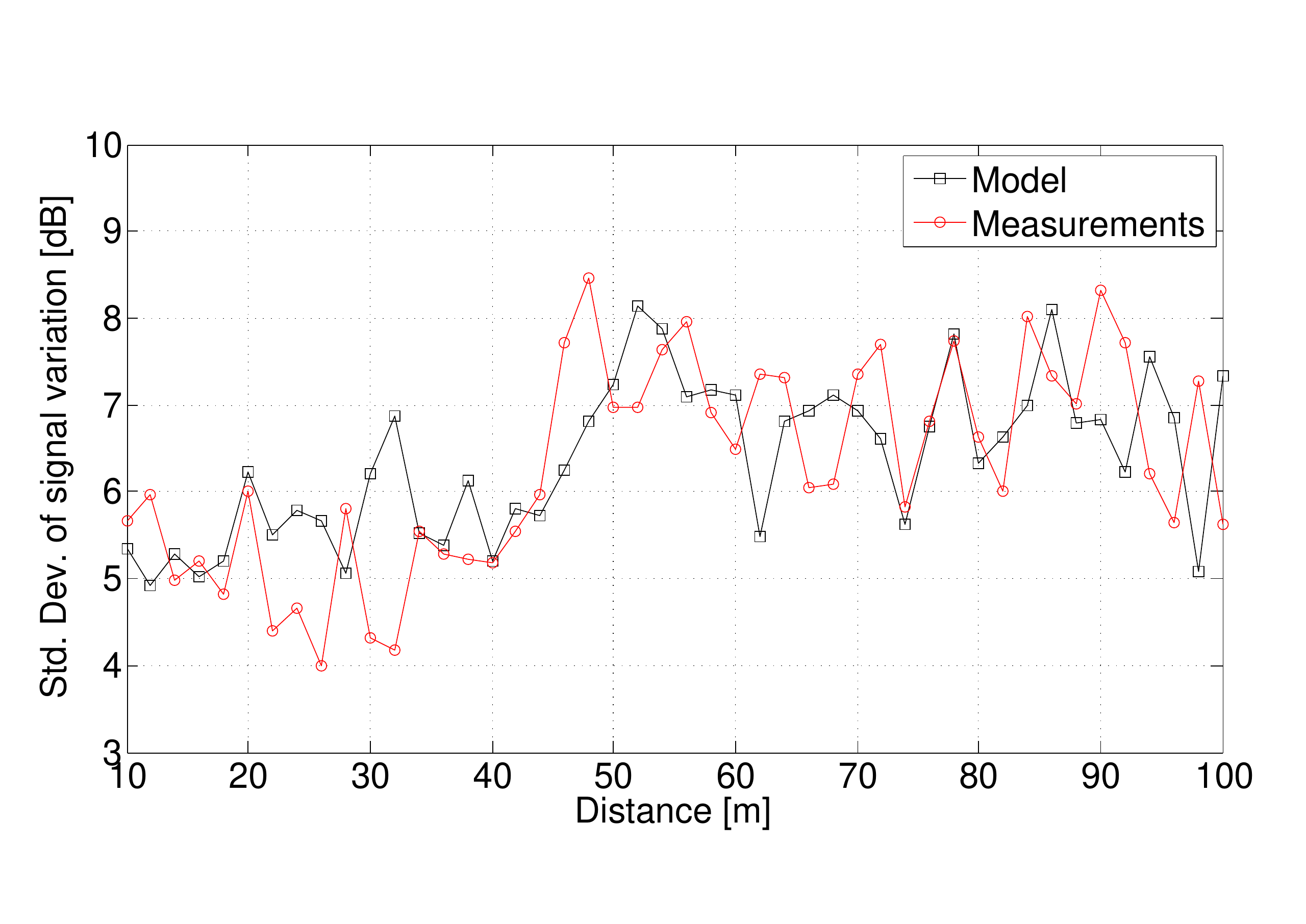}
     \caption[Comparison of fading generated by the model and measurements]
     {Standard deviation of signal variation generated by the model (additional small-scale variation included) and extracted from measurements in downtown Porto. All three link types (LOS, NLOSv, NLOSb) combined. Two-meter distance bins. Only bins with more than 40 data points are included.}
      \label{fig:fadingModelVsMeasurements}
   \end{center}
\end{figure}

Figure~\ref{fig:fadingModelVsMeasurements} shows the signal variation around the mean for the model and measurements. 
For each two-meter bin, the variation is a composite result of the variation generated by the large-scale model (that also includes a part of small-scale effects through single-interaction reflections and diffractions) with the addition of the zero-mean, normally distributed variable with standard deviation $\sigma$ determined using eq.~\ref{eq:FadingCalc} that represents the additional small-scale (multipath) variations. 
 The model generates the overall signal variation comparable to that obtained through measurements, with %
 the variation across the distance bins for both the model and the measurements of approximately 6.3~dB. %
 
\begin{figure}
  \begin{center}
    \includegraphics[width=0.45\textwidth]{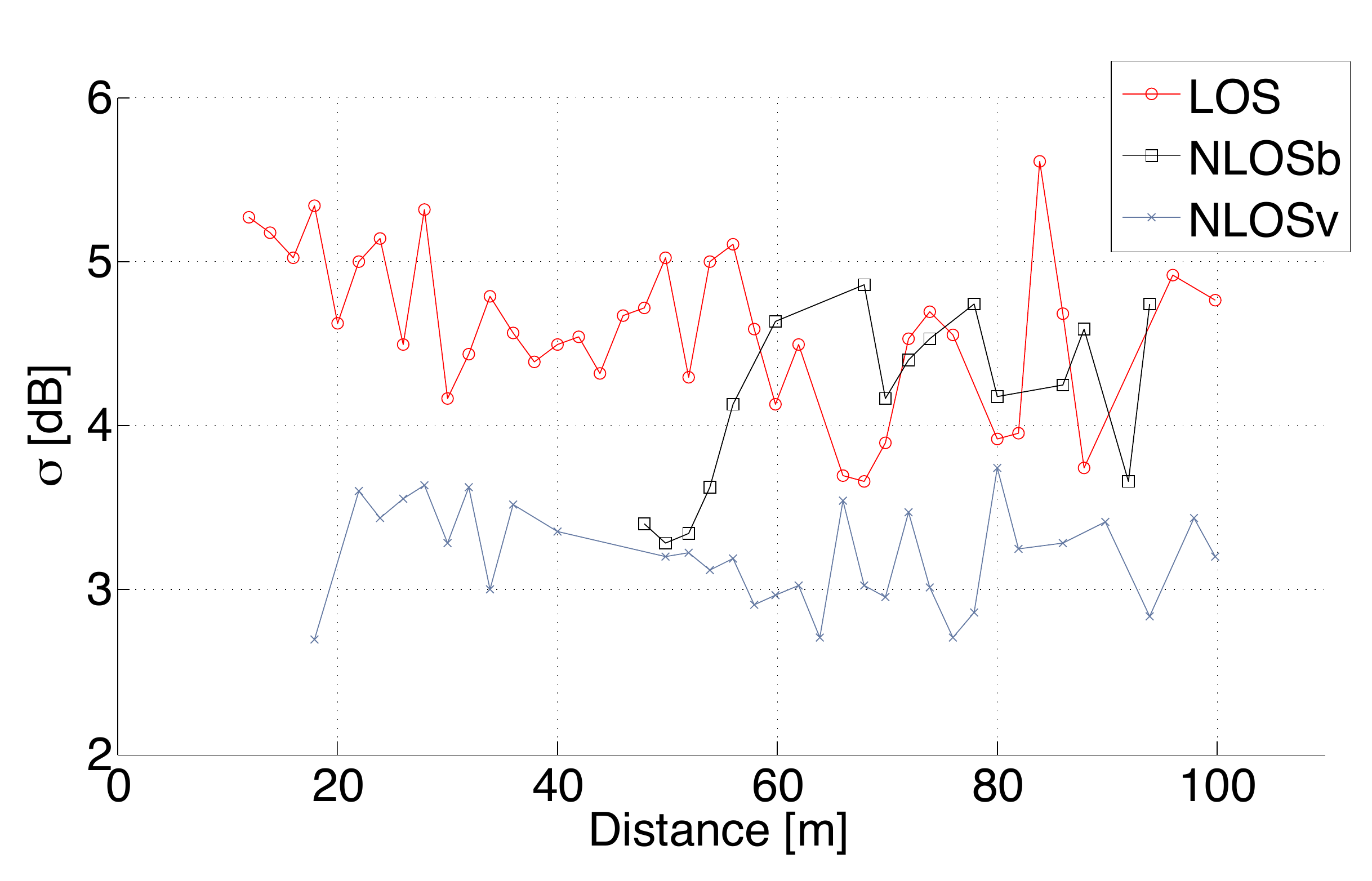}
     \caption[Values of $\sigma$ generated by the model for downtown Porto]{Values of $\sigma$ generated by the model for downtown Porto for different link types. Two-meter distance bins. Only bins with more than 40 data points are included.}
      \label{fig:sigmaLOSConditions}
   \end{center}
\end{figure}

 Figure~\ref{fig:sigmaLOSConditions} shows the value of $\sigma$, i.e., the deviation of the additional small-scale signal variation (eq.~\ref{eq:FadingCalc}) as generated by the model. %
The value of $\sigma$ is comparatively lower for the NLOSv and NLOSb links, since the variation generated by the most significant reflected and diffracted rays is already included in the large-scale model. Unlike the data generated by the model, in the measurement data we have no way of distinguishing the signal variation generated by the large-scale and small-scale effects (apart from splitting the links in LOS, NLOSv, and NLOSb categories). However, we know that for the LOS links, the only deviation from the theoretical two-ray ground reflection model is generated by $\sigma$, therefore the per-bin average $\sigma$ in Fig.~\ref{fig:sigmaLOSConditions} is higher than in case of NLOSv and NLOSb, where the signal variation is generated by both $\sigma$ and the large-scale model.

 The results show that the implemented small-scale model can capture the fast-varying signal changes in vehicular environment by considering the objects surrounding the communicating pair. %

\section{A Few Notes on the Performance of the Model}\label{sec:Performance}
\begin{figure*}
\centering

\subfigure[\scriptsize Processing time for different network sizes.]{\label{fig:totalSimTime}\includegraphics[trim=8cm 20cm 8cm 20cm,clip=true,width=0.4\textwidth]{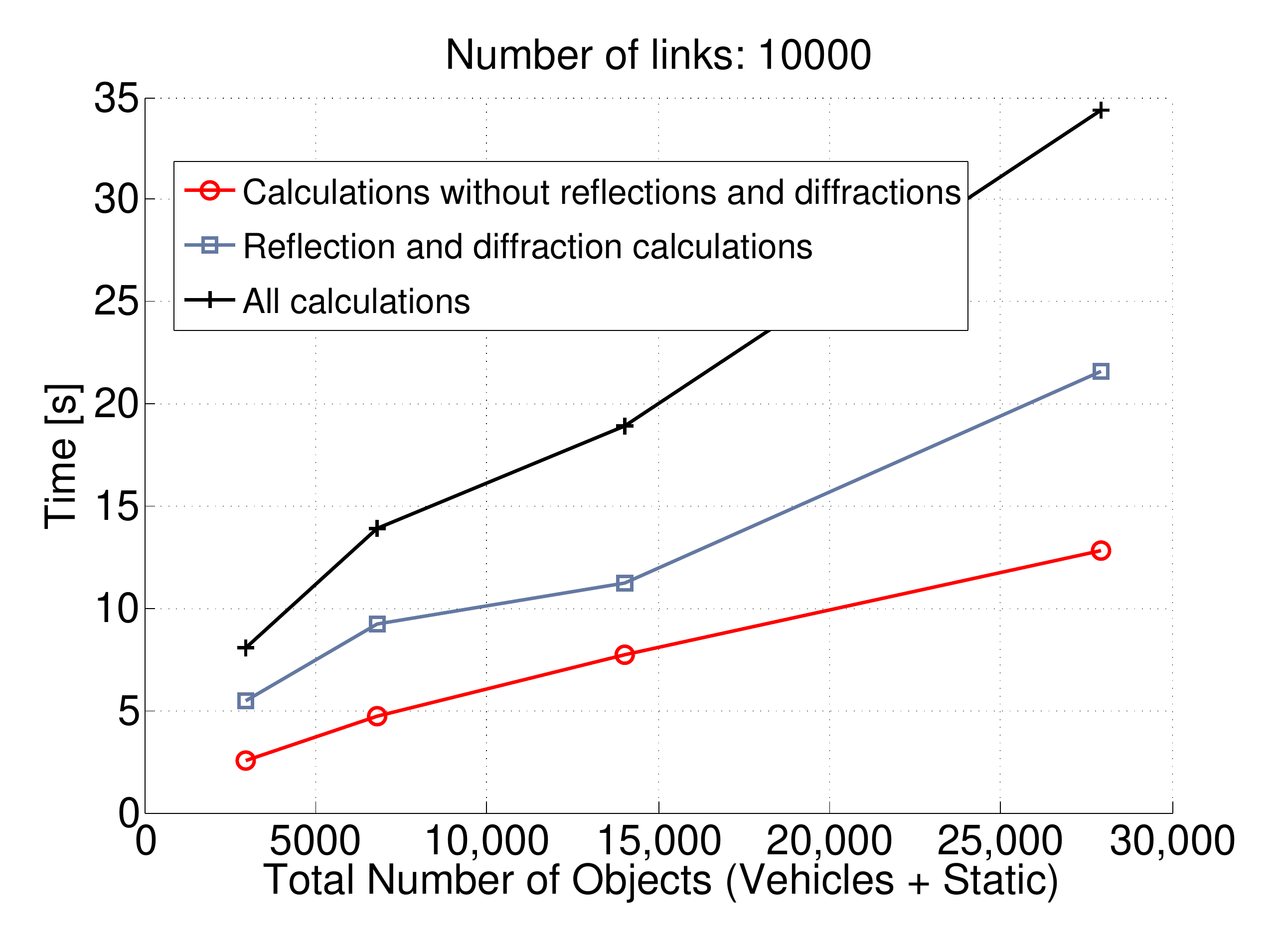}}
  \hspace{5mm}
\subfigure[\scriptsize R-tree construction times. Note that static R-tree needs to be constructed only once, whereas vehicle R-tree is constructed at each time step.]{\label{fig:RTreeTimes}\includegraphics[trim=8cm 20cm 8cm 20cm,clip=true,width=0.4\textwidth]{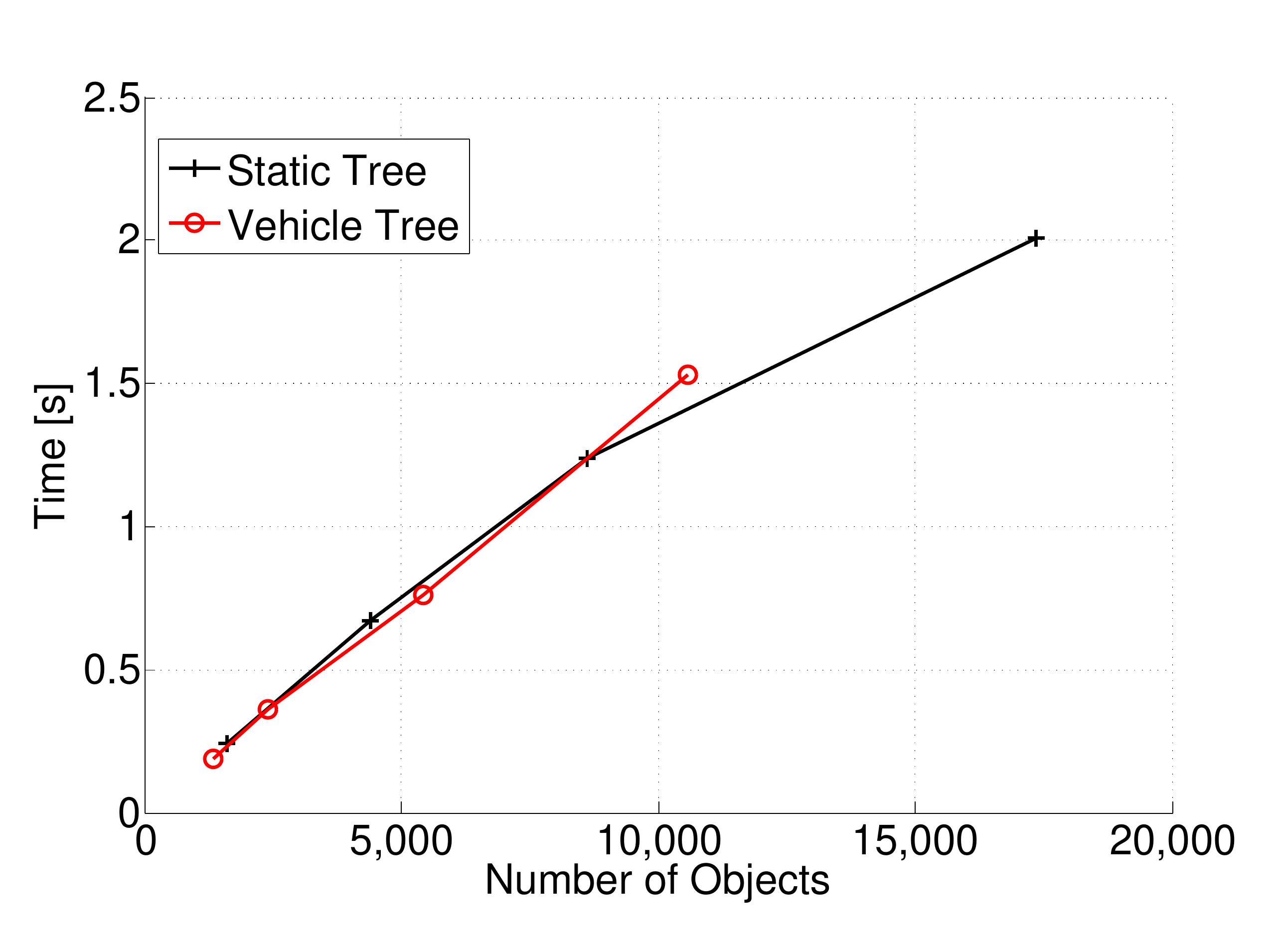}}

\subfigure[\scriptsize Time to classify fixed number of links (10000) for different network sizes.]{\label{fig:10kLinkClassification}\includegraphics[trim=8cm 20cm 8cm 20cm,clip=true,width=0.4\textwidth]{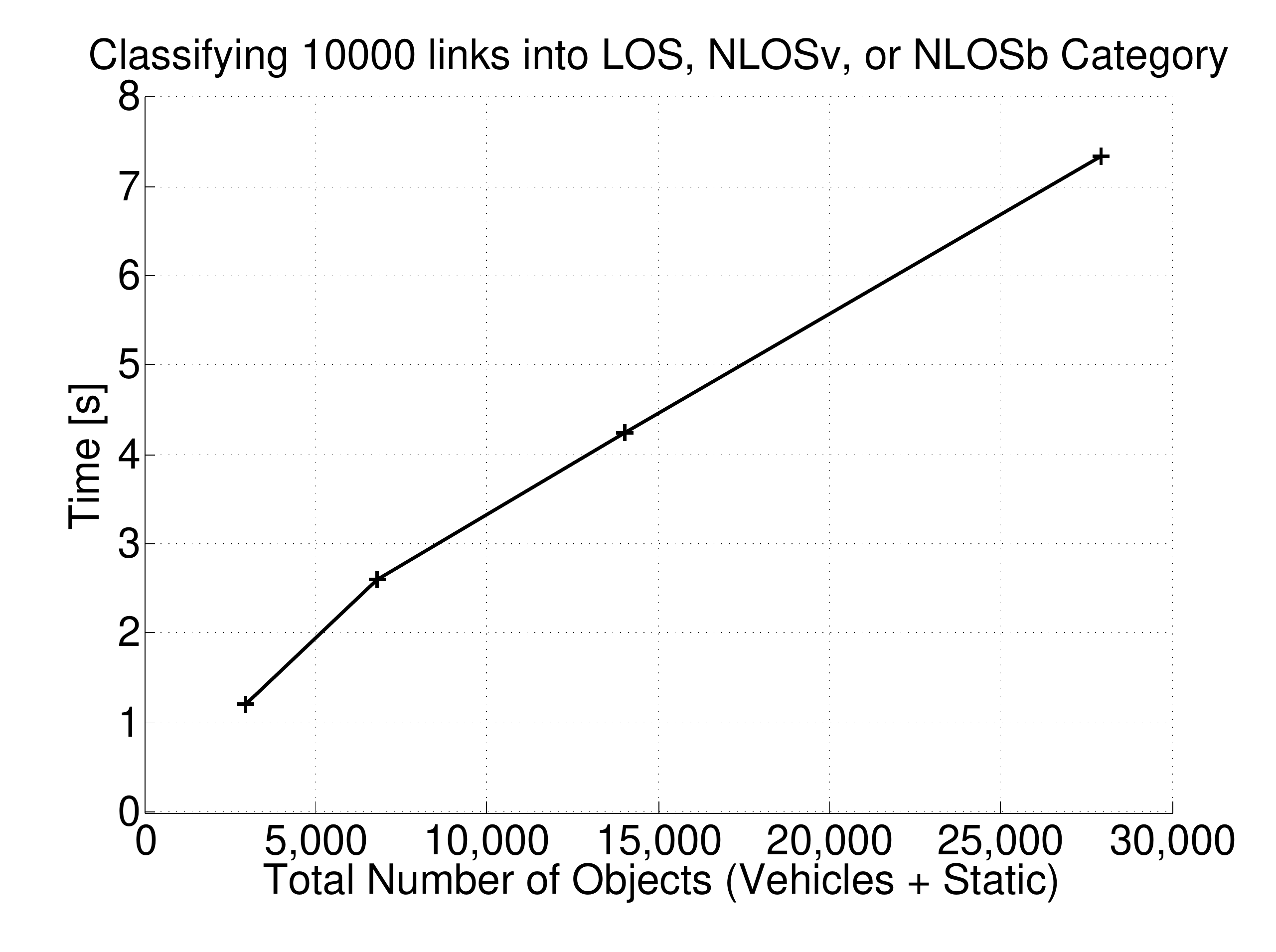}}
\hspace{5mm}
\subfigure[\scriptsize Time to classify different number of links for a fixed network size.]{\label{fig:linkClassificationTime}\includegraphics[trim=8cm 20cm 8cm 20cm,clip=true,width=0.4\textwidth]{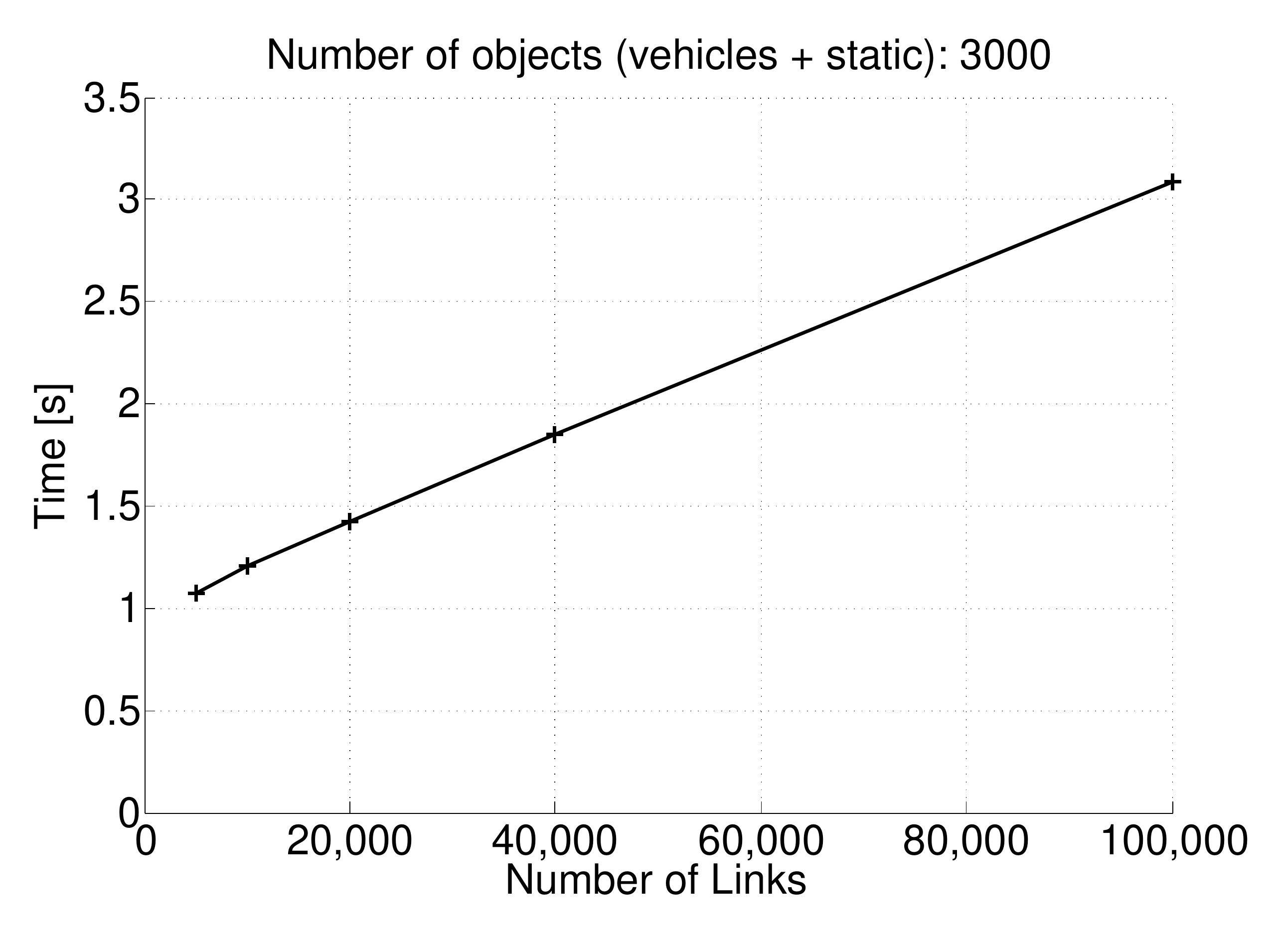}}

\caption[Performance of the Implemented Model]{Calculation times for various parts of the model on the downtown Porto dataset. We used the following hardware: 2011 MacBook Air, 1.7GHz Core i5, 4~GB RAM. The complete dataset contains 10566 vehicles and 17346 static objects over 41 km$^2$. For simulations on smaller networks (first three data points in subfigures (a), (b), and (c)), we used half, quarter, and eighth of the entire city area, which contained corresponding number of vehicles and static objects. All results are for single-core operation (i.e., no parallelization).  %
 }

\label{fig:modelPerformance}
\end{figure*}

We implemented the complete model in MATLAB. We were able to simulate the entire city of Porto, with an area of approximately 41~km$^2$ containing 10566 vehicles and 17346 static objects, on commodity hardware (2011 MacBook Air) and the communication ranges shown in Table~\ref{tab:rLOS}. %
Figure~\ref{fig:modelPerformance} shows the processing times for the most salient parts of the model. Figure~\ref{fig:totalSimTime} shows the complete time it takes to determine channel conditions for 10000 links and varying network sizes. By increasing the network size (i.e., number of objects in the scene), the processing time increases fairly linearly, even for the largest network size with 28000 objects. Figure~\ref{fig:totalSimTime} also shows separately the time to perform the most computationally intensive operation: calculating reflections and diffractions. Across different network sizes, calculating reflections and diffractions (which are calculated for NLOSb links only) accounts for two-thirds of the computation time. For this reason, we plan to explore if comparably realistic results can be obtained for NLOSb links without explicitly calculating the reflections and diffractions (e.g., by utilizing a log-distance path loss with appropriate exponent~\cite{Cheng2007}). 

Figure~\ref{fig:RTreeTimes} shows that the R-tree construction scales linearly with the number of objects that need to be stored in the tree. The results for constructing vehicle and static object R-trees are similar, since it takes only marginally more time to fit the more complex static objects (outlines of buildings/foliage) in the minimum bounding rectangles. After that stage, the calculations per object are identical. Figure~\ref{fig:10kLinkClassification} shows the increase in link classification time when the network size (and therefore, the vehicle and static R-tree size) increases. Again, the increase is linear with the size of the network.

Network simulators need to account for interference from neighboring communicating pairs. In order to calculate Signal to Interference plus Noise Ratio (SINR), signal contributions from all currently transmitting neighboring vehicles need to be taken into account. Depending on the employed medium access protocol, this might imply calculating a large number of parallel and interfering transmissions from neighbors. For example, time and frequency division protocols will on average result in low interference, whereas random access protocols (such as the default medium access scheme of IEEE 802.11p -- CSMA/CA~\cite{astm03}) will generate higher interference, and thus higher number of links that need to be taken into consideration. This implies that, to calculate the SINR for a link, the interference from a relatively large number of active neighboring links might also need to be calculated. However, with regards to per-link processing time, Fig.~\ref{fig:linkClassificationTime} shows that, for a fixed network size, increasing the number of links results in a linear increase of processing time with a mild slope (e.g., to classify 5000 links, it takes 1.1 second, whereas for 100000 links it takes 3.1 second). Therefore, the additional burden due to calculating the power from nearby interfering links is not overly high for the proposed model.

Furthermore, it has to be noted that, from the computational complexity point of view, selecting the correct communication range if quite important. Increasing the range results in quadratic increase in the number of objects neighboring a given communication pair. By design, in the extreme case, if the communication range is equal to the size of the simulated area, the number of neighboring objects (and therefore calculations) is quadratic with the number of communicating pairs. Therefore, the communication range for each of the LOS types needs to be carefully chosen so that it is minimized while accounting for potentially communicating pairs. 

With regards to the scalability of the model, the trends shown in Fig.~\ref{fig:modelPerformance} are far more important than the actual processing times. The result show linear behavior even for large networks comprising tens of thousands of objects and communicating pairs. We also point out that the operations required by the model (R-tree construction and classification of links through object querying and intersection tests) are inherently parallelizable. Since the model relies on geometric manipulations of the objects that impact the channel, analogies can be made to computer graphics domain, where parallel rendering techniques are utilized to perform occlusion/visibility and intersection testing~\cite{chazelle89, molnar94}. Parallelization techniques can be employed in both the object querying and intersection testing, as well as the R-tree construction. %
Since there is no dependency between different communication pairs (links), parallelizing the computations across different links is straightforward. Furthermore, recent advances in parallel R-tree construction, querying, and intersection testing (e.g., see Luo et.~al in~\cite{luo12}) indicate that significant speed increase can be obtained by using multicore graphics processing units.

\section{Related Work}\label{sec:relWorkComplete}

Several recent studies tackled efficient and realistic simulation of vehicle-to-vehicle channels in different VANET environments. Karedal et al.~in~\cite{karedal10} and Mangel et al.~in~\cite{mangel11_2} designed channel models focused on street intersections, where buildings create non-LOS (NLOSb) conditions. Both studies selected representative urban intersections and performed measurements which were then used to design channel models and calibrate the path loss and fading parameters. Karedal et al.~in~\cite{karedal09} designed a %
V2V channel model based on measurements performed in highway and suburban environment at the 5.2 GHz frequency band. The model distributes the vehicles and static objects randomly and analyzes four distinct signal components: LOS, discrete components from vehicles, discrete components from static objects, and diffuse scattering. Based on the measurements, the authors propose a set of model parameters for highway and suburban environment. While it enables modeling of different propagation characteristics (path loss, multipath, Doppler spread, etc.), the proposed model assumes the LOS component exists, therefore it does not specify how to determine the LOS conditions of the channel and the transitions between LOS, NLOSv, and NLOSb. %
Figure~\ref{fig:Outlet} shows that modeling the transitions between the LOS conditions is essential for obtaining realistic results, since the ensuing path loss is the most important component in determining the received power and, consequently, the decodability of the packet.

Sommer et al.~in~\cite{sommer11} perform measurements based on which they calibrate a computationally efficient path loss model aimed at distinguishing between the LOS and NLOSb conditions. The authors design the model based on the assumption that transmission through buildings is the main propagation mechanism, while diffractions and reflections are not modeled. This goes against previous experimental and modeling studies reported by Anderson in~\cite{anderson98} and Durgin et al.~in~\cite{durgin98}, which concluded that reflections and diffractions are the dominant propagation mechanisms in the 1.9 and 5.9~GHz frequency bands. Furthermore, the results we obtained in isolated experiments (Fig.~\ref{fig:Outlet}) show that the V2V channel obstructed by buildings for DSRC systems in the 5.9~GHz band is well represented by accounting for diffractions and reflections only, thus indicating that the transmission through buildings plays a minor role.  

In terms of channel modeling on a city-wide scale, studies reported by Giordano et al.~in~\cite{giordano10}, Martinez et al.~in~\cite{Martinez2009}, and Cozzetti et al.~in~\cite{cozzetti12} focus on computationally efficient channel modeling in grid-like urban environments, where streets are assumed to be straight and intersecting at a right angle. While such assumptions hold for certain urban areas, in others they might be overly simple (e.g., in the city of Porto -- Fig.~\ref{portoBuildings}).
With regards to improving the channel modeling using location-specific information, Wang et al.~in~\cite{wang12} utilize aerial photography to determine the density of scatterers in the simulated area. By processing the aerial data to infer the scatterer density, the authors determine the fading level for a given location on the road.  

A number of studies were performed in various VANET environments to estimate the channel by performing measurements and fitting the measured data using well-known models (e.g., log-distance path loss~\cite{00parsons}). Cheng et al.~in \cite{Cheng2007} performed measurements in suburban Pittsburgh, PA, USA, in the 5.9 GHz frequency band. The authors fit the measurement data to a dual slope piecewise log-distance path loss model (i.e., they use two path loss exponents and two fading deviations). They also make an observation that buildings and vehicles significantly impact the received power and that there exists a need for ``a multi-state model, with different states being applicable when a line of sight does and does not exist between the vehicles''. Similarly, Paschalidis et~al. in~\cite{paschalidis11} make measurements in different environments (urban, suburban, rural, highway) and fit the measurements data to the log-distance path loss model. Again, the path loss exponent varies considerably (between 1.83 and 3.59) for different locations and LOS conditions. The large range of path loss exponent reported in these studies goes to show that a single path loss exponent can not capture the characteristics of a channel, even for a single location/environment. Therefore, different LOS conditions (LOS, NLOSv, NLOSb) need to be distinguished and modeled separately.

When it comes to evaluating the impact of vehicular obstructions, several experimental studies emphasized the importance of obstructing vehicles. Sepulcre et al.~in~\cite{sepulcre12} and Gallagher et al.~in~\cite{gallagher06} quantified the impact of vehicular obstructions on different parameters, such as packet reception, throughput, and communication range. Interestingly, Gonzalvez et~al. in~\cite{gozalvez12} perform experiments where the impact of vehicular traffic and tall vehicles (buses) also heavily influenced the vehicle-to-infrastructure (V2I) links, where roadside units are placed at elevated positions (between 3 and 10~meters) next to or above the roads. Tall vehicles decreased the effective communication range by 40\%, whereas the dense traffic reduced the range by more than 50\%. Furthermore, other studies suggested that obstructing vehicles could be an important factor in channel modeling %
 (e.g.,~\cite{Otto2009,Jerbi2007,matolak05,takahashi03}). 
 
The studies above were aimed at measuring the channel characteristics and fitting the channel models to the already collected measurements. However, research aimed at incorporating the vehicles in the channel model and therefore \emph{predicting} their effect has been scarce. Apart from our previous work reported in~\cite{boban11}, to the best of our knowledge, there have only been two studies aimed at explicitly introducing vehicular obstruction in channel modeling. %
Abbas et al.~in~\cite{abbas12} performed V2V measurements and showed that a single vehicle can incur more than 10~dB attenuation, which is in line with the results reported in~\cite{boban11}. Based on the measurements, the authors design a stochastic channel model for highway environments that incorporates vehicular obstructions and determines the time duration of the LOS, NLOSv, and NLOSb states using the measured probability distributions of each state. 
Wang et al.~in~\cite{wang09} perform isolated (``parking lot'') measurements and characterize the loss due to vehicles obstructing the LOS. Furthermore, they model the loss due to vehicles by employing a three-ray knife-edge model, where diffraction loss is calculated over the vehicles and on the vehicle sides using the modified Bullington method~\cite{bullington77}. The results show a good agreement between the isolated measurement results and the proposed method.

\section{Conclusions}\label{sec:conclusionsComplete}
We proposed a computationally efficient channel model that can be used in large scale packet-level VANET simulators. Compared to the simple statistical channel models currently used in VANET simulators, the proposed model utilizes the geographic descriptors to enable location-specific per-link modeling of the channel. Furthermore, the time-dependent component of the channel is accounted for: depending on the density of the vehicles in an area, the channel between two vehicles can change considerably as the surrounding vehicles create LOS obstruction, reflections, and diffractions.
Compared to the ray-tracing methods, the model is beneficial in terms of: 1) computational complexity, since it performs only a subset of complex calculations required for full ray-tracing models; and 2) reduced requirements for geographical information -- the only information required are outlines and type of buildings and foliage, and locations and dimensions of vehicles -- these are freely available through geographical databases and mobility traces. 

Furthermore, with limited (and often imperfect) geographical description of the simulated area, at a certain level of simulation detail,
there is a point of diminishing returns, where a marginal improvement in the realism requires a large computational effort. For this reason, 
we used VANET-specific information (e.g., number of surrounding objects, dimensions of vehicles, different attenuation levels due to different object types, etc.) to limit the complexity of the model. 
To enable a more efficient channel modeling, we separated the links into three categories: 1) line of sight (LOS); 2) non-LOS due to vehicles (NLOSv); and 3) non-LOS due to buildings/foliage (NLOSb). The results regarding LOS and NLOSv conditions shown in figs.~\ref{fig:LOSFits} and~\ref{fig:NLOSvFits} indicate that, in order to correctly model LOS and NLOSv scenarios, it is sufficient to consider the main type of propagation mechanism for respective link type (two-ray for LOS and vehicles-as-obstacles for NLOSv), with added fading proportional to the number of vehicles and area of the static objects (buildings, foliage) in the communicating pair's ellipse (as defined in Fig.~\ref{NLOSExplanation}). This allows for an efficient implementation in the simulation environments, as the only operations required are LOS classification and determining the number and area of objects in the communication pair's ellipse (both of which can be performed efficiently using bounding volume hierarchies). On the other hand, for NLOSb cases, it is beneficial to consider reflections and diffractions, as they enable for a better estimation of the received power, particularly when vehicles are communicating ``around the corner'' (i.e., where vehicles are on two sides of a corner of a single building, as shown in Figs.~\ref{fig:OutletOverlayReflDiffr} and~\ref{fig:outletAnnotated}). In such scenarios, not accounting for reflected and diffracted rays would imply that the level of obstruction (i.e., the ``depth'' of the building corner with respect to LOS path) is not important. However, due to reflections and diffractions accounting for two thirds of the computation time, as part of our future work, we will look for alternative techniques for channel estimation for NLOSb links.  %

We implemented the model and showed that it can be used to simulate networks containing thousands of communicating vehicles across different environments (highway, suburban, urban). The model is able to simulate city-wide areas containing buildings and foliage. The results showed that the model behaves linearly with the increase of both the network size (i.e., number of objects in the simulation) and the number of communicating pairs.

\chapter{TVR -- Tall Vehicle Relaying} \label{ch:TVR}

In Chapters~\ref{ch:experiments} and~\ref{ch:vehModel}, we have shown that line of sight (LOS) conditions have a direct influence on V2V link quality. %
LOS is influenced by both mobile objects (vehicles) and static objects (buildings, foliage) located between and around the communicating nodes. %
We have shown experimentally that vehicles can obstruct the LOS communication up to 50\% of the time; furthermore, a single obstructing vehicle can reduce the power at the receiver by more than 20~dB. Similarly, buildings obstructing the LOS severely impede communication, with under 20\% of packets successfully delivered when the distance between the communicating pair is above 100~m.
Based on both experimental measurements (Ch.~\ref{ch:experiments}) and simulations performed using the proposed channel model (Ch.~\ref{ch:vehModel} and~\ref{ch:completeModel}), in this chapter we show that the elevated position of the antennas on tall vehicles significantly improves communication performance. Tall vehicles can increase the effective communication range, with an improvement of up to 50\% in certain scenarios~\cite{boban11_2,boban13}. Using these findings, we propose a new V2V relaying scheme called Tall Vehicle Relaying (TVR) that takes advantage of the better channel characteristics provided by tall vehicles. TVR distinguishes between tall and short vehicles and, where appropriate, chooses tall vehicles as next hop relays. We investigate TVR's system-level performance through a combination of small-scale experiments and large-scale simulations and show that it outperforms existing techniques.

\section{Motivation} \label{sec:Introduction}

The relatively low height of the antennas located on the vehicles makes V2V communication susceptible to line of sight (LOS) obstruction by non-communicating vehicles. The probability of having LOS communication decreases with distance, with less than a 50\% chance of LOS near the maximum V2V communication range~\cite{boban11}. Furthermore, the Dedicated Short-Range Communications (DSRC)~\cite{dsrc09} frequency band reserved for VANET communication is in the 5.9~GHz band. As noted by Parsons in~\cite{00parsons}, in this frequency band the ``propagation paths must have line of sight between the transmitting and receiving antennas, otherwise losses are extremely high''.
This has been empirically shown to be the case for V2V links in \cite{meireles10}, where a single large truck attenuated the received power between two passenger cars by 27~dB.  %
Consequently, obstructing vehicles cause a reduction of the effective communication range of up to 60\% and Packet Delivery Ratio (PDR) of up to 30\%, depending on the environment. %

Motivated by these findings, we explore how the adverse effects of vehicular obstructions can be ameliorated by opting for the taller vehicles as next hop relays. %
We distinguish between tall vehicles, such as commercial and public transportation vehicles (vans, buses, trucks, etc.) 
and short vehicles (passenger cars). We base this distinction on the analysis performed in \cite{boban11}, which showed that the dimensions of the most popular passenger cars differ significantly from the dimensions of commercial freight and public transportation vehicles. Specifically, it was observed that the latter are, on average, more than 1.5~meters taller than personal vehicles. %
By separating the vehicles in this manner, we showed in~\cite{boban11_2,boban13} that the antennas mounted on top of tall vehicles experience a significantly better communication channel, 
which is not as affected by obstruction from other vehicles as is the case for 
short vehicles (i.e., the probability of having LOS increases).

We perform small-scale experiments and large-scale simulations that provide insights into the end-to-end benefits of tall vehicle relaying. %
Based on the benefits we observed while performing experiments, %
we introduce the Tall Vehicle Relaying (\emph{TVR}) technique,
a paradigm shift from the farthest relay technique, which selects the farthest tall vehicle in the direction of message destination. We compare the performance of \emph{TVR} with two techniques: i) \emph{Farthest Neighbor}, which selects the farthest neighbor with which communication is possible; and ii) and \emph{Most New Neighbors}, which selects the vehicle with the largest number of new neighbors in the direction of message dissemination. The results show that \emph{TVR} matches these techniques in lower vehicle density scenarios and outperforms them in high density scenarios in terms of the number of hops required to reach a destination and end-to-end delay. %

The main contributions of this work can be summarized as follows:
\begin{itemize}
\item %
Leveraging previous results based on aerial photography and a validated channel model,
we quantify the benefits of using tall vehicles as next hops in terms of: 1) LOS communication; 2) received signal power; and 3) effective communication range; %
\item We perform real-world experiments to determine the benefits of using tall vehicles as relays; the results show that selecting tall vehicles is beneficial in terms of higher received power, smaller number of hops to reach the destination (thus decreasing end-to-end delay), and increased per-hop communication range;
\item We introduce the Tall Vehicle Relay (\emph{TVR}) technique, which matches existing techniques in low vehicle density scenarios and outperforms them in high density scenarios %
in terms of the number of hops needed to reach the destination, delay, and medium contention. %
\end{itemize}

The rest of the chapter is organized as follows. %
Model-based analysis of selecting tall vehicles as relays is presented in Section~\ref{sec:model}. %
The experimental setup and results are described in Section~\ref{sec:experiment}. Section~\ref{sec:largeScale} describes the TVR technique and the results of the large-scale simulations, %
whereas 
Section~\ref{sec:RelatedWork} describes related work. %
Finally, Section~\ref{sec:Discussion} %
provides conclusions and discusses possible future work. %

\section{Model-Based Analysis of the Benefits of Tall Vehicles as Relays}\label{sec:model}

In this section we analyze the effect of vehicle height on the probability of line of sight and on the received signal strength. The discussion on Packet Delivery Ratio (PDR) is deferred to the next section in order to facilitate comparison with the experimentally obtained data.

\subsection{Setup}

To assess the effect of tall vehicles on communication performance, apart from experimental evaluation, we also designed a large-scale simulative study. For the study to be credible, we require the following: 1) accurate information on vehicle positions and dimensions; and 2) a realistic channel model.

For accurate vehicle positioning, 
we leverage a dataset of real vehicle positions obtained from aerial photography of the A28 highway located near Porto, Portugal. The 404 vehicle dataset is described in Table~\ref{dataset}; more details on the dataset are available in~\cite{ferreira10}. In addition to vehicle location, this dataset specifies the heading and the length of each vehicle. %
To assign width and height %
to each vehicle, we used the empirically derived distributions of the dimensions of tall and short vehicles described in~\cite{boban11}. The heights of both types of vehicles are normally distributed, with a mean of 3.35 meters for tall and 1.5 meters for short vehicles. The standard deviation is 0.08 meters for both types.

\begin{table} 
	\centering
		\caption{ Aerial photography dataset (A28 highway)}
		\begin{tabular}{|c c c c c|}
\hline \textbf{Highway} & \textbf{Length} & \textbf{\# Vehicles} & \textbf{\# Tall Vehicles} & \textbf{Veh. Density} \\
		 	\hline 
\hline A28 & 12.5~km & 404 & 58 (14.36\%) & 32.3~veh/km\\ 
\hline
\end{tabular}
	\label{dataset}
\end{table}

To realistically model the received power, %
we use the channel model developed in Chapter~\ref{ch:vehModel}. Since it accounts for vehicles as three-dimensional obstructions to the transmitted signal, it allows to calculate both the vehicles' impact on LOS, as well as the received power. %

\subsection{Impact of Vehicles on Line of Sight}\label{sec:LOS}
We first set out to determine how often the LOS is blocked by non-communicating vehicles and the difference in LOS blocking between short and tall vehicles using the aerial dataset.
For this purpose, we define the per-vehicle ratio of LOS links as follows.
For each vehicle, we determine the number of neighbors it has a LOS with (a neighbor vehicle is any vehicle within radio range, i.e. it receives the signal above the sensitivity threshold, based on the employed channel model). Then, we divide that number by the total number of neighbors. This gives the ratio of LOS links for a specific vehicle. By doing the same calculation for each vehicle and by separating the tall and short vehicles, we obtain the distribution of the ratio of LOS links. %

\begin{figure}
  \begin{center}
    \includegraphics[width=0.435\textwidth]{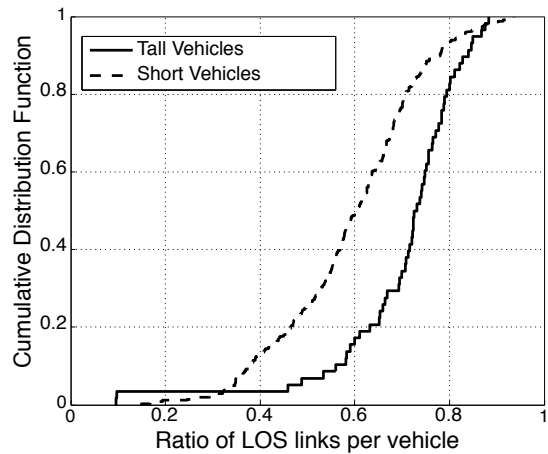}
     \caption[CDF of the per-vehicle ratio of LOS links]{\small Cumulative Distribution Function of the per-vehicle ratio of LOS links
     for tall and short vehicles based on the aerial photography dataset. %
}
      \label{PLOS-cars-trucks}
   \end{center}
\end{figure}

Figure~\ref{PLOS-cars-trucks} shows the difference in the ratio of LOS links for tall and short vehicles. %
The ratio of LOS links %
is notably higher for tall vehicles; 50\% of the short vehicles have more than 60\% of LOS links, whereas 
for tall vehicles, %
the value rises from 50\% to 80\%.

\begin{figure}
  \begin{center}
    \includegraphics[width=0.45\textwidth]{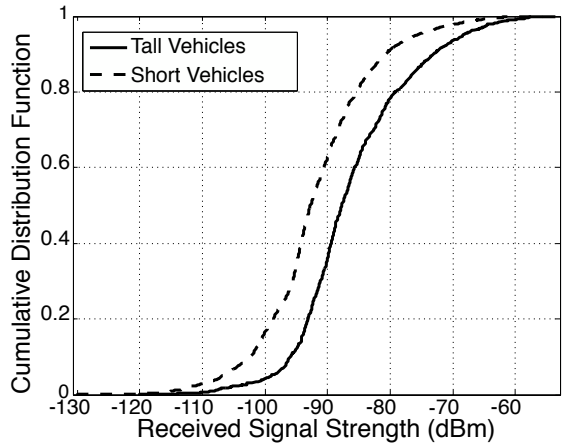}
     \caption[CDF of received signal strength for tall and short vehicles]{\small Cumulative Distribution Function of received signal strength for tall and short vehicles based on the aerial photography dataset. Different distance bins are equally represented for tall and short vehicle links.}
      \label{empirical-CDF}
   \end{center}
\end{figure}

\subsection{Difference Between Received Signal Strength for Tall and Short Vehicles}\label{sec:Power}

Figure~\ref{empirical-CDF} shows the cumulative distribution function of the received signal strength for tall and short vehicle links calculated using the described channel model and position information. Despite the fact that the average distance between the communicating vehicles for tall and short vehicle links is the same, the received signal strength for tall vehicle links is consistently higher by approximately 4~dB. 
Using the minimum sensitivity threshold of -85~dBm for a 3~Mb/s data rate as defined in the DSRC standard~\cite{astm03}, Fig.~\ref{empirical-CDF} shows that the 4~dB difference in received power results in more than 20 percentage points higher packet delivery ratio for links involving tall vehicles than that composed of only short vehicles. %

\section{Experimental Analysis of the Benefits of Tall Vehicles as Relays
}\label{sec:experiment}

\begin{figure}
  \begin{center}
    \includegraphics[width=0.5\textwidth]{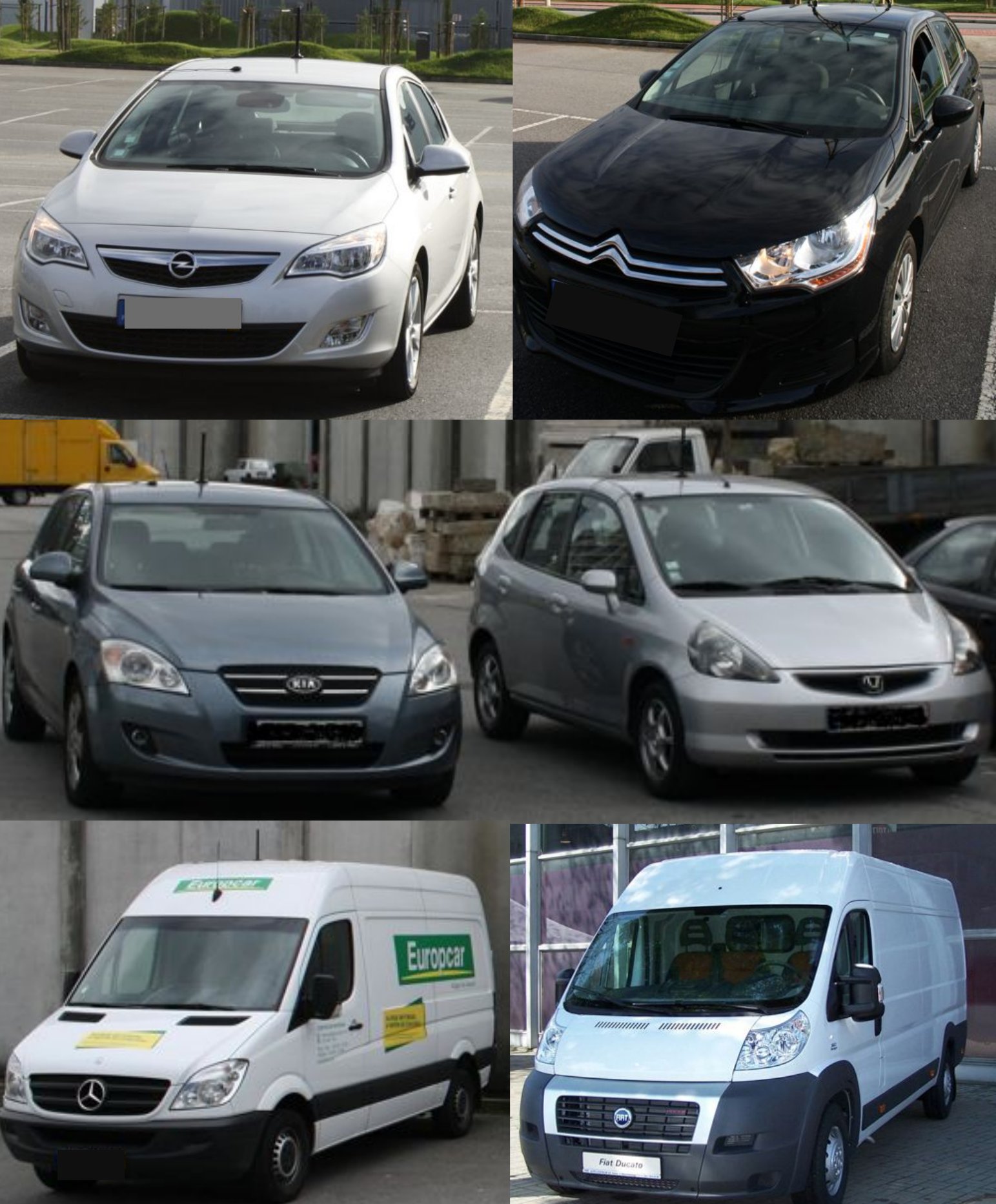}
     \caption[Vehicles used in the experiments]{\small  Vehicles used in the experiments. Clockwise from top left: Opel Astra, Citroen C4, Honda Jazz, Fiat Ducato, Mercedes Sprinter, and Kia Cee'd. The four cars have a height of approximately 1.5 meters, which coincides with the statistical mean height for personal vehicles~\cite{boban11}, %
whereas both vans are approximately 2.5~meters tall.}
      \label{vehicles}
   \end{center}
\end{figure}

We performed small-scale experiments to complement the model-based analysis by measuring the benefits of choosing a tall vehicle as a relay in a real-world scenario.  Using regular passenger cars to represent the short vehicle class and full-size vans to represent the tall vehicle class, we performed experiments comprising two-node and three-node networks. Vehicles used in the experiments are depicted in Fig.~\ref{vehicles}; their dimensions are listed in Table~\ref{tab:dimensions}. %

\begin{table}
	\centering
		\caption{ Dimensions of Vehicles Used in the Experiments} 
\begin{tabular}{|c c c c|} \hline
		 & \multicolumn{3}{c|}{\bf Dimensions (meters)} \\ %
		 \bf Vehicle & \bf Height & \bf Width & \bf Length \\ \hline \hline
		 \textbf{Passenger (short) vehicles}&&& \\ \hline
		2011 Citroen C4 & 1.491 & 1.789 & 4.329\\ \hline
		2011 Opel Astra & 1.510 & 1.814&  4.419\\ \hline
		2007 Kia Cee'd & 1.480 & 1.790 & 4.260\\ \hline
		2002 Honda Jazz & 1.525 & 1.676 & 3.845\\ \hline 
		\textbf{Commercial (tall) vehicles}&&& \\ \hline
		2010 Mercedes Sprinter & 2.591 & 1.989 & 6.680\\ \hline
		2010 Fiat Ducato & 2.524 & 2.025 & 5.943\\ \hline
		\end{tabular} 
\label{tab:dimensions}
\end{table}

\subsection{Experimental Scenarios}\label{subsec:scenarios}

\begin{figure}
  \begin{center}
    \includegraphics[width=0.65\textwidth]{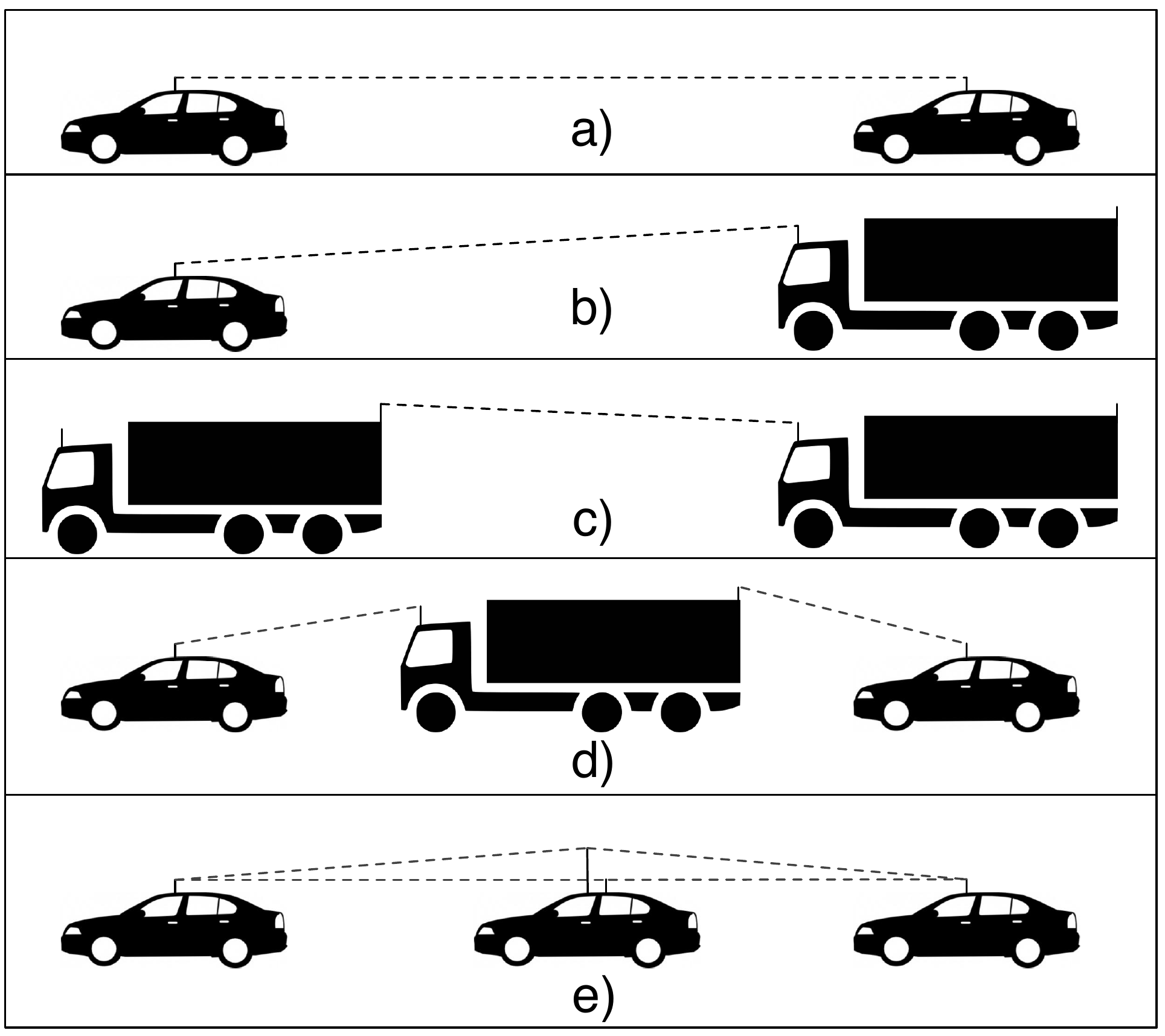}
     \caption[Types of experiments performed]{\small We performed the following experiments: %
      a) car-car; b) car-van; c) van-van; d) car-van-car; e) car-car-car (tall and short relay antenna).}
      \label{linksSingleMultiHop}
   \end{center}
\end{figure}

\begin{figure}
  \begin{center}
    \includegraphics[width=0.4\textwidth]{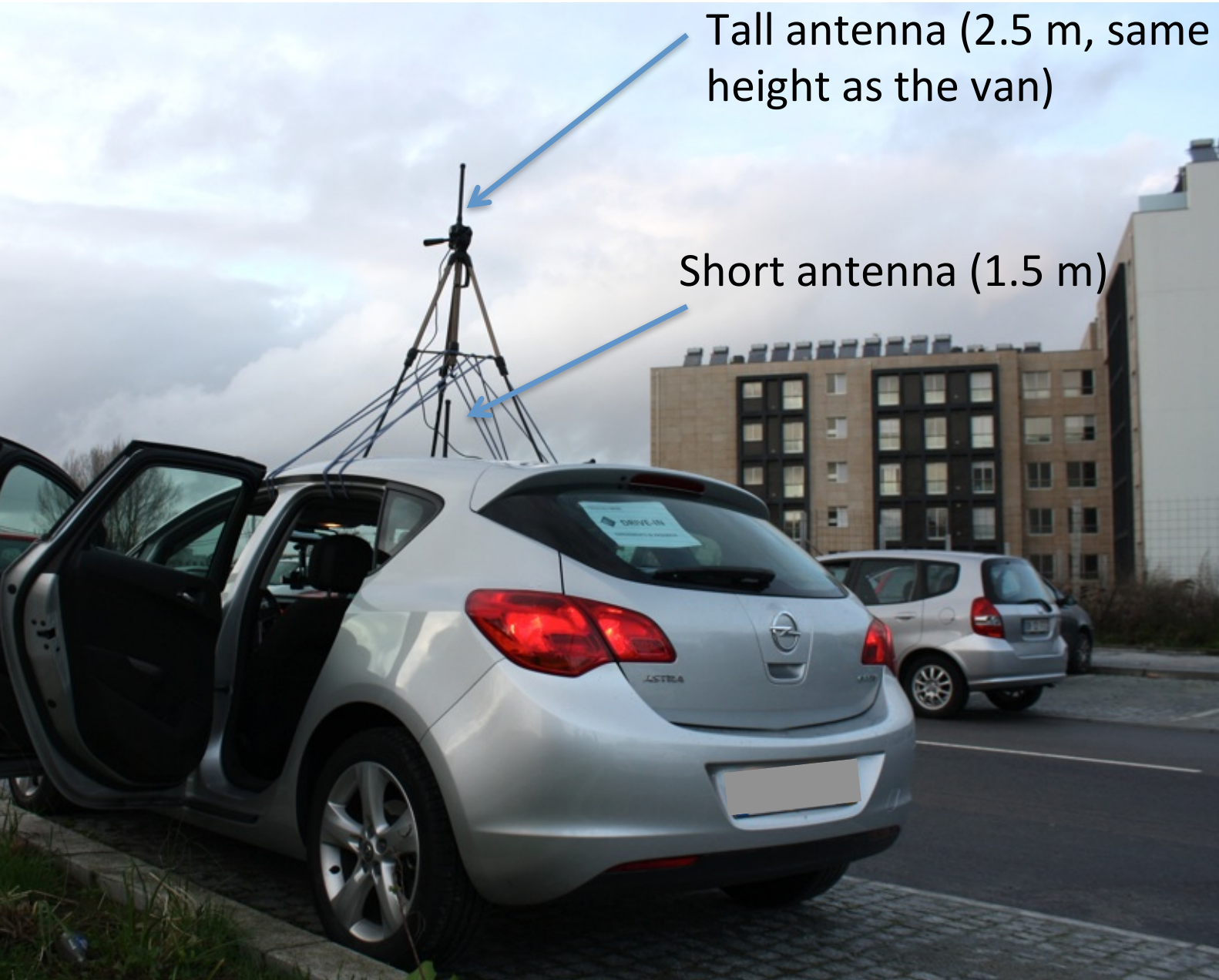}
     \caption[Tall and short antenna mounted on the relay vehicle]{\small Tall and short antenna mounted on the relay vehicle. The vehicle was used as a relay node between two other short vehicles and the experiments with both antennas as relays were performed simultaneously. This experimental setup isolated the antenna height as the only factor affecting the received power and Packet Delivery Ratio (PDR). %
     We made sure that the tripod holding the tall antenna does not interfere with the short antenna by isolating any metal parts and placing the tripod legs so that they do not block the LOS with front and rear vehicle.}
      \label{twoAntennas}
   \end{center}
\end{figure}

We consider the following five scenarios: %

\begin{itemize}
\item Single-hop experiments, where two vehicles drive in tandem:
\begin{enumerate}
\item \textbf{Car-car} (Fig.~\ref{linksSingleMultiHop}a) ---  A link between two passenger cars is used to establish a baseline for single-hop comparison.

\item \textbf{Car-van} (Fig.~\ref{linksSingleMultiHop}b) --- A link between a passenger car and a van is used to evaluate the channel between vehicles of different types. %

\item \textbf{Van-van} (Fig.~\ref{linksSingleMultiHop}c) --- A link between two vans is used to quantify the maximum potential benefit of tall relays. When both vehicles are tall, the likelihood of their LOS being obstructed is minimized. %

\end{enumerate}
\item Two-hop experiments, where three vehicles drive in tandem, the source and destination at the ends and a relay in the middle:
\begin{enumerate}\setcounter{enumi}{3}
\item  \textbf{Car-van-car} (Fig.~\ref{linksSingleMultiHop}d) --- A van is equipped with two antennas, one in the front, and one in the rear. A car drives in front of the van, exchanging messages with the van's front-mounted antenna. A second car drives behind the van, communicating with the rear-mounted antenna. This scenario quantifies the benefits of tall vehicle relays between two short vehicles.
\item  \textbf{Car-car-car} (Fig.~\ref{linksSingleMultiHop}e) --- Here we have a leading car, a trailing car and a relay car in the middle. The relay car is equipped with two radios and two antennas, one mounted directly on the roof and one mounted on a one meter tall tripod placed on top of the roof, as depicted in Fig.~\ref{twoAntennas}. This scenario enabled us to exclude the impact of all variables other than antenna height on the communication performance (i.e., the conditions in terms of terrain topography, vehicular density, and blocking vehicles were exactly the same for both tall and short antennas). %
\end{enumerate}
\end{itemize}

\begin{figure*}
\centering
\subfigure[13.5~Km section of the A28 highway used in our experiments.]{\label{fig:A28}\includegraphics[height=.6\textwidth]{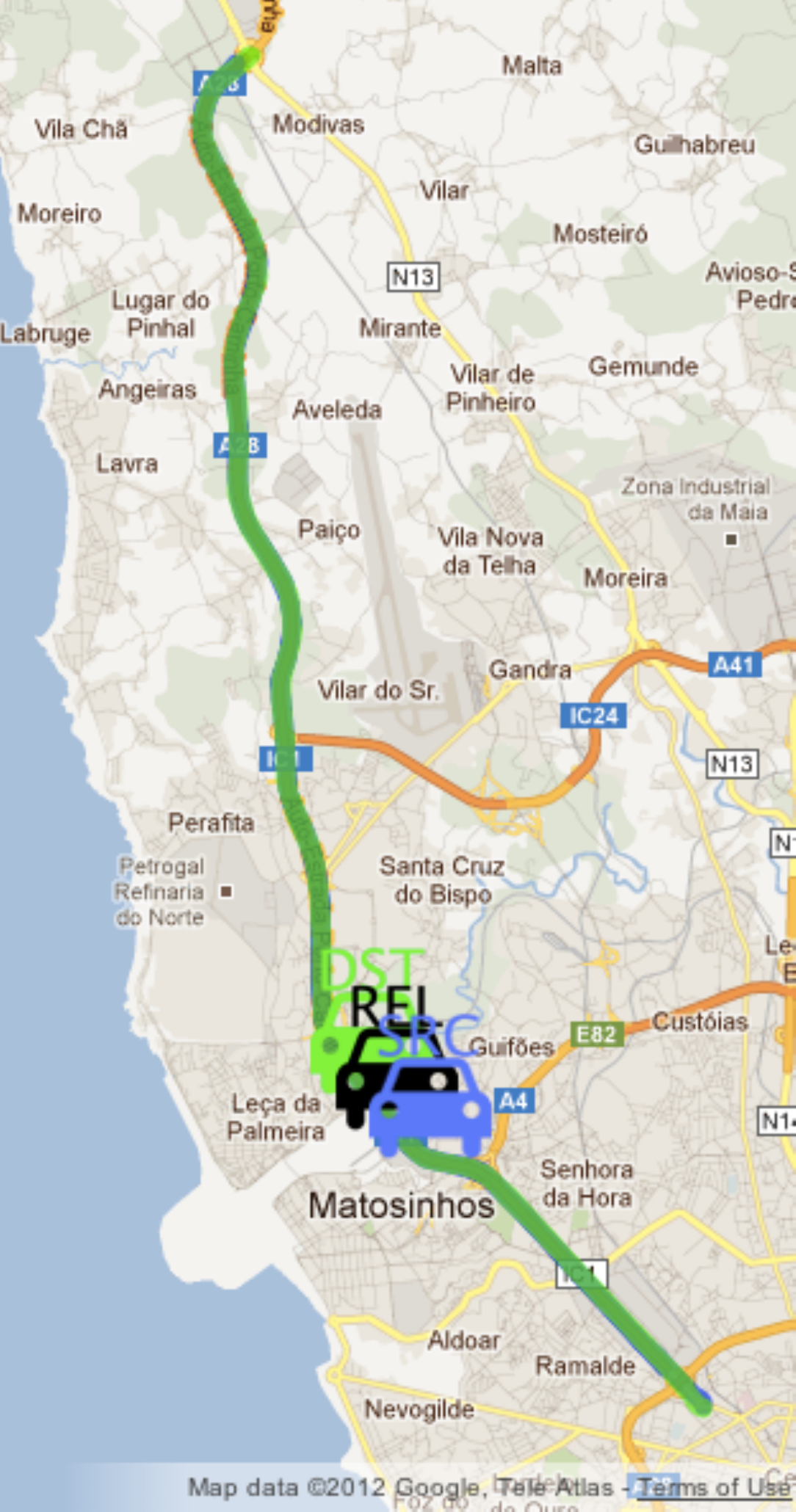}}
\subfigure[24~Km section of the VCI urban highway used in our experiments.]{\label{fig:VCI}\includegraphics[height=.6\textwidth]{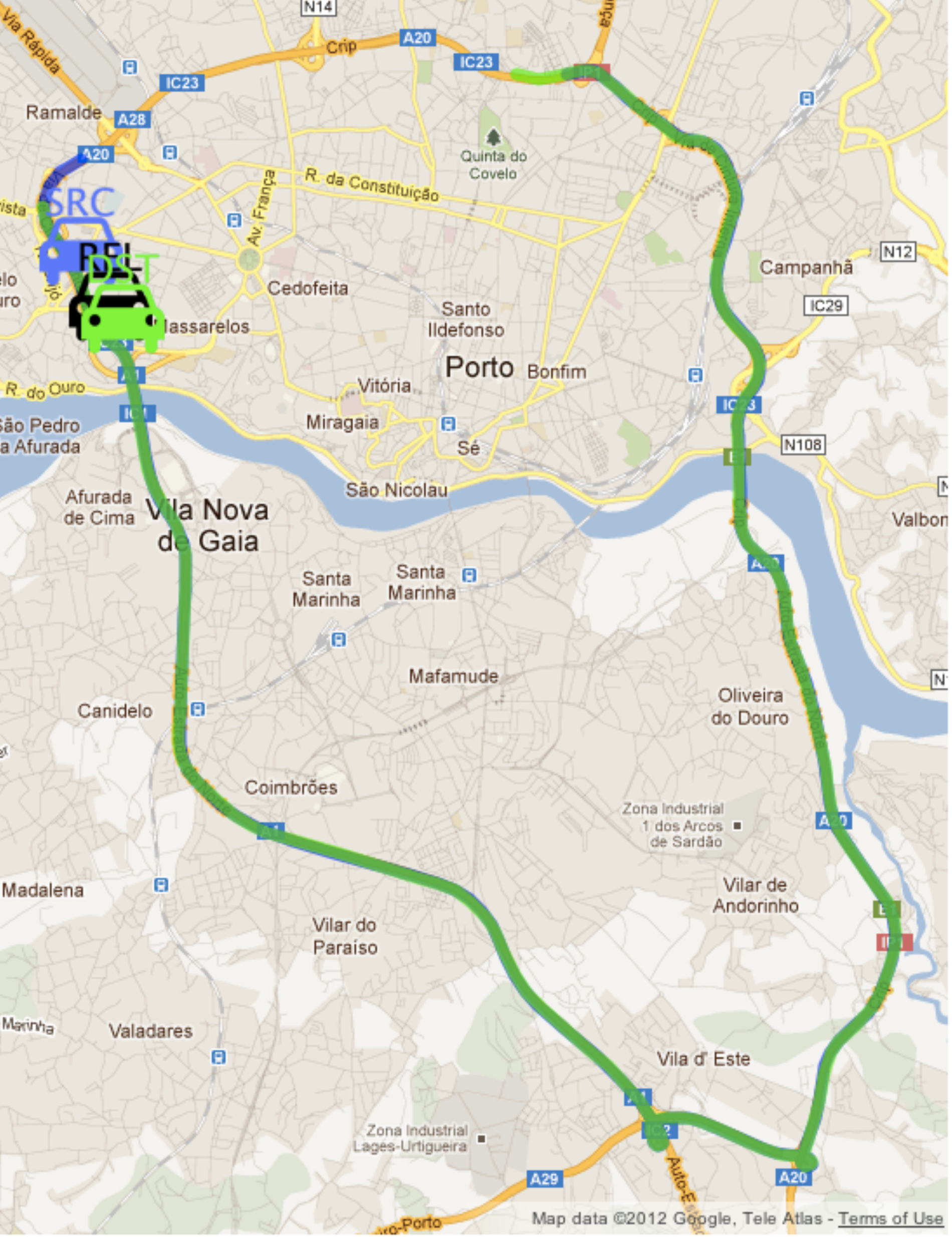}}\\
\caption[Highways where the experiments were performed]{\small Highways where the experiments were performed. The three test vehicles shown in subfigures (a) and (b) are: source (SRC), relay (REL), and destination (DST). The SRC and DST vehicles were always passenger cars (i.e., short vehicles). The relay vehicle was either a van (tall vehicle) or a passenger car with two antennas, one mounted at 1.5~m height and the other at 2.5~m height, as shown in Fig.~\ref{twoAntennas}. The One-hop experiments were performed only on the A28 highway, whereas the two-hop experiments were performed on both highways. %
}
\label{fig:highways}
\end{figure*}

Figure~\ref{fig:highways} shows the highways where we performed the experiments. The two highways, A28 and VCI, represent distinct scenarios. The A28 is a typical highway with little to no buildings near the road and occasional trees and other vegetation nearby (Fig.~\ref{fig:A28Img}). The VCI highway is an urban ring road that goes around the twin cities of Porto and Vila Nova de Gaia, with occasional buildings close to the road %
and portions of the road lined with concrete walls (Fig.~\ref{fig:VCIImg}). 
To make the results comparable to the model-based analysis described in the previous section, we performed the experiments on the same stretch of the A28 highway that was analyzed through aerial photography (Table~\ref{dataset}). %
On both highways, the experiments were performed in %
medium to moderately dense traffic during the 11~a.m.--9~p.m. period on weekdays and weekends in March, April, and December, 2011. Each experiment run was approximately one hour long, with the vehicles traversing the A28 highway south to north and vice versa and making an incomplete loop on the VCI highway as shown in Fig.~\ref{fig:highways}. Speeds ranged from 40 to 120~km/h, in accordance with traffic conditions. The single-hop experiments were performed on A28, whereas the two-hop experiments were performed on both A28 and VCI.

\subsection{Hardware Setup}
Vehicles were equipped with NEC LinkBird-MXs V3, a development platform for vehicular communications, described in section~\ref{subsec:NetworkConfiguration}. %
Each node was configured to send periodic position beacons that were then used to record Received Signal Strength Indicator (RSSI) and Packet Delivery Ratio (PDR) information during the experiments (PDR is defined as the the ratio between the number of received messages and the number of sent messages). The position information was obtained from an external GPS receiver connected to each LinkBird. The system parameters are shown in Table~\ref{tab:hw-config}.

\begin{figure*}
\centering

\subfigure[Experimental overall results]{\label{fig:pdr:overall}\includegraphics[width=0.4\textwidth]{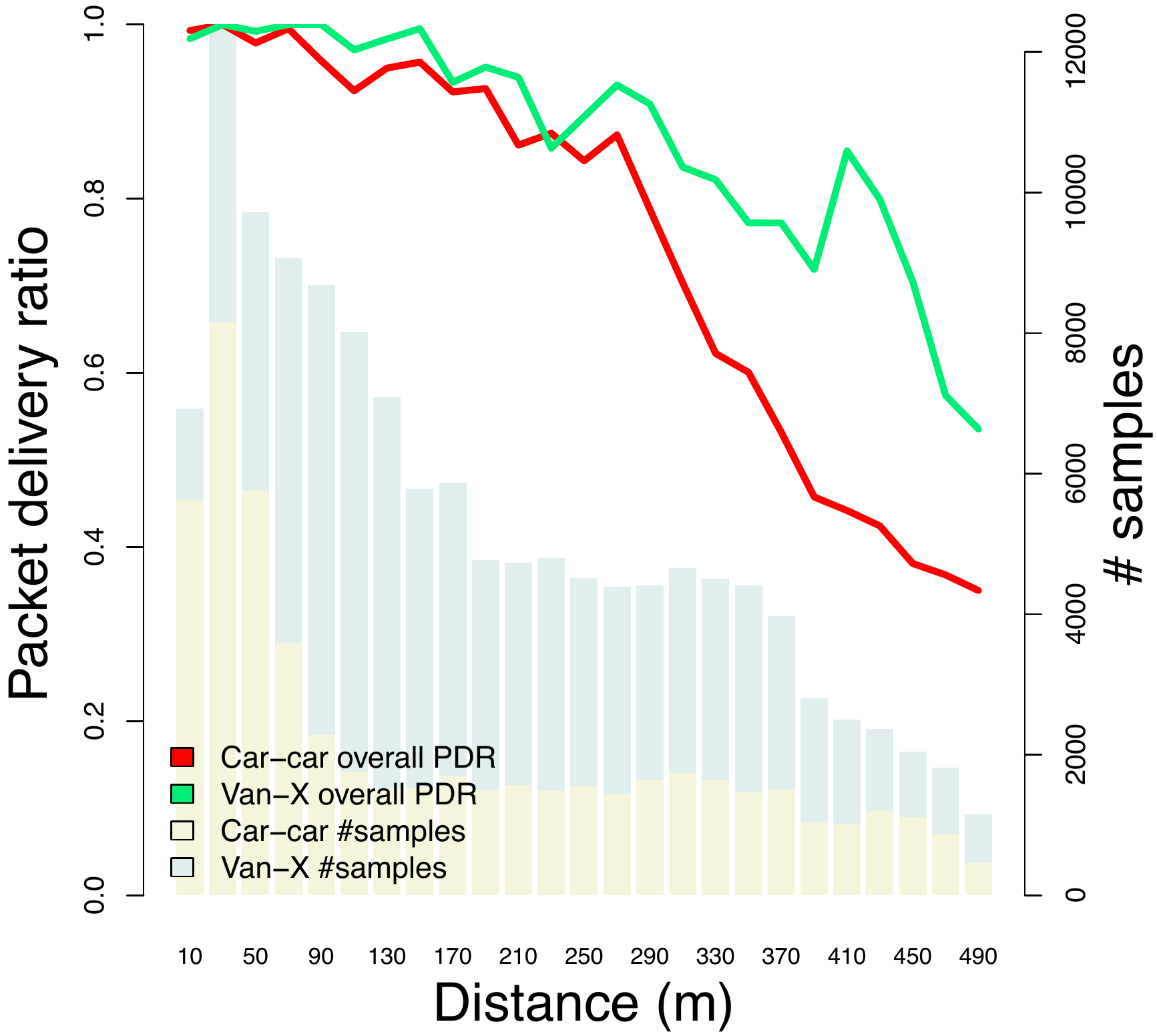}}
\subfigure[Model-based overall results]{\label{fig:pdr:overallM}\includegraphics[width=0.4\textwidth]{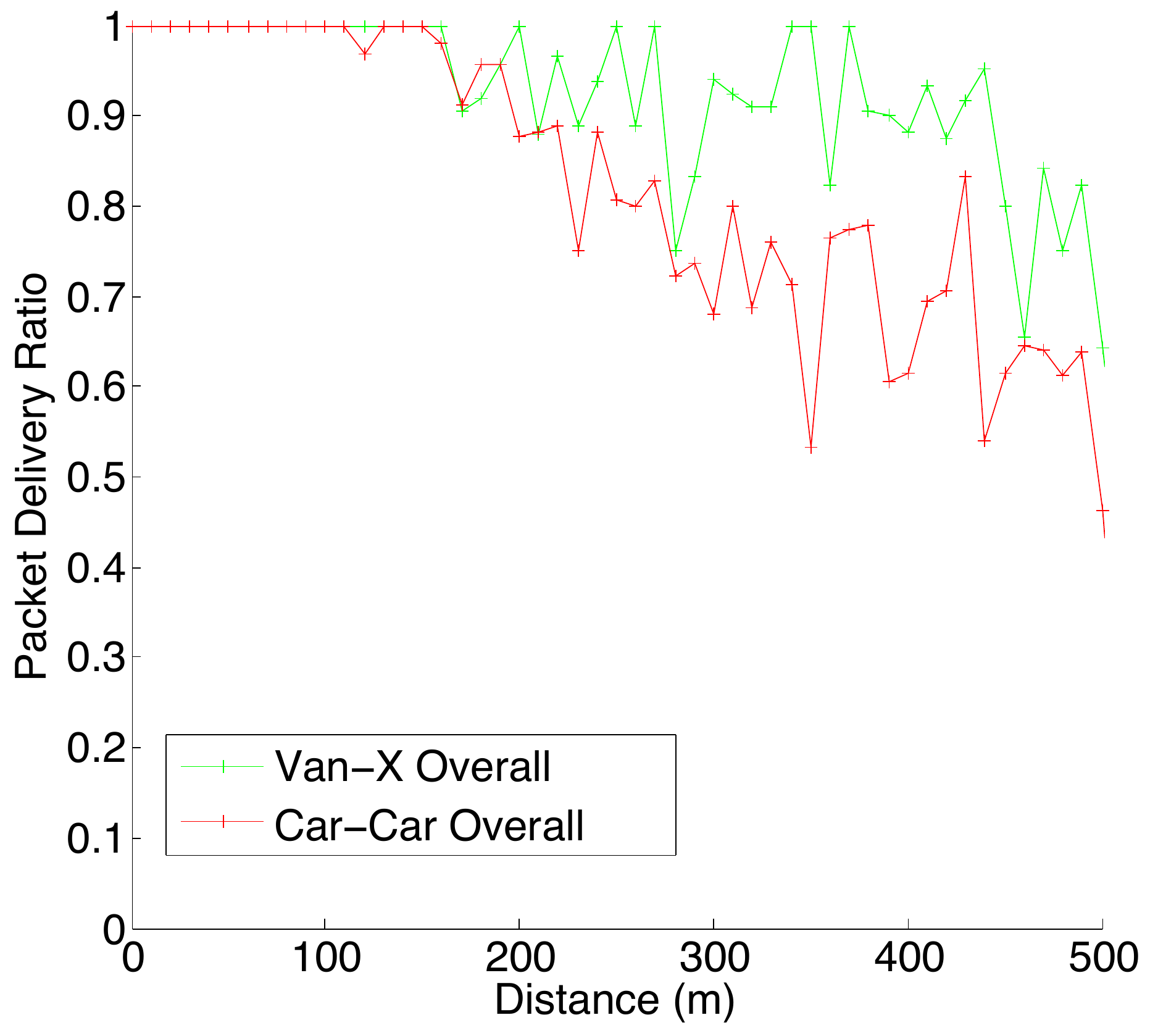}}

\subfigure[Experimental NLOSv results]{\label{fig:pdr:nlos}\includegraphics[width=0.4\textwidth]{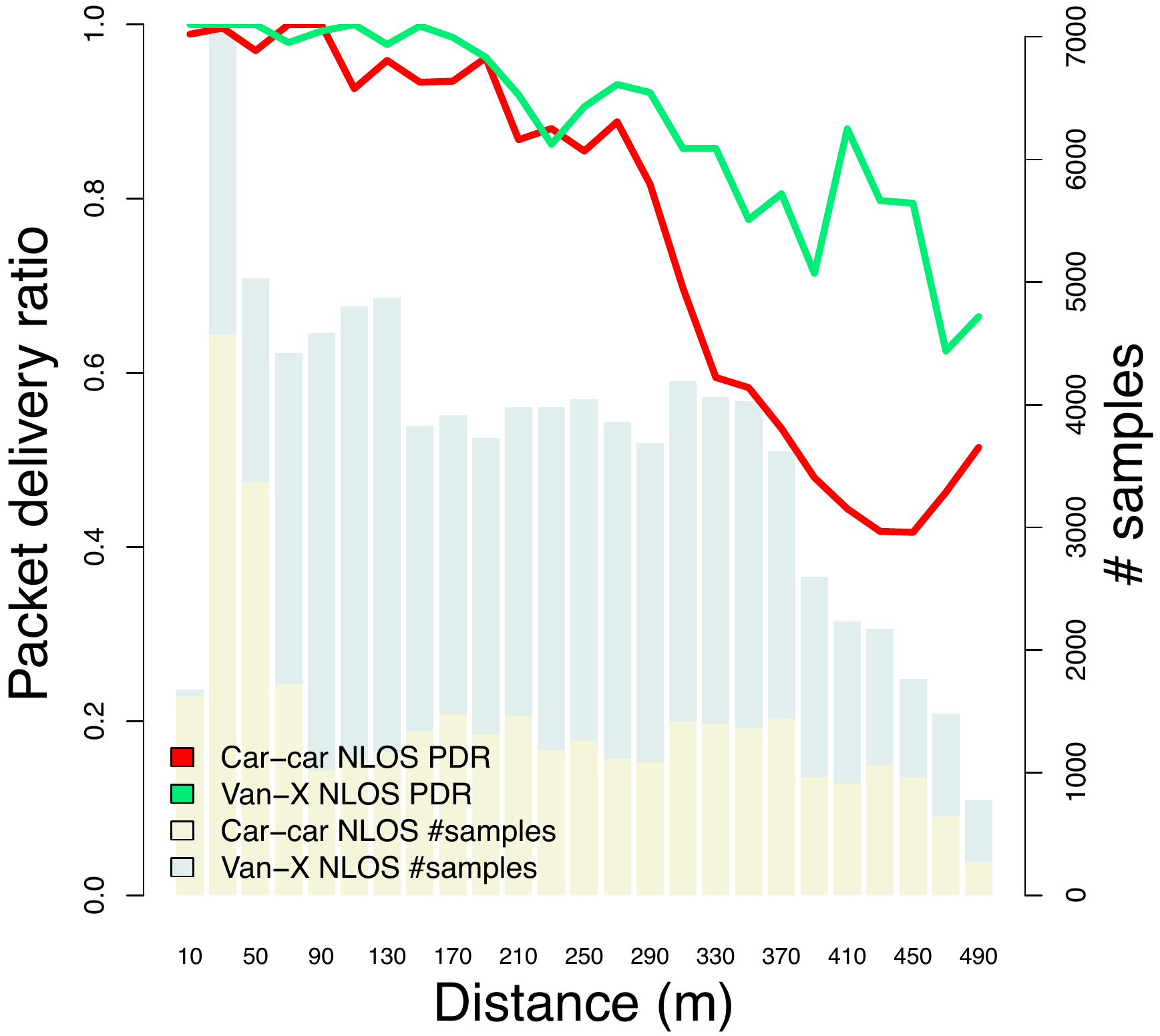}}
\subfigure[Model-based NLOSv results]{\label{fig:pdr:nlosM}\includegraphics[width=0.4\textwidth]{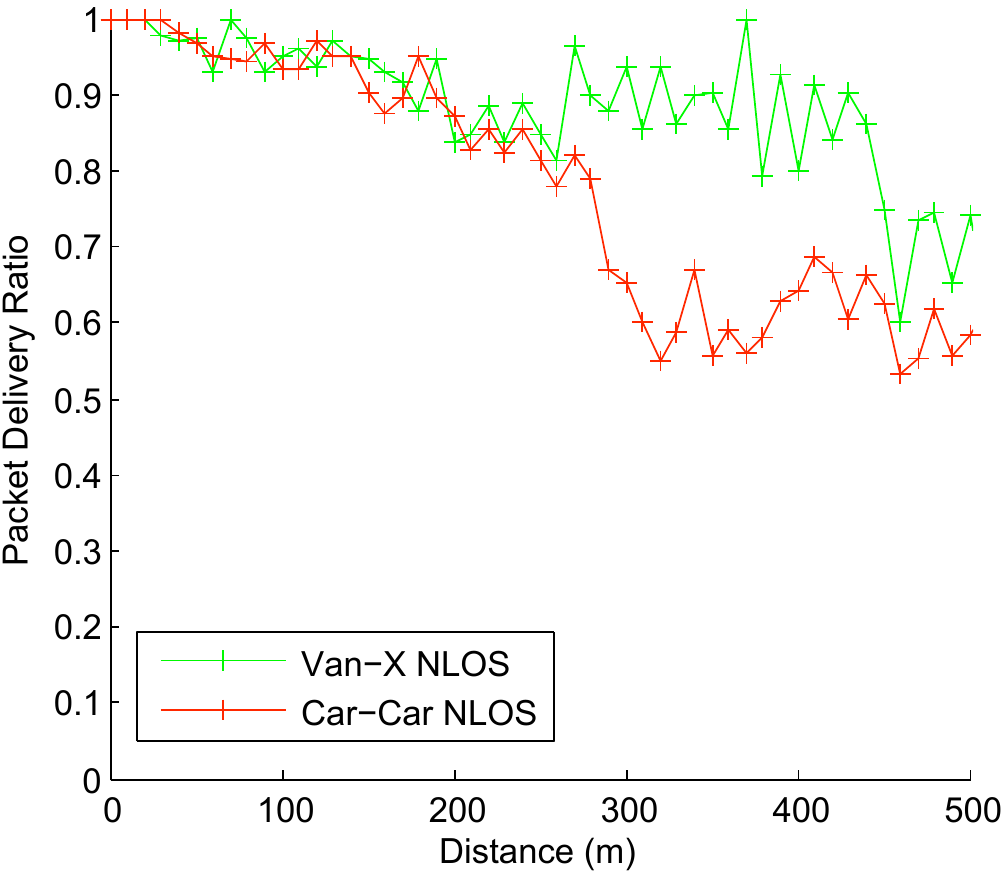}}
\caption[Packet Delivery Ratio -- experiments and model]{\small Packet Delivery Ratio (PDR) obtained through the experiments and the model for single-hop communication.} %
\label{fig:pdr}
\end{figure*}

On the passenger cars, the antenna was positioned at the center of the roof, which has been empirically shown to be the overall optimal position~\cite{kaul07}.
On the vans, we used two antennas: one at the front of the roof, and another at the back (shown in Fig.~\ref{linksSingleMultiHop}). %
This prevents the van itself from significantly deteriorating the channel characteristics by blocking the LOS path between its own antenna and the antenna of the vehicle it is communicating with. %

We distinguish two types of links: 1) line of sight (\textbf{LOS}); and 2) non-LOS due to vehicles (\textbf{NLOSv})\footnote{On highways, non-LOS due to buildings/foliage (NLOSb) did not occur often, thus we discard these links.}.
To help us distinguish between LOS and NLOSv conditions, we recorded videos of the experiments from the vehicle following in the rear in case of single-hop, and from both the leading and trailing vehicles in case of two-hop experiments (two videos were required in two-hop experiments to determine LOS conditions for each link). We then synchronized the videos to the experimental data using a custom web-based visualization suite~\cite{losexpsite} and classified each part of the experiment as LOS or NLOSv with a one second resolution. We classified the conditions as NLOSv when one or more vehicles, short or tall, were present between the two communicating vehicles. Given that the experiments were performed on highways, the number of static obstructions such as buildings was negligible and thus not considered.

\subsection{Experimental Results - One Hop Experiments}
\begin{figure}
  \begin{center}
    \includegraphics[width=0.65\textwidth]{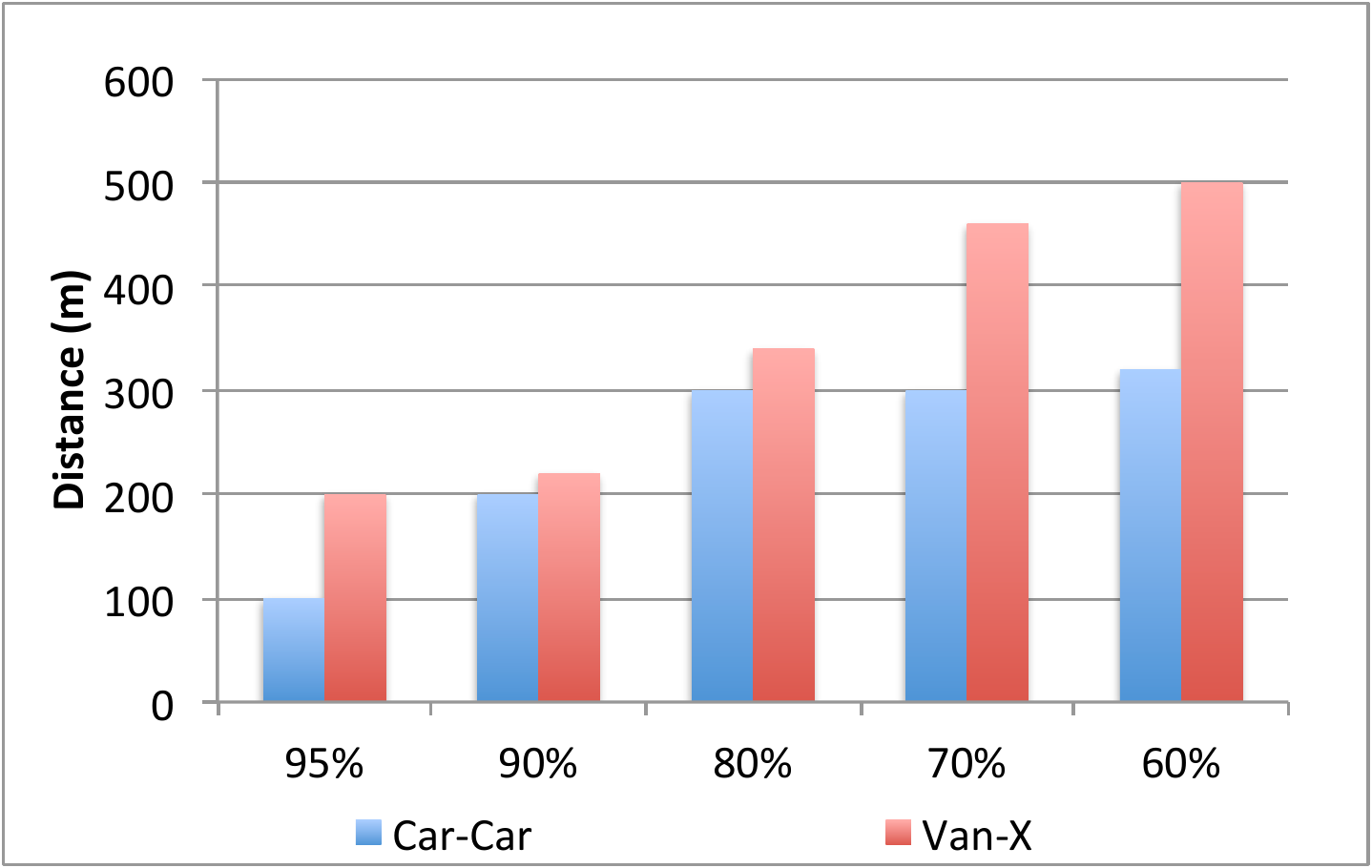}
     \caption[Experimental results on the effective communication range]{\small  Experimental results on the effective communication range as a function of desired packet delivery ratio for NLOSv conditions.}
      \label{fig:effective-comm-range}
   \end{center}
\end{figure}

\begin{figure*}[t!]
\centering
\subfigure[A28 Packet Delivery Ratio]{\label{fig:A28PDR}\includegraphics[width=0.4\textwidth]{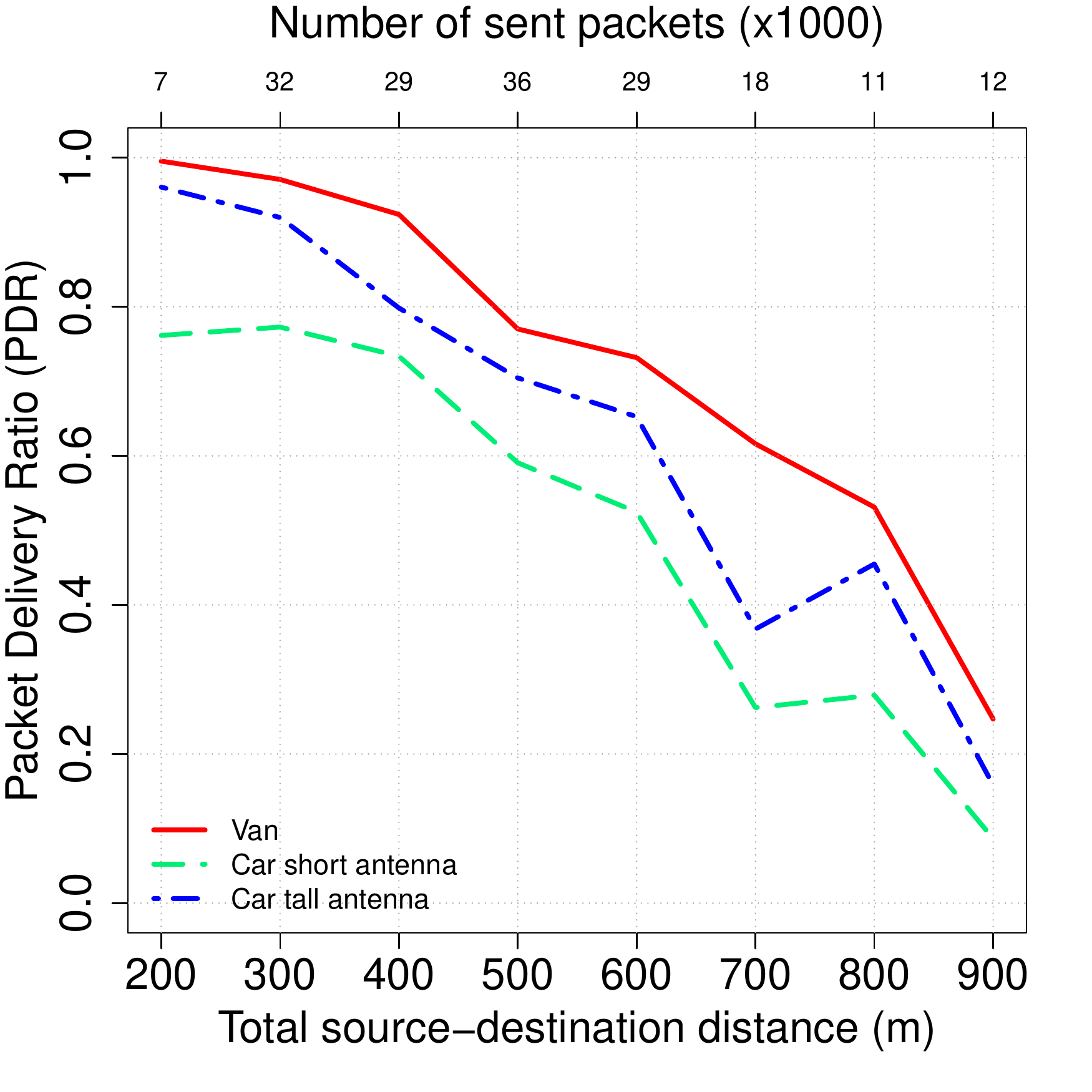}}
\subfigure[VCI Packet Delivery Ratio]{\label{fig:VCIPDR}\includegraphics[width=0.4\textwidth]{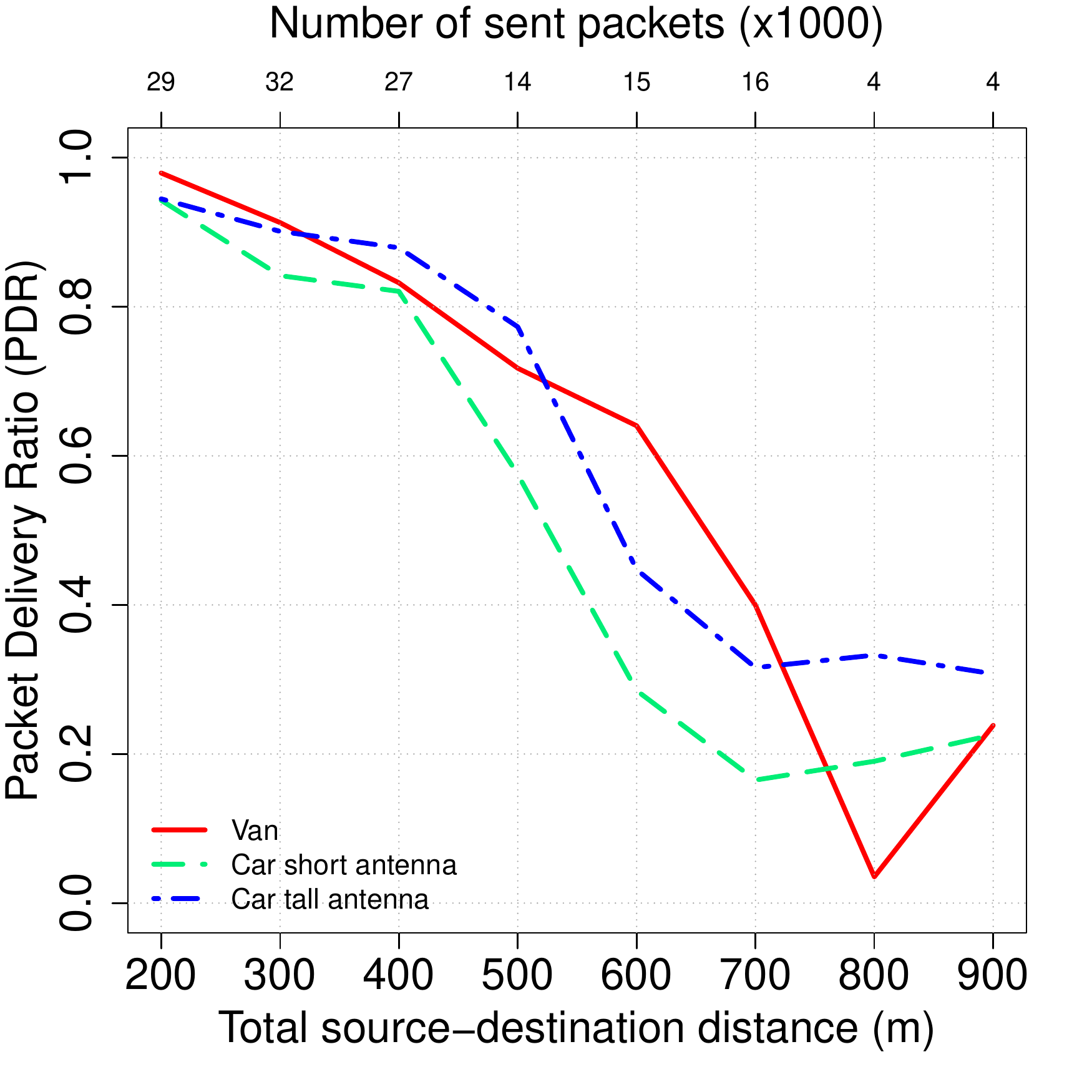}}
\caption[ Overall Packet Delivery Ratio results for the two-hop experiments]{\small Overall Packet Delivery Ratio (PDR) results for the two-hop experiments. The end-to-end PDR is computed by multiplying the PDR of the two individual links.}
\label{fig:2hopPDR}
\end{figure*}

We first present results for one-hop PDR as a function of distance, depicted in Fig.~\ref{fig:pdr}. The figure shows the PDR results obtained through both the experiments and the model described in the previous section. Similarly to the model-based results, we aggregate the van-van and van-car cases to analyze the benefit of tall vehicles regardless of the height of the other vehicle. We call this combined scenario Van-X. For each message sent, we check whether it was received or not and place that information in a distance bin with a 20 meter granularity based on the distance between the communicating vehicles. In addition to the PDR, for experimental data we plot the number of samples placed in each bin.

Figure~\ref{fig:pdr:overall} shows the overall experimentally obtained PDR for both Car-Car and Van-X scenarios, regardless of the LOS conditions. We can observe that the Van-X PDR is consistently better than the Car-Car PDR. Up to 280 meters, the difference is slight but after that it becomes quite significant, with Van-X offering an improvement of around 20 percentage points over Car-Car communication up to the maximum distance for the recorded data. Figure~\ref{fig:pdr:overallM} depicts the model-derived overall PDR, based on the aerial photography of the same A28 highway. The PDR exhibits a behavior similar to that of the experimentally collected data (Fig~\ref{fig:pdr:overall}). %

\begin{figure*}
\centering
\subfigure[A28 Overall (aggregated LOS and NLOSv) RSSI]{\label{fig:2hop-rssi-ExpVsModA28}\includegraphics[width=0.4\textwidth]{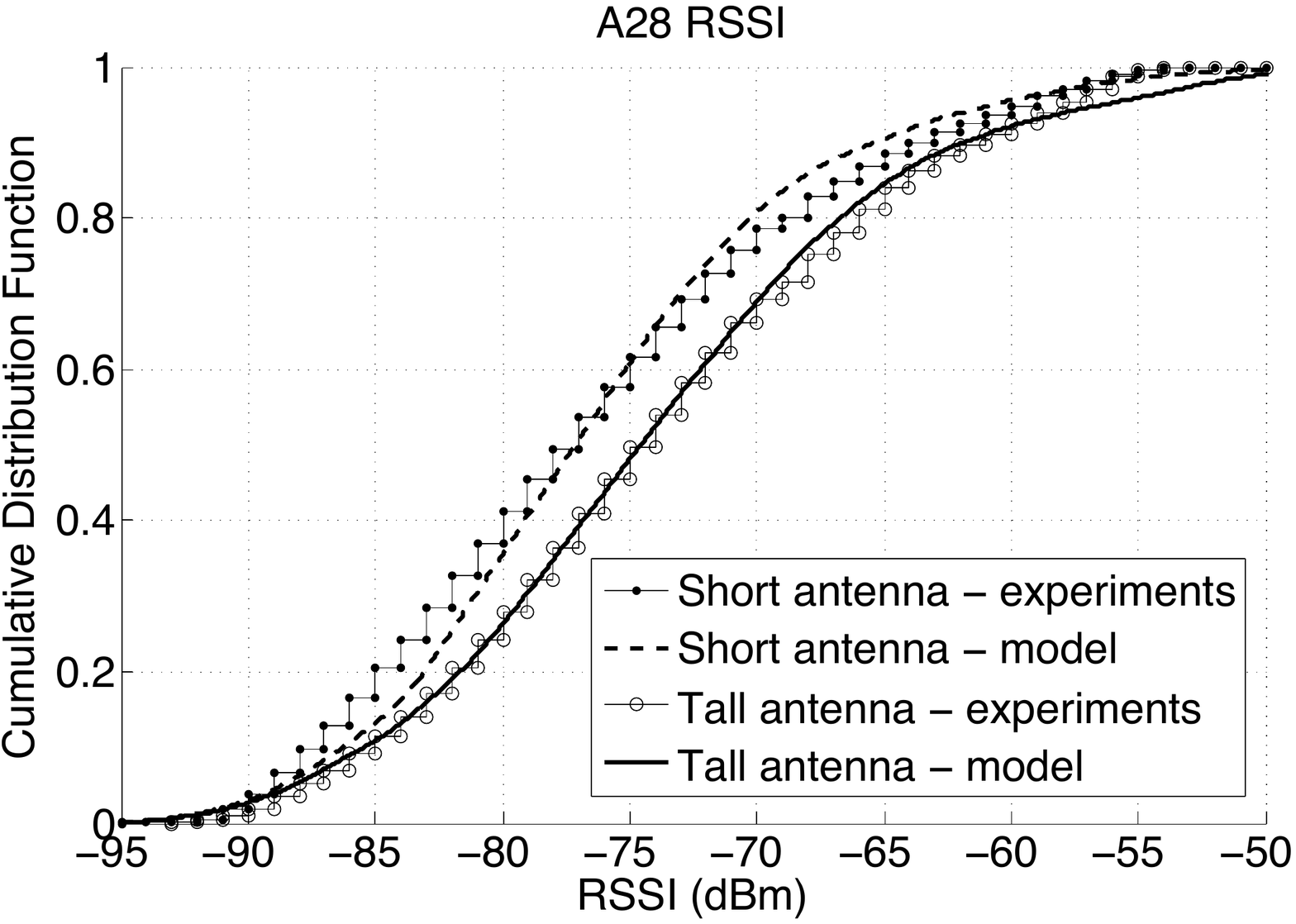}}
\subfigure[VCI Overall (aggregated LOS and NLOSv) RSSI]{\label{fig:2hop-rssi-ExpVsModVCI}\includegraphics[width=0.4\textwidth]{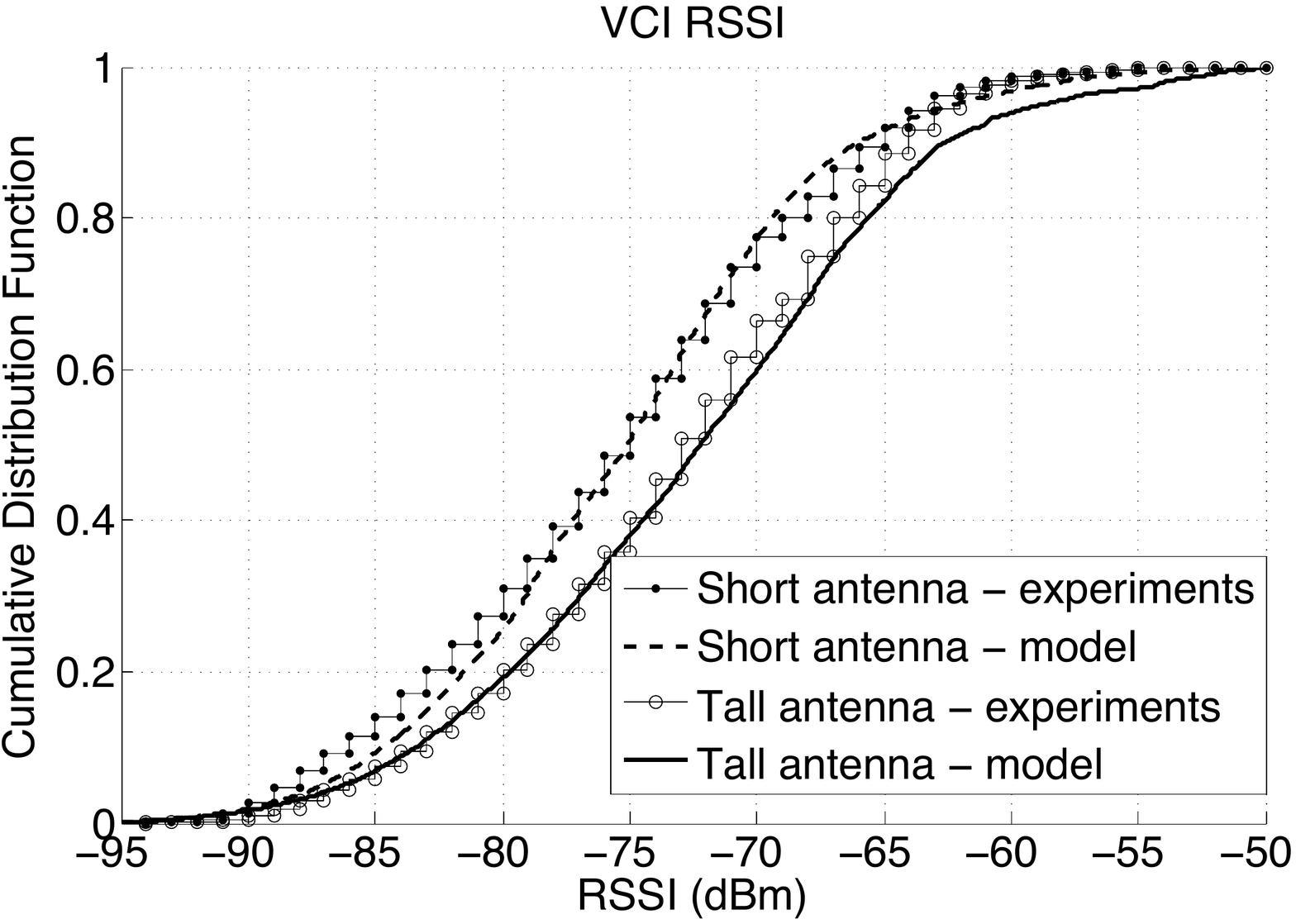}}\\
\caption[CDF of the RSSI for the tall and short relay antennas]{\small Cumulative Distribution Functions of the Received Signal Strength Indicator (RSSI) for the tall and short relay antennas, for both the car-car-car experiments (Fig.~\ref{linksSingleMultiHop}e) and the channel model. Both the LOS data (i.e., no obstruction) and non-LOS data (i.e., vehicle obstructions) is included. LOS data comprises 66\% of the total data, with the remaining 34\% being NLOSv.}
\label{fig:RSSICDF}
\end{figure*}

\begin{figure*}
\centering
\subfigure[NLOSv A28 RSSI gains]{\label{fig:2hop-rssi-gains-a28}\includegraphics[width=0.4\textwidth]{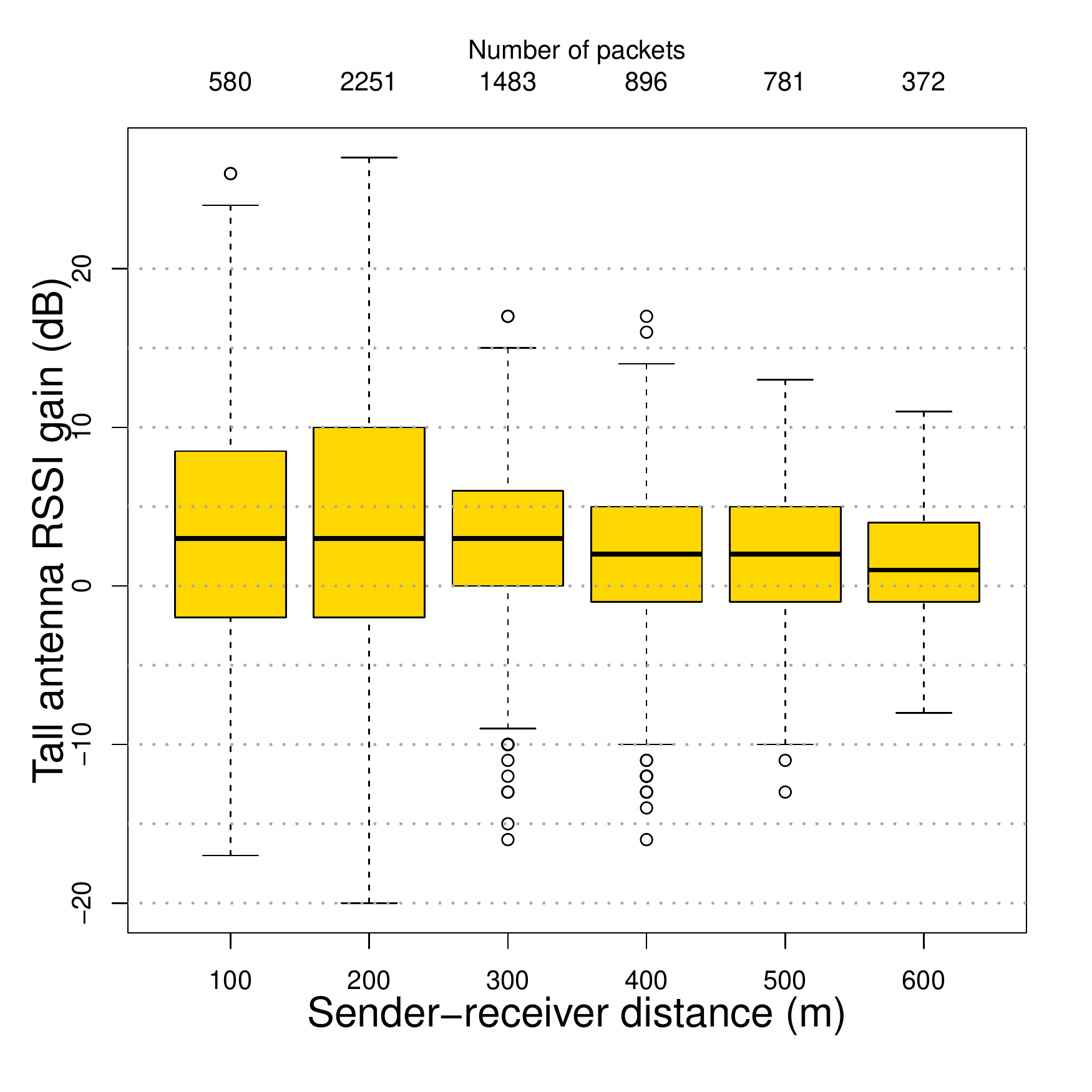}}
\subfigure[NLOSv VCI RSSI gains]{\label{fig:2hop-rssi-gains-vci}\includegraphics[width=0.4\textwidth]{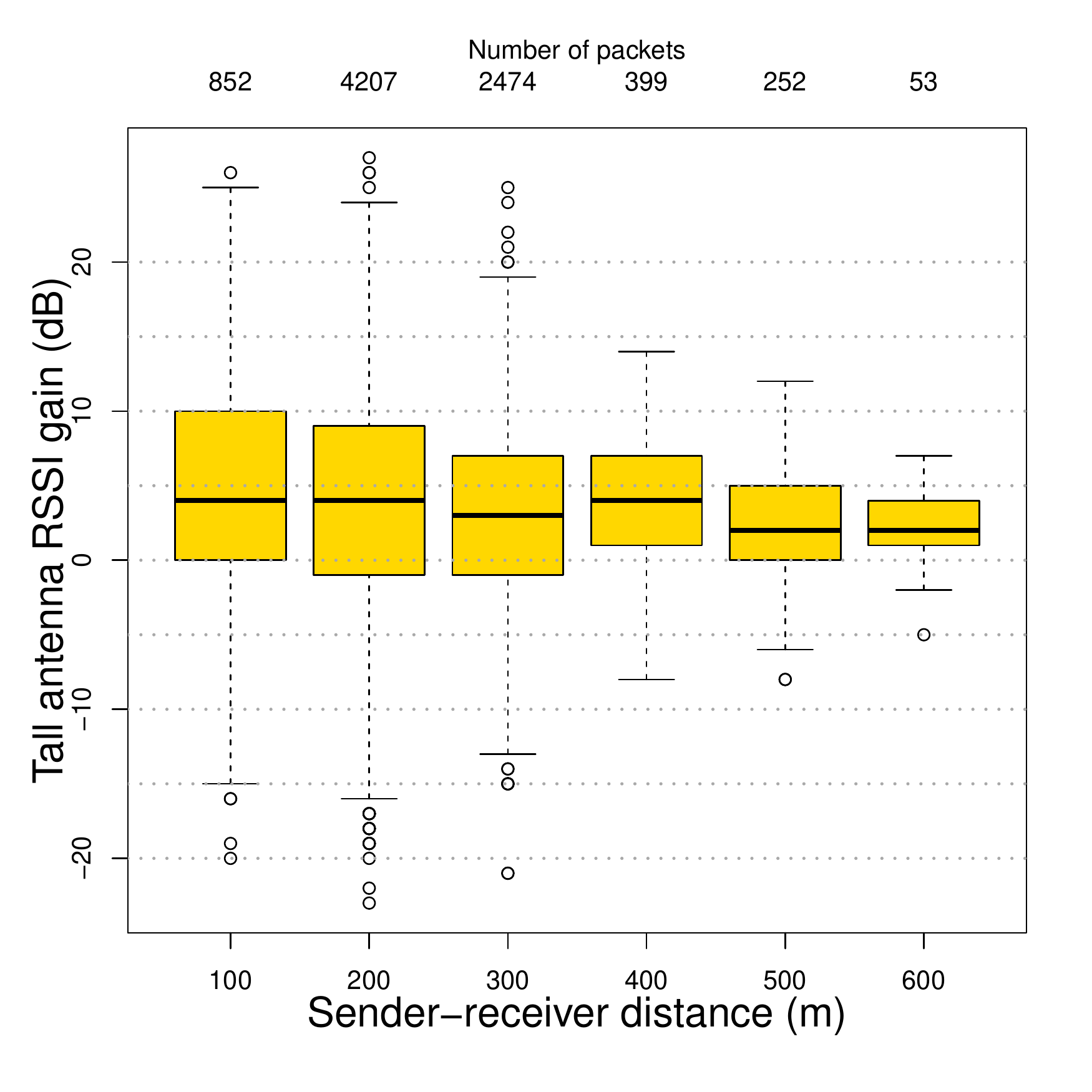}}
\caption[RSSI gains from the tall relay antenna relative to the short relay antenna]{\small %
RSSI gains from the tall relay antenna relative to the short relay antenna for the car-car-car experiments (Fig.~\ref{linksSingleMultiHop}e) under NLOSv conditions. Each box plot represents the median and lower and upper quartiles. The error bars represent the minimum and maximum ranges, except for outliers (more than 1.5 times the interquartile range), which are represented by small circles.}
\label{fig:2hop-rssi-gains}
\end{figure*}

Figure~\ref{fig:pdr:nlos} depicts the experimentally obtained PDR for NLOSv cases only, where there were other vehicles between the communicating vehicles that potentially obstructed the LOS. The shapes of the curves are similar to the overall case, %
with Van-X providing a clear advantage when compared to Car-Car communication at distances larger than 250 meters. %
When the received power is close to the reception threshold, the improved channel made possible by the use of tall vehicles often makes the difference between a decodable and a non-decodable packet. Figure~\ref{fig:pdr:nlosM} shows the PDR for NLOSv data as predicted by the model. As with the overall case, the results are similar to those obtained experimentally, thus validating the employed model.

From an application's point of view, the benefit of using tall vehicles as forwarders can be seen as an increase in the effective communication range given a certain delivery probability requirement. Figure~\ref{fig:effective-comm-range} shows the difference in communication range under NLOSv conditions, using the data derived from the graph in Fig.~\ref{fig:pdr:nlos}, as a function of the desired delivery ratio. Tall vehicles increased the effective communication range by a margin of up to 200 meters. %
The results show that significant benefits can be achieved by differentiating between different types of vehicles according to their height. Selecting tall vehicles allows for higher probability of LOS, increased network reachability and received signal power, all of which result in a higher PDR, which is of particular importance for effective implementation of safety applications~\cite{torrent06}.

\subsection{Experimental Results - Two Hop Experiments}

Figure~\ref{fig:2hopPDR} shows the overall (i.e. aggregated LOS and NLOSv) end-to-end
PDR results obtained for the two-hop experiments on the A28 and VCI highways (Figs.~\ref{linksSingleMultiHop}d and \ref{linksSingleMultiHop}e). There are three PDR curves for each highway: 1) for the car-van-car scenario (Fig.~\ref{linksSingleMultiHop}d); 2) for the car-car-car scenario using the low-mounted antenna as a relay (Fig.~\ref{linksSingleMultiHop}e), and 3) for the car-car-car scenario using the high-mounted antenna as the relay (Fig.~\ref{linksSingleMultiHop}e). Curves 2) and 3) share the exact same spatial and temporal conditions (vehicle density, surroundings, obstructing vehicles), whereas curve 1) was obtained by redoing the experiments with a van as a relay. 

The PDR results follow a trend similar to the one-hop results (Fig.~\ref{fig:pdr}), with both the van and the high-mounted antenna outperforming the low-mounted antenna as relays. The taller antenna results in an improvement of up to 20 percentage points when compared with the short antenna. Using a van results in an even more pronounced improvement of up to 40 percentage points at larger communication distances.

Fig.~\ref{fig:RSSICDF} shows the RSSI Cumulative Distribution Function (CDF) for both the car-car-car two-hop experiment (Fig.~\ref{linksSingleMultiHop}e), where the relay vehicle has both tall and short antennas, and the RSSI values generated by the channel model based on the vehicle location information and LOS conditions obtained during the experiments. The plots encompass the aggregated data for LOS and NLOSv. %
The tall relay antenna shows a consistent advantage over the short antenna, with up to 4~dB higher RSSI. Furthermore, there is a good agreement between the experimental and model-derived values.

To obtain a deeper insight into the benefits of a tall antenna in NLOSv conditions,  
Fig.~\ref{fig:2hop-rssi-gains} %
 shows the Received Signal Strength Indicator (RSSI) %
results in the form of a box plot for each 100~meter sender-receiver distance bin in the case of NLOSv communication. %
We computed RSSI difference for the pairs of packets that were received by both the high and the low-mounted antennas. %
The high-mounted antennas provide a median advantage between 2 and 4~dB in received signal strength on both experiments. As discussed earlier, the benefit is due to the higher-mounted antenna being less susceptible to blocking from the non-communicating vehicles.

\section{Large-scale Simulations} %
\label{sec:largeScale}

\begin{figure}
  \begin{center}
    \includegraphics[width=0.35\textwidth]{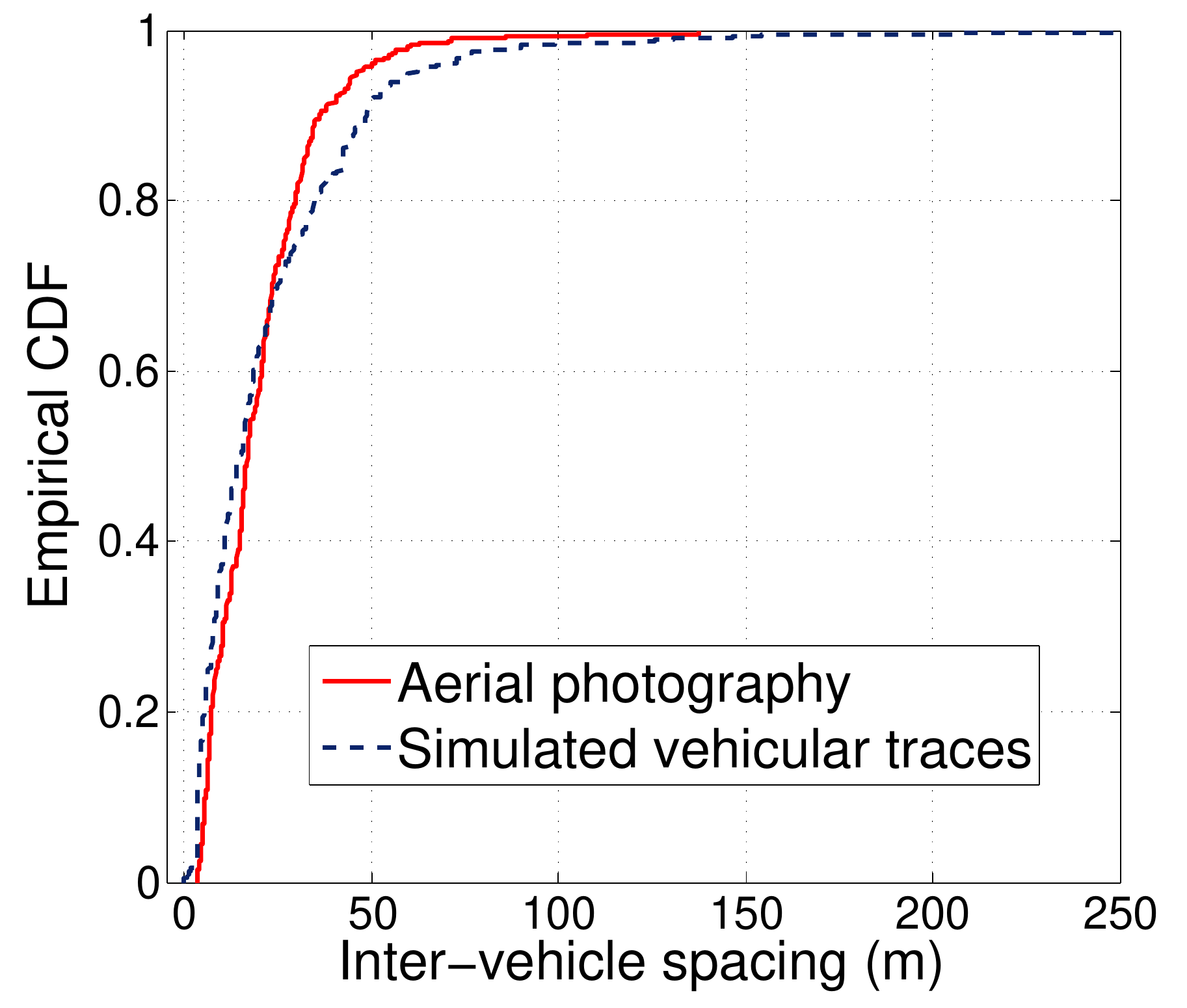}
     \caption[Inter-vehicle spacing for the simulated vehicular mobility trace]{\small Inter-vehicle spacing for the simulated medium density vehicular mobility trace and for the aerial photography of the A28 highway in Portugal.}
      \label{inter-veh-spacing}
   \end{center}
\end{figure}

We used the insights from the experiments to test the tall vehicle relaying concept on a system-wide scale. We generated vehicular traces using the STRAW mobility model~\cite{choffnes05} on a road of the same length (13.5~km), the same number of lanes (four), and similar shape to highway A28 where aerial imagery was acquired. We used three vehicular densities: 2.5, 7.5, and 10~vehicles/km/lane (designated in~\cite{naumov06} as low, medium, and high, respectively) while keeping the same percentage of tall vehicles of approximately 14\% as observed in the aerial dataset. This resulted in 135, 404, and 675 vehicles in the system for different vehicular densities. The medium density dataset was comparable to the A28 dataset (equal number of vehicles). We validated the traces against the aerial imagery by calculating the inter-vehicle distance (distance from each vehicle to its nearest neighbor for the generated medium density and the A28 dataset). Figure~\ref{inter-veh-spacing} shows a good agreement between the cumulative distribution function of the inter-vehicle spacing for the generated medium density traces and for the A28 highway, which also gives us confidence in drawing conclusions based on the generated vehicular traces for low and high densities. \\

\subsection{Relay Techniques Under Consideration}
To determine the impact of using tall vehicles as relays on end-to-end (i.e., multi-hop) communication, we implemented the following three techniques. In the subsequent text, we define a \emph{neighbor} as the vehicle which receives the signal from the current vehicle above the sensitivity threshold, based on the channel model from Chapter~\ref{ch:completeModel}.
 Furthermore, we define the \emph{best route} as the route that contains the least number of hops from the transmitter to the destination. 

\subsubsection{Most New Neighbors technique} 
This technique will select the neighbor that contributes the most new neighbors in the direction of the destination, which are not neighbors of the current sending node. The reasoning behind this technique is that a neighbor which contributes most new neighbors has the highest local connectivity (or, in other words, highest degree distribution) in the direction of the destination.
Consequently, the conjecture is that having more potential next hop relays increases the likelihood of delivering the message to the intended recipient. Referring to Fig.~\ref{MostNew-Tall-Farthest}, the selected vehicle (\emph{Most New}) has most new neighbors (three) that are not neighbors of the current sender (\emph{Source}). 
\subsubsection{Farthest Neighbor technique} 
This technique simply selects the farthest %
neighbor in the direction of the destination. 
Referring to Fig.~\ref{MostNew-Tall-Farthest}, the selected vehicle is designated as \emph{Farthest}. The intuition behind this technique is that maximizing the distance travelled in each individual hop will lead to a smaller number of hops to reach the destination. This technique has often been used in the literature (e.g., see~\cite{wisit07}). %
\subsubsection{TVR technique} 

\begin{figure*}
  \begin{center}
    \includegraphics[width=0.95\textwidth]{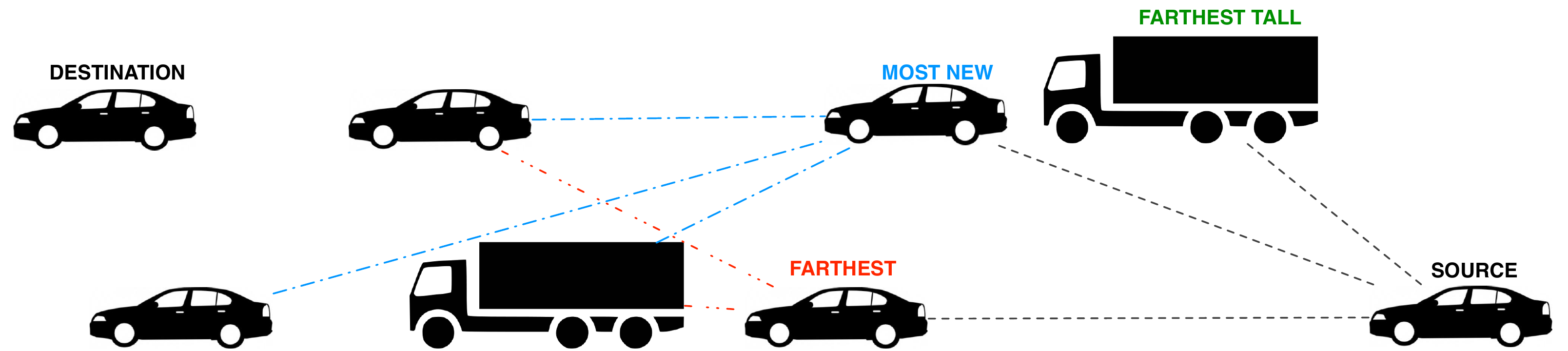}
     \caption[Relay selection for the three techniques]{\small Relay selection for the three techniques. In case of \emph{Most New Neighbors} relay technique, the vehicle designated \emph{Most New Neighbors} will be selected, as it has most new neighbors (three) in the direction of the destination (designated \emph{Destination}) that are not neighbors of the current sending node (designated \emph{Source}). In case of \emph{Farthest Neighbor} relay technique, vehicle designated \emph{Farthest Neighbor} will be selected, as it is farthest from the current sending node, and in the direction of the destination \emph{Dest}. In the case of \emph{TVR}, the tall vehicle designated \emph{Farthest Tall} will be selected. Note that a single vehicle can be selected by multiple techniques (e.g., farthest vehicle might have most new neighbors, and it can also be a tall vehicle, which would make it the best relay for all three techniques)}
      \label{MostNew-Tall-Farthest}
   \end{center}
\end{figure*}

Based on the benefits we observed when performing the tall vehicle experiments (see Figs.~\ref{fig:pdr},~\ref{fig:effective-comm-range},~\ref{fig:2hopPDR}), we implemented a technique that selects the farthest tall vehicle in the direction of the message destination (Fig.~\ref{MostNew-Tall-Farthest}), provided that $dist(Tx,Far_{Short})-dist(Tx,Far_{Tall})\leq x_{max}$, where $dist(x,y)$ is the Euclidean distance between points $x$ and $y$, $x_{max}$ is the maximum distance difference at which a tall vehicle is still a better relay, $Tx$ is the location of the transmitter, and $Far_{Short}$ and $Far_{Tall}$ are the locations of the farthest reachable short and tall neighbors, respectively. %
In other words, TVR selects a tall vehicle if the distance difference between the farthest tall vehicle and the current transmitter and the farthest short vehicle and the transmitter is less than a threshold $x_{max}$; otherwise, the farthest short node is selected. 

At each hop, we define the vehicle selected by a given technique as the \emph{best relay} for that technique. %

\subsection{Calculating $x_{max}$}
\begin{figure}
  \begin{center}
    \includegraphics[width=0.65\textwidth]{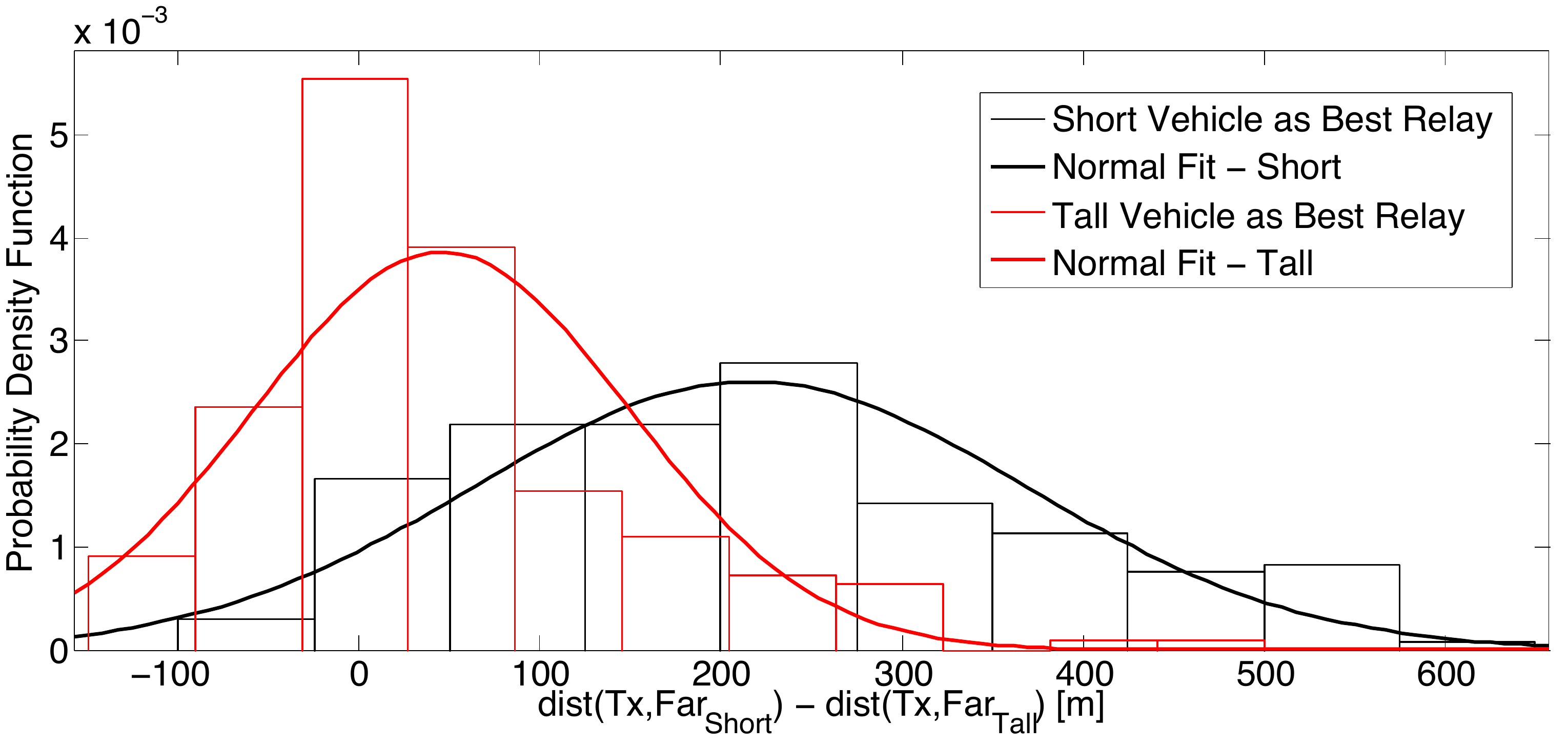}
     \caption[Probability distributions of distance from transmitter to farthest short and tall vehicle]{\small Probability distributions of the distance from the transmitter to the farthest short and farthest tall vehicle $dist(Tx,Far_{Short})-dist(Tx,Far_{Tall})$ for a transmit power of 10~dBm, tested on the aerial photography data of the A28 highway. %
 Negative distance implies that the tall vehicle is farther from the transmitter than the short vehicle. For the given transmit power, when a short vehicle is the best relay, it is on average 210~meters farther from the transmitter than the tall vehicle. When a tall vehicle is the best relay, it is on average 50~meters closer to the transmitter than the short vehicle.} %
      \label{tallVsShort}
   \end{center}
\end{figure}

In order to calculate  $x_{max}$, we first look at the distribution of distance difference $dist(Tx,Far_{Short})-dist(Tx,Far_{Tall})$, as shown in Fig.~\ref{tallVsShort}, which was derived from aerial photography.
The case when a tall vehicle is the best relay (in terms of least number of end-to-end hops) is the distribution colored red, whereas the case when a short vehicle is the best relay is colored black. To determine when a tall vehicle is more likely to be a better relay, let us define a binary random variable $\theta$ as being one when a tall vehicle is more likely to be a better relay, and zero otherwise: %

\begin{equation}
\theta = \left\{ \begin{array}{cl}1, & \textrm{when } \displaystyle \frac{\int_{-\infty}^{\textstyle x} f_T(t)dt}{\int_{\textstyle x}^{+\infty} f_S(s)ds}>1;\\
0, & \textrm{otherwise,}
\end{array} \right.
\label{eq:Ptheta}
\end{equation}
where $f_T(t)$ and $f_S(s)$ are probability distributions  of $dist(Tx,Far_{Short})-dist(Tx,Far_{Tall})$ %
for best tall vehicle and best short vehicle case, respectively. In other words, we can interpret eq.~\ref{eq:Ptheta} as $\theta=1$ when the cumulative distribution $F_T(t)$ for a given value $x$ is larger than the complementary cumulative distribution of $F_S(s)$ and $\theta=0$ otherwise.

In order to calculate the maximum distance difference $x_{max}$ at which a tall vehicle is still a better relay, we need to solve $F_T(t) = 1- F_S(s)$.
In the specific case of our collected data, %
for tractability purposes we approximate the distance difference distributions of $s$ and $t$ with normal distributions (normal fits shown in Fig.~\ref{tallVsShort}). In this case, $x_{max}$ can be calculated by solving

\begin{equation}
1 - Q\left(\frac{x_{max}-\mu_s}{\sigma_s}\right) = Q\left(\frac{x_{max}-\mu_t}{\sigma_t}\right),%
\label{eq:Qfn}
\end{equation}
where $\mu_s$, $\sigma_s$, $\mu_t$, and $\sigma_t$ are the means and variances of $s$ and $t$, respectively, and $Q(\cdot)$ is the $Q$-function, defined as $Q(x) = \frac{1}{\sqrt{2\pi}} \int_x^\infty \exp\Bigl(-\frac{u^2}{2}\Bigr) \, du$. 

Figure~\ref{tallVsShort} shows the distributions of $s$ and $t$ for a single transmit power (10~dBm); to analyze the behavior of $s$ and $t$ with different communication ranges, we vary the transmission power from 1 to 20~dBm.
Distributions of $s$ and $t$ are readily available in simulators by implementing an appropriate channel model (such as~\cite{boban11}), since the global network knowledge (``oracle'') is available. However, 
obtaining these distributions  %
is not straightforward without global knowledge, %
which means that %
the distributions of $s$ and $t$ will not be available to the routing protocols in the real world. Therefore, we %
set a fixed value for $x_{max}$. %
 We used a value of $x_{max}$ calculated based on the aerial photography dataset in Table~\ref{dataset} as follows. We choose $x_{max}$ to be the average value of $t$ across %
 transmission powers from 1 to 20~dBm (typical transmit powers for the DSRC standard). %
Formally, 
\begin{align}   
x_{max}&= \sum_{i=1}^{20}E[t|Pwr=i~dBm]\cdot P[Pwr=i~dBm] \\
\nonumber&=\frac{1}{20}\sum_{i=1}^{20}E[t|Pwr=i~dBm]\\
\nonumber&=\frac{1}{20}\sum_{i=1}^{20}\int_{-\infty}^{\infty}tf_T(t|Pwr=i~dBm)dt,
\label{eq:proradi}
\end{align}   
where $Pwr$ is the transmit power.
The calculated value is $x_{max} = 50$~meters (i.e., in the simulations, we use a tall vehicle %
as the next hop when $dist(Tx,Far_{Short})-dist(Tx,Far_{Tall})\leq 50$). Note that calculating $x_{max}$ based on specific values of $E[t|Pwr]$ yields better results for that specific transmission power. However, using different values of $x_{max}$ might be impractical for protocol implementation, as it may vary across different environments. %

\subsection{Comparing the Performance of the Relay Techniques}
\begin{figure}
  \begin{center}
    \includegraphics[width=0.65\textwidth]{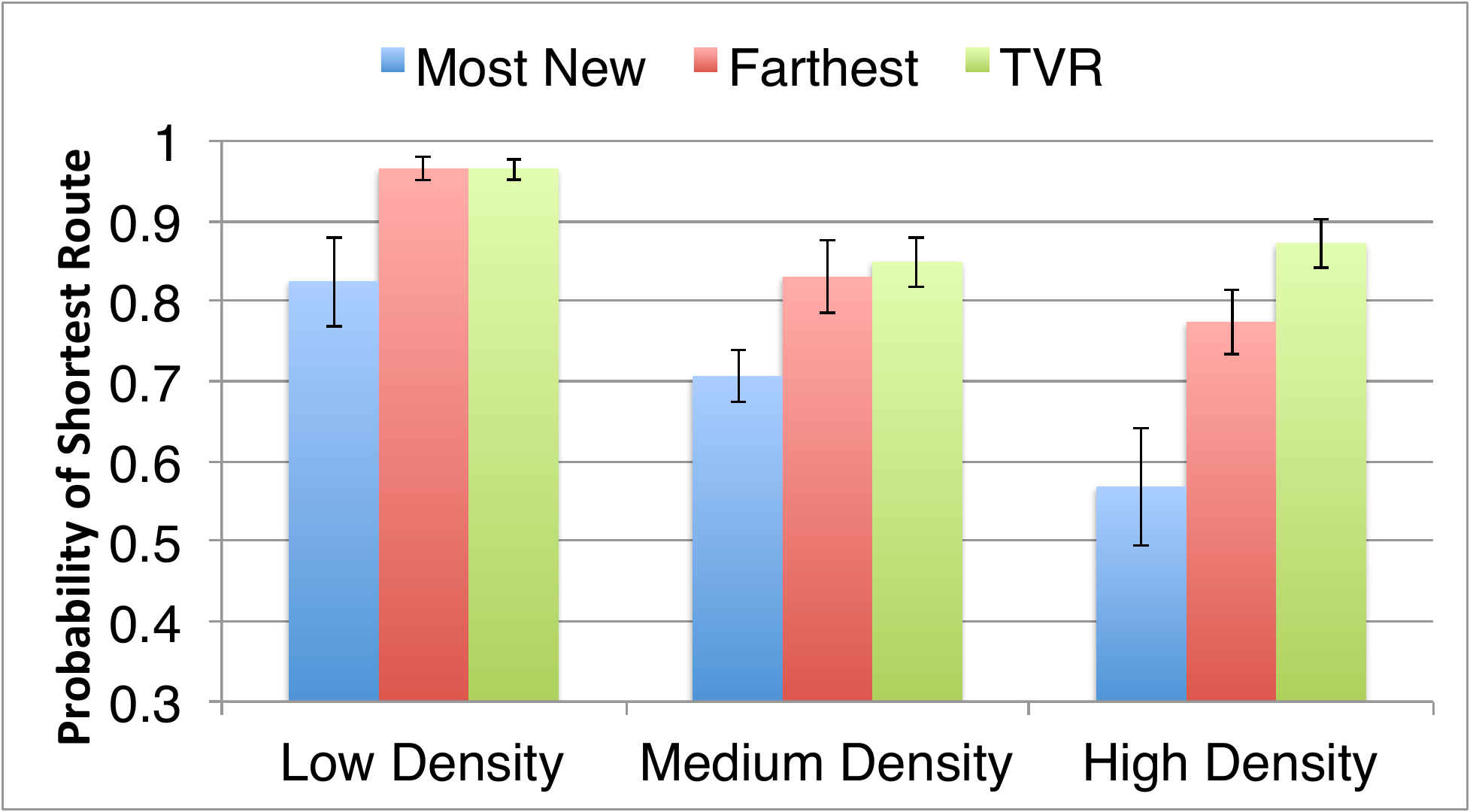}
     \caption[Performance of the three techniques]{\small Performance of the three techniques in terms of the percentage of minimum hop routes from source to destination. Error bars represent one standard deviation drawn from the 20 different power settings (from 1 to 20 dBm).}
      \label{techniquesBestRouteComparison}
   \end{center}
\end{figure}

To determine the performance of the three techniques,  %
in each generated vehicular mobility dataset (i.e., low, medium, and high),
we randomly selected a set of source-destination pairs such that the source and destination are not direct neighbors. The number of analyzed source-destination pairs for each transmit power was 10000. To have a fair comparison, we used the same set of randomly selected pairs to test all three techniques. %
The total number of source-destination pairs analyzed across different densities and transmit powers was $10^4\times3\times20=6\times10^5$. %
Figure~\ref{techniquesBestRouteComparison} shows the comparison of the three relaying techniques in terms of the probability of selecting a shortest (minimum-hop) route. Shortest route for a source-destination pair is defined as the least number of hops achieved by any of the three techniques. This was taken as a baseline: any of the techniques that had more than this number of hops did not choose the best route. %
 \emph{TVR} equals or outperforms the remaining two techniques, and as the density increases, its performance relative to the other two techniques improves. It is comparable to the \emph{Farthest Neighbor} technique at low density, on average 1.5 percentage points better than it at medium density, and 10 percentage points better at high density. Reduced number of hops exhibited by \emph{TVR} directly affects the end-to-end delay: fewer hops means a shorter time to get to the destination. Reducing hops is particularly important in dense vehicular networks, where broadcast storms can occur~\cite{wisit07_2}. Fewer hops results in a decrease of both the number of transmissions and the end-to-end delay. 

It is interesting to see that the ratio of best routes per technique decreases as the vehicular density increases; this is due to the inability of any particular technique to always find %
 the best next relay. When the vehicular density is low, there are fewer neighbors to choose from, therefore choosing the one with best properties is easier. As the density increases, the ability to choose that specific relay decreases. \\

\subsection{Properties of Selected Best-Hop Links}
\begin{figure}
  \begin{center}
    \includegraphics[width=0.65\textwidth]{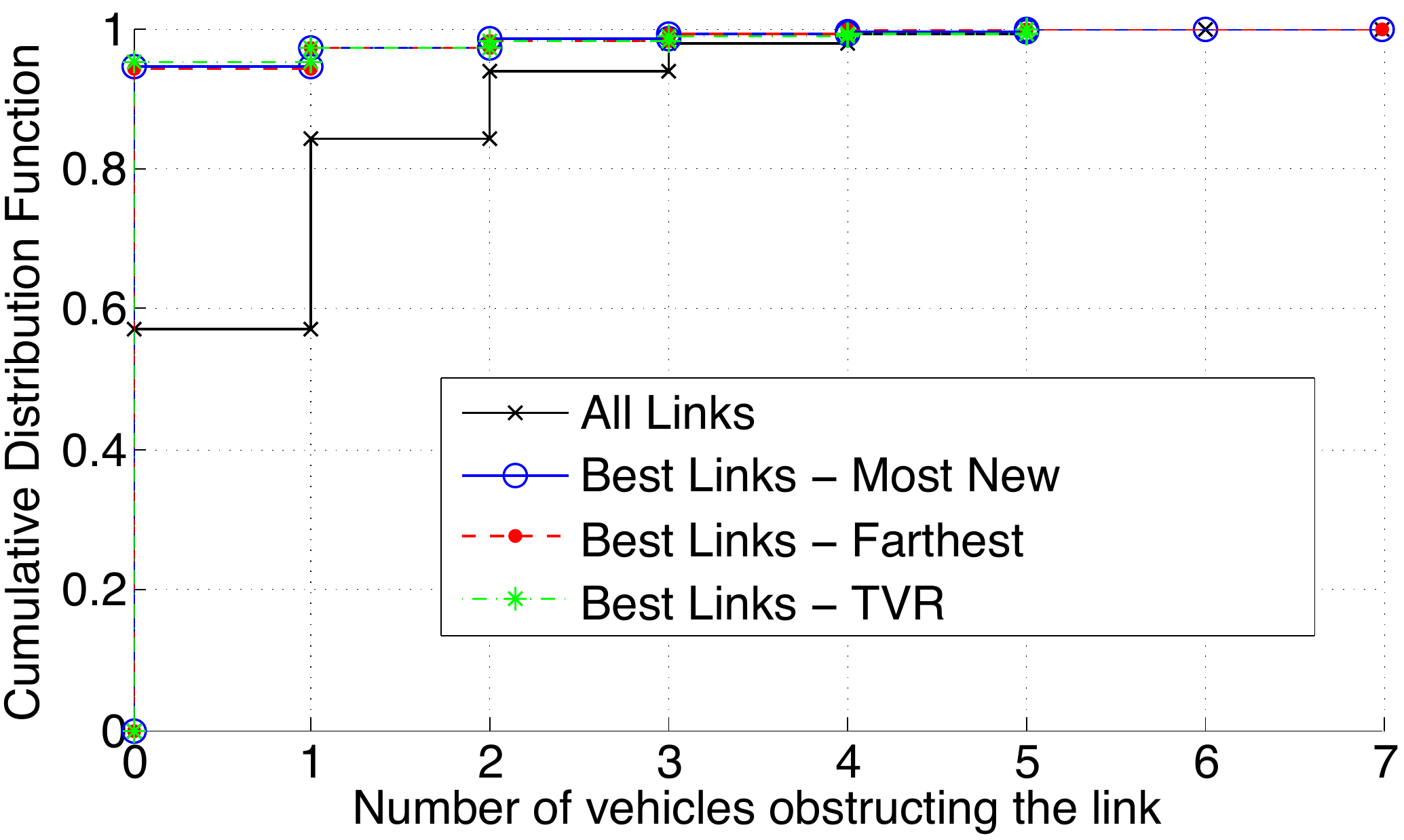}
     \caption[Difference between the number of obstructing vehicles: all links vs best links]{\small Difference between the number of obstructing vehicles in all links above the reception threshold in the system and the best links selected by the three employed techniques (\emph{Most New Neighbors}, \emph{Farthest Neighbor}, \emph{TVR}). Tested on the aerial photography data of the A28 highway. Power settings: Tx Power 10~dBm; Receiver sensitivity threshold: -90~dBm. Other power settings exhibit similar behavior. %
}
      \label{allSchemesVsTall}
   \end{center}
\end{figure}

Figure~\ref{allSchemesVsTall} shows the number of vehicles obstructing the LOS for the links selected by the three techniques (i.e., best-hop links) as well as all the links in the system. While system-wide only 58\% of links have LOS (i.e., zero obstructing vehicles), all three employed techniques select LOS links more than 92\% of the time. This result emphasizes the importance of having unobstructed (LOS) link conditions. %
All three techniques are implicitly preferring the LOS links: the next hop in the \emph{Most New Neighbors} technique will often have the most new neighbors due to privileged LOS conditions; with \emph{Farthest Neighbor} technique, the farthest neighbor most often has LOS with transmitter -- therefore, it receives the message above the threshold at a farther distance; and \emph{TVR} benefits from the increased height of the antennas on tall vehicles to reduce the chance of NLOSv.

\subsection{How Often is a Tall Vehicle Relay Available?}

The measurements %
described in~\cite{boban11}, \cite{wisit07}, and~\cite{bai09} %
show that the inter-vehicle spacing for free-flow traffic
follows an exponential distribution:
\begin{equation}
f_{K}(k)=\lambda_s e^{-\lambda_s k},
\label{eq:exp}
\end{equation}
where $\lambda_s$ is the inverse of the average inter-vehicle spacing %
in meters. For a certain ratio $\gamma$ of tall vehicles ($0\leq\gamma\leq1$), we have the following inter-vehicle spacing distribution for tall vehicles:
\begin{equation}
f_{K}(k)=\gamma\lambda_s e^{-\gamma\lambda_s k}.
\label{eq:expT}
\end{equation}
To calculate the probability $P_{T}$ of there being at least one tall vehicle relay within a certain average communication range $R$, we calculate the complement of the probability of having zero tall vehicles within R: %
\begin{align}
P_{T} %
& = 1-Pr(k\geq R) \\
\nonumber& =F_K(R) \\
\nonumber& = 1- e^{-\gamma\lambda_s R},
\label{eq:PrNoTallinho}
\end{align}   
where $F_K(\cdot)$ is the Cumulative Distribution Function (CDF) of the inter-vehicle spacing between tall vehicles. It has to be noted that, in real situations, $R$ is going to be a variable that is dependent on many factors (transmission power, road surroundings, etc.), including the vehicle density, since the increased vehicular density will decrease the transmission range, as shown in~\cite{meireles10}. Therefore, we consider $R$ as an average communication range for which the value can be determined from experimental data such as that in Fig.~\ref{fig:effective-comm-range}.
However, for the employed \emph{TVR} technique we are not interested in the existence of a tall vehicle within the entire $R$; rather, we are interested in a distance interval $[R-x_{max}, R]$, where $x_{max}$ is calculated %
as described in eq.~\ref{eq:Qfn} and eq.~3 %
and $x_{max}\leq R$. Therefore, we have the following probability %
of having at least one tall vehicle relay within $[R-x_{max}, R]$:

\begin{align}
P_{T[R-x_{max},R]} & =P_{T[0,x_{max}]} \\
\nonumber& = 1-Pr(k> x_{max}) \\
\nonumber& = F_K(x_{max}) \\
\nonumber& = 1-e^{-\gamma\lambda_s x_{max}},
\label{eq:PrNoTallinho}
\end{align} 
where the first step is a consequence of the memoryless property of the exponential distribution.
We analyze a fully connected network (i.e., at a certain point in time, each node has a route to all other nodes) with free-flow traffic\footnote{Free-flow traffic is defined as traffic where each vehicle is free to move at the desired speed~\cite{may90}, meaning the traffic volume is low enough so there are no traffic-induced decelerations. The converse of free-flow is high volume traffic near or in congestion. Arguably, in such a network, for the same ratio of tall vehicles, the probability distribution of tall vehicles, and therefore the probability of having a tall vehicle neighbor will be lower-bounded by %
eq.~\ref{eq:exp},~\ref{eq:expT}, and~\ref{eq:PrNoTallinho}.}. The converse setup would involve disconnected networks, which is beyond the scope of this study. %

\begin{figure}
  \begin{center}
    \includegraphics[width=0.65\textwidth]{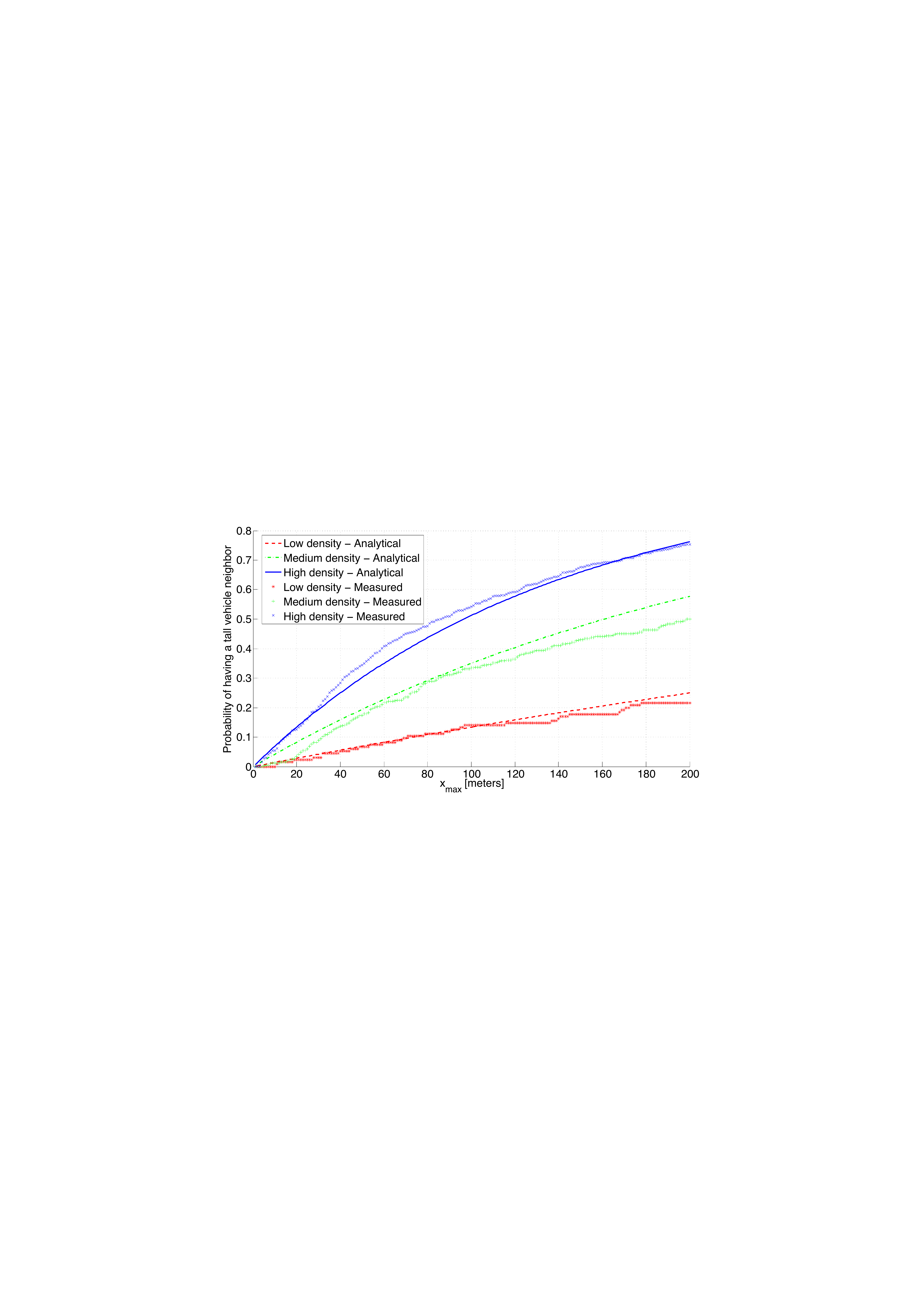}
     \caption{Probability of having a tall vehicle neighbor within $[R - x_{max}, R]$.}
      \label{ProbTallVehNeighbor}
   \end{center}
\end{figure}

Figure~\ref{ProbTallVehNeighbor} shows the analytical probability (eq.~\ref{eq:PrNoTallinho}) of having a tall vehicle neighbor within $[R - x_{max}, R]$ compared to that measured using aerial photography (medium density) %
and the generated vehicular traces (low and high density). %
There %
is a good match between the measured and analytical results; in both cases, the probability is approximately 35\% and 20\% when  $x_{max}=50$~meters for high and medium density, respectively. Only for low density the probability is below 20\% even with $x_{max}=150$~meters. This result explains why at higher densities \emph{TVR} performs better: the increase in the overall number of neighboring vehicles increases the probability of having a tall vehicle within the $[R - x_{max}, R]$ region, thus enabling the selection of shorter routes via tall vehicles. In low density scenarios, there simply are not enough tall vehicles to make a positive difference, therefore \emph{TVR} most often falls back to farthest neighbor relaying.

\begin{table}
	\centering
		\caption{ Percentage of Vehicles Used for Relaying} 
\begin{tabular}{|c c c c|} \hline
		 \bf Density & \bf Low & \bf Medium & \bf High \\ 
		 \bf Technique &&& \\ \hline \hline
		\bf Most New Neighbors & 44\% & 40\% & 31\%\\ \hline
		\bf Farthest  & 34\% & 28\%&  26\%\\ \hline
		\bf TVR & 33\% & 27\% & 21\%\\ \hline
		\end{tabular} 
\label{tab:percentageRelaying}
\end{table}

\subsection{Does TVR Create Bottlenecks on Tall Vehicles?}
In this study, we focused on the effects of tall vehicle relaying in terms of per-hop increase in the received power (i.e., ``physical layer'') and improvement in end-to-end relaying by reducing the number of hops (i.e., ``network layer''), thus directly decreasing both the delay and the overall number of messages that need to be exchanged in the system. %
For both of these metrics, \emph{TVR} was shown to perform better than other techniques. 
However, in our simulations, we assumed a perfect medium access scheme, which does not incur any contention or interference-induced losses.  %
Therefore, one question arises: if the majority of data traffic is relayed over tall vehicles, does this create bottlenecks -- situations where the tall vehicles cannot support the traffic being relayed over them? 
To answer this question, we analyzed the percentage of vehicles that are used for relaying as follows. For each technique, the same set of 10000 source-destination routes per vehicle density were taken into consideration, and the percentage of total number of vehicles used as relays by \emph{any} route has been reported in Table~\ref{tab:percentageRelaying} (results rounded to the nearest percentage point). As can be seen, the \emph{TVR} technique does use a smaller percentage of vehicles; however, the difference is at most five percentage points %
when compared to the \emph{Farthest Neighbor} technique. %
Furthermore, this result also implies that neither of the techniques uses all vehicles in the system; rather, those vehicles are selected that have strategically better positions for relaying (e.g., a vehicle connecting two otherwise disconnected clusters, a vehicle that has a clear LOS with the most neighbors, etc). %

\section{Related Work} \label{sec:RelatedWork}

A number of VANET studies have pointed out the importance of antenna height in different contexts. %
The benefits of vertical antenna diversity were explored by Oh et al.~in~\cite{oh09}, where antennas were vertically displaced by 0.4~meters on a passenger car (i.e., a short vehicle) by installing one antenna inside the passenger cabin and a number of antennas on the car's roof. %
Both parking lot and on-road experiments were conducted using IEEE 802.11a radios operating in the 5.2~GHz frequency band. While mainly focusing on mitigating the negative effects of ground reflections rather than dealing with vehicular obstructions, the results show that the vertical diversity increases the effective communication range by more than 100 meters in certain scenarios. Kaul et al. 
reported a similar study in~\cite{kaul07}, with a focus on determining the single best location for an antenna in a passenger car. By performing parking lot and on-road experiments using IEEE 802.11a radios operating in the 5.2~GHz frequency band, the center of the roof was found to be the best overall position, with significant variation in reception patterns when the antenna was displaced horizontally and vertically. On the other hand, two simulation studies based on detailed ray-optical channel models (Reichardt et al.~in~\cite{reichardt09} and Kornek et al.~in~\cite{kornek10}) indicate that antenna positions  other than those on the roof (e.g., on side mirrors) can be preferable in certain scenarios. %

With respect to Vehicle-to-Infrastructure (V2I) links and the impact of antenna placement, %
Paier et al.~in~\cite{paier10_2} performed experiments %
which showed significantly better results %
with a road-side unit (RSU) that was placed above the height of the tallest vehicles. %
Placing the RSUs higher up %
results in a more reliable communication channel, which is particularly important for safety related applications. Since the RSU radio design is similar to the on-board unit (OBU) radios in vehicles, this finding suggests that the same applies for V2V communication; i.e., placing the antennas on taller vehicles is likely to result in improved radio channel. A similar study was reported in \cite{paier10}, where the authors analyzed the performance of a downlink between an RSU and an OBU installed in a vehicle. Antenna heights and traffic had a severe impact on the downlink performance, and the authors pointed out that ``shadowing effects caused by trucks lead to a strongly fluctuating transmission performance, particularly for settings with long packet lengths and higher speeds.'' This reinforces the findings reported by Meireles et al.~in~\cite{meireles10}, where high losses were observed when obstructing vehicles were present between communicating vehicles.

Regarding the performance analysis and modeling of LOS and non-LOS (NLOS) %
Tan et al.~in~\cite{tan08} performed V2V and V2I measurements in urban, rural, and highway environments at 5.9~GHz. The results point out significant differences with respect to delay spread and Doppler shift in case of LOS and NLOS  channels (NLOS was often induced by trucks obstructing the LOS). The paper distinguishes LOS and NLOS communication scenarios by coarsely dividing the overall obstruction levels. Similarly, Otto et al.~in~\cite{Otto2009} performed V2V experiments in the 2.4~GHz frequency band in an open road environment and reported a significantly worse signal reception during a heavy traffic, rush hour period in comparison to a no traffic, late night period. In the WINNER project~\cite{baum05}, a series of 5.3 GHz wireless experiments were performed with a stationary base station and a moving node. The results were then used to derive channel models for use in simulation. Higher antenna heights were found to be beneficial to communication: the higher the antenna, the lower the path loss exponent.
Several other experimental studies and surveys discuss potential impact of vehicles %
on the channel quality:~\cite{abbas11,dhoutaut06,paier07,matolak05,matolak08, matolak09,usdot06_2, Jerbi2007}. %

Many relay selection metrics have been proposed for vehicular networks. The most common can be divided into: \textbf{1)} hop-count metrics (e.g., \cite{namboodiri04}); \textbf{2)} received power metrics (e.g., \cite{naumov06}); \textbf{3)} metrics based on geographic characteristics such as vehicle position, direction, or map information, etc. (e.g., \cite{naumov07}, \cite{lochert03}); and \textbf{4)} vehicular density based metrics (e.g., \cite{wisit07}). Combination of two or more of these metrics is also common in the literature. 
We have shown that relaying messages over tall vehicles is beneficial in terms of the hop count metrics (\emph{TVR} results in fewer hops, particularly in dense vehicular networks) and received power metrics (tall vehicles exhibit higher received power, PDR, and communication range). %
Apart from our preliminary study reported in~\cite{boban11_2}, to the best of our knowledge, none of the existing studies proposed utilizing the %
information about the type and height of vehicles to improve the performance of V2V communication. %

\section{Conclusions} \label{sec:Discussion} %
We have determined the benefits of utilizing the height of vehicles %
to enable more efficient V2V communication.
We have shown that using knowledge about vehicle type/height to appropriately select the next hop vehicle consistently results in increased effective communication range and larger per-hop message reachability. %
Through both experiments and simulations that use a validated model, we have shown that tall vehicles are significantly better relay candidates than short vehicles when tall vehicles are within a certain distance of the farthest vehicle. Selecting tall vehicles in such situations results in a higher received signal power, %
increased packet delivery ratio, and larger effective communication range. 

Furthermore, we characterized the properties of preferred next hops in an experimental setting and by evaluating three relay techniques through large-scale simulations. %
Both experiments and simulations showed that, when available, LOS links are preferred, regardless of the specific environment or relaying technique.
However,
since the distinction between LOS and NLOS links is not straightforward at the transmitter, %
we propose the tall vehicle relay (\emph{TVR}) technique, which increases the likelihood of having a LOS link. %
We have shown that by selecting tall instead of farthest vehicles, %
\emph{TVR} outperforms other techniques in terms of the number of hops to reach the destination, which in turn reduces the end-to-end delay. %
Therefore, the farthest neighbor metric might not be the best solution for selecting the next-hop relay where heterogeneous vehicle types exist (i.e., tall and short). The type of potential relay candidate can play an important role in deciding which next hop to select. %
Additionally, since TVR increases the received power level and reduces hop count, it can be used to improve performance of \emph{existing} routing protocols by adding binary information on the type of vehicle (tall or short).

It is important to note that our findings can be used to enhance different types of routing protocols, be it unicast~\cite{boban08, boban09}, broadcast~\cite{viriyasitavat11, tonguz10_2}, geocast~\cite{lochert03} or multicast~\cite{kihl07}. On highways, trucks and other tall commercial vehicles can be used as moving hotspots that relay the messages between the shorter vehicles. In urban environments, public transportation vehicles such as buses and streetcars can be used for the same purpose. %

\chapter{Conclusion} \label{ch:conclusion}

The main goals of this thesis were to realistically and efficiently model the vehicle-to-vehicle (V2V) channel and to design protocols that improve V2V communication. Towards this goal, we started by performing an extensive set of experiments in order to set up the ``ground truth'' regarding the following: 1) the impact of vehicular obstructions on inter-vehicle communication; 2) the impact of static objects (buildings and foliage); and 3) the performance characteristics of different vertical antenna displacements.  
The measurements confirmed our hypothesis that vehicles have a significant impact on signal propagation. Next, using the insights from measurements, we designed a computationally efficient model that enables realistic modeling of the impact of vehicles. Then, in order to enable simulating the channel in areas where static objects play an important role (e.g., urban and suburban areas), we introduced an efficient and simplified geometry-based model that incorporates the effects of both vehicles and static objects (namely, buildings and foliage) in terms of large-scale propagation effects (shadowing) as well as small-scale (multipath fading). The model exhibits realistic behavior as validated using the measurements. At the same time, the model is computationally efficient and requires minimum geographic descriptors (namely, outlines of the modeled objects) that are readily available. This makes it suitable for implementation in large scale, discrete-event VANET simulators. Finally, in order to alleviate the shadowing effects caused by vehicles, we proposed a technique that selects tall vehicles (e.g., buses, trucks) as next-hop relays where appropriate. The technique improves communication performance in terms of the number of hops, end-to-end delay, and effective communication range.

\section{Contributions}
The main contributions of this thesis are as follows. 
\begin{itemize}
\item \textbf{Experimental evaluation of the impact of vehicular obstructions on inter-vehicle communication}. %
We characterized in detail the impact that LOS-obstructing vehicles have on inter-vehicle communication in terms of received power, packet delivery rate, and effective communication range by performing extensive measurements in different environments (open space, highway, suburban, urban, parking lot). 

\item \textbf{Model for incorporating vehicles as obstacles in channel models.} We present a model that accounts for vehicles as three-dimensional obstacles and takes into account their impact on the line of sight obstruction, received signal power, and packet reception rate. %
The model is based on the empirically derived vehicle dimensions, accurate vehicle positioning, and realistic mobility patterns. We validate the model against measurements and it exhibits realistic propagation characteristics while maintaining manageable complexity.
  
\item \textbf{Computationally efficient, geometry-based, validated channel model for vehicular communications.} The model incorporates both static objects (buildings, foliage) and mobile (vehicles on the road). %
It requires minimum geographic information: 1) Information about the location of the vehicles on the road; and 2) Information about the outline of static objects.  %
We validate the model against measurements and show that it successfully captures both the large-scale propagation effects as well as fading in different environments (highway, urban, suburban, open space). 

\item \textbf{Tall Vehicle Relaying technique.} We performed an experimental study to determine the benefits of using tall vehicles as next-hop relays. The results showed that selecting tall vehicles is beneficial in terms of higher power at the receiver, smaller number of hops to reach the destination, and increased per-hop communication range. Based on these results, we proposed a technique that improves the inter-vehicle communication by selecting tall vehicles (e.g., buses, trucks) as next-hop relays. The technique improves communication performance by reducing the number of hops needed to reach the destination and increasing the effective communication range. 
\end{itemize}

\section{Future Work}

The following are possible directions in which the work done in this thesis could be extended.

\begin{itemize}	
\item Currently, we assume that the terrain is flat. %
With mainly flat terrains, this simplification does not have a significant bearing on the results. %
For locations with significant elevation changes, the model can be adapted so that the elevation is included, %
provided such data is sufficiently accurate. %
To this end, we will explore using publicly available topography data from sources such as  Global Digital Elevation Model (\url{http://asterweb.jpl.nasa.gov/gdem.asp}) and Shuttle Radar Topography Mission (\url{http://wiki.openstreetmap.org/wiki/SRTM}).

\item We limit the discussion to single antenna systems at both the transmitter and the receiver, with a note that the extension of the model to multiple antenna systems can be incorporated through  techniques such as the one presented by Reichardt et al.~in~\cite{reichardt11}.

\item Interference from other nearby communicating pairs can be introduced using our model. By simulating multiple communication sessions between nearby pairs of vehicles using the developed model, the following can be evaluated: 1) the signal to interference ratio (SIR); 2) the impact on channel access (e.g., channel availability, back-off times, etc.); and 3) the effect of interference on the upper layers of the protocol stack. %
\end{itemize}

\bibliographystyle{IEEEtran}
\bibliography{draftIII_tex,los-paper,refs}

\begin{thebibliography}{100}
\providecommand{\url}[1]{#1}
\csname url@samestyle\endcsname
\providecommand{\newblock}{\relax}
\providecommand{\bibinfo}[2]{#2}
\providecommand{\BIBentrySTDinterwordspacing}{\spaceskip=0pt\relax}
\providecommand{\BIBentryALTinterwordstretchfactor}{4}
\providecommand{\BIBentryALTinterwordspacing}{\spaceskip=\fontdimen2\font plus
\BIBentryALTinterwordstretchfactor\fontdimen3\font minus
  \fontdimen4\font\relax}
\providecommand{\BIBforeignlanguage}[2]{{%
\expandafter\ifx\csname l@#1\endcsname\relax
\typeout{** WARNING: IEEEtran.bst: No hyphenation pattern has been}%
\typeout{** loaded for the language `#1'. Using the pattern for}%
\typeout{** the default language instead.}%
\else
\language=\csname l@#1\endcsname
\fi
#2}}
\providecommand{\BIBdecl}{\relax}
\BIBdecl

\bibitem{dhoutaut06}
D.~Dhoutaut, A.~Regis, and F.~Spies, ``Impact of radio propagation models in
  vehicular ad hoc networks simulations,'' \emph{VANET 06: Proceedings of the
  3rd international workshop on Vehicular ad hoc networks}, pp. 69--78, 2006.

\bibitem{koberstein09}
J.~Koberstein, S.~Witt, and N.~Luttenberger, ``Model complexity vs. better
  parameter value estimation: comparing four topography-independent radio
  models,'' in \emph{Simutools '09: Proceedings of the 2nd International
  Conference on Simulation Tools and Techniques}, ICST, Brussels, Belgium,
  2009, pp. 1--8.

\bibitem{kotz04}
D.~Kotz, C.~Newport, R.~S. Gray, J.~Liu, Y.~Yuan, and C.~Elliott,
  ``Experimental evaluation of wireless simulation assumptions,'' in
  \emph{Proc. ACM MSWiM '04}.\hskip 1em plus 0.5em minus 0.4em\relax New York,
  NY, USA: ACM, 2004, pp. 78--82.

\bibitem{bai06}
F.~Bai, T.~Elbatt, G.~Hollan, H.~Krishnan, and V.~Sadekar, ``Towards
  characterizing and classifying communication-based automotive applications
  from a wireless networking perspective,'' \emph{1st IEEE Workshop on
  Automotive Networking and Applications ({AutoNet})}, 2006.

\bibitem{boban11_3}
M.~Boban and T.~T.~V. Vinhoza, ``Modeling and simulation of vehicular networks:
  Towards realistic and efficient models,'' in \emph{Mobile Ad-Hoc Networks:
  Applications}.\hskip 1em plus 0.5em minus 0.4em\relax InTech, January 2011.

\bibitem{cheng09}
C.-X. Wang, X.~Cheng, and D.~I. Laurenson, ``Vehicle-to-vehicle channel
  modeling and measurements: Recent advances and future challenges,''
  \emph{IEEE Communications Magazine}, vol.~47, no.~11, pp. 96--103, November
  2009.

\bibitem{molisch09}
A.~Molisch, F.~Tufvesson, J.~Karedal, and C.~Mecklenbrauker, ``Propagation
  aspects of vehicle-to-vehicle communications - an overview,'' in \emph{IEEE
  Radio and Wireless Symposium, 2009. RWS '09.}, Jan. 2009, pp. 179 --182.

\bibitem{matolak09}
D.~Matolak and Q.~Wu, ``Vehicle-to-vehicle channels: Are we done yet?'' in
  \emph{IEEE GLOBECOM Workshops}, 2009, pp. 1 --6.

\bibitem{matolak08}
D.~Matolak, ``Channel modeling for vehicle-to-vehicle communications,''
  \emph{IEEE Communications Magazine}, vol.~46, no.~5, pp. 76 --83, may 2008.

\bibitem{goldsmith05}
A.~J. Goldsmith, \emph{Wireless Communications}.\hskip 1em plus 0.5em minus
  0.4em\relax Cambridge University Press, 2006.

\bibitem{rappaport96}
T.~S. Rappaport, \emph{Wireless Communications: Principles and Practice}.\hskip
  1em plus 0.5em minus 0.4em\relax Prentice Hall, 1996.

\bibitem{maurer04}
J.~Maurer, T.~Fugen, T.~Schafer, and W.~Wiesbeck, ``{A new inter-vehicle
  communications (IVC) channel model},'' in \emph{IEEE Vehicular Technology
  Conference, 2004, VTC2004-Fall}, September 2004, pp. 9--13.

\bibitem{usdot06_2}
``Vehicle safety communications project, final report,'' U.S. Department of
  Transportation, NHTSA, Crash Avoidance Metrics Partnership, Tech. Rep. DOT HS
  810 591, 2006.

\bibitem{Otto2009}
J.~Otto, F.~Bustamante, and R.~Berry, ``Down the block and around the corner
  the impact of radio propagation on inter-vehicle wireless communication,'' in
  \emph{29th IEEE International Conference on Distributed Computing Systems,
  ICDCS '09.}, June 2009, pp. 605--614.

\bibitem{boban11}
M.~Boban, T.~T.~V. Vinhoza, J.~Barros, M.~Ferreira, and O.~K. Tonguz, ``Impact
  of vehicles as obstacles in vehicular ad hoc networks,'' \emph{IEEE Journal
  on Selected Areas in Communications}, vol.~29, no.~1, pp. 15--28, January
  2011.

\bibitem{meireles10}
R.~Meireles, M.~Boban, P.~Steenkiste, O.~K. Tonguz, and J.~Barros,
  ``{Experimental study on the impact of vehicular obstructions in VANETs},''
  in \emph{IEEE Vehicular Networking Conference (VNC 2010)}, Jersey City, NJ,
  USA, December 2010, pp. 338--345.

\bibitem{masui02}
H.~Masui, T.~Kobayashi, and M.~Akaike, ``Microwave path-loss modeling in urban
  line-of-sight environments,'' \emph{IEEE Journal on Selected Areas in
  Communications}, vol.~20, no.~6, pp. 1151--1155, Aug 2002.

\bibitem{jist}
\BIBentryALTinterwordspacing
``{JiST/SWANS} homepage.'' [Online]. Available:
  \url{http://jist.ece.cornell.edu/}
\BIBentrySTDinterwordspacing

\bibitem{ns2}
\BIBentryALTinterwordspacing
``{Network Simulator 2}.'' [Online]. Available:
  \url{http://nsnam.isi.edu/nsnam/index.php/Main_Page}
\BIBentrySTDinterwordspacing

\bibitem{ns3}
\BIBentryALTinterwordspacing
``{Network Simulator 3}.'' [Online]. Available: \url{www.nsnam.org}
\BIBentrySTDinterwordspacing

\bibitem{qualnet}
\BIBentryALTinterwordspacing
``{Qualnet}.'' [Online]. Available: \url{http://www.qualnet.com}
\BIBentrySTDinterwordspacing

\bibitem{klosovski98}
J.~Klosowski, M.~Held, J.~Mitchell, H.~Sowizral, and K.~Zikan, ``Efficient
  collision detection using bounding volume hierarchies of k-{DOP}s,''
  \emph{IEEE Transactions on Visualization and Computer Graphics}, vol.~4,
  no.~1, pp. 21 --36, January 1998.

\bibitem{dsrc09}
``{IEEE Draft Standard IEEE P802.11p/D9.0},'' Tech. Rep., July 2009.

\bibitem{Cheng2007}
L.~Cheng, B.~E. Henty, D.~D. Stancil, F.~Bai, and P.~Mudalige, ``Mobile
  vehicle-to-vehicle narrow-band channel measurement and characterization of
  the 5.9 ghz dedicated short range communication {(DSRC)} frequency band,''
  \emph{IEEE Journal on Selected Areas in Communications}, vol.~25, no.~8, pp.
  1501--1516, Oct. 2007.

\bibitem{andersen95}
J.~Andersen, T.~Rappaport, and S.~Yoshida, ``Propagation measurements and
  models for wireless communications channels,'' \emph{IEEE Communications
  Magazine}, vol.~33, no.~1, pp. 42--49, Jan 1995.

\bibitem{martinez09}
F.~J. Martinez, C.~K. Toh, J.-C. Cano, C.~T. Calafate, and P.~Manzoni, ``A
  survey and comparative study of simulators for vehicular ad hoc networks
  ({VANETs}),'' \emph{Wireless Communications and Mobile Computing}, vol.~11,
  no.~7, pp. 813 -- 828, July 2011.

\bibitem{carblueprints}
\BIBentryALTinterwordspacing
``{Car blueprints database}.'' [Online]. Available:
  \url{http://carblueprints.info}
\BIBentrySTDinterwordspacing

\bibitem{festag08}
A.~Festag, R.~Baldessari, W.~Zhang, L.~Le, A.~Sarma, and M.~Fukukawa,
  ``{CAR-2-X} communication for safety and infotainment in {Europe},''
  \emph{{NEC} Technical Journal}, vol.~3, no.~1, 2008.

\bibitem{80211p2010}
``{IEEE Draft Standard IEEE P802.11p},'' Tech. Rep., June 2010.

\bibitem{wireshark}
\BIBentryALTinterwordspacing
G.~Combs \emph{et~al.}, ``{Wireshark-network protocol analyzer}.'' [Online].
  Available: \url{http://www.wireshark.org}
\BIBentrySTDinterwordspacing

\bibitem{kaul07}
S.~Kaul, K.~Ramachandran, P.~Shankar, S.~Oh, M.~Gruteser, I.~Seskar, and
  T.~Nadeem, ``Effect of antenna placement and diversity on vehicular network
  communications,'' in \emph{4th Annual IEEE Communications Society Conference
  on Sensor, Mesh and Ad Hoc Communications and Networks, SECON '07}, june
  2007, pp. 112--121.

\bibitem{takahashi03}
S.~Takahashi, A.~Kato, K.~Sato, and M.~Fujise, ``Distance dependence of path
  loss for millimeter wave inter-vehicle communications,'' in \emph{Proc. IEEE
  58th Vehicular Technology Conference (VTC 2003-Fall)}, vol.~1, Oct. 2003, pp.
  26--30.

\bibitem{tan08}
I.~Tan, W.~Tang, K.~Laberteaux, and A.~Bahai, ``Measurement and analysis of
  wireless channel impairments in {DSRC} vehicular communications,'' in
  \emph{IEEE International Conference on Communications, ICC '08.}, May 2008,
  pp. 4882--4888.

\bibitem{paier07}
A.~Paier, J.~Karedal, N.~Czink, H.~Hofstetter, C.~Dumard, T.~Zemen,
  F.~Tufvesson, A.~Molisch, and C.~Mecklenbrauker, ``{Car-to-car radio channel
  measurements at 5 GHz: Pathloss, power-delay profile, and delay-Doppler
  spectrum},'' in \emph{4th International Symposium on Wireless Communication
  Systems, ISWCS 2007.}, Oct. 2007, pp. 224--228.

\bibitem{sen08}
I.~Sen and D.~Matolak, ``{Vehicle-Vehicle Channel Models for the 5-GHz Band},''
  \emph{IEEE Transactions on Intelligent Transportation Systems}, vol.~9,
  no.~2, pp. 235--245, June 2008.

\bibitem{Jerbi2007}
M.~Jerbi, P.~Marlier, and S.~M. Senouci, ``Experimental assessment of {V2V and
  I2V} communications,'' in \emph{Proc. IEEE Internatonal Conference on Mobile
  Adhoc and Sensor Systems (MASS 2007)}, Oct. 2007, pp. 1--6.

\bibitem{Wu2005}
H.~Wu, M.~Palekar, R.~Fujimoto, R.~Guensler, M.~Hunter, J.~Lee, and J.~Ko, ``An
  empirical study of short range communications for vehicles,'' in \emph{Proc.
  of the 2nd ACM International workshop on Vehicular ad hoc networks}, 2005,
  pp. 83--84.

\bibitem{matolak05}
D.~Matolak, I.~Sen, W.~Xiong, and N.~Yaskoff, ``{5 GHz wireless channel
  characterization for vehicle to vehicle communications},'' in \emph{Proc.
  IEEE Military Communications Conference (MILCOM 2005)}, October 2005, pp.
  3016--3022.

\bibitem{acosta07}
G.~Acosta-Marum and M.~Ingram, ``Six time- and frequency-selective empirical
  channel models for vehicular wireless {LANs},'' \emph{IEEE Vehicular
  Technology Magazine}, vol.~2, no.~4, pp. 4 --11, December 2007.

\bibitem{boban12}
M.~Boban, T.~T.~V. Vinhoza, O.~K. Tonguz, and J.~Barros, ``Seeing is believing
  -- enhancing message dissemination in vehicular networks through visual
  cues,'' \emph{IEEE Communications Letters}, vol.~16, no.~2, pp. 238--241,
  Feb. 2012.

\bibitem{nadeem04}
T.~Nadeem, S.~Dashtinezhad, C.~Liao, and L.~Iftode, ``Trafficview: traffic data
  dissemination using car-to-car communication,'' \emph{SIGMOBILE Mob. Comput.
  Commun. Rev.}, vol.~8, no.~3, pp. 6--19, 2004.

\bibitem{chen05}
W.~Chen and S.~Cai, ``Ad hoc peer-to-peer network architecture for vehicle
  safety communications,'' \emph{IEEE Communications Magazine}, vol.~43, no.~4,
  pp. 100--107, April 2005.

\bibitem{tonguz10}
O.~K. Tonguz and M.~Boban, ``Multiplayer games over vehicular ad hoc networks:
  A new application,'' \emph{Ad Hoc Networks}, vol.~8, no.~5, pp. 531 -- 543,
  2010.

\bibitem{choffnes05}
D.~R. Choffnes and F.~E. Bustamante, ``An integrated mobility and traffic model
  for vehicular wireless networks,'' in \emph{VANET '05: Proceedings of the 2nd
  ACM international workshop on Vehicular ad hoc networks}.\hskip 1em plus
  0.5em minus 0.4em\relax New York, NY, USA: ACM, 2005, pp. 69--78.

\bibitem{nctuns}
\BIBentryALTinterwordspacing
``{NCTUns 6.0 network simulator and emulator}.'' [Online]. Available:
  \url{http://nsl.csie.nctu.edu.tw/nctuns.html}
\BIBentrySTDinterwordspacing

\bibitem{glassner89}
A.~Glassner, Ed., \emph{An Introduction to Ray Tracing}.\hskip 1em plus 0.5em
  minus 0.4em\relax Academic Press, 1989.

\bibitem{acosta06}
G.~Acosta and M.~Ingram, ``{Model development for the wideband expressway
  vehicle-to-vehicle 2.4 GHz channel},'' in \emph{IEEE Wireless Communications
  and Networking Conference, 2006. WCNC 2006.}, vol.~3, April 2006, pp.
  1283--1288.

\bibitem{zang05}
Y.~Zang, L.~Stibor, G.~Orfanos, S.~Guo, and H.-J. Reumerman, ``An error model
  for inter-vehicle communications in highway scenarios at 5.9ghz,'' in
  \emph{PE-WASUN '05: Proceedings of the 2nd ACM international workshop on
  Performance evaluation of wireless ad hoc, sensor, and ubiquitous
  networks}.\hskip 1em plus 0.5em minus 0.4em\relax New York, NY, USA: ACM,
  2005, pp. 49--56.

\bibitem{kim09}
H.~Kim and H.-S. Lee, ``Accelerated three dimensional ray tracing techniques
  using ray frustums for wireless propagation models,'' \emph{Progress In
  Electromagnetics Research, PIER}, no.~96, pp. 595--611, 2009.

\bibitem{karedal09}
J.~Karedal, F.~Tufvesson, N.~Czink, A.~Paier, C.~Dumard, T.~Zemen,
  C.~Mecklenbrauker, and A.~Molisch, ``A geometry-based stochastic {MIMO} model
  for vehicle-to-vehicle communications,'' \emph{IEEE Transactions on Wireless
  Communications}, vol.~8, no.~7, pp. 3646--3657, July 2009.

\bibitem{paier07_2}
A.~Paier, J.~Karedal, N.~Czink, H.~Hofstetter, C.~Dumard, T.~Zemen,
  F.~Tufvesson, C.~Mecklenbrauker, and A.~Molisch, ``First results from
  car-to-car and car-to-infrastructure radio channel measurements at 5.2
  {GHz},'' in \emph{IEEE 18th International Symposium on Personal, Indoor and
  Mobile Radio Communications, 2007. PIMRC 2007.}, September 2007, pp. 1--5.

\bibitem{cheng09_3}
X.~Cheng, C.-X. Wang, D.~I. Laurenson, S.~Salous, and A.~V. Vasilakos, ``An
  adaptive geometry-based stochastic model for non-isotropic mimo
  mobile-to-mobile channels,'' \emph{IEEE Transactions on Wireless
  Communications}, vol.~8, no.~9, pp. 4824--4835, 2009.

\bibitem{rothery92}
R.~W. Rothery, ``Car following models,'' \emph{In Trac Flow Theory}, 1992.

\bibitem{tonguz09_2}
O.~K. Tonguz, W.~Viriyasitavat, and F.~Bai, ``Modeling urban traffic: a
  cellular automata approach,'' \emph{IEEE Communications Magazine}, vol.~47,
  no.~5, pp. 142--150, 2009.

\bibitem{shih11}
O.~Shih, H.~Tsai, H.~Lin, and A.~Pang, ``A rule-based mixed mobility model for
  cars and scooters (poster),'' in \emph{IEEE Vehicular Networking Conference
  (VNC)}, 2011, pp. 198--205.

\bibitem{itu07}
ITU-R, ``Propagation by diffraction,'' International Telecommunication Union
  Radiocommunication Sector, Geneva, Recommendation P.526, February 2007.

\bibitem{00parsons}
J.~D. Parsons, \emph{The Mobile Radio Propagation Channel}.\hskip 1em plus
  0.5em minus 0.4em\relax John Wiley \& Sons, 2000.

\bibitem{epstein53}
J.~Epstein and D.~W. Peterson, ``An experimental study of wave propagatin at
  850{MC},'' \emph{Proceedings of the IRE}, vol.~41, no.~5, pp. 595--611, 1953.

\bibitem{deygout66}
J.~Deygout, ``Multiple knife-edge diffraction of microwaves,'' \emph{IEEE
  Transactions on Antennas and Propagation}, vol.~14, no.~4, pp. 480--489,
  1966.

\bibitem{giovaneli84}
C.~L. Giovaneli, ``An analysis of simplified solutions for multiple knife-edge
  diffraction,'' \emph{IEEE Transactions on Antennas and Propagation}, vol.~32,
  no.~3, pp. 297--301, 1984.

\bibitem{mccasland65}
{McCasland, W T}, ``{Comparison of Two Techniques of Aerial Photography for
  Application in Freeway Traffic Operations Studies},'' \emph{Photogrammetry
  and Aerial Surveys}, 1965.

\bibitem{ferreira09}
M.~Ferreira, H.~Concei{\c{c}\~{a}}o, R.~Fernandes, and O.~K. Tonguz,
  ``Stereoscopic aerial photography: an alternative to model-based urban
  mobility approaches,'' in \emph{Proceedings of the Sixth ACM International
  Workshop on VehiculAr Inter-NETworking (VANET 2009)}.\hskip 1em plus 0.5em
  minus 0.4em\relax ACM New York, NY, USA, 2009.

\bibitem{wisit07}
N.~Wisitpongphan, F.~Bai, P.~Mudalige, V.~Sadekar, and O.~K. Tonguz, ``Routing
  in sparse vehicular ad hoc wireless networks,'' \emph{IEEE Journal on
  Selected Areas in Communications}, vol.~25, no.~8, pp. 1538--1556, Oct. 2007.

\bibitem{acap}
\BIBentryALTinterwordspacing
``{Associa\c c\~ao Autom\'ovel de Portugal}.'' [Online]. Available:
  \url{http://www.acap.pt/}
\BIBentrySTDinterwordspacing

\bibitem{carfolio}
\BIBentryALTinterwordspacing
``{Automotive Technical Data and Specifications}.'' [Online]. Available:
  \url{http://www.carfolio.com/}
\BIBentrySTDinterwordspacing

\bibitem{papoulis84}
A.~Papoulis, \emph{Probability, Random Variables and Stochastic Processes},
  2nd~ed.\hskip 1em plus 0.5em minus 0.4em\relax McGraw-Hill, 1984.

\bibitem{wang04}
Z.~Wang, E.~Tameh, and A.~Nix, ``Statistical peer-to-peer channel models for
  outdoor urban environments at 2 ghz and 5 ghz,'' in \emph{IEEE Vehicular
  Technology Conference, 2004. VTC2004-Fall. 2004}, vol.~7, Sept. 2004, pp.
  5101--5105 Vol. 7.

\bibitem{astm03}
``{Standard Specification for Telecommunications and Information Exchange
  Between Roadside and Vehicle Systems - 5GHz Band Dedicated Short Range
  Communications (DSRC) Medium Access Control (MAC) and Physical Layer (PHY)
  Specifications},'' \emph{ASTM E2213-03}, Sep. 2003.

\bibitem{boban11_2}
M.~Boban, R.~Meireles, J.~Barros, O.~K. Tonguz, and P.~Steenkiste, ``Exploiting
  the height of vehicles in vehicular communication,'' in \emph{IEEE Vehicular
  Networking Conference (VNC 2011)}, Amsterdam, The Netherlands, November 2011,
  pp. 284--291.

\bibitem{drivein}
\BIBentryALTinterwordspacing
``{DRIVE-IN (Distributed Routing and Infotainment through VEhicular
  Inter-Networking)}.'' [Online]. Available:
  \url{http://drive-in.cmuportugal.org}
\BIBentrySTDinterwordspacing

\bibitem{openstreetmap}
M.~Haklay and P.~Weber, ``{OpenStreetMap}: User-generated street maps,''
  \emph{IEEE Pervasive Computing}, vol.~7, no.~4, pp. 12--18, 2008.

\bibitem{wang12}
X.~Wang, E.~Anderson, P.~Steenkiste, and F.~Bai, ``Improving the accuracy of
  environment-specific vehicular channel modeling,'' in \emph{Proceedings of
  the seventh ACM international workshop on Wireless network testbeds,
  experimental evaluation and characterization, {WiNTECH '12}}.\hskip 1em plus
  0.5em minus 0.4em\relax New York, NY, USA: ACM, 2012, pp. 43--50.

\bibitem{blaschke10}
T.~Blaschke, ``Object based image analysis for remote sensing,'' \emph{ISPRS
  Journal of Photogrammetry and Remote Sensing}, vol.~65, no.~1, pp. 2--16,
  2010.

\bibitem{naip}
\BIBentryALTinterwordspacing
``{National Agriculture Imagery Program (NAIP)}.'' [Online]. Available:
  \url{http://www.fsa.usda.gov/FSA/apfoapp?area=home&subject=prog&topic=nai}
\BIBentrySTDinterwordspacing

\bibitem{haverkort04}
H.~J. Haverkort, ``Results on geometric networks and data structures,'' Ph.D.
  dissertation, Utrecht University, 2004.

\bibitem{guttman84}
A.~Guttman, ``R-trees: a dynamic index structure for spatial searching,''
  \emph{SIGMOD Rec.}, vol.~14, pp. 47--57, June 1984.

\bibitem{theodoridis96}
Y.~Theodoridis and T.~Sellis, ``A model for the prediction of {R-tree}
  performance,'' in \emph{Proceedings of the fifteenth ACM SIGACT-SIGMOD-SIGART
  symposium on Principles of database systems}, ser. PODS '96.\hskip 1em plus
  0.5em minus 0.4em\relax New York, NY, USA: ACM, 1996, pp. 161--171.

\bibitem{berg97}
M.~de~Berg, M.~van Kreveld, M.~Overmars, and O.~Schwarzkopf,
  \emph{Computational Geometry Algorithms and Applications}.\hskip 1em plus
  0.5em minus 0.4em\relax Springer-Verlag, 1997.

\bibitem{karedal10}
J.~Karedal, F.~Tufvesson, T.~Abbas, O.~Klemp, A.~Paier, L.~Bernad\'{o}~and, and
  A.~Molisch, ``Radio channel measurements at street intersections for
  vehicle-to-vehicle safety applications,'' in \emph{IEEE Vehicular Technology
  Conference (VTC 2010-Spring)}, May 2010, pp. 1 --5.

\bibitem{sommer2011using}
C.~Sommer and F.~Dressler, ``{Using the Right Two-Ray Model? A Measurement
  based Evaluation of PHY Models in VANETs},'' in \emph{17th ACM International
  Conference on Mobile Computing and Networking (MobiCom 2011), Poster
  Session}.\hskip 1em plus 0.5em minus 0.4em\relax Las Vegas, NV: ACM,
  September 2011.

\bibitem{mangel11}
T.~Mangel, M.~Michl, O.~Klemp, and H.~Hartenstein, ``Real-world measurements of
  {Non-Line-of-Sight} reception quality for 5.9 {GHz IEEE 802.11p} at
  intersections,'' in \emph{Proceedings of the Third international conference
  on Communication technologies for vehicles}, ser.
  Nets4Cars/Nets4Trains'11.\hskip 1em plus 0.5em minus 0.4em\relax Berlin,
  Heidelberg: Springer-Verlag, 2011, pp. 189--202.

\bibitem{durgin98}
G.~Durgin, T.~Rappaport, and H.~Xu, ``Measurements and models for radio path
  loss and penetration loss in and around homes and trees at 5.85 {GHz},''
  \emph{IEEE Transactions on Communications}, vol.~46, no.~11, pp. 1484--1496,
  November 1998.

\bibitem{goldhirsh98}
J.~Goldhirsh and W.~J. Vogel, ``Handbook of propagation effects for vehicular
  and personal mobile satellite systems -- overview of experimental and
  modeling results,'' The Johns Hopkins University, Applied Physics Laboratory
  and The University of Texas at Austin, Electrical Engineering Research
  Laboratory, Tech. Rep. A2A-98-U-0-021 (APL), EERL-98-12A (EERL), December
  1998.

\bibitem{ulaby90}
F.~Ulaby, M.~Whitt, and M.~Dobson, ``Measuring the propagation properties of a
  forest canopy using a polarimetric scatterometer,'' \emph{IEEE Transactions
  on Antennas and Propagation}, vol.~38, no.~2, pp. 251 --258, feb 1990.

\bibitem{benzair91}
B.~Benzair, H.~Smith, and J.~Norbury, ``Tree attenuation measurements at 1-4
  {GHz} for mobile radio systems,'' in \emph{Sixth International Conference on
  Mobile Radio and Personal Communications}, December 1991, pp. 16--20.

\bibitem{anderson98}
H.~Anderson, ``Building corner diffraction measurements and predictions using
  {UTD},'' \emph{IEEE Transactions on Antennas and Propagation}, vol.~46,
  no.~2, pp. 292 --293, February 1998.

\bibitem{abdi02}
A.~Abdi, J.~Barger, and M.~Kaveh, ``A parametric model for the distribution of
  the angle of arrival and the associated correlation function and power
  spectrum at the mobile station,'' \emph{IEEE Transactions on Vehicular
  Technology}, vol.~51, no.~3, pp. 425--434, 2002.

\bibitem{kunisch08}
J.~Kunisch and J.~Pamp, ``Wideband car-to-car radio channel measurements and
  model at 5.9 {GHz},'' in \emph{IEEE Vehicular Technology Conference, 2008.
  VTC 2008-Fall}, September 2008, pp. 1 --5.

\bibitem{karedal11TVT}
J.~Karedal, N.~Czink, A.~Paier, F.~Tufvesson, and A.~Molisch, ``Path loss
  modeling for vehicle-to-vehicle communications,'' \emph{IEEE Transactions on
  Vehicular Technology}, vol.~60, no.~1, pp. 323--328, January 2011.

\bibitem{landron96}
O.~Landron, M.~Feuerstein, and T.~Rappaport, ``A comparison of theoretical and
  empirical reflection coefficients for typical exterior wall surfaces in a
  mobile radio environment,'' \emph{IEEE Transactions on Antennas and
  Propagation}, vol.~44, no.~3, pp. 341--351, March 1996.

\bibitem{abbas11}
T.~Abbas, J.~Karedal, F.~Tufvesson, A.~Paier, L.~Bernado, and A.~Molisch,
  ``Directional analysis of vehicle-to-vehicle propagation channels,'' in
  \emph{73rd IEEE Vehicular Technology Conference (VTC Spring)}, May 2011, pp.
  1--5.

\bibitem{paier09_2}
A.~Paier, J.~Karedal, N.~Czink, C.~Dumard, T.~Zemen, F.~Tufvesson, A.~Molisch,
  and C.~Mecklenbr\"auker, ``Characterization of vehicle-to-vehicle radio
  channels from measurements at 5.2 {GHz},'' \emph{Wireless Personal
  Communications}, vol.~50, pp. 19--32, 2009.

\bibitem{cardote11}
F.~Neves, A.~Cardote, R.~Moreira, and S.~Sargento, ``Real-world evaluation of
  {IEEE} 802.11p for vehicular networks,'' in \emph{Proceedings of the Eighth
  ACM international workshop on Vehicular inter-networking (VANET '11)}.\hskip
  1em plus 0.5em minus 0.4em\relax New York, NY, USA: ACM, 2011, pp. 89--90.

\bibitem{giordano10}
E.~Giordano, R.~Frank, G.~Pau, and M.~Gerla, ``{CORNER}: a realistic urban
  propagation model for {VANET},'' in \emph{Proceedings of the 7th
  international conference on Wireless on-demand network systems and services
  (WONS)}, 2010, pp. 57--60.

\bibitem{paschalidis11}
P.~Paschalidis, K.~Mahler, A.~Kortke, M.~Peter, and W.~Keusgen, ``Pathloss and
  multipath power decay of the wideband car-to-car channel at 5.7 {GHz},'' in
  \emph{IEEE Vehicular Technology Conference (VTC Spring)}, May 2011, pp. 1--5.

\bibitem{abbas12}
T.~Abbas, F.~Tufvesson, and J.~Karedal, ``Measurement based shadow fading model
  for vehicle-to-vehicle network simulations,'' \emph{arXiv preprint
  arXiv:1203.3370v2}, 2012.

\bibitem{fernandez11}
J.~Fernandez, K.~Borries, L.~Cheng, B.~Kumar, D.~Stancil, and F.~Bai,
  ``Performance of the 802.11p physical layer in vehicle-to-vehicle
  environments,'' \emph{IEEE Transactions on Vehicular Technology}, vol.~61,
  no.~1, pp. 3 --14, January 2012.

\bibitem{chazelle89}
B.~Chazelle and L.~Guibas, ``Visibility and intersection problems in plane
  geometry,'' \emph{Discrete \& Computational Geometry}, vol.~4, no.~1, pp.
  551--581, 1989.

\bibitem{molnar94}
S.~Molnar, M.~Cox, D.~Ellsworth, and H.~Fuchs, ``A sorting classification of
  parallel rendering,'' \emph{IEEE Computer Graphics and Applications},
  vol.~14, no.~4, pp. 23--32, 1994.

\bibitem{luo12}
L.~Luo, M.~Wong, and L.~Leong, ``Parallel implementation of {R}-trees on the
  {GPU},'' in \emph{17th IEEE Asia and South Pacific Design Automation
  Conference (ASP-DAC)}, 2012, pp. 353--358.

\bibitem{mangel11_2}
T.~Mangel, O.~Klemp, and H.~Hartenstein, ``{A validated 5.9 GHz
  Non-Line-of-Sight path-loss and fading model for inter-vehicle
  communication},'' in \emph{11th International Conference on ITS
  Telecommunications (ITST)}, August 2011, pp. 75 --80.

\bibitem{sommer11}
C.~Sommer, D.~Eckhoff, R.~German, and F.~Dressler, ``A computationally
  inexpensive empirical model of {IEEE} 802.11p radio shadowing in urban
  environments,'' in \emph{Eighth International Conference on Wireless
  On-Demand Network Systems and Services (WONS)}, 2011, pp. 84 --90.

\bibitem{Martinez2009}
F.~J. Martinez, C.-K. Toh, J.-C. Cano, C.~T. Calafate, and P.~Manzoni,
  ``Realistic radio propagation models {(RPMs)} for {VANET} simulations,'' in
  \emph{Wireless Communications and Networking Conference, 2009. WCNC 2009.
  IEEE}, April 2009, pp. 1--6.

\bibitem{cozzetti12}
H.~Cozzetti, C.~Campolo, R.~Scopigno, and A.~Molinaro, ``Urban {VANETs} and
  hidden terminals: evaluation through a realistic urban grid propagation
  model,'' in \emph{IEEE International Conference on Vehicular Electronics and
  Safety (ICVES)}, July 2012, pp. 93 --98.

\bibitem{sepulcre12}
M.~Sepulcre and J.~Gozalvez, ``Experimental evaluation of cooperative active
  safety applications based on {V2V} communications,'' in \emph{Proceedings of
  the ninth ACM international workshop on Vehicular inter-networking, systems,
  and applications}, ser. VANET '12.\hskip 1em plus 0.5em minus 0.4em\relax New
  York, NY, USA: ACM, 2012, pp. 13--20.

\bibitem{gallagher06}
B.~Gallagher, H.~Akatsuka, and H.~Suzuki, ``Wireless communications for vehicle
  safety: radio link performance and wireless connectivity methods,''
  \emph{IEEE Vehicular Technology Magazine}, vol.~1, no.~4, pp. 4 --24,
  December 2006.

\bibitem{gozalvez12}
J.~Gozalvez, M.~Sepulcre, and R.~Bauza, ``{IEEE} 802.11p vehicle to
  infrastructure communications in urban environments,'' \emph{IEEE
  Communications Magazine}, vol.~50, no.~5, pp. 176 --183, May 2012.

\bibitem{wang09}
P.-J. Wang, C.-M. Li, and H.-J. Li, ``Influence of the shadowing on the
  information transmission distance in inter-vehicle communications,'' in
  \emph{IEEE International Symposium on Personal, Indoor and Mobile Radio
  Communications (PIMRC)}, September 2009, pp. 3015 --3019.

\bibitem{bullington77}
K.~Bullington, ``Radio propagation for vehicular communications,'' \emph{IEEE
  Transactions on Vehicular Technology}, vol.~26, no.~4, pp. 295 -- 308, nov
  1977.

\bibitem{boban13}
M.~Boban, R.~Meireles, J.~Barros, P.~A. Steenkiste, and O.~K. Tonguz, ``{TVR} -
  tall vehicle relaying in vehicular networks,'' \emph{IEEE Transactions on
  Mobile Computing}, vol.~99, p.~1, 2013.

\bibitem{ferreira10}
M.~Ferreira, H.~Conceicao, R.~Fernandes, and O.~Tonguz, ``Urban connectivity
  analysis of {VANETs} through stereoscopic aerial photography,'' in \emph{IEEE
  Vehicular Technology Conference Fall (VTC 2009-Fall)}, Sept. 2009, pp. 1--3.

\bibitem{losexpsite}
\BIBentryALTinterwordspacing
``{802.11p Line of Sight Experiment website}.'' [Online]. Available:
  \url{http://drive-in.cmuportugal.org/los}
\BIBentrySTDinterwordspacing

\bibitem{torrent06}
M.~Torrent-Moreno, P.~Santi, and H.~Hartenstein, ``Distributed fair transmit
  power adjustment for vehicular ad hoc networks,'' in \emph{3rd Annual IEEE
  Communications Society Conference on Sensor and Ad Hoc Communications and
  Networks, SECON '06}, vol.~2, sept. 2006, pp. 479 --488.

\bibitem{naumov06}
V.~Naumov, R.~Baumann, and T.~Gross, ``An evaluation of inter-vehicle ad hoc
  networks based on realistic vehicular traces,'' in \emph{MobiHoc '06:
  Proceedings of the 7th ACM international symposium on Mobile ad hoc
  networking and computing}, New York, NY, USA, 2006, pp. 108--119.

\bibitem{wisit07_2}
N.~Wisitpongphan, O.~K. Tonguz, J.~Parikh, P.~Mudalige, F.~Bai, and V.~Sadekar,
  ``Broadcast storm mitigation techniques in vehicular ad hoc networks,''
  \emph{IEEE Wireless Communications}, vol.~14, no.~6, pp. 84--94, Dec. 2007.

\bibitem{bai09}
F.~Bai and B.~Krishnamachari, ``Spatio-temporal variations of vehicle traffic
  in {VANETs}: facts and implications,'' in \emph{VANET '09: Proceedings of the
  sixth ACM international workshop on VehiculAr InterNETworking}.\hskip 1em
  plus 0.5em minus 0.4em\relax New York, NY, USA: ACM, 2009, pp. 43--52.

\bibitem{may90}
A.~D. May, \emph{Traffic flow fundamentals}.\hskip 1em plus 0.5em minus
  0.4em\relax Prentice Hall, 1990.

\bibitem{oh09}
S.~Oh, S.~Kaul, and M.~Gruteser, ``Exploiting vertical diversity in vehicular
  channel environments,'' in \emph{IEEE 20th International Symposium on
  Personal, Indoor and Mobile Radio Communications}, Sept. 2009, pp. 958 --962.

\bibitem{reichardt09}
L.~Reichardt, T.~Fugen, and T.~Zwick, ``Influence of antennas placement on car
  to car communications channel,'' in \emph{3rd European Conference on Antennas
  and Propagation, EuCAP 2009.}, march 2009, pp. 630--634.

\bibitem{kornek10}
D.~Kornek, M.~Schack, E.~Slottke, O.~Klemp, I.~Rolfes, and T.~KŸ~andrner,
  ``Effects of antenna characteristics and placements on a vehicle-to-vehicle
  channel scenario,'' in \emph{IEEE International Conference on Communications
  Workshops}, May 2010, pp. 1--5.

\bibitem{paier10_2}
A.~Paier, D.~Faetani, and C.~Mecklenbr\"auker, ``Performance evaluation of
  {IEEE} 802.11p physical layer infrastructure-to-vehicle real-world
  measurements,'' in \emph{Proceedings of {ISABEL} 2010}, Rome, Italy, November
  2010.

\bibitem{paier10}
A.~Paier, R.~Tresch, A.~Alonso, D.~Smely, P.~Meckel, Y.~Zhou, and N.~Czink,
  ``Average downstream performance of measured {IEEE} 802.11p
  infrastructure-to-vehicle links,'' in \emph{IEEE International Conference on
  Communications Workshops, 2010}, may 2010, pp. 1 --5.

\bibitem{baum05}
{D. S. Baum et al.}, ``{IST-2003-507581 WINNER I, D5.4, Final report on link
  level and system level channel models},'' Information Society Technologies,
  Tech. Rep., 2005.

\bibitem{namboodiri04}
V.~Namboodiri, M.~Agarwal, and L.~Gao, ``A study on the feasibility of mobile
  gateways for vehicular ad-hoc networks,'' in \emph{VANET '04: Proceedings of
  the 1st ACM international workshop on Vehicular ad hoc networks}.\hskip 1em
  plus 0.5em minus 0.4em\relax New York, NY, USA: ACM, 2004, pp. 66--75.

\bibitem{naumov07}
V.~Naumov and T.~Gross, ``Connectivity-aware routing {(CAR)} in vehicular
  ad-hoc networks,'' in \emph{26th IEEE International Conference on Computer
  Communications, INFOCOM 2007.}, May 2007, pp. 1919--1927.

\bibitem{lochert03}
C.~Lochert, H.~Hartenstein, J.~Tian, H.~Fussler, D.~Hermann, and M.~Mauve, ``A
  routing strategy for vehicular ad hoc networks in city environments,''
  \emph{Proceedings of the IEEE Intelligent Vehicles Symposium, 2003.}, pp.
  156--161, June 2003.

\bibitem{boban08}
M.~Boban, G.~Misek, and O.~K. Tonguz, ``What is the best achievable {QoS} for
  unicast routing in {VANETs?}'' in \emph{IEEE GLOBECOM, 3rd Workshop on
  Automotive Networking and Applications (AutoNet)}, New Orleans, LA, USA,
  November 2008, pp. 1--10.

\bibitem{boban09}
M.~Boban, O.~K. Tonguz, and J.~Barros, ``Unicast communication in vehicular ad
  hoc networks: a reality check,'' \emph{IEEE Communications Letters}, vol.~13,
  no.~12, pp. 995--997, December 2009.

\bibitem{viriyasitavat11}
W.~Viriyasitavat, O.~K. Tonguz, and F.~Bai, ``{UV-CAST: an urban vehicular
  broadcast protocol},'' \emph{IEEE Communications Magazine}, vol.~49, no.~11,
  pp. 116--124, November 2011.

\bibitem{tonguz10_2}
O.~K. Tonguz, N.~Wisitpongphan, and F.~Bai, ``{DV-CAST: A distributed vehicular
  broadcast protocol for vehicular ad hoc networks},'' \emph{IEEE Wireless
  Communications}, vol.~17, no.~2, pp. 47--57, April 2010.

\bibitem{kihl07}
M.~Kihl, M.~Sichitiu, T.~Ekeroth, and M.~Rozenberg, ``Reliable geographical
  multicast routing in vehicular ad-hoc networks,'' in \emph{Wired/Wireless
  Internet Communications}, ser. Lecture Notes in Computer Science, F.~Boavida,
  E.~Monteiro, S.~Mascolo, and Y.~Koucheryavy, Eds.\hskip 1em plus 0.5em minus
  0.4em\relax Springer Berlin / Heidelberg, 2007, vol. 4517, pp. 315--325.

\bibitem{reichardt11}
L.~Reichardt, J.~Pontes, W.~Wiesbeck, and T.~Zwick, ``Virtual drives in
  vehicular communication,'' \emph{IEEE Vehicular Technology Magazine}, vol.~6,
  no.~2, pp. 54 --62, june 2011.

\end{thebibliography}
\clearpage

\newpage
\newpage
\newpage
\newpage

\end{document}